\documentclass[10pt,twopage]{article}         
\usepackage{graphicx}      
\usepackage[small]{caption}
\usepackage{hyperref}
\usepackage{subfigure}
\usepackage{epsfig}

\newcommand{\be}{\begin{equation}}
\newcommand{\ee}{\end{equation}}
\newcommand{\ba}{\begin{eqnarray}}
\newcommand{\ea}{\end{eqnarray}}
\newcommand{\gtrsim}
	{\mathrel{\raisebox{0.4ex}{\hbox{$>$}}\kern-0.75em
	\raisebox{-0.5ex}{\hbox{$\sim$}}}}
\newcommand{\lesssim}
	{\mathrel{\raisebox{0.4ex}{\hbox{$<$}}\kern-0.75em
	\raisebox{-0.5ex}{\hbox{$\sim$}}}}

\newcommand{\ket}[1]{|#1\rangle}

\newcommand{\deu}{$D$}
\newcommand{\tro}{$^3He$}
\newcommand{\qua}{$^4He$}

\newcommand{\sep}{$^{7}Li$}

\newcommand{\centeron}[2]{{\setbox0=\hbox{#1}\setbox1=\hbox{#2}\ifdim
                             \wd1>\wd0\kern.5\wd1\kern-.5\wd0\fi \copy0
                             \kern-.5\wd0\kern-.5\wd1\copy1\ifdim\wd0>\wd1
                             \kern.5\wd0\kern-.5\wd1\fi}}

\newcommand{\ltap}{\>\centeron{\raise.35ex\hbox{$<$}}
                     {\lower.65ex\hbox{$\sim$}}\>}
\newcommand{\gtap}{\>\centeron{\raise.35ex\hbox{$>$}}
                     {\lower.65ex\hbox{$\sim$}}\>}

\newcommand{\gsim}{\mathrel{\gtap}}
\newcommand{\lsim}{\mathrel{\ltap}}

\newcommand{\B}{B^1}

\newcommand{\qR}{q^1_R}

\def\bone{B^{(1)}}
\def\erone{e^{(1)}_R}
\def\etal{{\it et al.~}}
\def\eg{{\it e.g.~}}
\def\ie{{\it i.e.~}}

\def\GC{Galactic center~} 
\def\susy{SUSY}

\begin{document}
%
%
%

%
%
\title{Particle Dark Matter: Evidence, Candidates and Constraints}
\author{Gianfranco Bertone$^1$, Dan Hooper$^2$ and Joseph Silk$^2$ \\ \\
$^1$ NASA/Fermilab Theoretical Astrophysics Group, Batavia, IL 60510\\
$^2$ University of Oxford, Astrophysics Dept., Oxford, UK OX1 3RH}

\maketitle

\begin{abstract}
In this review article, we discuss the current status of particle dark
matter, including experimental evidence and theoretical motivations. We
discuss a wide array of candidates for particle dark matter, but focus on
neutralinos in models of supersymmetry and Kaluza-Klein dark matter in
models of universal extra dimensions. We devote much of our attention to
direct and indirect detection techniques, the constraints placed by these
experiments and the reach of future experimental efforts.
\end{abstract}

\vspace{-11.0cm}
\hspace{7.0cm}
FERMILAB-Pub-04/047-A \

\hspace{7.0cm}
hep-ph/0404175

\clearpage

\tableofcontents

\clearpage

\section{Introduction}

\subsection{Overview}
A great deal of effort has been made since 1687, the year of publication of 
Issac Newton's classic work {\it``Philosophiae Naturalis Principia 
Mathematica''}, towards explaining the motion of astrophysical
objects in terms of the laws of gravitation. Since then, 
the deviations of observed motions from expected trajectories 
have proved very effective in deepening our understanding of the 
Universe. Whenever anomalies were observed in the motion 
of planets in the Solar system, the question arose: should such anomalies be regarded as a refutation of the laws of gravitation 
or as an indication of the existence of unseen (today we would say 
``dark'') objects? 

The second approach proved to be correct in the case of the 
anomalous motion of Uranus, which led the French astronomer
U.~Le Verrier and the English astronomer John Couch Adams to conjecture the existence of Neptune, eventually
discovered in 1846 by J.G.~Galle. Conversely, the attempt  
to explain the anomalies in the motion of Mercury as due to
the  existence of a new planet, called Vulcan, failed, and the 
final solution had to wait for the advent of Einstein's theory of 
general relativity, \ie the introduction of a more refined
description of the laws of gravitation.

The modern problem of dark matter is conceptually very similar 
to the old problem of unseen planets. We observe in large astrophysical 
systems, with sizes ranging from galactic to cosmological scales, 
some ``anomalies'' that can only be explained either by assuming 
the existence of a large amount of unseen, {\it dark}, matter, or 
by assuming a deviation from the known laws of gravitation and the theory of general relativity. 

About ten years ago, Jungman, Kamionkowski and Griest wrote a review of supersymmetric dark matter for Physics Reports~\cite{Jungman:1995df}. This article, although incredibly useful, complete and popular, has gradually become outdated over the last decade. With this in mind, we have endeavored to write a new review of particle dark matter. As with the Jungman {\it et al.} article, our review is intended to be suitable for a wide range of readers. It could be used as an introduction for graduate students interested in this subject or for more experienced scientists whose research focuses in other areas. It is also intended to be a useful reference in day-to-day research for particle physicists and astrophysicists actively working on the problem of dark matter. Unlike the review by Jungman {\it et al.}, we do not limit our discussion to supersymmetric dark matter.

The article is
organized as follows: we first present, in this chapter, a brief review
of the Standard Model of particle physics and cosmology, and review 
our present understanding of the history of the Universe. We focus
in particular on the freeze-out of dark matter particles and on the calculation of 
their relic abundance, and discuss the possible relationship between 
dark matter and physics beyond the Standard Model of particle physics. 

Chapter 2 is devoted to the compelling evidence for dark matter at
all astrophysical length scales. We review the key observations
and discuss the theoretical predictions (from N-body simulations) for 
the distribution of dark matter, focusing in particular on the innermost 
regions of galaxies, and discuss how they compare with observations.
Particular attention is devoted to the galactic center, where the presence
of a supermassive black hole could significantly modify the dark matter 
distribution.

Dark matter candidates are presented in chapter 3. We start with an
introduction to the ``dark matter zoo'', \ie a description of the many candidates
that have been proposed in the literature. We then focus on two particularly interesting dark matter candidates: the supersymmetric neutralino and Kaluza-Klein dark matter. For each
of these candidates, we give a brief introduction to the physical motivations and underlying
theories. We conclude chapter 3 with a review of the constraints put 
on dark matter from collider experiments, and discuss the prospects for future experiments.

The second part of this review is devoted to the astrophysical constraints on 
particle dark matter. We begin in chapter 4 with a review of  existing and 
next-generation experiments that will 
probe the nature of dark matter. This chapter is propedeutical to chapter 5, 
which discusses the many possible direct and indirect searches of dark matter
and which constitutes the heart of this review. We give our conclusions in chapter 6. Some useful particle physics details 
are given in the appendices.

\subsection{Standard Cosmology}

Although the exact definition of the {\it Standard} cosmological 
model evolves with time, following the progress of experiments in measuring the
cosmological parameters, most cosmologists agree
on a fundamental picture, the so-called {\it Big Bang} scenario, which
describes the Universe as a system evolving from a highly 
compressed state existing around $10^{10}$~years ago. 

This picture has its roots in the discovery
of Hubble's law early in  the past century,
and has survived all sorts of cosmological observations,
unlike alternative theories such as the ``steady state cosmology'',
with continuous creation of baryons, which, among other problems,
failed to explain the existence and features of the
cosmic microwave background.

We now have at our disposal an extremely sophisticated
model, allowing us to explain in a satisfactory way
the thermal history, relic background radiation, abundance of elements, large scale structure and many other properties of the Universe.
Nevertheless, we are aware that our understanding is 
still only partial. It is quite clear that 
new physics is necessary to investigate the first 
instants of our Universe's history (see section \ref{beyond}). 

To ``build'' a cosmological model, in a modern sense, three 
fundamental ingredients are needed:
\begin{itemize}
\item {\it Einstein equations}, relating the geometry of the
Universe with its matter and energy content
\item {\it metrics}, describing the symmetries of
the problem
\item {\it Equation of state}, specifying the physical properties
of the matter and energy content
 \end{itemize}

The Einstein field equation can be derived almost from first principles, 
assuming that: 1) the equation is invariant under general coordinate
transformations, 2) the equation tends to Newton's law in the limit
of weak fields, and 3) the equation is of second differential order
and linear in second derivatives~\cite{Ohanian:uu}. The resulting equation
reads\index{Einstein equations}
\be
R_{\mu \nu}- \frac{1}{2}g_{\mu \nu}R = -\frac{8 \pi G_N}{c^4}T_{\mu \nu}+
\Lambda g_{\mu \nu},
\ee
where $R_{\mu \nu}$ and $R$ are, respectively, the Ricci 
tensor and scalar 
(obtained by contraction of the Riemann curvature tensor). $g_{\mu \nu}$ is the metric tensor, $G_N$ is Newton's
constant, $T_{\mu \nu}$ is the
energy-momentum tensor, and $\Lambda$ is the 
so--called cosmological constant.

Ignoring for a moment the term involving the cosmological
constant, this equation is easily understood. We learn that {\it the geometry of the Universe}, described by the terms
on the left-hand-side, {\it is determined by its energy content}, described 
by the energy-momentum tensor on the right-hand-side. 
This is the well known relationship
between the matter content and geometry of the Universe, 
which is the key concept of general relativity.

The addition of the cosmological constant term, initially
introduced by Einstein to obtain a stationary solution
for the Universe and subsequently abandoned when the
expansion of the Universe was discovered, represents 
a ``vacuum energy'' associated with space-time itself, 
rather than its matter content, and is a source
of gravitational field even in the absence of matter.
The contribution of such ``vacuum energy'' to the
total energy of the Universe can be important, if one
believes recent analyses of type Ia supernovae and
parameter estimates from the cosmic microwave background (for further discussion
see section~\ref{cmb}).

To solve the  Einstein equations one has to specify the
symmetries of the problem. Usually one assumes
the properties of statistical {\it homogeneity} and {\it isotropy} 
of the Universe, 
which greatly simplifies the mathematical analysis. 
Such properties, made for mathematical convenience, 
are confirmed by many observations. In
particular, observations of the Cosmic Microwave
Background (CMB) have shown remarkable isotropy (once the
dipole component, interpreted as due to the Earth
motion with respect to the CMB frame, and the contribution from the
galactic plane were  subtracted). 
Isotropy alone, if combined with the Copernican 
principle, or ``mediocrity'' principle, would imply
homogeneity. Nevertheless, direct evidence of homogeneity
comes from galaxy surveys, suggesting a homogeneous
distribution at scales in excess of  $\sim 100$ Mpc. More specifically, 
spheres with diameters larger than $\sim 100$ Mpc centered in any place of 
the Universe should contain, roughly, the same amount of matter.

The properties of isotropy and homogeneity imply a specific 
form of the metric: the line element can in fact be expressed as
\be
ds^2=-c^2\mbox{d}t^2+a(t)^2\left({\mbox{d}r^2\over 1-kr^2}+r^2\mbox{d}
\Omega^2\right),
\label{ds2}
 \ee
where $a(t)$ is the so-called {\it scale factor} and the constant 
$k$, describing the spatial curvature, can take the values $k= -1,\,0,\,+1$. 
For the simplest case, $k=0$, the spatial part of Eq.~\ref{ds2} 
reduces to the metric of ordinary (flat) Euclidean space.

The Einstein equations can be solved with this metric, one of its
components leading to the Friedmann equation \index{Friedmann equation}
\be
\left({\dot a\over a}\right)^2+{k\over a^2}={8\pi G_N\over 3}\rho_{tot},
\label{eq:friedmann}
\ee
where $\rho_{tot}$ is the total average energy density of the 
universe. It is common to introduce the Hubble parameter
\be
H(t) ={\dot a(t)\over a(t)} \;\;.
\ee

A recent estimate~\cite{Contaldi:2003hi} of the present 
value of the Hubble parameter, $H_0$, (also referred to as the 
Hubble {\it constant}) is $H_0=73 \pm 3$ km s$^{-1}$ Mpc$^{-1}$. 
We see from Eq.~\ref{eq:friedmann} that the universe is flat 
($k=0$) when the energy density equals the {\em critical density}, $\rho_{c}$:
\be
\rho_{c}\equiv {3H^2\over 8\pi G_N} \;\;.
\label{rhocrit}
\ee
In what follows we will frequently express the abundance of
a substance in the Universe (matter, radiation or vacuum energy),
in units of $\rho_{c}$. We thus define the quantity $\Omega_i$
of a substance of species $i$ and density $\rho_i$ as
\be
\Omega_i \equiv {\rho_i \over \rho_{c}} \;\;.
\label{omega}
\ee
It is also customary to define 
\be
\Omega =\sum_i \Omega_i \equiv \sum_i {\rho_i \over \rho_{c}},
\ee
in terms of which the Friedmann equation (Eq.~\ref{eq:friedmann}) can be written
\be
  \Omega-1={{k}\over {H^2 a^2}} \ .
\ee
The sign of $k$ is therefore determined by whether $\Omega$ is
greater than, equal to, or less than one (see table 1). 

Following Ref.~\cite{Bergstrom:2000pn}, we note
 that the various  $\Omega_i$
evolve with time differently, depending on the equation of 
state of the component. A general
expression for the expansion rate is
\ba
{H^2(z)\over H_0^2}=\left[
\Omega_X\left(1+z\right)^{3(1+\alpha_X)}
+\Omega_K\left(1+z\right)^2\right. \nonumber \\
\left. +\Omega_M\left(1+z\right)^3
+\Omega_R\left(1+z\right)^4
\right]
\label{eq:general}
\ea
where $M$ and $R$ are labels for matter and radiation,
$\Omega_K={-k\over a_0^2H_0^2}$ and X refers to a 
generic substance with equation of state $p_X=\alpha_X\rho_X$
(in particular, for the cosmological constant, $\alpha_\Lambda=-1$). $z$ is the redshift.

We discuss in Sec.~\ref{cmb} recent estimates of cosmological 
parameters using CMB measurements, combined with various 
astrophysical observations.

\begin{center}
\begin{table}
\centering
\begin{tabular}{|c|ccc|}
\hline 
$\rho<\rho_{c}$ & $\Omega < 1$ &$k=-1$ & ${\rm open}$\\
$\rho=\rho_{c}$ & $\Omega = 1$ &$k=0$ & ${\rm flat}$\\
$\rho>\rho_{c}$ & $\Omega > 1$ &$k=1$ & ${\rm closed}$\\
\hline
\end{tabular}
\label{tabUniv}
\caption{Classification of cosmological models based on the value of
the average density, $\rho$, in terms of the critical density, $\rho_{c}$.}
\vskip 0.3cm
\end{table}
\end{center}

\subsection{The Standard Model of Particle Physics}
\label{sec:SM}
\index{Standard Model of particle physics}
The Standard Model (SM) of particle physics has, for many years, accounted for all observed particles and interactions \footnote{It is a matter of definition whether one considers neutrino masses as part of the SM or as physics beyond the SM.}. Despite this success, it is by now clear that a more fundamental 
theory must exist, whose low-energy realization should
coincide with the SM.

In the SM, the fundamental constituents of matter are fermions: {\it quarks} and 
{\it leptons}. Their interactions are 
mediated by integer spin particles called {\it gauge bosons}. Strong 
interactions are mediated by gluons $G_a$, electroweak
interaction by  $W^{\pm}$, $Z_0$, $\gamma$ and the Higgs boson 
$H^0$. 
The left-handed leptons and quarks are arranged 
into three generations of $SU(2)_L$ doublets
\begin{equation}
\label{su2doublets}
\left(\begin{array}{c}
\nu_e \\
e^-
\end{array}\right)_L\qquad
\left(\begin{array}{c}
\nu_\mu \\
\mu^-
\end{array}\right)_L\qquad
\left(\begin{array}{c}
\nu_\tau \\
\tau^-
\end{array}\right)_L
\end{equation}
\begin{equation}\label{2.66}
\left(\begin{array}{c}
u \\
d^\prime
\end{array}\right)_L\qquad
\left(\begin{array}{c}
c \\
s^\prime
\end{array}\right)_L\qquad
\left(\begin{array}{c}
t \\
b^\prime
\end{array}\right)_L       
\end{equation}
with the corresponding right-handed fields transforming as singlets
under $ SU(2)_L $. Each generation contains two flavors of quarks
with {\it baryon number} $B=1/3$ and {\it lepton number} 
$L=0$ and two leptons with $B=0$ and $L=1$. Each particle also
has a corresponding antiparticle with the same mass and 
opposite quantum numbers. 

The quarks which are primed are {\it weak eigenstates} related to
{\it mass  eigenstates} by the Cabibbo-Kobayashi-Maskawa (CKM)
matrix
\be
\label{ckm}
\left(\begin{array}{c}
d^\prime \\ s^\prime \\ b^\prime
\end{array}\right)=
\left(\begin{array}{ccc}
V_{ud}&V_{us}&V_{ub}\\
V_{cd}&V_{cs}&V_{cb}\\
V_{td}&V_{ts}&V_{tb}
\end{array}\right)
\left(\begin{array}{c}
d \\ s \\ b
\end{array}\right)=\hat V_{\rm CKM}\left(\begin{array}{c}
d \\ s \\ b
\end{array}\right).
\end{equation}

Gauge symmetries play a 
fundamental role in particle physics. It is in fact in terms 
of symmetries and using the formalism of gauge theories 
that we describe electroweak and strong interactions.
The SM is based on the  $ SU(3)_C\otimes SU(2)_L\otimes U(1)_Y$
gauge theory, which undergoes the spontaneous breakdown:
\be
SU(3)_C\otimes SU(2)_L\otimes U(1)_Y \to
 SU(3)_C\otimes U(1)_Q
\label{brkdown}
\ee
where $Y$ and $Q$ denote the weak hypercharge and the electric charge
generators, respectively, and $SU(3)_C$ describes the strong (color) 
interaction, known as Quantum ChromoDynamics (QCD). This spontaneous symmetry breaking results in the generation of the massive $W^{\pm}$ and $Z$ gauge bosons as well as a massive scalar Higgs field.

\subsection{A very brief history of the Universe}

\begin{figure}[t!]
\centering
\includegraphics[width=0.7\textwidth,clip=true]{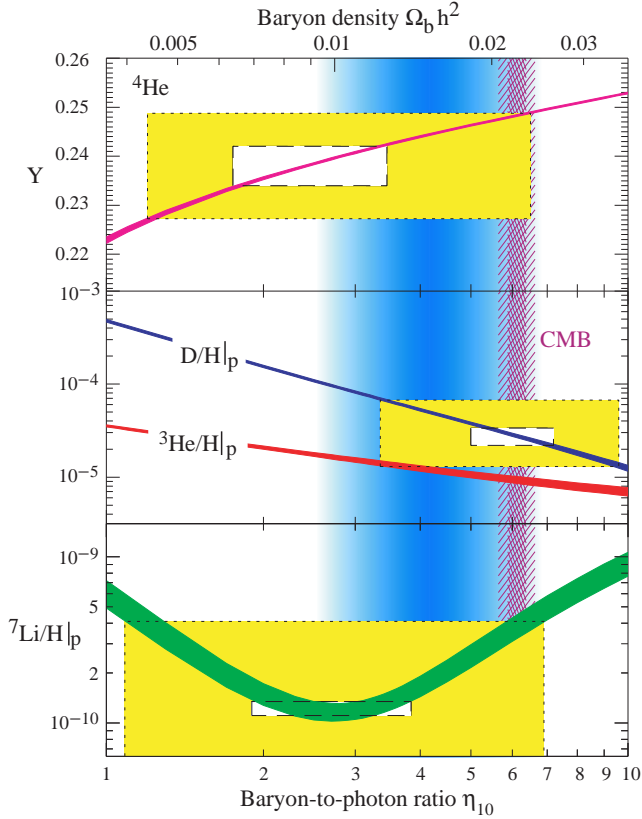}
\caption[Big Bang nucleosynthesis predictions for the 
abundances of \qua , \deu, \tro\ and \sep]
{Big Bang nucleosynthesis predictions for the 
abundances of light elements as a function of the baryon over photon ratio 
$\eta$ or $\Omega_b h^2$~\cite{Coc:2003ce}. From Ref.~\cite{subirbbn}.}
\label{bbn}
\end{figure}

Our description of the early Universe is based on an
extrapolation of known physics back to the Planck epoch,
when the Universe was only $t=10^{-43}$seconds old, or 
equivalently up to energies  at which the gravitational interaction becomes strong (of the order of the 
Planck mass, $M_{Pl}=10^{19}$~GeV). Starting at this epoch 
we take now a brief tour through the evolution of
the Universe:

\begin{itemize} 
\item $T \sim 10^{16}$~GeV. It is thought that at this scale,  
some (unknown) grand unified group, $G$, breaks down into the 
Standard Model gauge group, $SU(3)_C\otimes SU(2)_L\otimes U(1)_Y$. Little is known about this transition, however. 

\item $T \sim 10^{2}$~GeV. The Standard Model gauge symmetry breaks into 
$SU(3)_C \otimes U(1)_Q$ (see Eq.~\ref{brkdown}). This transition, called electroweak symmetry breaking, could be the origin of 
baryogenesis (see \eg Ref.~\cite{Affleck:1984fy}) and
possibly of primordial magnetic fields 
(\eg Ref.~\cite{Joyce:1997uy}).

\item $T \sim 10^1 - 10^3$~GeV. Weakly interacting
dark matter candidates with GeV-TeV scale masses {\it freeze-out}, as discussed in next section.
This is true in particular for the {\it neutralino}
and the $\bone$ Kaluza-Klein excitation that we discuss in 
Chapter~\ref{candidates}.

\item $T \sim 0.3$~GeV. The QCD phase transition occurs, 
which drives the confinement of quarks and
gluons into hadrons. 
 
\item $T \sim 1$~MeV. Neutron freeze-out occurs.

\item $T \sim 100$~keV. Nucleosynthesis: protons 
and neutrons fuse into light elements (D, $^3$He, $^4$He, Li).
The standard Big Bang nucleosynthesis (BBN) provides by far
the most stringent constraints to the Big Bang theory, and 
predictions remarkably agree with observations (see Fig.~\ref{bbn}).

\item $T \sim 1$~eV. The matter density becomes equal to 
that of the radiation, allowing for the formation of structure to begin.

\item $T \sim 0.4$~eV. Photon decoupling produces the 
cosmic background radiation (CMB), discussed in Sec.~\ref{cmb}.  

\item $T = 2.7 K \sim 10^{-4}$~eV. Today.
\end{itemize}

\subsection{Relic Density}
\label{sec:relic}
\index{relic density}

We briefly recall here the basics of the calculation of the density of a thermal relic. The discussion is based on Refs.~\cite{Kolb:vq, 
Gondolo:1990dk,Servant:2002hb} and we refer to them 
for further comments and details.

A particle species in the early Universe has to 
interact sufficiently or it will fall out of local thermodynamic
equilibrium. Roughly speaking, when its interaction rate drops below the expansion rate of the Universe, the equilibrium can no longer be maintained and the 
particle is said to be {\it decoupled}. 

\subsubsection{The standard calculation}
The evolution of the phase space distribution function,
$f({\bf p},{\bf x})$, is governed by the Boltzmann equation
\be
{\bf L}[f] = {\bf C}[f],
\ee
where ${\bf L}$ is the Liouville operator, and 
${\bf C}$ is the collision operator, describing the
interactions of the particle species considered.

After some manipulation, the Boltzmann equation 
can be written as an equation for the particle number
density $n$: 
\be
\frac{dn}{dt}+3Hn=-\langle\sigma v\rangle\left(n^2-({n^{eq}})^2\right) ,
\label{BE}
\ee
where $\sigma v$ is the total annihilation cross section multiplied by velocity,
brackets denote thermal average, $H$ is Hubble constant,
and $n^{eq}$ is the number density at thermal equilibrium. For
massive particles, \ie in the non-relativistic limit, and in
the Maxwell-Boltzmann approximation, one has
\be
n^{eq}=g\left(\frac{mT}{2\pi}\right)^{3/2}e^{-m/T} ,
\label{exponentialdecrease}
\ee
where $m$ is the particle mass and $T$ is the temperature.
We next introduce the variables
\be
Y \equiv \frac{n}{s} \;\;\;\; , \;\;\;\; Y^{eq} \equiv \frac{n^{eq}}{s} 
\ee
where $s$ is the entropy density $s={2\pi^2 g_* T^3}/{45}$ and
$g_*$ counts the number of relativistic degrees of freedom.
Using the conservation of entropy per co-moving volume
($sa^3=$constant), it follows that $\dot{n}+3Hn=s\dot{Y}$
and Eq.~\ref{BE} reads
\be
s\dot{Y}=-\langle\sigma v\rangle s^2\left(Y^2-({Y^{eq}})^2\right) .
\label{BE2}
\ee
If we further introduce the variable $x \equiv \frac{m}{T}$, 
Eq.~\ref{BE2} can be expressed as
\be
\label{Boltz}
\frac{dY}{dx}=-\frac{\langle\sigma v\rangle s}{H x} \left(Y^2-({Y^{eq}})^2\right) .
\ee
For heavy states, we can approximate $\langle\sigma v\rangle$ with the
non-relativistic expansion in powers of $v^2$
\be
\langle\sigma v\rangle= a+b \langle v^2 \rangle 
+ {\cal O}(\langle v^4 \rangle ) \approx a+6 \ b/x \; ,
\label{approsig}
\ee
which leads to our final version of Eq.~\ref{Boltz}
in terms of the variable $\Delta=Y-Y^{eq}$:
\be
\label{delta}
\Delta^{\prime}=-{Y^{eq}}^{\prime}-f(x)\Delta(2Y^{eq}+\Delta) ,
\ee
where prime denotes $d/dx$ and
\be
f(x)=\sqrt{\frac{\pi g_*}{45}}\  m \ M_{Pl}\ (a+6 \ b/x) \  x^{-2} .
\ee

Following Ref.~\cite{Kolb:vq} we introduce the quantity $x_F \equiv m/T_F$, where $T_F$ is the freeze-out temperature of the relic particle,
and we notice that Eq.~\ref{delta} can be solved analytically in 
the two extreme regions $x \ll x_F$ and $x \gg x_F$ 
\ba
\Delta =& -\frac{{Y^{eq}}^{\prime}}{2f(x)Y^{eq}} \; \mbox{for } x \ll x_F \\
\Delta^{\prime} =& -f(x) \Delta^{2} \; \mbox{for } x \gg x_F.
\label{latetimes}
\ea
These regions correspond to long before freeze-out and long after freeze-out, respectively. Integrating the last equation between $x_F$ and $\infty$ 
and using $\Delta_{x_F} \gg \Delta_{\infty}$, we can derive the
value of $\Delta_{\infty}$ and arrive at
\be
\label{Y_infty}
Y_{\infty}^{-1}=\sqrt{\frac{\pi g_*}{45}}M_{Pl} \ m \ x_F^{-1}(a+3b/x_F). 
\ee
The present density of a generic relic, $X$, is simply given by
$\rho_X=m_X 
n_X=m_X s_0 Y_{\infty}$, where $s_0=2889.2$ cm$^{-3}$ is the present entropy density (assuming three Dirac
neutrino species).
The relic density can finally be expressed in 
terms of the critical density (see Eq.~\ref{omega}) 
\be
\label{eq:relic}
\Omega_X h^2 \approx \frac{1.07 \times 10^9 \, \rm{GeV}^{-1}}
{M_{Pl}}\frac{x_F}{\sqrt{g_*}}\frac{1}{(a+3b/x_F)} ,
\ee
where $a$ and $b$ are expressed in GeV$^{-2}$ and 
$g_*$ is evaluated at the freeze-out temperature. It is conventional to write the relic density in terms of the Hubble parameter, $h = H_0 / 100 \, \rm{km} \, \rm{s}^{-1}  \, \rm{Mpc}^{-1}$.

To estimate the relic density, one is thus left with
the calculation of the annihilation cross sections (in all of the possible channels) and the extraction of the parameters 
$a$ and $b$, which depend on the particle mass. 
The freeze-out temperature $x_F$ can be estimated
through the iterative solution of the equation
\be
\label{XF}
x_F=\ln \left[c (c+2) \sqrt{\frac{45 }{8}}
\frac{g}{2\pi^3} \frac{m \ M_{Pl} (a+6b/x_F)}{g^{1/2}_* x^{1/2}_F}\right],
\ee
where $c$ is a constant of order one determined by matching the late-time 
and early-time solutions.

It is sometimes useful to perform an order-of-magnitude estimate using an approximate version of Eq.~\ref{eq:relic}~\cite{Jungman:1995df}:
\be
\Omega_X h^2 \approx \frac{3 \times 10^{-27} \mbox{cm}^3 \mbox{s}^{-1}}
{\langle\sigma v\rangle} \;\; .
\ee 

We note that the approximation introduced in Eq.~\ref{approsig}
is not always justified (see \eg Ref.~\cite{Jungman:1995df}). For example, Ref.~\cite{Salati:2002md} suggests a scenario where the presence of a scalar field in the early Universe could
significantly affect the value of the relic density. Furthermore,
a dramatic change in the relic density can be induced by resonance enhancements or so-called {\it coannihilations}. We discuss the effects of coannihilations in the next section.

\subsubsection{Including coannihilations}
\index{coannihilations}
Following earlier works (see Ref.~\cite{Binetruy:1983jf}),
Griest and Seckel~\cite{Griest:1990kh} noticed that 
if one or more particles have a  mass similar to the relic 
particle and share a quantum number with it, the standard
calculation of relic density fails.

Let us consider $N$ particles $X_i$
($i=1,\ldots,N$) with masses $m_i$ and internal degrees of freedom
(statistical weights) $g_i$.  Also assume that $m_1 \leq m_2 \leq
\cdots \leq m_{N-1} \leq m_N$, and that the lightest particle is 
protected against decay thanks to some symmetry 
({\it i.e.} R-parity or KK-parity, for neutralinos or Kaluza-Klein particles, respectively. See section~\ref{candidates}). We will also denote the lightest 
particle by $X_1$.

In this case, Eq.~\ref{BE} becomes
\begin{equation}
  \frac{dn}{dt} = -3Hn - \sum_{i,j=1}^N \langle \sigma_{ij} v_{ij} \rangle 
  \left( n_{i}n_{j} - n_{i}^{\rm{eq}}n_{j}^{\rm{eq}} \right),
\end{equation}
where $n$ is the number density of the relic particle 
and $n= \sum_{i=1}^N n_{i}$, due to the fact that the decay rate 
of particles, $X_i$, other than the lightest is much faster than the 
age of the Universe. Here,
\begin{eqnarray}
  \sigma_{ij}  & = & \sum_X \sigma (X_i X_j \rightarrow X_{SM})
\end{eqnarray}
is the total annihilation rate for  $X_i X_j$ annihilations
into a Standard Model particle. Finally,  
\begin{equation}
  v_{ij} = \frac{\sqrt{(p_{i} \cdot p_{j})^2-m_{i}^2 m_{j}^2}}{E_{i} E_{j}}
\end{equation}
is the relative particle velocity, with $p_{i}$ and $E_{i}$ being the four-momentum and energy of 
particle $i$.

The thermal average $\langle\sigma_{ij}v_{ij}\rangle$ is defined
with equilibrium distributions and is given by
\begin{equation}
  \langle \sigma_{ij}v_{ij} \rangle = \frac{\int d^3{\bf
      p}_{i}d^3{\bf p}_{j} 
  f_{i}f_{j}\sigma_{ij}v_{ij}}
  {\int d^3{\bf p}_{i}d^3{\bf p}_{j}f_{i}f_{j}},
\end{equation}
where $f_{i}$ are distribution functions in the Maxwell-Boltzmann
approximation.

The scattering rate of supersymmetric particles off particles in the
thermal background is much faster than their annihilation rate.
We then obtain
\begin{equation} \label{eq:Boltzmann2}
  \frac{dn}{dt} =
  -3Hn - \langle \sigma_{\rm{eff}} v \rangle 
  \left( n^2 - n_{\rm{eq}}^2 \right),
\end{equation}
where
\begin{equation} \label{eq:sigmaveffdef}
  \langle \sigma_{\rm{eff}} v \rangle = \sum_{ij} \langle
  \sigma_{ij}v_{ij} \rangle \frac{n_{i}^{\rm{eq}}}{n^{\rm{eq}}}
  \frac{n_{j}^{\rm{eq}}}{n^{\rm{eq}}}.
\end{equation}

Edsjo and Gondolo \cite{Edsjo:1997bg} reformulated 
the thermal average into the more convenient expression
\begin{equation} \label{eq:sigmavefffin2}
  \langle \sigma_{\rm{eff}}v \rangle = \frac{\int_0^\infty
  dp_{\rm{eff}} p_{\rm{eff}}^2 W_{\rm{eff}} K_1 \left(
  \frac{\sqrt{s}}{T} \right) } { m_1^4 T \left[ \sum_i \frac{g_i}{g_1}
  \frac{m_i^2}{m_1^2} K_2 \left(\frac{m_i}{T}\right) \right]^2},
\end{equation}
where  $K_{i}$ are the modified Bessel functions of the second kind and of 
order $i$. The quantity $W_{\rm{eff}} $ is defined as
\begin{equation} \label{eq:weff}
  W_{\rm{eff}} = \sum_{ij}\frac{p_{ij}}{p_{11}}
  \frac{g_ig_j}{g_1^2} W_{ij} = 
  \sum_{ij} \sqrt{\frac{[s-(m_{i}-m_{j})^2][s-(m_{i}+m_{j})^2]}
  {s(s-4m_1^2)}} \frac{g_ig_j}{g_1^2} W_{ij},
\end{equation}
where $ W_{ij} = 4 E_{i} E_{j} \sigma_{ij} v_{ij}$ and $p_{ij}$ 
is the momentum of the particle $X_i$ (or $X_j$) in the
center-of-mass frame of the pair $X_i X_j$, and 
$s  =  m_{i}^2+m_{j}^2 + 2E_{i}E_{j}-2 |{\bf p}_{i}| |{\bf
    p}_{j}| \cos \theta$, with the usual meaning of the symbols.

The details of coannihilations in the framework 
of supersymmetric models are well established (see \eg
the recent work of Edsjo \etal~\cite{Edsjo:2003us}),
and numerical codes now exist including coannihilations
with all supersymmetric particles, \eg MicrOMEGAs 
\cite{Belanger:2002sq} and the new version
of DarkSusy \cite{darksusynew,Gondolo:2002tz},
  publicly released in 2004. The case of coannhilations with 
a light top squark, such as the one required for the realization 
of the electroweak baryogenesis mechanism, has been discussed in 
Ref.\cite{Balazs:2004bu}.

\subsection{Links with Physics Beyond the Standard Model}
\label{beyond}

The concepts of dark energy and dark matter do not
find an explanation in the framework of the Standard Model
of particle physics. Nor are they understood 
in any quantitative sense in terms of astrophysics.
It is interesting that also in the
realm of particle physics, evidence is accumulating for the
existence of physics {\it beyond} the Standard Model, based on theoretical and perhaps experimental arguments. 

On the experimental side, there is strong evidence for
oscillations of atmospheric neutrinos (originating from
electromagnetic cascades initiated by cosmic rays in the 
upper atmosphere) and solar neutrinos. The oscillation 
mechanism can be explained under the hypothesis that
neutrinos {\it do} have mass, in  contrast to the zero mass
neutrinos of the Standard Model (see Ref.~\cite{Maltoni:2003da}
for a recent review).

On the theoretical side, many issues make the 
Standard Model unsatisfactory, for example the 
{\it hierarchy problem}, \ie the enormous difference 
between the weak and Planck scales in the presence of the Higgs field
(this will be discussed in some detail in 
Sec.~\ref{whysusy}), or the problem of unification
addressing the question of whether there exists a unified description of all known forces, possibly including gravity.

The list of problems could be much longer, and it is 
natural to conjecture that our Standard
Model is the low-energy limit of a more fundamental 
theory. Two examples of popular extensions of the Standard Model include: 
\begin{itemize} 
\item \underline{{\it Supersymmetry.}} 
As a complete symmetry between fermions and bosons, supersymmetry's theoretical appeal is very great \cite{wesszumino}. So great, in fact, is this appeal, that it appears to many as a necessary
ingredient of future extensions of the Standard Model. Many interesting features make it 
attractive, including its role in understanding the 
fundamental distinction between bosons and fermions, and the problems of hierarchy and unification discussed 
above. Last, but not least, it provides an excellent
dark matter candidate in terms of its lightest stable particle, the {\it neutralino}. We will present the 
basics of supersymmetry and the properties of the neutralino 
in Sec.~\ref{sec:susy}.

\item \underline{{\it Extra dimensions.}} 
In the search of a fundamental theory with a unified
description of all interactions, physicists developed theories with extra spatial dimensions, following an early idea
of Kaluza~\cite{Kaluza:tu}, who extended to four
the number of space dimensions to include electromagnetism
into a ``geometric'' theory of gravitation. 
In theories with {\it unified} extra dimensions, 
in which all particles and fields of the Standard Model
can propagate in the extra dimensions, 
the lightest Kaluza-Klein particle, \ie the lightest 
of all the states corresponding to the first excitations
of the particles of the Standard Model, is a viable 
dark matter candidate, as we discuss in Sec.~\ref{sec:kk}.

\end{itemize}

Despite the fact that neutrinos are thought to be massive,
they are essentially ruled out as dark matter candidates (see 
Sec.~\ref{sec:zoo}). Consequently, the Standard Model does not provide
a viable dark matter candidate. This is further supported by the fact that most
of the dark matter is non-baryonic (see section~\ref{cmb}). Dark matter is therefore
a motivation to search for physics beyond the Standard Model
(others might  say that this is {\it evidence} for physics
beyond the Standard Model). 

This is a typical example of the strong interplay between 
particle physics, theoretical physics, 
cosmology and astrophysics. From one
side, theoretical particle physics stimulates the formulation
of new theories predicting new particles that turn out to
be excellent dark matter candidates. On the other side, cosmological
and astrophysical observations constrain the properties 
of such particles and consequently the parameters of the 
new theories.


\section{Evidence and Distribution}
\label{obs}
\subsection{The Galactic Scale}

The most convincing and direct evidence for dark matter on 
galactic scales comes from the observations of the {\it rotation curves} of galaxies, namely the graph
of circular velocities of stars and gas as a function of their
distance from the galactic center. 

\begin{figure}
\centering
\includegraphics[width=0.5\textwidth,clip=true]{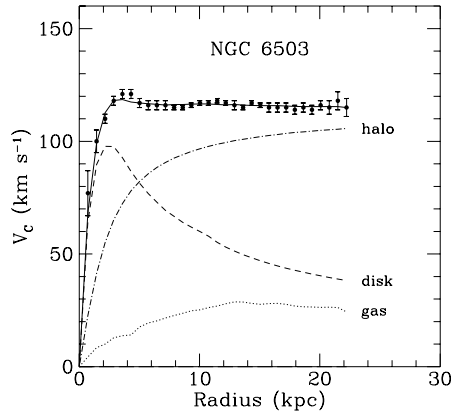}
\caption[Rotation curve of NGC 6503]{Rotation curve of NGC 6503. The dotted, dashed 
and dash-dotted lines are the contributions of 
gas, disk and dark matter, respectively. From Ref.~\cite{rot}.}
\label{rot}
\end{figure}

Rotation curves are usually obtained by combining observations
of the $21$cm line with optical surface photometry.
Observed rotation curves usually exhibit a characteristic {\it flat} behavior
at large distances, \ie out towards, and even far beyond, 
the edge of the visible disks (see a typical example in
Fig.~\ref{rot}). 

In Newtonian dynamics the circular velocity is expected to be
\be
v(r) = \sqrt{\frac{G M(r)}{r }}\,,
\ee
where, as usual, $M(r)\equiv 4 \pi \int 
\rho(r) r^2 \mbox{d}r$, and $\rho(r)$ is the mass density profile,
 and should be falling $\propto 1/\sqrt{r}$ beyond the 
optical disc. The fact that $v(r)$ is approximately constant implies 
the existence of an halo with $M(r) \propto r$ and 
$\rho \propto 1/r^2$. 

Among the most interesting objects, from the point
of view of the observation of rotation curves, are
the so--called Low Surface Brightness (LSB) galaxies, 
which are probably everywhere dark matter-dominated, with the observed 
stellar populations making only a small 
contribution to rotation curves. 
Such a property is extremely
important because it allows one to avoid the difficulties associated
with the deprojection and 
disentanglement of the dark and visible contributions to the rotation curves. 

Although there is a consensus about the shape of dark matter halos at large distances, it is unclear whether
galaxies present cuspy or shallow profiles in their
innermost regions, which is an issue of crucial 
importance for the effects we will be discussing in 
the following chapters.
\begin{figure}[t]
\begin{center}
$\begin{array}{c@{\hspace{0.5in}}c}
\includegraphics[width=0.4\textwidth]{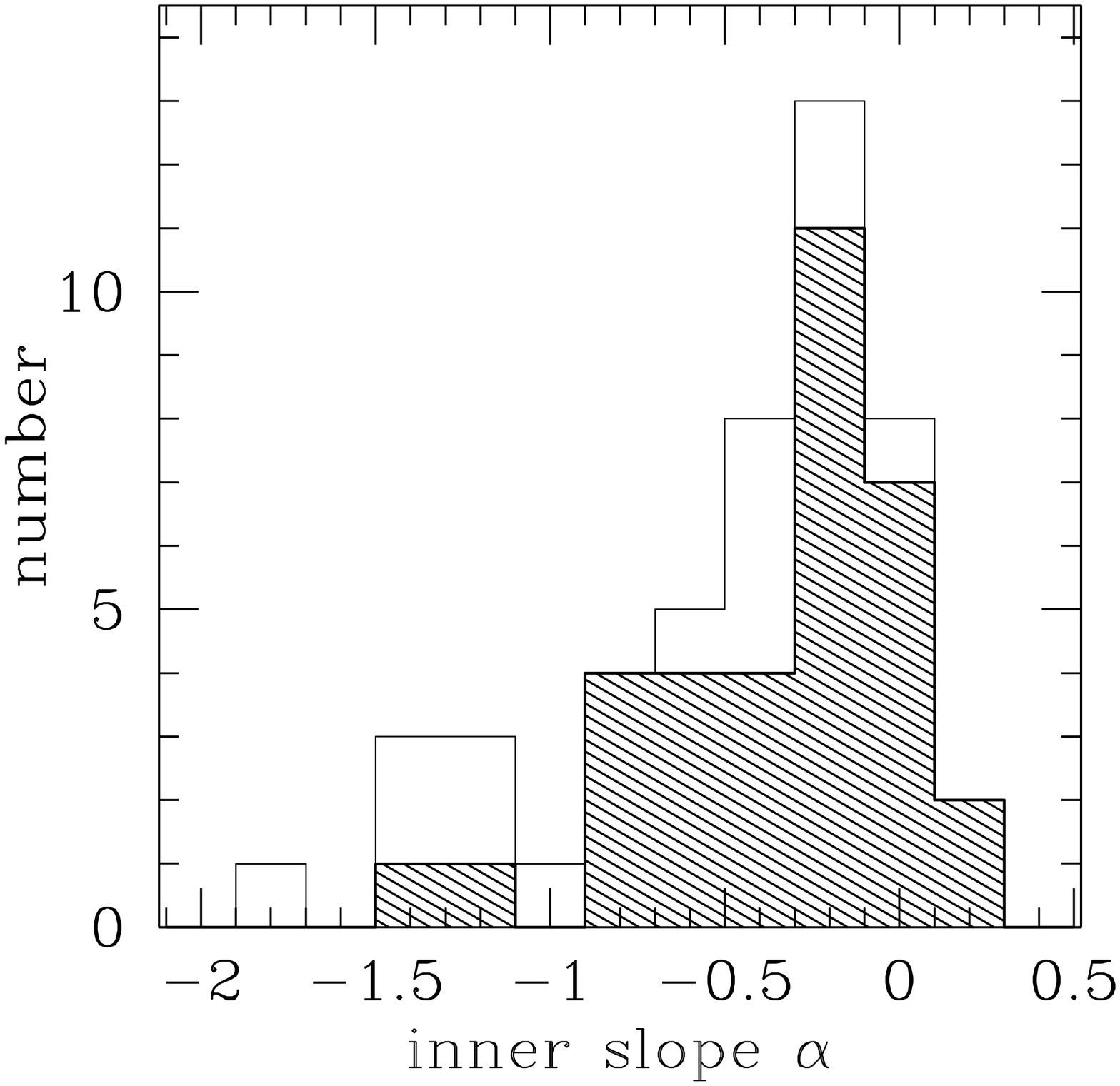} &
\includegraphics[width=0.4\textwidth]{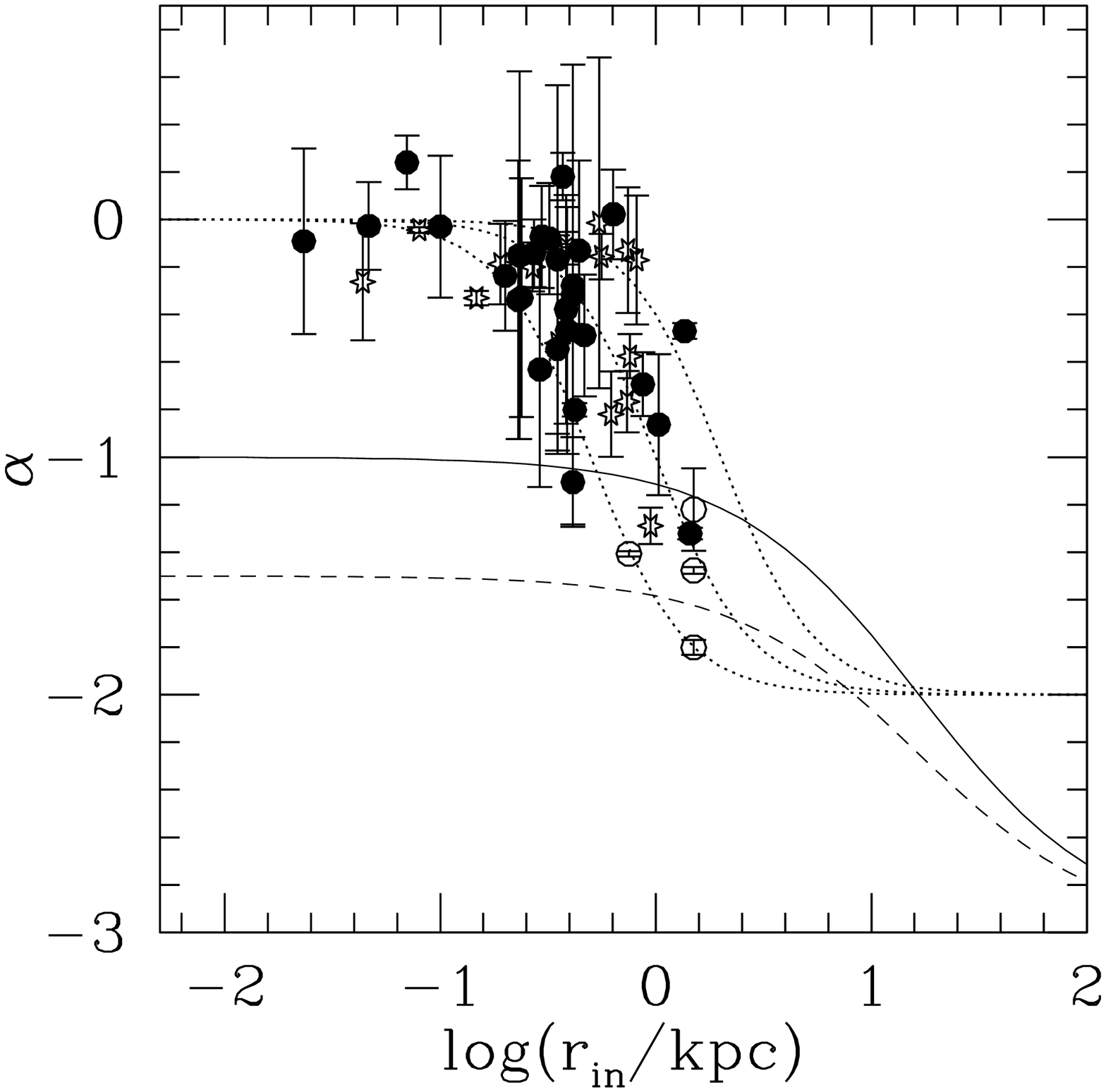} \\ [0.4cm]
\end{array}$
\end{center}
\caption[Parameters of dark matter density profiles in innermost regions]
{Left panel: The distribution of inner slopes, $\alpha$, of 
dark matter density profiles in LSB galaxies. The hatched 
(blank) histogram represents well--resolved (unresolved) galaxies. 
Right panel: The value of $\alpha$ as a function of the 
radius of the innermost point. From Ref.~\cite{deBlok:2001fe}.}
\label{deblok}
\end{figure}

Using high--resolution data of 13 LSB galaxies, 
de Blok \etal~\cite{deBlok:2001fe}
recently showed, that the distribution of inner slopes,
\ie the power--law indices of the density profile in the
innermost part of the galaxies,
suggests the presence of shallow, or even flat, 
cores (see Fig.~\ref{deblok}). Furthermore, the 
highest values of the power--law index are obtained 
in correspondence to galaxies with the poorest resolution, 
as can be seen from the right panel of the same
figure.

Following Salucci and Borriello~\cite{Salucci:2002jg},
rotation curves of both low and high surface luminosity galaxies 
appear to suggest a universal density
profile, which can be expressed as the sum of an
exponential thin stellar disk, and a spherical 
dark matter halo with a flat core of radius $r_0$ and 
density $\rho_0 = 4.5 \times 10^{-2} (r_0/\mbox{kpc})
^{-2/3} M_\odot \mbox{pc}^{-3}$ (here, $M_\odot$ denotes a solar mass, $2 \times 10^{30}\, \rm{kg}$).  
In a similar way the analysis of Reed {\it et al.}~\cite{Reed:2003hp}
leads to the
conclusion that simulated halos have significantly steeper 
density profiles than are inferred from observations.

Nevertheless, claims have been 
made in the literature about the possibility of  
reconciling these results with the steep profiles 
predicted by numerical simulations (see section~\ref{nbody}
for a discussion on the state of art of N-body simulations and for further 
discussions, see Refs.~\cite{deBlok:2001fe,vandenBosch:2000rz,Rhee:2003vw}).
In particular, Hayashi \etal~\cite{Hayashi:2003sj} have claimed consistency between 
most observations and their simulated profiles and
have argued that the remaining discrepancies could be explained
by taking into account the difference between the circular velocity 
and gas rotation speed, likely to arise in gaseous disks 
embedded within realistic, triaxial cold dark matter halos.

Another area of contention is that of the dark matter content in the
inner halos of massive disk galaxies. It has been argued that  barred
galaxies cannot contain substantial amounts of dark matter out to the
outermost extent of the observed bars, otherwise the rapidly rotating
bars would have slowed down due to dynamical friction on the dark matter \cite{sellwood1,sellwood2}.
 One counterargument is the contention that bars may be
 dynamically young systems that formed by  secular evolution of
 unstable cold disks and hence poor dynamical probes \cite{Combes:2004vu}. Another is that
 the slowing down of bars, perhaps in an earlier phase of the forming
 galaxy,  actually  heated the dark matter  and generated a core.

Despite the uncertainties of the slope in the
innermost regions of galaxies, rotation curves of disk galaxies provide
strong evidence for the existence of a spherical
dark matter halo. The total amount of dark matter present is difficult
to quantify, however, as we do
not  know to what distances halos extend. Additional
evidence for dark matter at galactic scales
comes from mass modelling of the detailed rotation curves, including
spiral arm features. Submaximal disks are often, although not always,
required \cite{Slyz:2003qh}.

Some elliptical galaxies show evidence for dark matter  via 
strong gravitational lensing~\cite{Koopmans:2002qh}.  X-ray evidence 
reveals the presence of extended atmospheres of hot gas that fill the
  dark halos of isolated ellipticals and whose  hydrostatic support
  provides evidence for dark matter.
In at least one case, an elliptical contains a cold gas disk whose HI
rotation curve is flat out to about 5 half light radii. In contrast,
however, planetary nebula studies to a similar distance for other
ellipticals can be explained only with a constant mass-to-light ratio.
There may be some dark matter in these cases, but its relative
dominance does not appear to increase with increasing galactocentric
distance.
Rather, it is asociated with the stellar distribution.

Other arguments for dark matter, both on subgalactic 
and inter--galactic
scales, also comes from a great 
variety of data. Without attempting to be complete, we cite among them:

\begin{itemize}

\item {\it Weak modulation of strong lensing} around individual
massive elliptical  galaxies. This provides evidence for substructure
on scales of $\sim 10^6\rm M_\odot$ \cite{Metcalf:2003sz,Moustakas:2002iz}.

\item The so--called {\it Oort discrepancy} in the
disk of the Milky Way (see \eg Ref.~\cite{bahcall:1992}). 
The argument follows an early suggestion of
Oort, inferring the existence of unobserved matter
from the inconsistency between the amount of stars, 
or other tracers in the solar neighborhood,
and the gravitational potential implied by their 
distribution.

\item {\it Weak gravitational lensing} of distant galaxies 
by foreground structure (see, \eg Ref.~\cite{Hoekstra:2002nf}).

\item The {\it velocity dispersions of dwarf spheroidal
galaxies} which imply
mass--to--light ratios larger than those observed in our
``local'' neighborhood.
While the profiles of individual dwarfs show scatter, there is no
doubt about the overall dark matter content (see Refs.~\cite{vogt,Mateo:1998wg}).

\item The {\it velocity dispersions of spiral galaxy satellites} which 
suggest the existence of dark halos around spiral galaxies,
similar to our own, extending at galactocentric radii
$\gtrsim 200$~kpc, \ie well behind the optical disc.
This applies in particular to the Milky Way, where both dwarf galaxy satellites
and globular clusters  probe the outer rotation curve (see Refs.~\cite{zari,Azzaro:2003hp}).
\end{itemize}

\subsection{The Scale of Galaxy Clusters}

A
cluster of galaxies gave  the first hints of dark matter (in the modern sense). In 1933,  
F.Zwicky \cite{zwicky33} inferred, from measurements of 
the velocity dispersion of galaxies in the Coma cluster, 
a mass--to--light ratio of around $400$ solar 
masses per solar luminosity, thus exceeding the ratio in the solar
neighborhood by two 
orders of magnitude.
Today, most dynamical estimates \cite{nbahcall,sasha,carlberg}
are consistent with a value $\Omega_M\sim 0.2-0.3$ on cluster 
scales. A convenient calibration is $\Omega_M=(M/L)/1000.$ 

The mass of a cluster can be determined via several
methods, including application of the virial theorem to the observed
distribution of radial velocities, by weak gravitational lensing, and 
by studying the profile of X--ray emission that 
traces the distribution of hot emitting gas in rich clusters. \index{X-ray
emission from clusters}
\begin{figure}
\centering
\includegraphics[width=\textwidth]{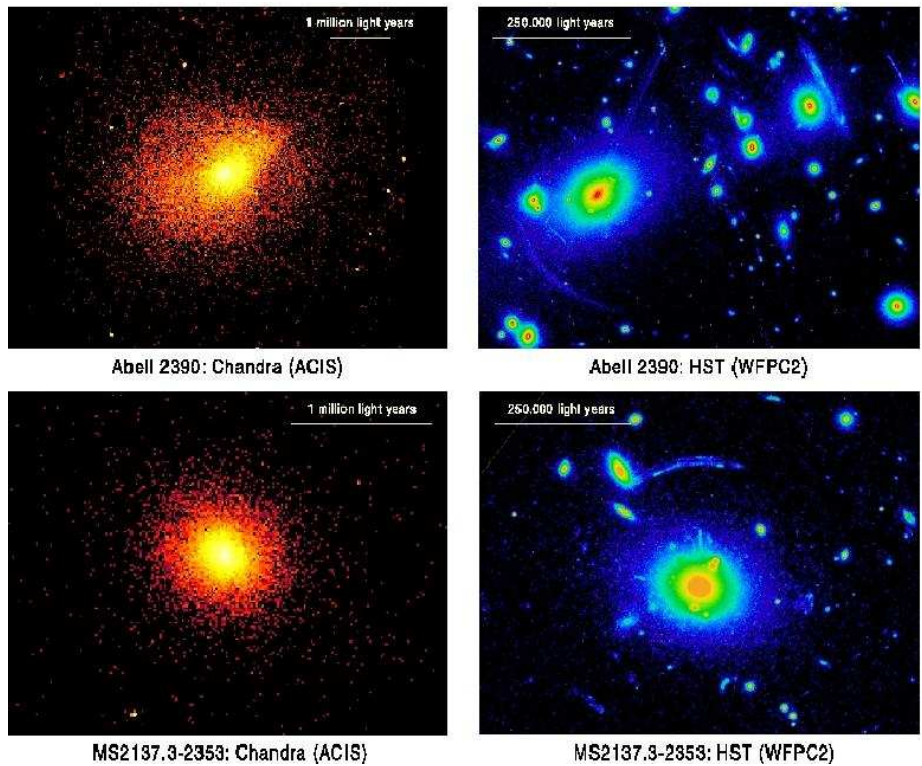}
\caption[X-ray and optical images of Abell 2390]
{Chandra X-ray (left) and Hubble Space Telescope Wide Field
Planetary  Camera 2 optical (right) images of Abell 2390 ($z=0.230$) 
and  MS2137.3-2353 ($z=0.313$). 
Note the clear gravitational arcs in the Hubble images. 
From Ref.~\cite{Fabian:2003ex}.}
\label{montage} 
\end{figure}

Consider the equation of hydrostatic equilibrium 
for a system with spherical symmetry
\be
\frac{1}{\rho} \frac{dP}{dr} = -a(r) \;,
\ee
where $P$, $\rho$, and $a$ are, respectively, the pressure, 
density, and gravitational acceleration of the gas,
at radius $r$. For an ideal gas, this can be rewritten in 
terms of the temperature, $T$, and the average molecular 
weight, $\mu\approx 0.6$, 
\be
\frac{d \log \rho}{d \log r} +\frac{d \log T}{d \log r} 
= -\frac{r}{T} \left( \frac{\mu m_p}{k} \right) a(r) \;,
\ee  
where $m_p$ is the proton mass. The temperature
of clusters is roughly constant outside of their 
cores and the density profile of the observed gas 
at large radii roughly follows a power--law
with an index between $-2$ and $-1.5$. We then find
that the temperature should obey the relation 
\be
kT \approx (1.3 - 1.8) \mbox{keV} \left( \frac{M_r}
{10^{14} M_\odot}\right) \left( \frac{1 \mbox{Mpc}}
{r}\right) 
\label{T}
\ee
for the baryonic mass of a typical cluster, where $M_r$
is the mass enclosed within the radius $r$.
The disparity between the temperature obtained using 
Eq.~\ref{T} and the corresponding observed temperature, 
$T \approx 10\,$keV, when  $M_r$
is identified with the baryonic mass,
suggests the existence of a substantial
amount of dark matter in clusters.

These conclusions can be checked against estimates
from gravitational lensing data (see Fig.~\ref{montage}).
Following Einstein's theory of general
relativity, light propagates along geodesics which
deviate from straight lines when passing near intense 
gravitational fields. The distortion of the images of
background objects due to the gravitational
mass of a cluster can be used to infer the shape 
of the potential well and thus the mass of the 
cluster (see \eg Ref.~\cite{tyson} for a spectacular 
demonstration of gravitational lensing in clusters).

The fraction of baryons inside a cluster, crucial
to disentangle the contributions of ordinary (visible) and dark matter,
can also be inferred through the so--called Sunyaev-Zel'dovich effect by which the cosmic microwave
background (see section~\ref{cmb}) gets spectrally 
distorted through Compton scattering on hot electrons.

Despite  general agreement between dark matter density profiles
at large radii and numerical simulations (see section~\ref{nbody}),
it is unclear whether there is agreement with the
predicted profiles in the cores of clusters. 
Gravitational lensing measurements appear to
be in conflict with cuspy profiles, excluding
at the 99\% confidence level cusps with power--law indices of about
$-1$ (see \eg Ref.~\cite{Sand:2002cz}).

This argument is strengthened by use of radial arcs which probe the
mass gradient, but is weakened  if the cluster is not spherically
symmetric. Indeed an asymmetry of a few percent alows the cluster
profiles to be consistent with NFW.
Moreover,  recent {\it Chandra} 
observations of X--ray emission from Abell 2029
suggest a full compatibility of dark matter distributions
with cuspy profiles (see Ref.~\cite{Lewis:2002mf}). For a critique of gravitational lensing constraints on dark matter halo profiles, see Ref.~\cite{Dalal:2003jk}.

\subsection{Cosmological Scales}
\label{cmb}
We have seen in the previous sections that, on distance 
scales of the size of galaxies and clusters of galaxies, 
evidence of dark matter appears to be compelling. Despite this, the observations discussed do not allow us to determine the {\it total} amount of dark matter in the Universe. We discuss in this section how such information can be
extracted from the analysis of the Cosmic Microwave 
Background (CMB).\index{CMB}

Excellent introductions to CMB theory exist in the literature
\cite{cmbreview1,cmbreview2}. 
Here, we limit ourselves to a brief review of the implications
of recent CMB data on the determination of cosmological
parameters. In particular, we discuss the stringent constraints on the abundances
of baryons and matter in the Universe placed by the 
Wilkinson Microwave Anisotropy Probe (WMAP) data.

The existence of background radiation originating
from the propagation of photons in the early Universe (once they decoupled from matter) was predicted by
George Gamow and his collaborators in 1948 and inadvertently
discovered by Arno Penzias and Robert Wilson in 1965.
After many decades of experimental effort,
the CMB is known to be isotropic at the  $10^{-5}$ level and to
follow with extraordinary precision the spectrum of
a black body corresponding to a temperature $T=2.726$ K.

Today, the analysis of CMB anisotropies enables accurate
testing of cosmological models and puts stringent
constraints on cosmological parameters.

\begin{figure}
\centering
\includegraphics[width=\textwidth]{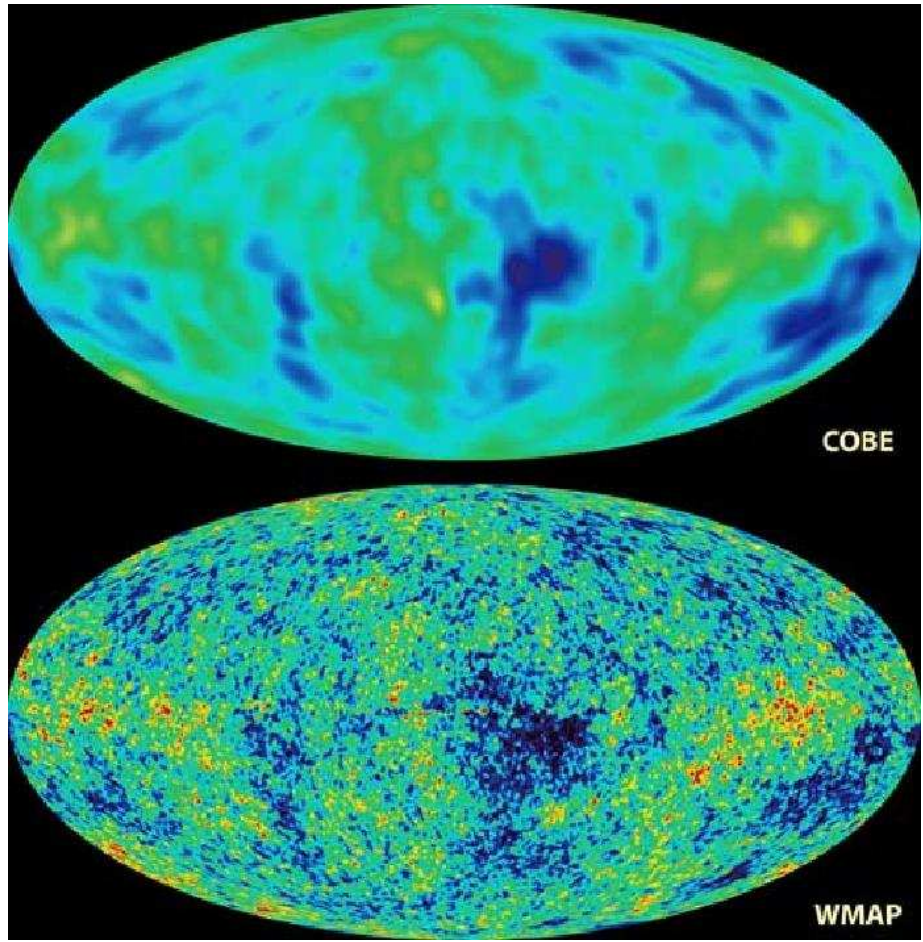}
\caption[Comparison between WMAP and COBE]
{CMB Temperature fluctuations: A comparison between COBE 
and WMAP. Image from http://map.gsfc.nasa.gov/.}
\label{fig:cmb} 
\end{figure}
\begin{figure}
\centering
\includegraphics[width=0.6\textwidth,angle=0]{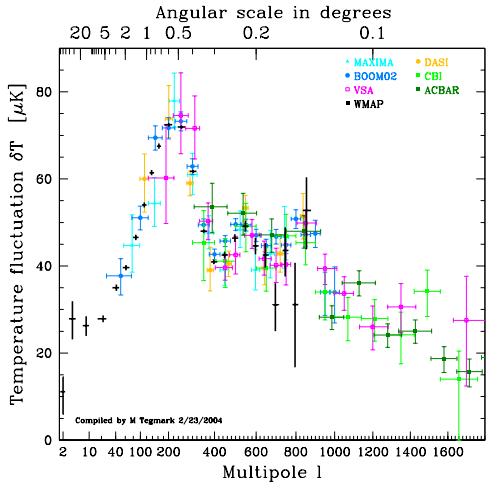}
\caption[WMAP power spectrum of CMB anisotropies]
{The observed power spectrum of CMB anisotropies. From Ref.~\cite{tegmarkhome}.}
\label{fig:cmbps} 
\end{figure}

The observed temperature anisotropies in the sky are 
usually expanded as
\be
\frac{\delta T}{T}(\theta, \phi) = \sum_{\ell = 2}^{+\infty}\sum_{m=-\ell}^{+\ell} a_{\ell m}Y_{\ell m}(\theta, \phi) 
\ee
where $Y_{\ell m} (\theta, \phi)$ are spherical harmonics.
The variance $C_\ell $ of $a_{\ell m}$ is given by
\be 
C_\ell \equiv < |a_{\ell m}|^2 > \equiv \frac{1}{2 \ell +1}
\sum_{m=-\ell}^{\ell}|a_{\ell m}|^2.
\ee
If the temperature fluctuations are assumed to be Gaussian,
as appears to be the case,
all of the information contained in CMB maps can be 
compressed into the power spectrum, essentially
giving  the behavior of $C_\ell $ as a function of 
$\ell$. Usually plotted is $\ell (\ell +1) C_\ell /2\pi$ (see Fig.~\ref{fig:cmbps}).

The methodology, for extracting information from CMB
anisotropy maps, is simple, at least in principle.
Starting from a cosmological model with a fixed number 
of parameters (usually 6 or 7), the best-fit parameters
are determined from the peak of the N-dimensional 
likelihood surface.

From the analysis of the WMAP data alone, the following
values are found for the abundance of baryons and matter 
in the Universe
\be
\Omega_b h^2= 0.024\pm 0.001 \,\,\,\,\,\,\,\,\, \Omega_M h^2=0.14 \pm 0.02.
\ee
Taking into account data from CMB experiments studying 
smaller scales (with respect to WMAP), such as  ACBAR~\cite{Kuo:2002ua} and 
CBI~\cite{Pearson:2002tr}, 
and astronomical measurements of the power spectrum 
from large scale structure (2dFGRS, see Ref.~\cite{Percival:2001hw}) 
and the Lyman $\alpha$ forest (see \eg Ref.\cite{Croft:2000hs}), the 
constraints become ~\cite{Spergel:2003cb}
\be
\Omega_b h^2= 0.0224\pm 0.0009 \,\,\,\,\,\,\,\,\,\rm{and}
\,\,\,\,\,\,\,\,\,  \Omega_M h^2=0.135^{+0.008}_{-0.009}.
\ee
The value of $\Omega_b h^2$ thus obtained is consistent
with predictions from Big Bang nucleosynthesis (\eg \cite{Olive:2003iq})
\be
0.018 < \Omega_b h^2 < 0.023.
\ee

Besides those provided by CMB studies, the most reliable cosmological measurements
are probably those obtained by Sloan Digital Sky Survey (SDSS) team,
which has recently 
measured the three-dimensional power spectrum, $P(k)$, using over 
200,000 galaxies. An estimate of the cosmological parameters  
combining the SDSS and WMAP measurements can be found 
in Ref.~\cite{Tegmark:2003ud}.

\subsection{N-Body Simulations}
\label{nbody}
\index{N-body simulations}
Our understanding of large scale structure is still far 
from a satisfactory level. The description
of the evolution of structures from seed inhomogeneities,
\ie primordial density fluctuations, is complicated by the 
action of many physical processes like gas dynamics, radiative cooling, 
photoionization, recombination and radiative transfer. 
Furthermore, any theoretical
prediction has to be compared with the observed luminous
Universe, \ie with regions where dissipative effects are 
of crucial importance. 

The most widely adopted approach to the problem of large-scale
structure formation involves the use of N-body simulations.
The first simulation of interacting galaxies was performed 
by means of an analog optical computer (Holmberg 1941 \cite{Holmberg})
using the flux from 37 light-bulbs, with photo-cells and
galvanometers to measure and display the inverse square 
law of gravitational force.
Modern, high resolution simulations make full use of the tremendous
increase in computational power over the last few decades.  

The evolution of structure is often approximated with 
non--linear gravitational clustering from specified initial 
conditions of dark matter particles and can be refined
by introducing the effects of gas dynamics, chemistry, 
radiative transfer and other astrophysical processes. 
The reliability of an N-body simulation is measured by 
its mass and length resolution. The mass resolution is 
specified by the mass of the smallest (``elementary'') 
particle considered, being the scale below which fluctuations
become negligible. Length resolution is limited by the so-called
softening scale, introduced to avoid infinities in the
gravitational force when elementary particles collide.

Recent N-body simulations suggest the existence of 
a {\it universal} dark matter profile, with the same shape
for all masses, epochs and input power spectra \cite{Navarro:1995iw}. 
The usual parametrisation for a dark matter halo density is
\begin{equation}
  \rho(r)= \frac{\rho_0}{(r/R)^{\gamma}
  [1+(r/R)^{\alpha}]^{(\beta-\gamma)/\alpha}} \;\;.
\label{profile} 
\end{equation}
Various groups have ended up with different results for 
the spectral shape in the innermost regions of galaxies
and galaxy clusters. In particular, several groups have failed to reproduce the initial results of
Navarro, Frenk and White \cite{Navarro:1995iw}, which
find a value for the power--law index in the innermost
part of galactic halos of $\gamma=1$. 
In table 2, we give the values of the parameters 
$(\alpha, \beta, \gamma)$ for some of the most widely used 
profile models, namely the Kravtsov et al. (Kra, \cite{kra}), 
Navarro, Frenk and White (NFW, \cite{Navarro:1995iw}), Moore et al. 
(Moore, \cite{Moore:1999gc}) and modified isothermal (Iso, e.g. Ref.~\cite{Bergstrom:1997fj}) profiles. \index{Dark matter density profiles}
\begin{center}
\begin{table}
\centering
\vskip 0.3cm
\begin{tabular}{|c|cccc|}
\hline 
&$\alpha$&$\beta$&$\gamma$&R (kpc)\\
\hline 
Kra& 2.0& 3.0&0.4 & 10.0\\
NFW& 1.0& 3.0& 1.0& 20.0\\
Moore& 1.5& 3.0& 1.5& 28.0\\
Iso& 2.0& 2.0& 0& 3.5\\ 
\hline 
\end{tabular}
\label{tab1}
\caption{Parameters of some widely used profile models for 
the dark matter density in galaxies (See Eq.~\ref{profile}). Values of $R$ can vary from system to system.}
\end{table}
\end{center}

Although it is definitely clear that the slope of the density
profile should increase as one moves from the center of a galaxy to 
the outer regions, the precise value of the power-law index in the
innermost galactic regions is still under debate. Attention should
be paid when comparing the results of different groups, as they
are often based on a single simulation, sometimes at very 
different length scales. 

Taylor and Navarro~\cite{taynava, Navarro:2001ij}
studied the behaviour of the phase-space density (defined
as the ratio of spatial density to velocity dispersion cubed,
$\rho/\sigma^3$) as a function of the radius, finding excellent agreement with a power-law extending over several 
decades in radius, and also with the the self-similar 
solution derived by Bertschinger~\cite{Bertschinger:pd}
for secondary infall onto 
a spherical perturbation. The final result of their analysis is
a ``critical'' profile, following a NFW profile in the outer
regions, but with a central slope converging to the value 
$\gamma_{TN}=0.75$, instead of $\gamma_{NFW}=1$.

The most recent numerical simulations (see Navarro \etal~\cite{Navarro:2003ew}, Reed \etal~\cite{Reed:2003hp} and Fukushige \etal~\cite{Fukushige:2003xc})
appear to agree on a
new paradigm, suggesting that density profiles do not converge
to any specific power-law at small radii. The logarithmic slope
of the profile continuously flattens when moving toward the 
galactic center. The slope at the innermost resolved radius
varies between 1 and 1.5, \ie between the predictions of the
NFW and Moore profiles. It is important to keep in mind that 
predictions made adopting such profiles probably overestimate 
the density near the Galactic center and should be used cautiously.  

Recently, Prada \etal~\cite{prada} have suggested that the effects of adiabatic
compression on the dark matter profile near the Galactic center could play an important role, possibly enhancing the dark matter density by an order of magnitude in the inner parsecs of the Milky Way.

The extrapolations of cuspy profiles at small radii have appeared 
in the past (and still appear to some) to be in disagreement with 
the flat cores observed in astrophysical systems, such as low surface brightness galaxies mentioned earlier. Such discrepancies prompted
proposals to modify the properties of dark matter particles, to make
them self-interacting, warm {\it etc}. Most of such 
proposals appear to create more problems than they solve
and will not be discussed here. 

Today, the situation appear less problematic, in particular after 
the analysis of Hayashi \etal~\cite{Hayashi:2003sj}. 
Our approach, given the uncertainties regarding observed and simulated halo
profiles, will be to consider the central slope of the galactic
density profile as a free parameter and discuss the 
prospects of indirect detection of dark matter for the different models 
proposed in literature.

\subsection{The Case of the Milky Way}
\label{milky}

Since the Milky Way is prototypical of the galaxies that contribute most to the cosmic luminosity density, it is 
natural to ask how the results discussed in the previous 
section compare with the  wide range of observational 
data available for our galaxy.

One way to probe the nature of matter in our neighborhood 
is to study microlensing events in the direction of the
galactic center. In fact, such events can only be due to 
compact objects, acting as lenses of background sources,
and it is commonly believed that dark matter is simply too weakly 
interacting to clump on small scales\footnote{It was noticed
by Berezinsky \etal~\cite{Berezinsky:1996eg} that if 
microlensing was due to neutralino stars (see the definition
of ``neutralino'' in the chapter on dark matter candidates), 
\ie self-gravitating systems of dark matter particles, then the gamma-ray
radiation originated by annihilations in these object would 
exceed the observed emission.}.

Binney and Evans (BE)~\cite{Binney:2001wu} recently 
showed that the number of observed microlensing events 
implies an amount of baryonic matter within the Solar circle
greater than about $3.9 \times 10^{10}$M$_\odot$. 
Coupling this result with estimates of the local dark matter 
density, they exclude cuspy profiles with power--law
index $\gamma \gtrsim 0.3$. 

Nevertheless, Klypin, Zhao and Somerville
(KZS)~\cite{Klypin:2001xu} find a good agreement between
NFW profiles ($\gamma=1$) and observational
data for our galaxy and M31. The main difference between
these analyses is the value of the microlensing optical depth towards the
Galactic center used. Observations of this quantity disagree by a factor of $\sim3$ and a low value within this range permits the presence of a dark matter cusp. Another difference arises from the modeling of the galaxy: KZS claim
to have taken into account dynamical effects neglected by BE 
and to have a ``more realistic'' description of the galactic 
bar.  

An important addition is adiabatic compression of the dark matter by
baryonic dissipation.
 This results in a dark matter density that is enhanced in the core by
 an order of magnitude. This result can be reconciled with modelling
 of the rotation curve if the lower value of the microlensing optical
 depth found by the EROS collaboration is used rather than that of the
 MACHO collaboration. In the latter case, little dark matter is
 allowed in the central few kpc. The microlensing result constrains
 the stellar contribution to the inner rotation curve, and hence
 to the total allowed density.
   
\begin{figure}
\centering
\includegraphics[angle=-90,width=0.6\textwidth,clip=true]{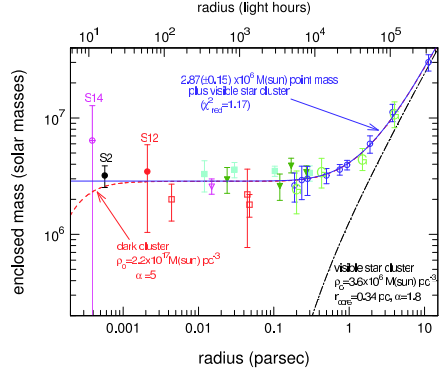}
\caption[Enclosed mass vs galactocentric coordinate]{
The mass distribution in the galactic center, as derived by
different observations, down to a $10^{-4}$pc scale. 
Lines represent fits under different assumptions, as
specified by the text in the figure. In particular,
the solid line is the overall best fit model: a   
$2.87\pm0.15 \times 10^{6} M_{\odot}$ central object,
plus a stellar cluster distributed with a power-law
of index 1.8. For more details see Ref.~\cite{Schoedel:2003gy}.
}
\label{enclosed}
\end{figure}
\subsubsection{The Galactic center}
\label{smbh}
\index{SMBH at the galactic center} \index{spike in DM density profile} 
The dark matter profile in the inner region of the Milky 
Way is even more uncertain. Observations of the velocity 
dispersion of high proper motion stars suggest the existence 
of a Super Massive Black Hole (SMBH) lying at the center 
of our galaxy, with a mass, $M_{\mbox{SMBH}} \approx 2.6 
\times 10^6 M_\odot$ ~\cite{Ghez:1998ab}\footnote{The 
existence of a SMBH at the center of the galaxy
is not surprising. There is, in fact, mounting evidence 
for the existence of $10^6$--$10^8\,M_\odot$ black 
holes in the centers of most galaxies with mass amounting to approximately 0.1\% of the stellar spheroid (see, \eg 
Ref.~\cite{Kormendy:2000cf}).}. 

Recently, near-infrared high-resolution imaging and 
spectroscopic observations 
of individual stars, as close as a few light days from
the galactic center,  were carried out at Keck~\cite{Ghez:2003qj} and ESO/VLT telescopes (see Ref.~\cite{Schoedel:2003gy}, for an excellent and updated
discussion of the stellar dynamics in the galactic center, based on 
the most recent observations at ESO/VLT).
The analysis of the orbital parameters of such
stars suggest that the mass of the SMBH could possibly 
be a factor of two larger with respect to the
above cited estimate from the velocity dispersion. 
In Fig.~\ref{enclosed} we show a plot of the enclosed
mass as a function of the galactocentric distance,
along with a best-fit curve, which corresponds to a dark object
with a mass of $2.87\pm0.15 \times 10^{6} M_{\odot}$.

It has long been argued (see \eg Peebles, Ref.~\cite{peebles72})
that if a SMBH exists at the galactic center, the process 
of adiabatic accretion of dark matter on it would 
produce a ``spike'' in the dark matter density profile. 
Gondolo and Silk~\cite{Gondolo:1999ef} have recently 
applied such a process to study the enhancement of the
the annihilation signal from the galactic center.

If we consider an initial power-law type profile of 
index $\gamma$, similar to  those discussed in Sec.~\ref{nbody},
the corresponding dark matter profile, $\rho'(r)$, after this accretion process is, 
following Ref.~\cite{Gondolo:1999ef},
\begin{equation}
\rho' = \left[ \alpha_{\gamma} \left( \frac{M}{\rho_D
D^3}\right)^{3-\gamma} \right]^{\gamma_{sp}-\gamma} \rho_D \; g(r) \left(
\frac{D}{r} \right)^{\gamma_{sp}},
\label{modaccre}
\end{equation}
where $\gamma_{sp}=(9-2\gamma)/(4-\gamma)$, $D \simeq 8\,\rm{kpc}$ is the solar distance from
the Galactic center and $\rho_D \simeq 0.3 \rm{GeV}/c^2/\rm{cm}^3$ is the density in the
solar neighborhood. The factors $\alpha_\gamma$ and $g_\gamma(r)$ cannot
be determined analytically (for approximate expressions and numerical
values see Ref.~\cite{Gondolo:1999ef}). Eq.~\ref{modaccre} is only valid in a
central region of size $R_{sp}=\alpha_\gamma D (M/\rho_D
D^3)^{1/(3-\gamma)}$, where the central black hole dominates the
gravitational potential.

It is easy to understand the basics of adiabatic accretion
under the assumptions of circular orbits. Assuming
an initial power--law distribution, $\rho \propto r^{-\gamma}$,
and a final distribution, $\rho \propto r^{-\gamma_{sp}}$, the 
equations of conservation of mass and angular momentum
can be expressed, respectively, as
\begin{equation}
 \rho_i r_i^2 \mbox{d}r_i = \rho_f r_f^2 \mbox{d}r_f 
\end{equation}
and
\begin{equation}
r_i M_i(r) = r_f M_f(r) \approx r_f M_{BH},
\end{equation}
which imply, respectively,
\begin{equation}
r_i \propto r_f^{(3-\gamma_{sp})/(3-\gamma)}
\end{equation}
and
\begin{equation}
r_i \propto r_f^{1/(4-\gamma)} \;.
\end{equation}
The final distribution will thus have a power--law index
\be
\gamma_{sp}=\frac{9-2\gamma}{4-\gamma},
\ee
which assumes values in the range of 2.25 to 2.5 as $\gamma$ varies in the interval of 0 to 2.

If we take into account the annihilation of dark matter particles, the
density cannot grow to arbitrarily high values, the maximal density being
fixed by the value 
\begin{equation}
\rho_{core}=\frac{m}{\sigma v\, t_{BH}},
\label{roco}
\end{equation}
where $t_{BH}\approx 10^{10}$~yr is the age of the central black hole.
The final profile, resulting from the adiabatic accretion of annihilating
dark matter on a massive black hole is
\begin{equation}
\rho_{dm}(r)= \frac{\rho'(r) \rho_{core}}{\rho'(r)+\rho_{core}}
\end{equation}
following a power-law for large values of $r$, and with a flat core of
density, $\rho_{core}$, and  dimension, 
\begin{equation}
R_{core}=R_{sp} \left( \frac{\rho(R_{sp})}{ \rho_{core}} \right)
^{(1/\gamma_{sp})}.
\end{equation}

We will use these equations when discussing the prospects
for indirect detection of dark matter in the presence of a spike.
We recall, nevertheless, that they have been derived under 
the simplifying assumption that the SMBH formed at a position 
coinciding exactly with the center of the galactic potential 
well, and neglecting all dynamical effects. 
\begin{figure}
\centering
\includegraphics[width=0.7\textwidth,clip=true]{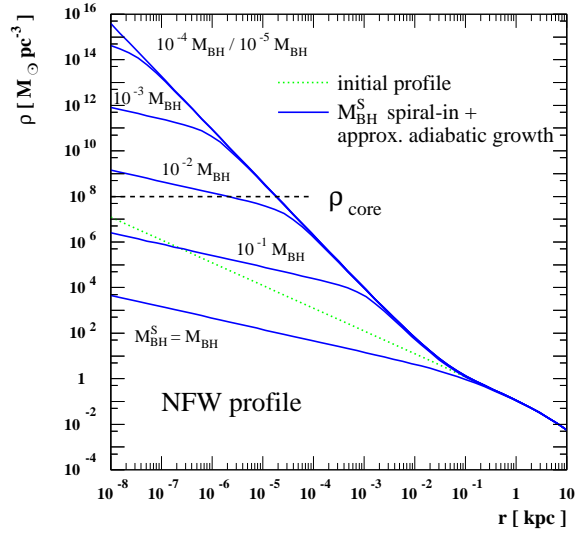}
\caption[Modification of a NFW profile due to a seed black hole]
{Modification of an NFW profile due to the off-center formation 
of a black-hole seed of mass $M_{BH}^S$, its spiral-in the center 
of the dark matter system and its adiabatic growth to the present-day 
mass, $M_{\mbox{SMBH}}$ (called $M_{\mbox{SM}}$ in the figure). From
Ref.~\cite{Ullio:2001fb}.}
\label{fig:ullio}
\end{figure}

It was shown by Ullio, Zhao and Kamionkowski~\cite{Ullio:2001fb} that if the black hole forms 
from a low--mass seed, then its spiral-in to reach
the exact center of the galaxy could take a length of time
longer than the age of the galaxy. If, conversely, 
the seed black hole is massive, the back-reaction
to the spiral-in of the black hole leads to the 
formation of a weak-density cusp, with 
$\rho \propto r^{-0.5}$. Fig.~\ref{fig:ullio} 
shows the modification of an NFW dark matter profile due to 
the off-center formation of the seed black hole.
The solution found by Gondolo and Silk would
be indistinguishable from the case of an initial light seed
of roughly $10^{-2}\,M_{\mbox{SMBH}}$, starting its
growth very near to the galactic center.

The spike could also be destroyed by hierarchical 
mergers, as discussed by Merritt \etal~\cite{Merritt:2002vj},
but such mergers are unlikely to have occurred 
in the recent history of the Milky Way. 
What can be stated with considerable 
confidence is that the Milky Way galaxy underwent one 
significant merger about 12 billion years ago. This resulted in the 
formation of the bulge, and therefore presumably of the 
SMBH, and of the thick disk. 
The chemical evidence for a unique merger origin in the 
case of our Milky Way's thick disk is compelling \cite{wyse1,wyse2},
as the continuity between thin disk, thick disk, and bulge 
would have been destroyed had anything significant 
happened more recently in the way of a merger (see also the
discussion of Bertone, Sigl and Silk~\cite{Bertone:2002je}). 

Furthermore, the scattering of dark matter particles 
by stars in the dense stellar cusp observed around 
the SMBH could substantially lower the dark matter 
density near the Galactic center over $10^{10}$ years, due both to 
kinetic heating, and to capture of dark matter 
particles by the SMBH~\cite{Merritt:2003qk}.

The existence of such spikes would produce a dramatic 
enhancement of the annihilation radiation from the galactic center.
The implications for indirect detection of dark matter particles
have been discussed in Refs.~\cite{Gondolo:1999ef, Gondolo:2000pn, 
Bertone:2001jv, Bertone:2002je}.

\subsubsection{The local density}

Very important to the prospects for direct and indirect detection is the density of dark matter in the region of our solar system.  Although this quantity is considerably more well known than the density near the galactic center, there are still uncertainties associated with the local density, which we will discuss here.

The local density of dark matter is determined by observing the rotation curves of the Milky Way.  This is somewhat difficult to do from our location within the galaxy.  Furthermore, rotation curves measure the total mass within an orbit, thus the density distributions of the galactic bulge and disk are needed to accurately calculate the dark matter profile. 

In addition to the local density, the velocity distribution of dark matter in the local region is needed to accurately calculate direct and indirect detection rates. This is also best inferred from observed rotation curves.

\begin{figure}
\centering
\includegraphics[width=0.6\textwidth,clip=true]{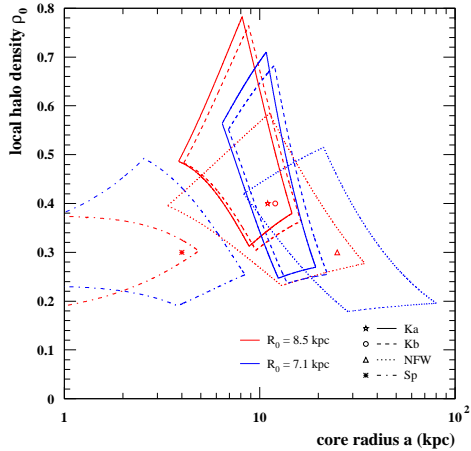}
\caption[Modification of a NFW profile due to a seed black hole]
{The range of local dark matter densities acceptable with observations of rotation curves for a variety of halo profiles and galactocentric distances. Densities in the range of $0.2-0.8\,$GeV/cm$^3$ are shown to be acceptable.  From Ref.~\cite{Bergstrom:1997fj}.}
\label{fig:local}
\end{figure}

Different groups have come to somewhat different conclusions regarding the local density and velocity distribution of dark matter. For example, Bahcall {\it et al.} finds a best-fit value of $\rho_0=0.34\,$GeV/cm$^3$ \cite{Bahcall:1981nv}, Caldwell and Ostriker find $\rho_0=0.23\,$GeV/cm$^3$ \cite{caldwellostriker} while Turner calculates $\rho_0=0.3-0.6\,$GeV/cm$^3$ \cite{turner1986}. In figure~\ref{fig:local} we show the range of local dark matter densities found to be acceptable by Bergstrom, Ullio and Buckley~\cite{Bergstrom:1997fj} for various choices of halo profile and galactocentric distance. They find local dark matter densities acceptable in the range of about $0.2-0.8\,$GeV/cm$^3$.

The velocity distribution of dark matter is typically described only by its average velocity, $\bar{v}=<v^2>^{1/2} \cong 270\,$km/s.

For more discussion on the local dark matter distribution, see section 2.4 of Ref.~\cite{Jungman:1995df}.

\section{Candidates}
\label{candidates}

As we have seen in the previous section, the evidence
for non-baryonic dark matter is compelling at all observed astrophysical 
scales. It is therefore natural to ask {\it what is the dark matter
  made of?} In this section,
 we present some of the candidates 
discussed  in  the
literature, and focus our attention especially on
two popular candidates:  the supersymmetric {\it neutralino},
probably the most widely studied candidate,
and the $\bone$ particle, the first Kaluza--Klein excitation of the $B$ boson 
in theories with universal extra dimensions. We will 
also briefly discuss ``superheavy'' candidates, also 
referred to as {\it wimpzillas}.

\subsection{The Non-Baryonic Candidate Zoo}
\label{sec:zoo} \index{non-baryonic Dark Matter}
There is no shortage of candidates for non-baryonic dark matter.
In this section we briefly describe some of these candidates.

\begin{itemize}
\item {\it Standard Model neutrinos} 

Neutrinos have been considered, until recently,
excellent dark matter candidates for their ``undisputed virtue of being 
known to exist''\cite{Bergstrom:2000pn}. However,
a simple calculation shows that, if we call $m_i$ the mass of
the i-th neutrino, their total relic density is predicted to be
\be
\Omega_{\nu} h^2 = \sum_{i=1}^{3} \frac{m_i}{93\mbox{~eV}} \; .
\label{omenu}
\ee
The best laboratory constraint on neutrino masses comes from 
tritium $\beta$-decay experiments at Troitsk and Mainz
\cite{Weinheimer:2003bh}, 
pointing to the following upper limit on the neutrino mass
\be
m_\nu  < 2.05 \mbox{~eV} \;\; \mbox{(95\% C.L.)} \;,
\ee
while next-generation experiments are expected to reach a 
sensitivity of approximately $0.2$~eV (see Ref.~\cite{Weinheimer:2003bh} 
and references therein). The above upper limit applies to all three
mass eigenvalues~\cite{Beacom:2002cb}, since the mass differences
among them must be very small to explain solar ($\Delta m^2 \approx 7 
\cdot 10^{-5}$~eV$^2$)
and atmospheric ($\Delta m^2 \approx 3 \cdot 10^{-3}$~eV$^2$) 
neutrino anomalies (see \eg Ref.~\cite{Gonzalez-Garcia:2002dz}).
This implies an upper bound on the total neutrino relic density of
\be
\Omega_{\nu} h^2 \lesssim 0.07 \;, 
\ee
which means that neutrinos are simply not abundant enough to 
be the dominant component of dark matter. A more stringent constraint
on the neutrino relic density comes from the analysis of CMB
anisotropies, combined with large-scale structure data, 
suggesting $\Omega_{\nu} h^2 < 0.0067$ (95\% confidence limit).
For three degenerate neutrino species this implies 
$m_\nu < 0.23$~eV. If extra neutrino interactions are allowed, 
e.g., the coupling of neutrinos to a light boson, 
the neutrino mass limits arising from large scale structure
can be evaded ~\cite{Beacom:2004yd}.

Being relativistic collisionless particles,
neutrinos erase (moving from high to low density regions)
fluctuations below a scale of $\sim 40$~Mpc ($m_{\nu}$/30 eV),
called the {\it free-streaming} length~\cite{bond}.
This would imply a {\it top-down} formation history 
of structure in the Universe, where big structures form 
first. The fact that our galaxy appears to be older
than the Local Group~\cite{peebles84}, and the discrepancy between the
predicted late formation of galaxies,
at redshift $z\lesssim 1$, against observations of 
galaxies around $z >4$ ~\cite{bond2}, is a further argument
against neutrinos as a viable dark matter candidate.

\item {\it Sterile neutrinos} 

These hypothetical particles are similar to Standard Model neutrinos, 
but without Standard Model weak interactions, apart from mixing.
They were proposed as dark matter 
candidates in 1993 by Dodelson and Widrow~\cite{Dodelson:1993je}.
Stringent cosmological and astrophysical constraints on sterile 
neutrinos come from the analysis of their cosmological abundance 
and the study of their decay products (see Ref.~\cite{Abazajian:2001nj}
and references therein). 

Light neutrinos, with masses below a 
few keV, would be ruled out as dark matter candidates. In fact,
if the WMAP result for the reionization optical depth is correct,
then dark matter structures were in place to form massive 
stars prior to redshift $z > 20$, which is is simply not possible 
if the dark matter particle mass is smaller than $\sim10$~keV 
~\cite{Yoshida:2003rm}.
An alternative explanation for the WMAP optical depth is 
reionization by decaying particles, such as sterile neutrinos
(see Ref.~\cite{Hansen:2003yj} and references therein).
Sterile neutrinos could also be cold dark matter, if there is 
a very small lepton asymmetry, in which case they are 
produced resonantly with a non-thermal spectrum ~\cite{Shi:1998km}.


\item {\it Axions} 

Introduced in an attempt
to solve the problem of CP violation in particle physics, axions have also 
often been discussed as a dark matter candidate.

Laboratory searches, stellar cooling and the dynamics of 
supernova 1987A constrain axions to be very light 
 ($\lesssim 0.01$~eV). Furthermore, they are expected to be extremely 
weakly interacting with ordinary particles, which implies 
that they were not in thermal equilibrium in the early universe.

The calculation of the axion relic density is uncertain,
and depends on the assumptions made regarding the production 
mechanism. Nevertheless, it is possible to find an acceptable
range where axions satisfy all present-day constraints
and represent a possible dark matter candidate (see \eg 
Ref.~\cite{Rosenberg:wb}).

\item {\it Supersymmetric candidates}
\begin{itemize}
\item {\it Neutralinos} 

Neutralinos in models of R-parity conserving supersymmetry are by far the most
widely studied dark matter candidates. We devote Sec.~\ref{sec:susy} to their
presentation.

\item {\it Sneutrinos} 

The superpartners of the Standard Model neutrinos in supersymmetric
models have long been considered as dark matter candidates. It has
been shown that sneutrinos will have a cosmologically interesting
relic density if their mass is in the range of 550 to 2300 GeV.
However,
the
scattering cross section of a sneutrino with nucleons is easily
calculated and is much larger than the limits found by direct dark
matter detection experiments \cite{falksneutrino}.

\item {\it Gravitinos} 

Gravitinos are the superpartners of the graviton in supersymmetric
models. In some supersymmetric scenarios, gauge mediated
supersymmetry for example, gravitinos can be the lightest
supersymmetric particle and be stable. Gravitinos are thus very strongly theoretically motivated. With only gravitational
interactions, however, gravitinos are very difficult to observe
\cite{Feng:2003xh}. 

It has been known for some time that long lived gravitinos can pose
problems for cosmology
\cite{gprob1,gprob2,gprob3,gprob4,gprob5,gprob6}. In particular, their
presence can destroy the abundances of primordial light elements in
some scenarios
\cite{reheating,lightelements1,lightelements2,lightelements3}. Gravitinos
may also be overproduced in the early universe if the temperature of
the reheating epoch is not sufficiently low \cite{reheating}. In some
scenarios, however, these problems can be circumvented
\cite{noreheating1,noreheating2,lightelementsrviolation}.

\item {\it Axinos} 

Axinos, the superpartner of the axion, were believed until recently to only be capable of acting as a warm, or hot, dark matter candidate \cite{axinowarm1,axinowarm2}. It has been shown, however, that for quite low reheating temperatures, cold axino dark matter may be possible \cite{axinocold1,axinocold2,Covi:2004rb,lythaxino}. In many ways, axinos and gravitinos share similar phenomenological properties.

\end{itemize}

\item {\it Light scalar dark matter}

Considering fermionic dark matter candidates with standard Fermi
interactions, Lee and Weinberg concluded that relic density arguments
preclude  such a WIMP with a mass less than a few GeV
\cite{leeweinberg} (see also Hut 1977~\cite{Hut:1977zn}). 
If the dark matter is made up other types of
particles, however, this limit could be evaded. For example, a 1-100
MeV scalar candidate has been proposed
\cite{mevproposed1,mevproposed2}.

Such a candidate, although somewhat ad hoc from a particle physics
perspective, has recently become experimentally motivated. In
Ref.~\cite{mevdetect}, it has been suggested that the 511 keV
gamma-ray line emission observed by the INTEGRAL satellite from the galactic bulge could be the
product of light dark matter particles annihilating into positrons
which then annihilate producing the observed gamma-ray line. To
confirm this hypothesis, more tests are needed. In particular, a
similar signature could be expected from dwarf spheroidal galaxies
\cite{mevdwarf}.

Very recently, light decaying dark matter particles such as axinos with R-parity violation \cite{Hooper:2004qf} or sterile neutrinos \cite{Picciotto:2004rp} have been suggested as the source of the observed 511 keV emission.

\item{\it Dark matter from Little Higgs models}

As an alternative mechanism (to supersymmetry) to stabilize the weak
scale, the so-called ``little Higgs'' models have been proposed and
developed
\cite{littlehiggspropose1,littlehiggspropose2,littlehiggspropose3,littlehiggspropose4}. In
these models, the Standard Model Higgs is a pseudo-Goldstone boson
with its mass protected by approximate non-linear global
symmetries. The divergences to the Higgs mass which remain are present
only at the two-loop level and, therefore, the weak scale can be
stabilized in an effective field theory which is valid up to $\sim 10\,$ TeV. Recall that in supersymmetry, the divergences to the Higgs mass are exactly cancelled at all orders.

At least two varieties of little Higgs models have been shown to
contain possible dark matter candidates. One of these classes of
models, called ``theory space'' little Higgs models, provide a
possibly stable, scalar particle which can provide the measured
density of dark matter \cite{lhdmrelic}. In Ref.~\cite{lhdmdetection},
the detection prospects for such a candidate were found to be not
dissimilar to WIMPs predicted in models of supersymmetry or universal
extra dimensions.

Cheng and Low \cite{littlehierarchy} have developed another variety of
little Higgs model, motivated by the problem of the hierarchy between
the electroweak scale and the masses of new particles constrained by
electroweak precision measurements. They solve this problem by
introducing a new symmetry at the TeV scale which results in the
existence of a stable WIMP candidate with a $\sim$TeV mass.

For a potential dark matter candidate from a little Higgs model to be stable, we must assume that the discrete symmetry which protects it from decay is fundamental and is not broken by the operators in the UV completion.

\item {\it Kaluza-Klein states}
 
Kaluza-Klein excitations of Standard Model fields which appear in models of universal
extra dimensions have also been discussed a great deal recently as a candidate for dark matter. They are 
discussed in Sec.~\ref{sec:kk}. Additionally, a dark matter candidate has been
proposed in the framework of ``warped'' universal extra-dimensions: an exotic 
particle with gauge quantum numbers of a right-handed neutrino, but carrying 
fractional baryon-number~\cite{Agashe:2004ci}.

\item {\it Superheavy dark matter}

Superheavy dark matter particles, sometimes called {\it Wimpzillas}, have interesting phenomenological consequences, including a possible solution to the problem of cosmic rays observed above the GZK cutoff. These are discussed in Sec.~\ref{sec:wzilla}.

\item {\it Q-balls \cite{Kusenko:1997si,Kusenko:1997vp}, mirror 
particles \cite{Hodges:yb,Foot:2003iv,Ignatiev:2003js,Mohapatra:2001sx,Foot:2004kd}, CHArged Massive Particles (CHAMPs) \cite{DeRujula:1989fe}, self interacting dark matter \cite{Spergel:1999mh,Dave:2000ar}, D-matter~\cite{shiuwang}, cryptons \cite{Ellis:1990iu,Ellis:1990nb}, superweakly interacting dark matter~\cite{Feng:2003xh}, brane world dark matter \cite{Cembranos:2003mr}, heavy fourth generation neutrinos \cite{Roy:1995ty,Kainulainen:2002pu}, etc.} 

Although some of these candidates or classifications present some intriguing 
features we will not discuss them  here. We refer the interested reader to 
the wide literature on the subject, \eg the reviews of non-baryonic 
candidates by Ellis~\cite{Ellis:1998gt} and Bergstrom~\cite{Bergstrom:2000pn}.

\end{itemize}

We stress that it is by no means assured that the dark matter
is made of a single particle species. On the contrary, 
we already know that Standard Model neutrinos contribute
to dark matter, but cannot account for all of it. 
Even in supersymmetry
models for dark matter, $N=2$ supersymmetry allows the possibility
of two
stable dark matter relics (see, for example, Ref.~\cite{Boehm:2003ha}).

In what
follows, we will assume that the abundance of our candidates
satisfy the limits provided by the analysis of the CMB
discussed in Sec.\ref{cmb}, but we stress that, although the 
upper bound is a strict limit, the lower bound can be 
relaxed, assuming that our candidate is a sub-dominant
component of dark matter. The interested reader will find in Ref.~\cite{Duda:2002hf} a detailed discussion on the detection
prospects of a subdominant density component of dark matter.

\subsection{Supersymmetry}
\label{sec:susy}
\label{sec:basics}
\index{Supersymmetry}
It would be impossible to review in only a few pages the 
history and theory of Supersymmetry (SUSY). Instead, we prefer
here to review the motivations that led to its 
introduction and to briefly present the concepts and
the notations that we will  use in the 
following chapters. Furthermore, we present a few of the
supersymmetric models discussed in 
the literature
(we reserve the word ``scenario'' for a specific choice
of parameters in the framework of a given model)
and discuss the consequences of various assumptions, involved in the process of model-building, on SUSY phenomenology.
For further discussions of supersymmetry, we refer the interested reader to Refs.~\cite{binetruy02,Collins:kn, Wess:cp,
Djouadi:1998di, Jungman:1995df,Martin:1997ns, Olive:2003iq,Chung:2003fi}.

\subsubsection{Basics of supersymmetry}
\label{whysusy}

As we saw in Sec.\ref{sec:SM}, in the Standard Model of particle
physics there is a fundamental distinction 
between bosons and fermions:
while bosons are the mediators of interactions, fermions
are the constituents of matter. It is therefore natural
to ask whether a symmetry exists which relates them, 
thus providing a sort of ``unified'' picture of matter
and interactions. 

Another way to state the problem is to ask whether
a Lie group exists mixing internal (Isospin, {\it etc.})
and space-time (Lorentz) symmetries \cite{Haag:1974qh}. Although apparently
uncorrelated to the differing behaviour of bosons and fermions,
this problem led to the study of the same algebraic
structures. Early attempts
to find a broad Lie group including the Poincar\'{e} 
and internal symmetry groups had to face the limitations
imposed by the so--called {\it no-go} theorem of Coleman and 
Mandula. Such limitations were finally circumvented with 
the introduction of {\it graded} Lie algebras, \ie  algebras
involving fermionic generators satisfying anticommutation 
relations (see below). 

For those who are not convinced by these symmetry arguments, there are other major reasons for interest in supersymmetry. One reason is
its role in understanding the {\it hierarchy problem}. The hierarchy problem
is linked to the enormous difference between the electroweak and Planck
energy scales. This problem arises in the radiative corrections to the mass of the Higgs boson. 

All particles get radiative corrections to 
their mass, but while fermion masses increase only logarithmically, scalar masses increase quadratically with energy, giving corrections at 1-loop of
\be
\delta m_s^2 \sim \left( \frac{\alpha}{2 \pi} \right) \Lambda^2,
\label{deltam2}
\ee
where $\Lambda$ is a high-energy cut-off where new physics
is expected to play an important role. The radiative corrections to the Higgs mass (which is expected to be of the
order of the electroweak scale $M_W \sim 100$ GeV) will destroy the stability of the electroweak scale if $\Lambda$ is higher than $\sim \rm{TeV}$, \eg if $\Lambda$ is near the Planck mass.

An appealing, though not the only, solution to this problem is to postulate
the existence of new particles with similar masses but with
spin different by one half. Then, since the contribution
of fermion loops 
to $\delta m_s^2$ have opposite sign to the corresponding bosonic loops,
at the 1-loop level, Eq.~\ref{deltam2} becomes
\be
\delta m_s^2 \sim \left( \frac{\alpha}{2 \pi} \right) 
\left( \Lambda^2 + m_B^2 \right)  -
\left( \frac{\alpha}{2 \pi} \right) 
\left( \Lambda^2 + m_F^2 \right) =
\left( \frac{\alpha}{2 \pi} \right) 
\left(  m_B^2 - m_F^2 \right) .
\ee
\begin{figure}[t]
\begin{center}
$\begin{array}{c@{\hspace{0in}}c}
\includegraphics[width=0.5\textwidth]{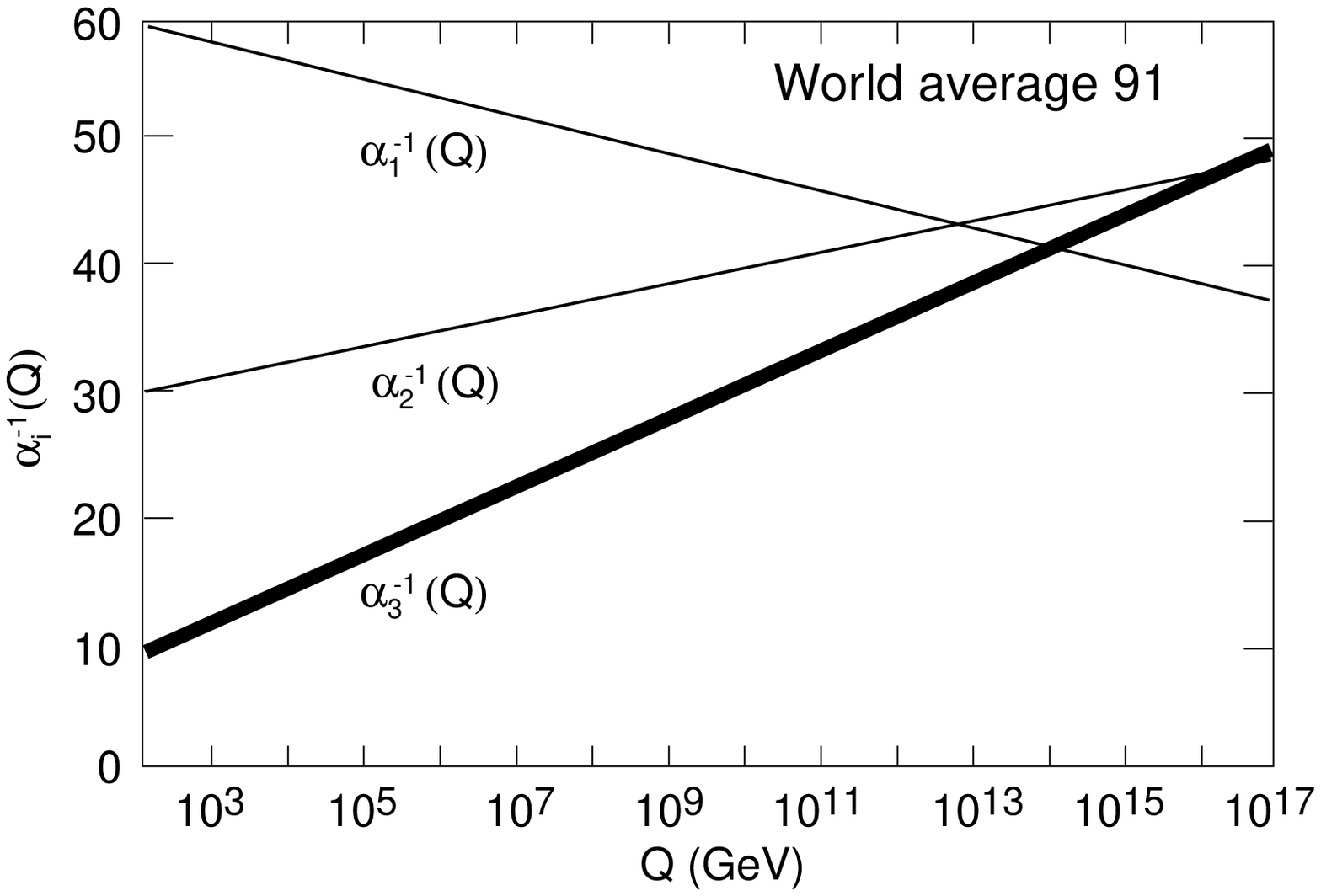} &
\includegraphics[width=0.5\textwidth]{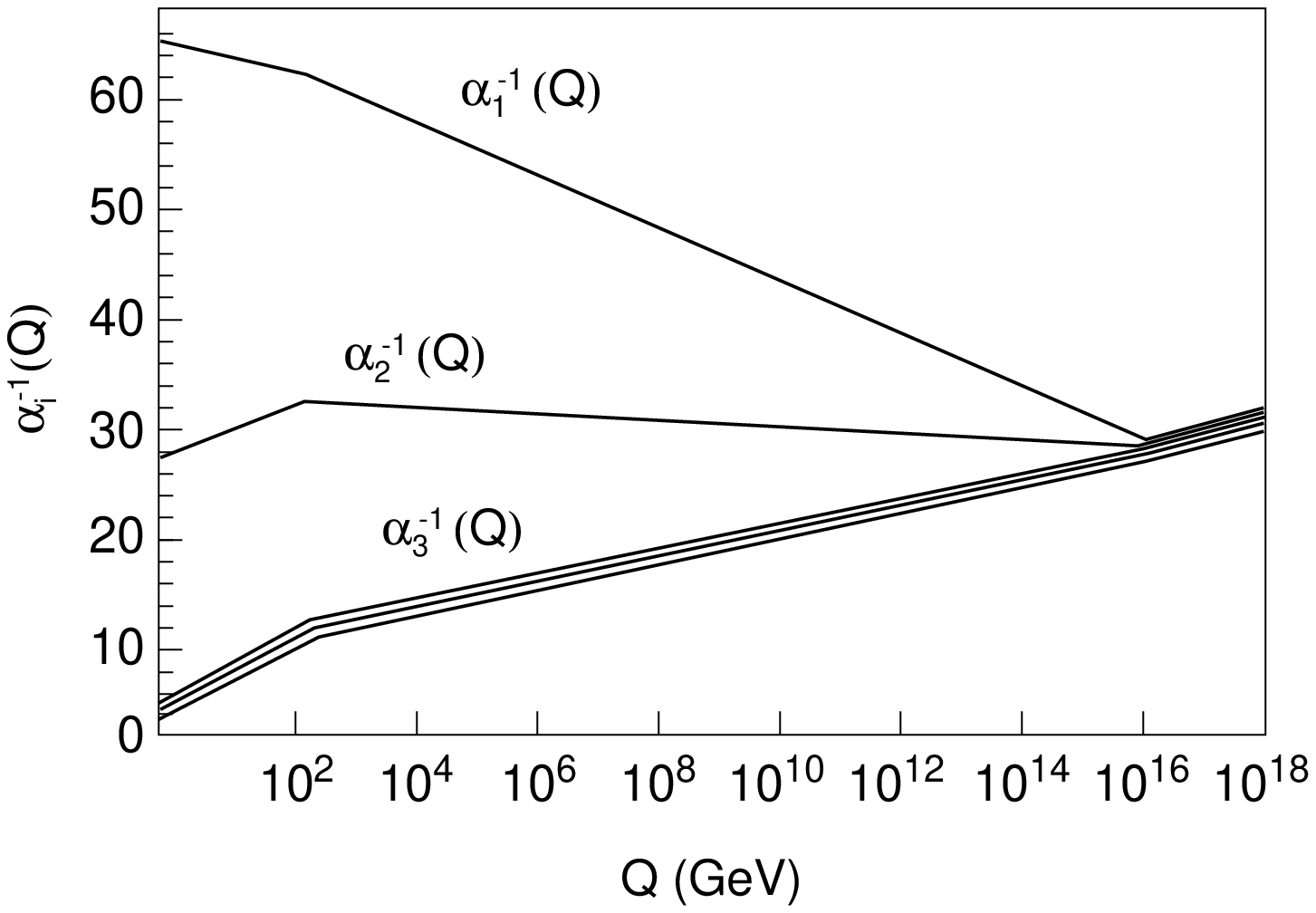} \\ [0.0cm]
\end{array}$
\end{center}
\caption[Running of gauge couplings]
{The measurements of the gauge coupling strengths at LEP do not (left)
evolve to a unified value if there is no supersymmetry but do (right) if
supersymmetry is included~\cite{deboergut,ADF}.} 
\label{runningcoup}
\end{figure}
Furthermore, the supersymmetric algebra insures that (provided $ |m_B^2 - m_F^2| \lesssim 1$ TeV) the quadratic
divergence to the Higgs mass is cancelled at all orders of perturbation theory. The algebra of supersymmetry 
naturally guarantees the existence of new particles,
with the required properties, associating to all of the 
particles of the Standard Model superpartners with the same mass
and opposite spin-type (boson or fermion).

Another reason for interest in supersymmetric theories
comes from the unification of gauge couplings at a scale $M_U \sim
2 \times 10^{16}$ GeV (see Fig.~\ref{runningcoup}). Although extrapolation of the coupling constants using only Standard Model particles fails to unify them to a common value (left frame of Fig.~\ref{runningcoup}), by introducing supersymmetry at the TeV scale, it was shown \cite{deboergut} that these forces naturally unify at a scale $M_U \sim 2 \times 10^{16}$ GeV (right frame of Fig.~\ref{runningcoup}). This has been  taken as a strong hint in favor of a Grand Unified Theory (GUT) which predicts gauge coupling
unification below the Planck scale.
 
\index{Supersymmetry}
The new generators introduced with supersymmetry change fermions into bosons and vise versa, \ie 
\be
Q\ket{\mbox{fermion}}=\ket{\mbox{boson}}; \;\;\;\; 
Q\ket{\mbox{boson}}=\ket{\mbox{fermion}}.
\ee

Because of their fermionic nature, the operators Q
must carry spin 1/2, which implies that supersymmetry
must be a spacetime symmetry. The question then arises 
of how to extend the Poincar\'{e} group of 
spatial translations and Lorentz transformations
to include this new boson/fermion symmetry.
The structure of such a group is highly 
restricted by the Haag-Lopuszanski-Sohnius extension 
of the Coleman and Mandula theorem cited above. 
For realistic theories, the operators, Q, which we choose 
by convention to be Majorana spinors, must satisfy
\ba
&&\{ Q_a, \overline{Q}_b \} = 2 \gamma^\mu_{ab} P_\mu  \label{susy1}
\\
&&\{ Q_a, P_\mu \}  = 0 \label{susy2}
\\
&&[ Q_a , M^{\mu \nu}  ] = \sigma^{\mu \nu}_{ab} Q^b \label{susy3}
\ea
where 
\be
\overline{Q}_a \equiv \left( Q^\dagger \gamma_0 \right)_a
\ee
and
\be
\sigma^{\mu \nu}= \frac{i}{4} [\gamma^\mu, \gamma^\nu]
\ee
are the structure constants of the theory. 

Just as Lorentz invariance is manifest in Minkowski
space-time, supersymmetry is manifest in the so-called 
{\it superspace} formalism, where a superspace is 
defined as a set of coordinates $\{x,\theta,\overline{\theta}\}$,
where $x=x^\mu$ are the usual coordinate of Minkowski 
spacetime, and $ \theta,\overline{\theta}$ are 
anti-commuting Weyl spinors. 

A {\it superfield} is then a function, 
$\Phi(x,\theta,\overline{\theta})$, defined on a superspace;
it is common to introduce {\it chiral fields} representing
matter and {\it vector fields} representing gauge fields.

\subsubsection{Minimal supersymmetric Standard Model}
\label{sec:minimal}

To continue our brief introduction to \susy, we consider
the {\it minimal} supersymmetric extension of the Standard Model \index{MSSM}
(MSSM, for Minimal Supersymmetric Standard Model). The MSSM is minimal in the 
sense that it contains the smallest possible field content
necessary to give rise to all the  fields of the Standard Model.
This can be done as follows:
\begin{itemize}
\item
We associate fermionic superpartners to all gauge fields.
Gluons, $W^\pm$ and $B$ bosons then get fermionic partners called
{\it gluinos} ($\tilde{g}$), {\it winos} ($\tilde{W}^i$) and 
{\it binos} ($\tilde{B}$), respectively. 
The common name for all partners of gauge fields is the {\it gaugino}.
\item
We associate scalar partners to the fermions, \ie quarks and leptons get scalar partners called {\it squarks} and {\it sleptons}. 
\item
We introduce one additional Higgs field (for a total of two Higgs doublets, corresponding to five physical Higgs states) and associate one
spin $1/2$ {\it Higgsino} to each Higgs boson. This is done to give masses to both up and down-type quarks upon electroweak symmetry breaking and also preserve supersymmetry (therefore, we cannot use the conjugate of the Higgs as is done in Standard Model). Introducing another Higgs doublet also makes the theory anomaly free.

\end{itemize}
The resulting particle content of the theory is shown in 
tables 3 and 4.
\begin{center}
\begin{table}
\footnotesize
\centering
\vskip 0.3cm
\begin{tabular}{|c|cc|cc|}
\hline 
Superfield& SM particles & Spin & Superpartners & Spin \\ [0.2cm]
\hline 
${Q}$ & $ \pmatrix{u_L \cr d_L \cr}$ &$1/2$& $ \pmatrix{\tilde{u_L} \cr \tilde{d_L} \cr}$ & 0 \\
${U}^c$ & $\bar{u}_R$ & $1/2$ & $\tilde u_R^*$ &0 \\
${D}^c$ & $\bar{d}_R$ & $1/2$ & $\tilde d_R^*$ &0 \\
 \hline 
${L}$ & $ \pmatrix{\nu_L \cr e_L \cr}$ &$1/2$& $ \pmatrix{\tilde{\nu_L} \cr \tilde{e_L} \cr}$ & 0 \\
${E}^c$ & $\bar{e}_R$ & $1/2$ & $\tilde e_R^*$ &0 \\
\hline 
${H}_1$ & $H_1$ & $0$ & $\tilde H_1$ &1/2 \\
${H}_2$ & $H_2$ & $0$ & $\tilde H_2$ &1/2 \\
 \hline
 ${G}^a$ & $g$ & $1$ & $\tilde g$ &1/2 \\
 ${W}_i$ & $W_i$ & $1$ & $\tilde W_i$ &1/2 \\
 ${B}$ & $B$ & $1$ & $\tilde B$ &1/2 \\ \hline 
\end{tabular} 
\label{tab:susy}
\caption{Field content of the MSSM.}
\end{table}
\end{center}
The MSSM is then specified through the
{\it superpotential}, defined as 
\be
W = \epsilon_{ij} \bigl[ y_e H_1^j  L^i E^c + y_d H_1^j Q^i D^c + y_u H_2^i
Q^j U^c \bigr] + \epsilon_{ij} \mu H_1^i H_2^j
\label{superw}
\ee
where $i$ and $j$ are SU(2) indices, and $y$ are Yukawa
couplings. Color and generation indices have been suppressed in
the above expression. The superpotential represents a supersymmetrization of the Standard Yukawa couplings plus a bilinear Higgs term. The superpotential enters the Lagrangian of
the theory through the terms
\be
{\cal L}_{\mbox{SUSY}}= - {1 \over 2} (W^{ij} \psi_i \psi_j + 
W_{ij}^* {\psi^i}^\dagger {\psi^j}^\dagger) - W^i W_i^*
\ee
where we have used $W^i \equiv \partial W / \partial
\phi_i$, $W_i^* \equiv \partial W / \partial {\phi^i}^*$, 
and $W^{ij} \equiv \partial^2
W / \partial \phi_i \partial \phi_j$. $\phi_i$ and $\psi_i$ are scalar and fermion fields, respectively.

\begin{table}
\footnotesize
\begin{center}
\begin{tabular}{lllllll} \hline 
  \multicolumn{2}{c}{Standard Model particles and fields} & \multicolumn{5}{c}{Supersymmetric partners} \\
  & & \multicolumn{3}{l}{Interaction eigenstates} & \multicolumn{2}{l}{Mass 
  eigenstates} \\
  Symbol & Name & Symbol & Name & & Symbol & Name \\ \hline
  $q=d,c,b,u,s,t$ & quark & $\tilde{q}_{L}$, $\tilde{q}_{R}$ & 
  squark & & $\tilde{q}_{1}$, $\tilde{q}_{2}$ & squark \\
  $l=e,\mu,\tau$ & lepton & $\tilde{l}_{L}$, $\tilde{l}_{R}$ & slepton & 
  & $\tilde{l}_{1}$, $\tilde{l}_{2}$ & slepton \\
  $\nu = \nu_{e}, \nu_{\mu}, \nu_{\tau}$ & neutrino & $\tilde{\nu}$ & 
  sneutrino & & $\tilde{\nu}$ & sneutrino \\
  $g$ & gluon & $\tilde{g}$ & gluino & & $\tilde{g}$ & gluino \\
  $W^\pm$ & $W$-boson & $\tilde{W}^\pm$ & wino & & & \\
  $H^-$ & Higgs boson & $\tilde{H}_{1}^-$ & higgsino & 
  \raisebox{-.25ex}[0ex][0ex]{$\left. \raisebox{0ex}[-3.3ex][3.3ex]{}
  \right\}$} &  $\tilde{\chi}_{1,2}^\pm$ & chargino \\
  $H^+$ & Higgs boson & $\tilde{H}_{2}^+$ & higgsino & & & \\
  $B$ & $B$-field & $\tilde{B}$ & bino & & & \\
  $W^3$ & $W^3$-field & $\tilde{W}^3$ & wino & & & \\
  $H_{1}^0$ & Higgs boson & 
  \raisebox{-1.75ex}[0ex][0ex]{$\tilde{H}_{1}^0$} & 
  \raisebox{-1.75ex}[0ex][0ex]{higgsino} & 
  \raisebox{.25ex}[0ex][0ex]{$\left. \raisebox{0ex}[-5.25ex][5.25ex]{}
  \right\}$} & \raisebox{0.5ex}[0ex][0ex]{$\tilde{\chi}_{1,2,3,4}^0$} & 
  \raisebox{.5ex}[0ex][0ex]{neutralino} \\[0.5ex]
  $H_{2}^0$ & Higgs boson & 
  \raisebox{-1.75ex}[0ex][0ex]{$\tilde{H}_{2}^0$} & 
  \raisebox{-1.75ex}[0ex][0ex]{higgsino} & & & \\[0.5ex]
  $H_{3}^0$ & Higgs boson & & & & & \\[0.5ex] \hline
\end{tabular}
\label{tab:susyparticles}
\caption[Standard Model particles and their superpartners in the MSSM]
{Standard Model particles and their superpartners in the MSSM
(adapted from Ref.~\cite{Edsjo:1997hp}).}
\end{center}
\end{table}
One additional ingredient of the MSSM is the conservation of \index{R-parity}$R$-parity. $R$-parity is a multiplicative quantum number defined as 
\be
R \equiv (-1)^{3B+L+2s} \;.
\ee
All of the Standard Model particles have $R$-parity $R=1$ and all sparticles (\ie superpartners) have $R=-1$. Thus, as a consequence of $R$-parity conservation, sparticles can only decay into an odd number of sparticles (plus Standard Model particles). The lightest sparticle (dubbed the LSP, for Lightest
Supersymmetric Particle) is, therefore, stable and can only
be destroyed via pair annihilation, making it an excellent dark matter
candidate~\cite{goldberglsp,ellislsp}. Note that this not the original motivation for R-parity. In fact, R-parity was first introduced to suppress the rate of proton decay~\cite{rparity1,rparity2,rparity3,rparity4,rparity5}.

\index{Lightest supersymmetric particle}

The nature of the LSP in the MSSM is constrained by many
observations. It cannot have a non-zero electric charge or color, or
it would have condensed with baryonic matter to produce heavy
isotopes, in conflict with observations. Among the neutral candidates,
a possibile LSP could be the sneutrino. Sneutrino LSPs have, however,
been excluded by direct dark matter detection experiments (see
sections~\ref{directexp} and \ref{direxp}).  Although axinos and
gravitinos cannot be {\it a prori} excluded, they arise only in a
subset of supersymmetric scenarios and have some unattractive
properties (see section~\ref{sec:zoo}). In particular, gravitinos and
axinos have very weak interactions and would be practically impossible
to detect, making them less interesting from a phenomenological
perspective. The lightest {\it neutralino} remains an excellent dark
matter candidate, and is further discussed in the next section.

To determine the identity of the LSP (or other characteristics) in a
given supersymmetric scenario, we have to specify how supersymmetry is
{\it broken}. If supersymmetry were not broken, then each superpartner
would have a mass identical to its Standard Model counterpart, which
is clearly not the case. Thus, new terms  which break supersymmetry
must be added to the
Lagrangian. These terms, however, should be
added carefully, in order not to destroy the hierarchy between Planck and
electroweak scales.  The possible forms for such terms are \ba {\cal
L}_{soft} & = & -{1\over 2} M^a_\lambda \lambda^a \lambda^a -{1\over
2} ({m^2})^{i}_{j} \phi_i {\phi^j}^* \nonumber \\ & & -{1\over 2}
{(BM)}^{ij} \phi_i \phi_j - {1\over 6} {(Ay)}^{ijk} \phi_i \phi_j
\phi_k + h.c., \ea
\noindent where the $M^a_\lambda$ are gaugino masses, $m^2$ are soft
scalar masses, $B$ is a bilinear mass term, and $A$ is a trilinear
mass term. We will discuss some specific supersymmetry breaking
scenarios later in this section.

\subsubsection{The lightest neutralino}
\label{sec:neutralino}

In the MSSM, the superpartners of the $B$, $W_3$ gauge bosons (or the photon and $Z$, equivalently) and the neutral Higgs bosons, $H_1^0$ and $H_2^0$, are called binos ($\tilde{B}$), winos ($\tilde{W_3}$), and higgsinos ($\tilde{H_1^0}$ and $\tilde{H_2^0}$), respectively. These states mix into four Majorana fermionic mass eigenstates, called neutralinos. The four neutralino mass eigenstates are typically labelled
$\tilde\chi^0_1$, $\tilde\chi^0_2$, $\tilde\chi^0_3$ and $\tilde\chi^0_4$, ordered with increasing mass\index{neutralino}. In the following we
will refer to $\tilde\chi^0_1$, \ie the lightest of the four 
neutralinos, as {\it the} neutralino, and denote it simply  
as, $\chi \equiv \tilde\chi^0_1$.

In the basis 
$(\tilde{B},\tilde{W}_{3},\tilde{H}_{1}^0,\tilde{H}_{2}^0)$, the 
neutralino mass matrix can be expressed as
\begin{equation}
\label{eq:neumass}
\arraycolsep=0.01in
{\cal M}_N=\left( \begin{array}{cccc}
M_1 & 0 & -M_Z\cos \beta \sin \theta_W^{} & M_Z\sin \beta \sin \theta_W^{}
\\
0 & M_2 & M_Z\cos \beta \cos \theta_W^{} & -M_Z\sin \beta \cos \theta_W^{}
\\
-M_Z\cos \beta \sin \theta_W^{} & M_Z\cos \beta \cos \theta_W^{} & 0 & -\mu
\\
M_Z\sin \beta \sin \theta_W^{} & -M_Z\sin \beta \cos \theta_W^{} & -\mu & 0
\end{array} \right)\;,
\end{equation}
where $M_1$ and $M_2$ are the bino and wino mass parameters, respectively, $\theta_W$ is the Weinberg angle and $\tan \beta$ is the ratio of the vacuum expectation values of the Higgs bosons. $\mu$ is the higgsino mass parameter. As we have seen, the (lightest) neutralino is a linear combination of $\tilde B$,
%
$\tilde W_3$, $\tilde H_1^0$ and  $\tilde H_2^0$,
\be
\chi = N_{11} \tilde B+  N_{12}\tilde W_3 + N_{13} \tilde H_1^0 + N_{14} \tilde H_2^0.
\ee
We then define the {\it gaugino fraction}, $f_G$,  and the {\it higgsino 
fraction}, $f_H$, as 
\be
f_G=N_{11}^2 + N_{12}^2
\ee
and
\be
f_H=N_{13}^2 + N_{14}^2\;.
\ee
For the analytic expressions used to diagonalize the neutralino mass matrix, see Appendix~\ref{diagonal}.

The neutralino interactions most relevant for the purposes of dark matter are 
self annihilation and elastic scattering off of nucleons. 
Neutralinos are expected to be extremely non-relativistic
in the present epoch, allowing us to safely keep only the $a$-term in the usual expansion of the annihilation cross section,
\be
\sigma v = a + b v^2 + {\cal O} \left( v^4 \right).
\ee
The $b$-term must be included in performing calculations of the neutralino relic density, however.

At low velocities, the leading channels for neutralino annihilation are annihilations to
fermion-antifermion pairs (primarily heavy fermions, such as top, bottom and charm quarks and tau leptons), gauge bosons pairs ($W^+ W^-$ and $Z^0 Z^0$) and final states containing Higgs 
bosons. In appendix~\ref{annappendix}, we give the most important neutralino annihilation diagrams, amplitudes and cross sections (in the low velocity limit). For a complete list of all tree level processes, diagrams, amplitudes and cross sections, see the excellent review of Jungman, Kamionkowski and Griest~\cite{Jungman:1995df}. 

\subsubsection{Supersymmetric models}
\label{sec:models}

Although relatively simple in many respects, the MSSM has a huge number 
of free parameters. Most of these parameters represent masses and mixing angles, much as in the case of the Standard Model. To allow for the practical phenomenological study of the MSSM, the number of parameters which are considered must be reduced. This can be done by making (theoretically well motivated) assumptions which reduce the free parameters from more than 100 to a more tractable quantity. Depending on the assumptions used, one obtains different 
supersymmetric models. In the following section, we will describe 
a few of the most widely considered supersymmetric scenarios, including mSUGRA (often called the constrained MSSM) and a phenomenologically simplified MSSM (called the phenomenological, or, pMSSM). We also discuss the phenomenological features of the MSSM in anomaly, gauge and gaugino mediated scenarios. 

\paragraph{mSUGRA}
\label{cMSSM}
\index{cMSSM} \index{mSUGRA}

The mSUGRA, or constrained MSSM, scenario is a simple phenomenological model based on a series of theoretical assumptions (see \eg Kane \etal~\cite{Kane:1993td}). The number of free parameters is reduced in this scenario by assuming that the MSSM parameters obey 
a set of boundary conditions at the Grand Unification scale:
\begin{itemize} 
\item Gauge coupling unification
\be
\alpha_{1} (M_U) = \alpha_2 (M_U) = \alpha_3 (M_U) \equiv \alpha_U
\ee
with $\alpha_i=g_i^2/4\pi$
\item  Unification of the gaugino masses
\be
M_1 (M_U)=M_2(M_U)=M_3(M_U) \equiv m_{1/2}
\label{Mis}
\ee
\item Universal scalar [sfermion and Higgs boson] masses
\ba
M_{\tilde{Q}}  (M_U)=M_{\tilde{u}_R}  (M_U)=M_{\tilde{d}_R} (M_U)=M_{\tilde{L}} (M_U)= M_{\tilde{l}_R} (M_U) &  \nonumber \\
=M_{H_u}  (M_U)=M_{H_d}  (M_U)\equiv  m_0&
\ea

\item Universal trilinear couplings: 
\be
A_u (M_U) = A_d (M_U) = A_l (M_U) \equiv  A_0
\ee
\end{itemize} 

\begin{figure}[t]
\begin{center}
$\begin{array}{c@{\hspace{0.01in}}c}
\includegraphics[width=0.43\textwidth]{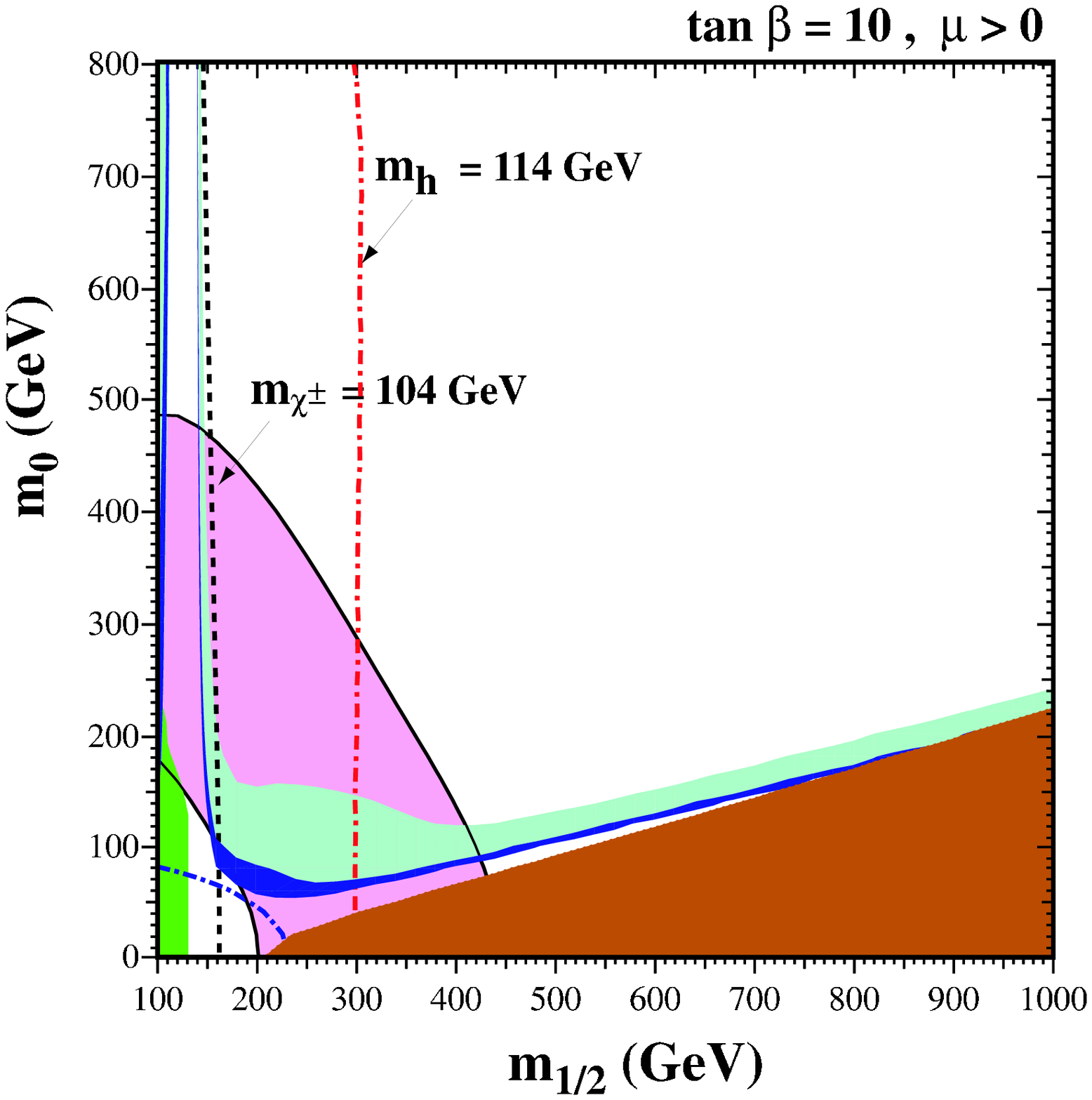} &
\includegraphics[width=0.43\textwidth]{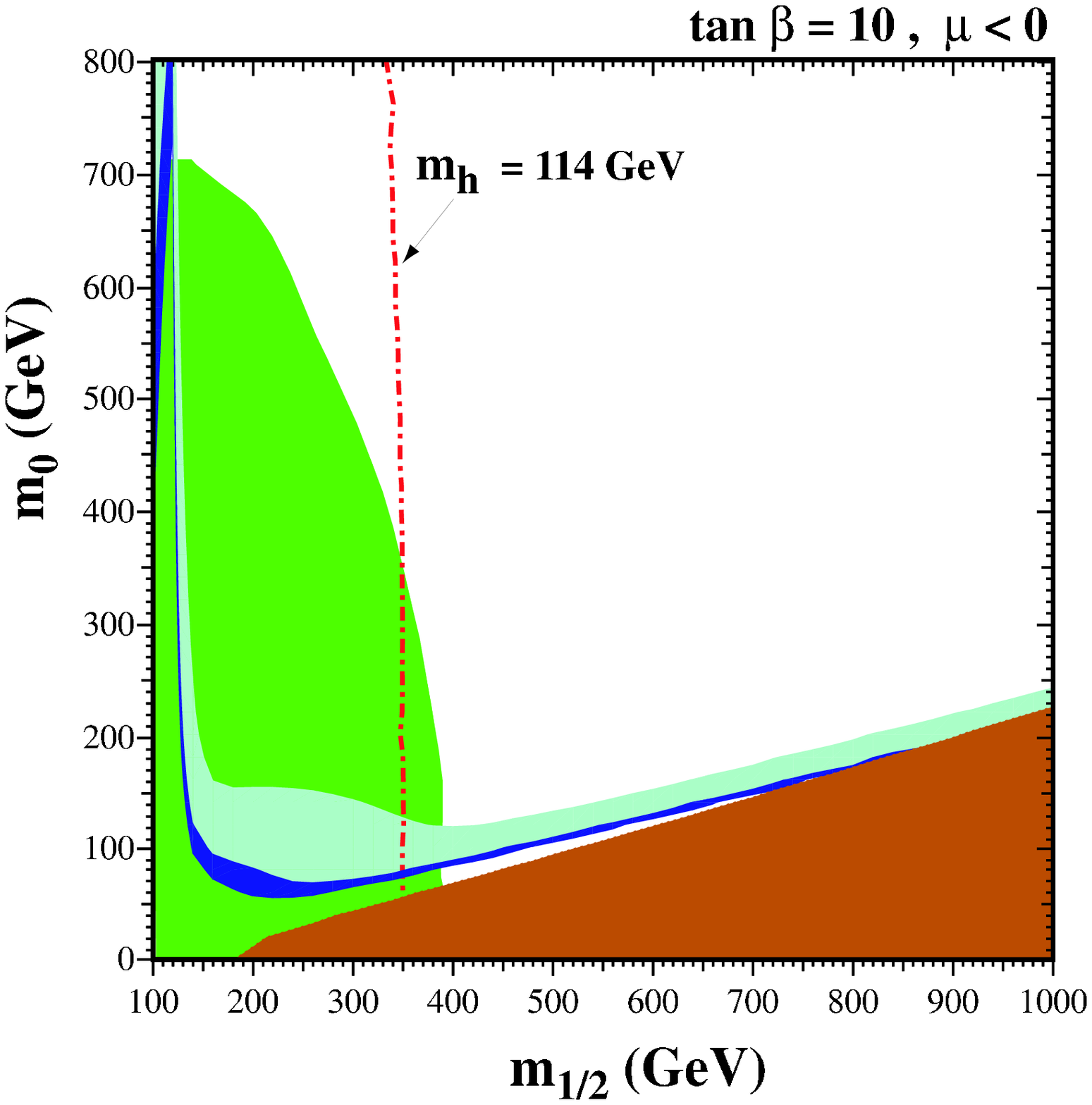} \\ 
\includegraphics[width=0.43\textwidth]{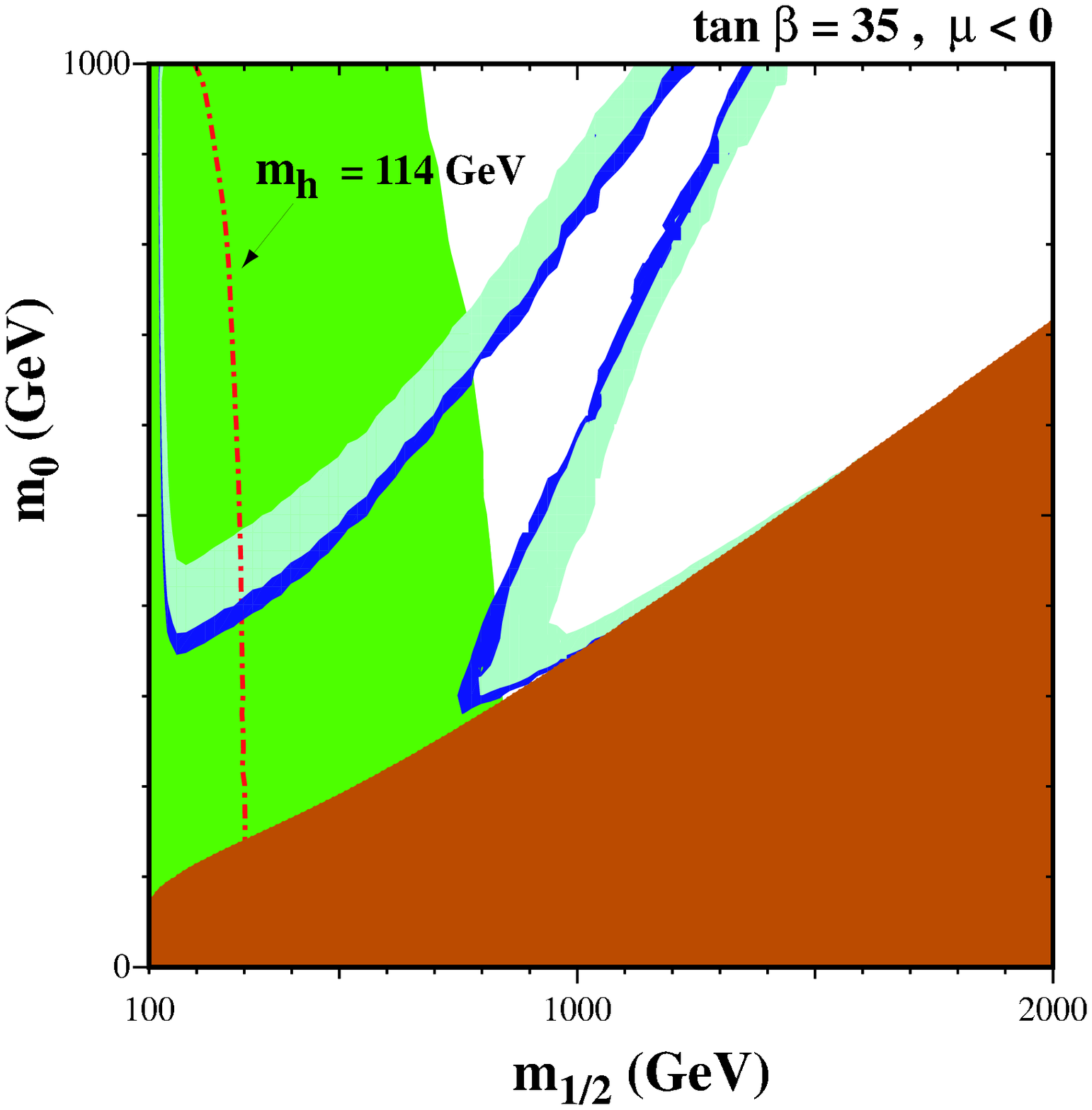} &
\includegraphics[width=0.43\textwidth]{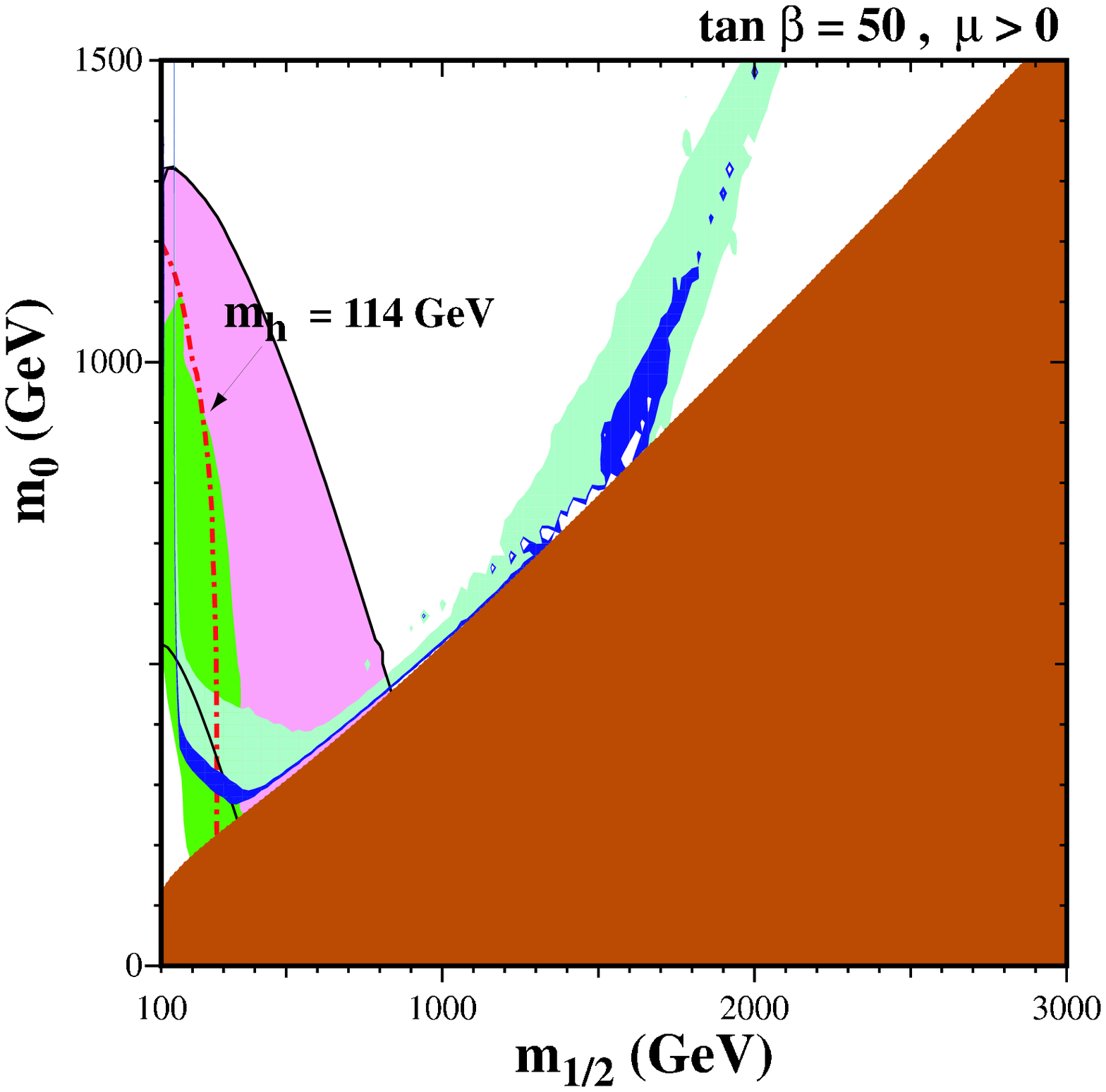} \\
\end{array}$
\end{center}
\small
\caption[Influence of CMB data on mSUGRA parameter space]{The $(m_{1/2}, m_0)$ planes for mSUGRA with (a) $\tan\beta = 10, \mu > 0$, (b)
$\tan\beta = 10, \mu < 0$, (c) $\tan\beta = 35, \mu < 0$, and (d)
$\tan\beta = 50, \mu > 0$. In each panel, the
region allowed by the older cosmological constraint $0.1 \le \Omega_\chi
h^2 \le 0.3$ has cyan shading, and the region allowed by the newer
cosmological constraint $0.094 \le \Omega_\chi h^2 \le 0.129$ has 
dark blue shading. For more details, see Ref.~\cite{Ellis:2003cw}.
}
\label{fig:Omega}
\end{figure}

\begin{figure}[t]
\begin{center}
\includegraphics[width=0.5\textwidth]{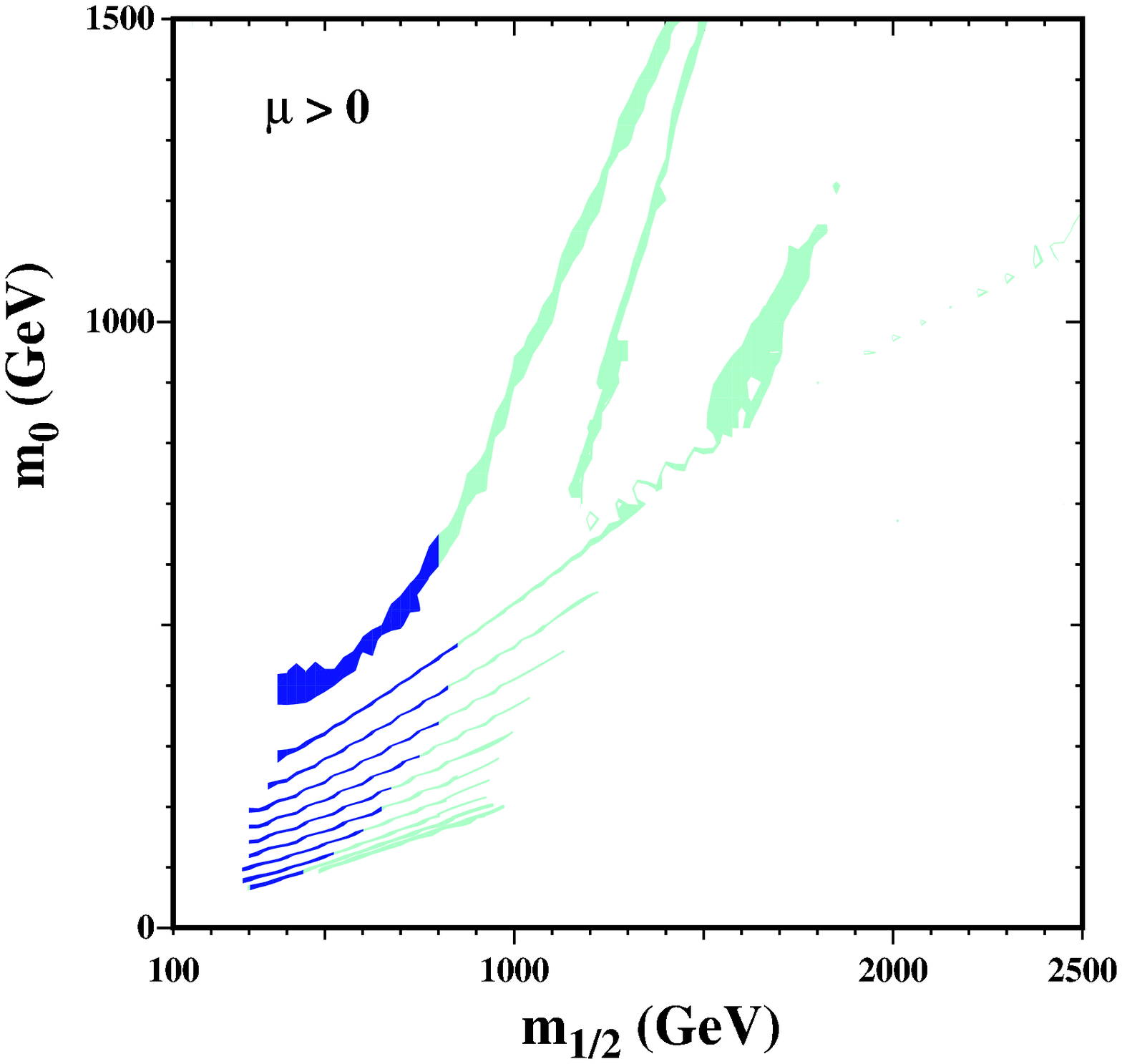} 
\end{center}
\caption[Allowed regions of the mSUGRA parameter space]{Regions of the $(m_{1/2}, m_0)$ plane in mSUGRA that are
compatible with $0.094 < \Omega_\chi h^2 < 0.129$ and laboratory
constraints for $\mu > 0$ and $\tan \beta = 5, 10, 15, 20, 25, 30,
35, 40, 45, 50, 55$. The parts of the strips compatible with $g_\mu - 2$ 
at the 2-$\sigma$ level have darker shading. From Ref.~\cite{Ellis:2003cw}.
}
\label{fig:strips}
\end{figure}

By requiring the minimization of the Higgs potential (in order to recover electroweak 
symmetry breaking), we are left with five (four continuous 
and one discrete) free parameters:
\be
\tan \beta \ , \ m_{1/2} \ , \ m_0 \ , \ A_0 \ , \ \ {\rm sign}(\mu),
\ee
where $\tan \beta$ is the ratio of the vacuum expectation values 
of the two Higgs fields and $\mu$ is the higgsino mass parameter.

A recent study of mSUGRA parameter space in light of the WMAP
measurement of the dark matter relic density can be found in 
Ref.~\cite{Ellis:2003cw}. We show in Fig.~\ref{fig:Omega}
and Fig.~\ref{fig:strips} the regions of the 
$(m_{1/2}, m_0)$ plane consistent with CMB and accelerator 
data. It is worth mentioning that neutralino models with 
relic densities lower than the WMAP measurement are not 
ruled out, although  evidently they cannot make up all the dark matter.

In addition to constraints on models in mSUGRA which come from the
WMAP measurements, strong constraints can also be placed by collider
data. In particular, 
constraints arise from the absence of new particles at LEP below $\approx 100$~GeV
and the agreement of $b \rightarrow s\gamma$ decays with predictions
of the Standard Model. Measurements of the anomalous magnetic momentum 
of the muon, $g_{\mu}-2$, also provide a possible constraint. These constraints have been studied in the context of mSUGRA in great detail \cite{Baer:2003yh,Ellis:2003si,Roszkowski:2001sb}. The interested reader will find a discussion of mSUGRA parameters and the definition
of SUSY {\it benchmarks} points in Ref.~\cite{Battaglia:2003ab}. For more on collider constraints, see section~\ref{sec:collider}.


\paragraph{The phenomenological MSSM}
\label{pMSSM}
\index{pMSSM}
The scenario we present in this section is not necessarily motivated by any theoretical arguments, but rather is justified by focusing on the aspects of supersymmetric phenomenology which are the most interesting for neutralino dark matter. The phenomenological MSSM, or pMSSM, is an adaptable framework which can be described by as many as tens of parameters, or as few as five or seven. It is NOT a model, but rather a convenient description of the phenomenology most relevant to supersymmetric dark matter. Common choices in defining a phenomenological MSSM include a) no new sources of CP violation (all the phases in the SUSY breaking terms are set to zero), b) no flavor-changing neutral currents and c) first and second generation universality.

One example of a phenomenological MSSM is used in the DarkSusy program package \cite{Gondolo:2002tz}\index{DarkSUSY}. In this scheme, in addition to the common features described above, gaugino unification is assumed (similar to Eq.~\ref{Mis}). The remaining inputs are defined by seven free parameters:
\be
\mu, M_2, \tan \beta, \  M_A, \  \ m_0, \  A_{b} \ \mbox{and} \ A_{t}
\ee
where $M_A$ is the mass of the pseudo-scalar Higgs boson,
$m_0$ is the common scalar mass, and $A_{b,t}$ are 
trilinear couplings appearing in SUSY breaking terms.
Unlike in the case of the mSUGRA scenario, the input parameters
are chosen at the electroweak scale without making
use of renormalization group equations. The inputs used in DarkSusy can be expanded beyond these seven to include other parameters, thus representing a more general MSSM.  

We note that these scenarios are less 
theoretically motivated in comparison to mSUGRA. Various combinations of theoretically and phenomenologically-based descriptions for supersymmetry are often considered in the literature, often maintaining some of the theoretically motivated constraints of mSUGRA while relaxing other requirement (for example, see Refs.~\cite{Bertin:2002sq,Birkedal-Hansen:2002am}).

\paragraph{The focus point region of mSUGRA}
\label{focuspt}
\index{focuspt}

In most of the parameter space of mSUGRA or other similar scenarios, the lightest neutralino is a gaugino-like neutralino with a mass of a couple hundred GeV or less. In the so-called ``focus point'' region of mSUGRA, however, the lightest neutralino can have a considerable higgsino content, and be significantly more heavy \cite{focuspoint1,focuspointdrees,focuspoint2,focuspoint3,focuspoint4}.
 
In the focus point region, very large scalar masses are
possible without violating naturalness constraints. This occurs because the soft masses squared of the Higgs
boson, $m^2_{H_u}$, have pseudo fixed-point behavior, and can start with a wide range of input values and run to a
similar negative value at the low scale. This is interesting because it indicates that, in the
focus point region, electroweak symmetry breaking does not require fine-tuning in the high
energy input values. 

A typical feature of the focus point region are large scalar masses (usually $\sim$ TeV). The main reason for a larger higgsino content in the LSP is the larger input value of the soft scalar mass. The tree level 
electroweak symmetry breaking condition gives 
\begin{equation}
\frac{1}{2}m_Z^2 \sim - m_{H_u}^2 -\mu^2.
\end{equation}
In the typical mSUGRA scenarios, $m_{H_u}^2$ is driven to some large
negative value due to the running of the renormalization group equations. This requires a large value of
$\mu$ to give the correct Z mass. In the focus point region, however, it
is possible that the large input value of the scalar soft mass makes
$m_{H_u}^2$ less negative. Hence, a smaller value of $\mu$ is possible, which leads to a larger higgsino content in the LSP.

\paragraph{Anomaly mediated SUSY breaking}
\label{AMSB}
\index{AMSB}
Anomaly Mediated Supersymmetry Breaking (AMSB) is an attractive alternative to general gravity mediated scenarios as it provides an elegant solution to the so-called flavor problem through an elegant decoupling mechanism. The resulting soft parameters are ``UV insensitive''. In this scenario, the SUSY parameters can all be expressed in terms of low energy parameters such as the Yukawa and gauge couplings. Although the details of AMSB are quite technical, and are beyond the scope of this work, we will here describe some of the phenomenological features of this scenario which are most relevant for dark matter.

In AMSB, the gaugino spectrum is given by

\begin{eqnarray}
M_{\lambda^a} &=& \frac{\beta_{g_a}}{g_a} m_{3/2}, 
\end{eqnarray}
where $\beta$ are beta-functions computed in the
supersymmetric limit \cite{anomalymediation1,anomalymediation2,amsb}. The proportionality of the $\beta$-function to the low energy masses leads to a specific relationship between the gaugino masses: $M_1:M_2:M_3=2.8:1:7.1$. This is very different than would be predicted by the GUT relations used in mSUGRA, for example, with the wino and bino mass hierarchy reversed. When the neutralino mass matrix is diagonalized in AMSB, these ratios result in an LSP which is almost purely (neutral) wino \cite{amsbdark1,amsbdark2}. Additionally, a charged wino, with a mass only a few hundred MeV heavier than the LSP, is predicted. This leads to a long lived chargino with distinctive collider signatures. Also, in AMSB, with such large values of $M_3$, the gluino and squarks are predicted to be considerably heavier than in mSUGRA or other similar scenarios.

\paragraph{The heterotic orbifold model}

The weakly coupled heterotic string with orbifold compactification is
among the earliest and best understood string models that 
can accomodate in four dimensions the Standard
Model gauge group, three generations of squarks and a coherent mechanism of 
supersymmetry breaking. These models show a behavior that interpolates 
between the phenomenology of mSUGRA and models dominated by superconformal 
anomalies (AMSB)~\cite{Binetruy:2003ad}.

Recently, the full one loop soft supersymmetry breaking terms in a large 
class of superstring effective theories have been calculated~\cite{Binetruy:2000md}
based on 
orbifold compactifications of the weakly coupled heterotic string 
(including the so-called anomaly mediated contributions). 
The parameter space in this class of models has already been severely 
constrained by taking into account accelerator and relic density constraints,
as well as direct and indirect searches (see Refs.~\cite{Bertone:2004ps,
Binetruy:2003ad, Binetruy:2003yf, Kane:2002qp,Birkedal-Hansen:2001is}.

\paragraph{Gauge mediated SUSY breaking}
\label{GMSB}
\index{GMSB}
Another alternative SUSY breaking mechanism is mediated by gauge interactions \cite{gaugemediation1,gaugemediation2,gaugemediation3,gaugemediation4,gaugemediation5,gaugemediation6}. In Gauge Mediated Supersymmetry Breaking (GMSB), we have the following approximate relationship between the low energy SUSY masses
and the gravitino mass 
\begin{equation}
\frac{m_{3/2}}{m_{\rm{SUSY}}} \sim
\frac{1}{\alpha_a}\frac{M_S}{M_{Pl}} \ll 1, 
\end{equation}
where $M_S$ is some typical supersymmetry breaking scale. Therefore,
generically, we will have a very light gravitino as the LSP \cite{gaugemediationdark}. Such a scenario provides a dark matter candidate which is very difficult to observe. We will not discuss gravitino dark matter or GMSB further for this reason.

\paragraph{Gaugino mediated SUSY breaking}
\label{GoMSB}
\index{GoMSB}

Gaugino mediated supersymmetry breaking \cite{kap1,kap2} represents another class of SUSY breaking mediation motivated by the brane-world scenario. In this scenario, unwanted supersymmetry breaking effects, such as
the flavor violating couplings, are suppressed by the separation of the observable and hidden sectors via the separation of their respective branes. Gauginos are allowed to propagate off of the branes (in the bulk) in this scenario, communicating SUSY breaking from the hidden sector.

The most important phenomenological feature of this mechanism is that the sfermion masses are suppressed relative to the gaugino masses. This is because sfermion masses can only be generated from the 1-loop diagrams in which a gaugino is
emitted, travels through the bulk to the supersymmetry breaking brane,
gets the information of SUSY breaking and then returns to join the
sfermion propagator again. Generically, these masses are suppressed relative to the gaugino mass by a loop factor, $m^2_{\tilde{f}} \sim M_{\lambda}^2/(16 \pi^2)$.

\subsection{Extra Dimensions}
\label{sec:kk}
\index{extra dimensions}
Although our world appears to consist of 3+1 (three space and one time) dimensions, it is possible that other dimensions exist and appear at higher energy scales.

From the physics point-of-view, the concept 
of extra dimensions received great attention after
the idea of Kaluza, in 1921, to unify electromagnetism
with gravity by identifying the extra components
of the metric tensor with the usual gauge fields. More recently, it has been realized that the 
hierarchy problem (see section~\ref{whysusy}) could be addressed,
and possibly solved, by exploiting the geometry of
spacetime.  

In many extra-dimensional models, the 3+1 dimensional spacetime we experience is a structure called a {\it brane}, which is embedded in a $(3+\delta+1)$ spacetime
called the {\it bulk}. The hierarchy problem can then 
addressed by postulating that all of the 
extra dimensions are compactified on circles (or other topology) of some size, $R$,
as has been done in the Arkani-Hamed, Dimopoulos and Dvali (ADD) scenario~\cite{Arkani-Hamed:1998rs}, thus lowering the fundamental Planck scale to an energy near the electroweak scale. Alternatively, this could be accomplished by introducing extra dimensions with large curvature (warped extra dimensions) as has been suggested by Randall and Sundrum~\cite{Randall:1999ee}. The extra dimensional scenario which we will focus on throughout the remainder of this review (universal extra dimensions) does not share the features of the ADD or RS scenarios. Rather, it introduces flat extra dimensions which are much smaller than those in the ADD framework. 

In addition to the hierarchy problem, motivation for the study of theories
with extra dimensions comes from {\it string theory}
and {\it M-theory}, which today appear to be the best candidates for a 
consistent theory of quantum gravity and a unified
description of all interactions. It appears that such theories may require the presence of six or seven extra-dimensions.

A general feature of extra-dimensional theories is that upon 
compactification of the extra dimensions, all of the fields propagating in the bulk have their momentum quantized in units of $p^2 \sim 1/R^2$. The result is that for each bulk field, a set of Fourier expanded modes, called Kaluza--Klein (KK) states, appears. From our point of view in the four dimensional world, these KK states appear as a series (called a tower) of states with masses $m_n= n/R$, where $n$ labels the mode number. Each of these new states contains the same quantum numbers, such as charge, color, etc.

In many scenarios, the Standard Model fields are assumed to be confined
on the brane, with only gravity allowed to propagate
in the bulk. Nevertheless, if the extra-dimensions are small,
it would be possible for all fields to freely propagate in the extra dimensions.  Such is the case in models with universal extra dimensions, which we discuss in the next section.

\index{unified extra dimensions}
\subsubsection{Universal extra dimensions }
\label{sec:ued}

Scenarios in which all fields are allowed to 
propagate in the bulk are called Universal Extra Dimensions (UED)~\cite{Appelquist:2000nn}.
Following Ref.~\cite{Servant:2002aq}, we note that there is 
significant phenomenological motivation to have all 
Standard Model fields propagate in the bulk, including: 
\begin{itemize}
\item Motivation for three families from anomaly cancellation
\item Attractive dynamical electroweak symmetry breaking 
\item Prevention of rapid proton decay  
\item {\it Provides a viable dark matter candidate.}
\end{itemize}

In the case of one extra dimension, the 
constraint on the compactification scale in UED models from precision
electroweak measurements is as low as $R^{-1}\gtrsim 300$ GeV 
\cite{Appelquist:2000nn}. Recently, it was shown that this
bound can be weakened to $R^{-1}\gtrsim 280$ GeV if one allows a Higgs mass 
as heavy as $m_H\gtrsim 800$ GeV \cite{Appelquist:2002wb}.
This is to be contrasted with another class of models 
where Standard Model bosons propagate in extra dimensions while fermions are localized
in 4 dimensions. In such cases, the constraint on
the compactification scale is much stronger, requiring $R^{-1}\gtrsim$ several 
TeV \cite{Cheung:2001mq}.

The prospect of UED models providing a viable dark matter candidate is indeed what motivates 
us in our discussion here. The existence of a viable dark matter candidate
can be seen as a consequence of the conservation of momentum in higher dimensional space. Momentum conservation in the compactified dimensions leads to the conservation of KK number. This does not stabilise the lightest KK state, however. To generate chiral fermions at the zero mode, the extra dimensions must be moded out by an orbifold, such as $S/Z_2$ for one extra dimension or $T^2/Z_2$ for two. This orbifolding results in the violating of KK number, but can leave a remnant of this symmetry called KK-parity (assuming that the boundary terms match). All odd-level KK particles are charged under this symmetry, thus ensuring that the lightest (first level) KK state is stable. In this way, the Lightest Kaluza-Klein Particle (LKP) is stabalized in a way quite analogous to the LSP in R-parity conserving supersymmetry. 

In the next section, we will discuss some of the characteristics of the LKP in models of UED.

\subsubsection{The lightest Kaluza--Klein particle}
\index{Lightest Kaluza--Klein particle}
\label{lkp}
The study of the Lightest Kaluza-Klein Particle (LKP) as a dark matter candidate dates
back to the work of Kolb and Slansky in 1984~\cite{Kolb:fm},
where the KK excitations were referred to as 
{\it pyrgons}, from the Greek $\pi \upsilon \rho \gamma
o  \varsigma$ for ``scale'' or ``ladder''. 
The LKP has since been reconsidered in the framework of universal extra dimensions,
in which it is likely to be associated with the first KK 
excitation of the photon, or more precisely the first KK excitation 
of the hypercharge gauge boson \cite{Cheng:2002iz}. We will refer to this state as $\bone$. \index{$\bone$ particle}

A  calculation of the $\bone$ relic density was
performed by Servant and Tait~\cite{Servant:2002aq}, 
who found that if the LKP is to account for the observed quantity of dark matter, its mass (which is inversely proportional to the compactification radius $R$) should lie in 
the range of 400 to 1200 GeV, well above any current experimental 
constraint.

\begin{figure}[t]
\begin{center}
\includegraphics[width=0.6\textwidth]{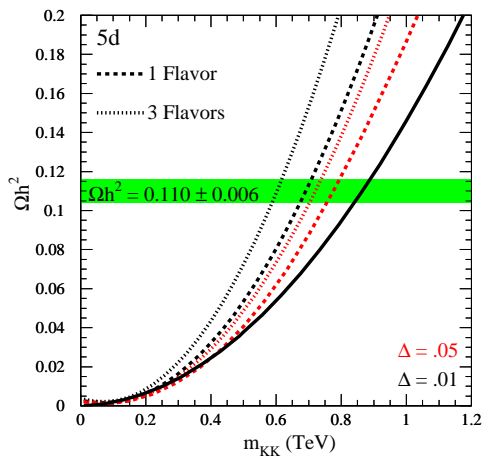}
\end{center}
\caption[Relic density versus mass of the $\bone$ (5D)]{
\footnotesize Relic density versus mass of the $\bone$.  The solid
line is the case for $\bone$ alone, dashed and dotted lines 
are for one (three) flavors of nearly degenerate 
$\erone$.  For each case, black curves (upper of each pair) 
are for $\Delta=0.01$ and red curves (lower of each pair) 
for $\Delta=0.05$. Figure kindly provided by G.~Servant.}
\normalsize
\label{bone}
\end{figure}

We show in Fig.~\ref{bone} the relic density of the
 $\bone$ particle versus its mass, including coannihilations
(see section~\ref{sec:relic}) with the next-to-lightest KK particle,
which in the case shown is $\erone$, the first KK excitation of the right-handed
electron. This figure is a new version of Fig.~3 in 
Ref.~\cite{Servant:2002aq}, updated to
include the new WMAP constraints on the cold dark matter relic density \footnote{see section~\ref{cmb} for a 
discussion of the CMB and, in particular, the recent
WMAP data}.

Note that the results of the LKP relic density calculation can vary
depending on the spectrum of other first level KK states. Unlike in
the case of supersymmetry, the density of KK dark matter is {\it
  increased} through coannihilations with other KK particles.

This
is due to the fact that in the case of neutralinos, the cross section 
for the interaction between neutralinos and 
the NLSP is much larger than the neutralino self-annihilation cross 
section, which implies that DM particles are kept longer in 
thermodynamic equlibrium, thus decoupling with a lower relic 
density. In constrast, the interactions between the $\bone$ and $\erone$
are comparable with the $\bone$ self-interaction. Decoupling in
presence of coannihilations thus happens essentially at the 
same time as in the case with no coannihilations, and the $\bone$
relic density becomes larger since the $\erone$, after 
decoupling at the same time, decays in the $\bone$.

The spectrum of first level KK states has been calculated to one
loop by Cheng, Matchev and Schmaltz~\cite{Cheng:2002iz}, although 
higher dimensional operators localized at the boundary may change the
details of the spectrum (without affecting KK parity). Variations in 
this spectrum can result in variations for the predicted LKP relic abundance. 

The $\bone$ annihilation cross section has been studied in 
Ref.~\cite{Servant:2002aq}, and is given by
\be
\sigma v =   \frac{95 g_1^4}{324 \pi m^2_{\bone}} \simeq \frac{0.6 \,\mbox{pb}}{m_{\bone}^2 [\mbox{TeV}]} \;.
\ee
The branching ratios for $\bone$ annihilation (see table 5)
are almost independent of the particle mass. Unlike in the case of
supersymmetry, the bosonic nature of the LKP means that there will be no
chirality suppression in its annihilations, and thus can annihilate efficiently
to fermion-fermion pairs. In particular, since the annihilation cross section
is proportional to hypercharge$^4$ of the final state, a large fraction of 
LKP annihilations produce charged lepton pairs.

\begin{table}[t]
\centering
\begin{tabular}{|lc|}
\hline 
Channel & Branching ratio \\
\hline
quark pairs & 35\% \\
charged lepton pairs & 59\% \\
neutrino pairs & 4\% \\ 
Higgs bosons & 2\% \\
\hline 
\end{tabular}
\caption[Branching ratios for the annilation of the $\bone$ particle]
{Branching ratios for the annihilation of the $\bone$ particle. Note that small variations from these results can occur with variation in the KK spectrum.}
\label{tab:br}
\end{table}

Direct detection of the LKP via its elastic scattering with nuclei was investigated in Refs.~\cite{Cheng:2002ej,Servant:2002hb}.
It was emphasized in Ref.~\cite{Servant:2002hb} that a one-ton detector is 
needed to probe the expected heavy masses as indicated by the
relic density calculation \cite{Servant:2002aq} of the LKP. One must, therefore, wait for the next generation of direct detection experiments such
as GENIUS \cite{Klapdor-Kleingrothaus:2000eq}
or XENON \cite{Aprile:2002ef} (see section~\ref{directexp}). Simultaneously, the LHC 
should probe most of the relevant KK mass parameter space (up to
$R^{-1}\sim 1.5$ TeV \cite{Cheng:2002ab})
and confirm or rule out UED at the TeV scale.

\subsection{Superheavy Candidates}
\label{sec:wzilla} \index{wimpzilla}
Dark matter particles are usually assumed to be relatively ``light'', meaning lighter than a few hundred TeV. This ``limit'' is a consequence of the 
existence of a maximum annihilation cross section, $\sigma v$,
for a particle of a given mass, $m_{DM}$, set by the so-called 
{\it unitarity bound} (see \eg Ref.~\cite{weinberg}).
Griest and Kamionkowski~\cite{Griest:1989wd} applied
this bound and the constraint on the relic density to infer an upper limit on the dark matter particle
mass:
\be
m_{DM} \lesssim 340 \;\; \mbox{TeV}\;\;. 
\label{c1}
\ee
We note that nowadays, using the WMAP constraint on
$\Omega_{DM} h^2$, such a constraint can be made ten times stronger,
\be
m_{DM} \lesssim 34 \;\; \mbox{TeV}\;. 
\label{c2}
\ee
The assumption behind this argument is that the dark matter 
particle is a thermal relic of the early Universe, otherwise
we could not have applied the relation between $\Omega_{DM} h^2$
and $\sigma v$. In this section, we consider 
superheavy dark matter candidates, defined as candidates with mass 
$m_{DM}  >10^{10}$~GeV, that we call generically {\it wimpzillas}
~\cite{Kolb:1998ki,Chang:1996vw}. Thus the first condition for this scenario
is that {\it wimpzillas must not have been in thermal equilibrium during freeze-out}. Since they are not in thermal equlibrium during freeze-out, their relic abundance does not depend on their annihilation cross section, but rather is a function of the wimpzilla's production cross section. Furthermore, we want them to be sufficiently stable 
against decay and annihilation to significantly contribute to 
the present day matter density.

There are many ways to produce wimpzillas in the early Universe. Among the 
most studied is
{\it gravitational production} at the end of inflation,
resulting from the expansion of the background spacetime  
(for details on this and other scenarios see \eg Ref.~\cite{Kolb:1998ki} and references therein).

Natural mass scales for wimpzillas include the inflaton or grand unified masses, which
are usually assumed to be roughly $10^{11} $~GeV or $10^{16}$~GeV, respectively. Alternatively, D-matter provides a good candidate for wimpzillas with a somewhat larger mass \cite{shiuwang}. The
interaction cross sections with ordinary matter for such particles can
vary from very weak to strong (in the latter case supermassive
particles are sometimes called {\it simpzillas}).

A common motivation for superheavy dark matter comes from the observation of cosmic rays at ultra-high energies\cite{agasahires}, above the so-called GZK (Greisen-Zatsepin-Kuzmin) cutoff \cite{gzk1,gzk2}. Above this cutoff, which occurs at $\sim 5 \times 10^{19}\,\mbox{eV}$, protons interact at resonance with CMB photons with a center-of-mass energy nearly equal to the mass of the $\Delta$-hadron (1.232 GeV). The cross section for this interaction is quite large, thus making the Universe opaque to ultra-high energy protons over cosmological distances ($\gsim 50$~Mega-parsecs). Since no astrophysical sources of ultra-high energy protons are known within this range, more exotic scenarios have been developed to account for these observed events. Such scenarios include ultra-high energy cosmic-ray production via the decay or annihilation of superheavy dark matter particles, called top-down cosmic-ray models (see, for example, Refs.~\cite{topdown1,topdown2,topdown3,Sarkar:1999bf,topdown4,topdown5,topdown6,topdown7}).


\subsection{Collider Constraints}
\label{sec:collider}

The constraints which can be placed on a dark matter candidate from collider experiments are highly model dependent in nature. It is, unfortunately, impossible to completely or simply describe the reach of colliders in their search for dark matter in any kind of general way. We will here, rather, review several of the most important collider searches which have been carried out for dark matter particles and for particles associated with a dark matter candidate.

\subsubsection{Current collider constraints}

\begin{itemize}
\item {\it Invisible $Z$ width}

If a dark matter candidate is sufficiently light, $Z$ bosons may decay invisibly to such particles with a non-zero branching fraction. Of course, there is a substantial background to such events, namely $Z \rightarrow \nu \bar{\nu}$ decays. Presently, to contribute less than one standard deviation to the measured neutrino contribution, the analysis of LEP2 finds that a decay width of $\Gamma_{Z \rightarrow XX} < 4.2 \,$ MeV is required ($X$ denotes a dark matter particle).

Similarly, single photon events can be an interesting search channel for light dark matter particles. At LEP2, the Standard Model background process for this signature is $e^+ e^- \rightarrow Z \rightarrow \nu \bar{\nu}$ with an additional photon radiated off of the initial state. The total cross section for $\gamma Z$ production at LEP2 is less than 31 pb with a minimum 1 GeV transverse momentum cut for the photon. The contribution to this final state (single photon) from particle dark matter, in addition to $Z \rightarrow XX$ with an additional photon radiated off of the initial state, are t-channel dark matter producing processes in which a photon is radiated off of a charged propagator (a selectron in supersymmetry, for example).

\item {\it Searches for new charged particles}

LEP2 has placed very stringent bounds on charged particles lighter than about 100 GeV. In $e^+ e^-$ colliders, cross sections for the direct pair production of charged particles are quite large, allowing for limits to be placed at or slightly below half of the center-of-mass energy of the collision. For LEP2, which reached a center-of-mass energy of 209 GeV, limits of 87-103 GeV have been placed for such particles; in particular for charginos ($m_{\chi^{\pm}} > 103\,$ GeV) and charged sleptons ($m_{\tilde{e}} > 99 \,$ GeV, $m_{\tilde{\mu}} > 96 \,$ GeV, $m_{\tilde{\tau}} > 87 \,$ GeV) in supersymmetric models. If the LSP is only slightly less massive than the charged particle, however, this limit may be substantially lower \cite{lepcharginos,lepsleptons1,lepsleptons2,lepsleptons3}. 

Limits on charged particles can only indirectly constrain dark matter, however. In supersymmetry, chargino masses and neutralino masses are, in some models, related by the unification of gaugino masses. Although such a relationship is often assumed, it is quite possible that the pattern of gaugino masses is not so simple. If gaugino mass unification is assumed chargino limits can translate to neutralino mass limits of about half of the chargino mass limit ($m_{\chi^0} \gsim 50 \,$ GeV). Without such a relationship, the LSP could be much lighter \cite{lightlsp1,lightlsp2}. 

\item {\it Sneutrino limits}

Limits for charged sleptons can be used to indirectly limit the possible masses for sneutrinos beyond the invisible $Z$ width constraints. Such a bound is the result of a basic SU(2) symmetry between the supersymmetry breaking masses of the left handed slepton and the sneutrino of a given lepton flavor. Limits somewhat lower than for charged particles ($m_{\tilde{\nu}} \gsim 85 \,$ GeV) can be placed on sneutrinos if such theoretical assumptions are made \cite{lepsleptons1,lepsleptons2,lepsleptons3}.

\item {\it Searches for colored particles}

Hadron colliders, such as the Tevatron, can place the strongest limits on colored particles (squarks and gluinos in supersymmetry or KK excitations of quarks and gluons in models with universal extra dimensions, for example). Such particles would most likely undergo a series of cascades upon decaying, possibly producing dark matter candidates among other particles. Combinations of squarks and gluinos are searched for using jets and missing energy signatures. This leads to exclusion contours in the squark and gluino mass plane.

In supersymmetry, the spectrum of neutralinos, charginos and sleptons lighter than the decaying squark/gluino is very important in placing limits on squark and gluino masses. Similar ambiguities are present in other models as well, such as universal extra dimensions, etc. Typically, limits of $\sim$200 GeV are obtained for new colored particles, unless there exists an invisible final state particle with a mass close to the new colored particle's mass \cite{tevcolor1,tevcolor2,tevcolor3,tevcolor4,tilmancode}.

\item {\it New gauge bosons}

Heavy gauge bosons appear in many models of particle physics beyond the Standard Model. Heavy charged gauge bosons (called $W^{'}$'s) and heavy neutral gauge bosons (called $Z^{'}$'s) have been excluded below about 600-800 GeV, depending on the details of the analysis \cite{heavybosons111,heavybosons222}. These limits assume that these particles have couplings equal to their Standard Model counterparts. If their couplings were smaller, the resulting limits could be considerably weaker. Electroweak precision measurements can also constrain heavy gauge bosons considerably (see below).

\item {\it Higgs searches}

In supersymmetric models, the Higgs mass is increased from its tree level mass (below $m_Z$) by loop processes involving superparticles, most importantly top squarks. Current bounds on the (lightest) Higgs mass, therefore, constrain the masses of top squarks and other superparticles. Furthermore, if supersymmetry is manifest below 1 TeV, as is normally expected, the Higgs mass must be less than about 130 GeV, not very far above current limits from LEP2 ($m_h < 114.1 \,$ GeV) \cite{lephiggs1,lephiggs1b,lephiggs2a,lephiggs2b,lephiggs3,lephiggs4}. Note that this bound is somewhat lower for cases with very large values of $\tan \beta$.

Searches for charged Higgs bosons can also provide constraints on models of physics beyond the Standard Model. More sensitive, however, may be the impact of charged Higgs bosons in the branching fraction of $b \rightarrow s \gamma$.

\item {\it Flavor changing neutral currents}

Many models of physics beyond the Standard Model introduce flavor changing neutral currents, often at tree level. To avoid the corresponding flavor constraints, either the masses of new particles involved must be quite large, or symmetries must be imposed to solve the ``flavor problem''. For example, the squark and slepton mass matrices are flavor diagonal in the constrained MSSM (mSUGRA) scenario, thus suppressing such processes. Flavor changing neutral currents in models with universal extra dimensions have also been explored \cite{kkfcnc}.

\item {\it $b \rightarrow s \gamma$}

The branching fraction for $b \rightarrow s \gamma$ \cite{Buras:2002tp,bsgammameasure1,bsgammameasure2,bsgammameasure3,bsgammameasure4}, measured at CLEO and BELLE, is of particular interest for supersymmetry and other beyond the Standard Model phenomenology. In many scenarios, the contributions to this process from new physics can add substantially to the Standard Model prediction. In particular, light charged Higgs bosons and/or charginos can be quite important for this decay \cite{Carena:2000uj,Okumura:2002wa,Okumura:2003hy}. In supersymmetry, the constraint is considerably stronger if $\mu < 0$, but also relevant for $\mu > 0$, especially for large values of $\tan \beta$. $b \rightarrow s \gamma$ is also an important constraint in models of universal extra dimensions \cite{kkfcnc}.

\item {\it $B_s \rightarrow \mu^+ \mu^-$}

The branching fraction for $B_s \rightarrow \mu^+ \mu^-$ is quite small in the Standard Model ($\simeq 3.5 \times 10^{-9}$) \cite{Anikeev:2001rk}. The contribution from supersymmetry scales as $\tan^6 \beta$, and thus becomes quite large for models with large values of $\tan \beta$. In run I of the Tevatron, a value consistent with the Standard Model was found. The sensitivity of run II of the Tevatron to this quantity will be considerably greater.

\item  {\it The anomalous magnetic moment of the muon, $g_{\mu}-2$}

In 2001, the E821 experiment at the Brookhaven National Laboratory reported a
measurement of the muon's magnetic moment which was 2.6 standard deviations from the Standard Model prediction \cite{gminus2announce}. Since then, however, an important error in the theoretical calculation was discovered which reduced the significance of this anomaly to about 1.6 standard deviations \cite{gminus2correct1,gminus2correct2,gminus2correct2b,gminus2correct3a,gminus2correct3b}. With the reduction of statistical error which has been achieved more recently, the deviation from the Standard Model prediction of this measurement is again about 3$\sigma$ using $e^+ e^-$ data (although the significance is somewhat less using $\tau$ decay data) \cite{gminus2current1,gminus2current2,gminus2current3,gminus2current4}. These measurements, although somewhat difficult to interpret, can be used to constrain TeV-scale particle physics beyond the Standard Model.

\item{\it Electroweak precision measurements}

In addition to the useful direct particle searches at LEP2, the Tevatron and other experiments, impressively accurate electroweak measurements have been made. Various limits on the scale of new physics and associated particle masses have been inferred from these measurements. Given these constraints, models with universal extra dimensions are limited to the scale $\sim 300 \,$ GeV or higher \cite{Appelquist:2000nn}. These measurements also yield particularly important bounds for models without a custodial SU(2) symmetry, such as many little Higgs models \cite{ewplittlehiggs}.

\end{itemize}

\begin{figure}
\centering
\includegraphics[width=0.7\textwidth,clip=true]{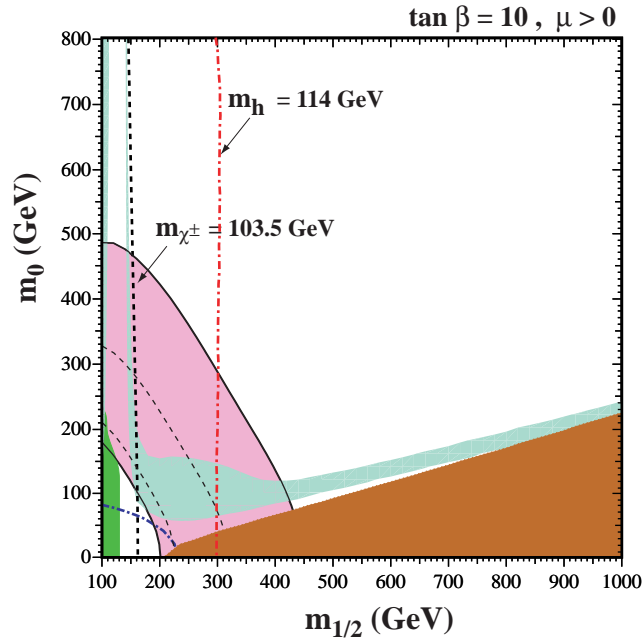}
\caption[]{\label{fig:collider1} 
An example of the impact of collider (and cosmological) constraints on a model of particle dark matter \cite{Olive:2003iq}. The model shown is the constrained Minimal Supersymmetric Standard Model (cMSSM or mSUGRA) with $\tan \beta = 10, \mu > 0$ and $A_0 = 0$. The almost vertical lines represent the limits on chargino (left) and Higgs (right) masses from LEP2. The blue dot-dash curve in the bottom left corner follows the 99 GeV selectron mass contour, excluded by LEP2. In the dark red region in the lower right, the LSP is a stau and is not, therefore, a viable dark matter candidate. The green region in the lower left corner is excluded by the $b \rightarrow s \gamma$ constraint. The long and often narrow turquoise region provides a relic density of $0.1\leq \Omega h^2 \leq 0.3$, near the observed quantity. The pink region extending over much of the lower left is the region within the 2$\sigma$ range for $g_{\mu}-2$. The two dashed curves within this region are the 1$\sigma$ bounds.}
\end{figure}

Together, these constraints can be very powerful, often providing very tight bounds for specific models. For example, in Fig.~\ref{fig:collider1}, we show the impact of collider and cosmological constraints on the constrained Minimal Supersymmetry Standard Model (or mSUGRA). We find that over the parameter space shown, constraints from LEP2 searches (Higgs, charginos and selectrons), along with $g_{\mu}-2$ and relic density constraints, leave only a small region near $m_{1/2}\cong 300-400$ GeV and $m_0 \cong 80-150$ GeV. Although the power of these and other constraints is quite model dependent, they are often very useful in supersymmetry and other classes of models.

\subsubsection{The reach of future collider experiments}

\begin{figure}
\centering
\includegraphics[width=0.8\textwidth,clip=true]{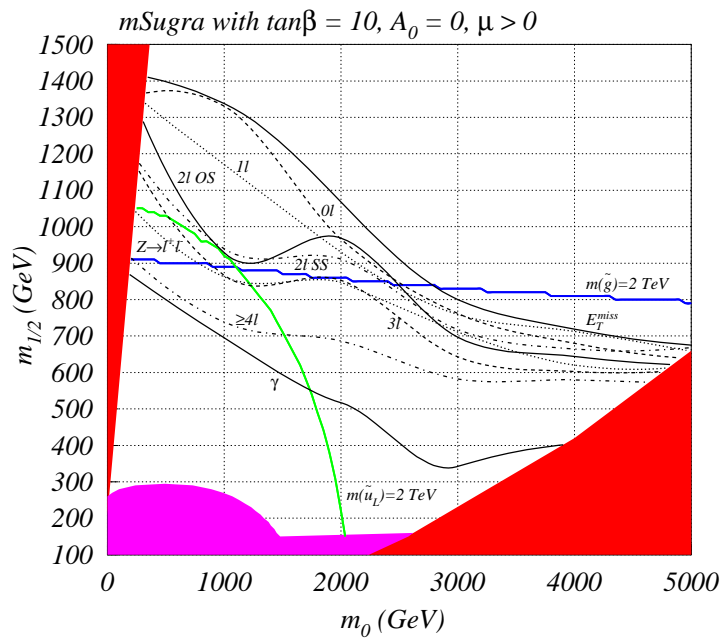}
\caption[]{\label{fig:collider2} 
An example of the reach of the Large Hadron Collider (LHC) to new TeV-scale physics \cite{lhcreach1}. As a function of $m_0$ and $m_{1/2}$ in the mSUGRA (or constrained MSSM) scenario, with $\tan \beta = 10$, $A_0=0$ and positive $\mu$, the reach is shown for a variety of channels: zero leptons (0$l$), one lepton (1$l$), leptons with opposite charge ($OS$), leptons with the same charge ($SS$), three leptons (3$l$), four or more leptons ($\ge 4 l$), any number of leptons plus a photon ($\gamma$), at least two opposite sign leptons with the invariant mass within an optimized interval around the $Z$ mass ($Z \rightarrow l^+ l^-$) and an ``inclusive'' missing transverse energy channel. Also shown are the 2 TeV up squark and 2 TeV gluino mass contours. The red regions are excluded by theoretical constraints, while the magenta region is excluded experimentally. $100 \rm{fb}^{-1}$ of integrated luminosity is assumed.}
\end{figure}

\begin{itemize}

\item{\it Future reach of the Tevatron}

The reach of the Tevatron extends to higher energies than any other accelerator until the time at which the Large Hadron Collider (LHC) becomes operational. The range of masses which can be searched for colored particles (squarks, gluinos and KK quarks, for example), heavy gauge bosons and other new physics will be increased significantly at the Tevatron IIb \cite{tevatronreach,tevcolor4}. 

\item {\it The Large Hadron Collider}

The Large Hadron Collider (LHC) is expected to begin operation around
2007 with proton-proton collisions at 14 TeV center-of-mass
energy. A luminosity of 300 inverse femtobarns is expected to be achieved, making the prospects for discovering new physics at the LHC excellent. Numerous classes of models which provide interesting dark matter candidates will be tested at this very important experiment, searching at scales of up to several TeV. In addition to the Higgs boson(s), the LHC will be sensitive to most supersymmetry scenarios, models with TeV-scale universal extra dimensional, little Higgs models, etc.

For a few examples of studies which discuss the sensitivity of the LHC to new physics, see Refs.\cite{atlas,cms,Dawson:1983fw,Allanach:2004ub,Allanach2,lhcreach1,lhcreach4,lhcreach6,lhcreach7}. In Fig.~\ref{fig:collider2}, an example of such a study is shown \cite{lhcreach1}. It is interesting to note that in the region of the MSSM which is the most difficult to probe at the LHC, direct dark matter detection rates are very high \cite{directwherenolhc}.

\item {\it Beyond the LHC}

After the LHC, other collider experiments are likely to follow. Although no specific post-LHC program is certain at this time, a 500-1000 GeV linear collider is a possibility, perhaps followed by a Very Large Hadron Collider (VLHC). These or other post-LHC colliders will, of course, have great value to particle dark matter studies \cite{weigleing,Baur:2002ka}.

\end{itemize}

\section{Experiments}

\subsection{Direct Detection Experiments}
\label{directexp}

Direct detection experiments appear today as one of the most 
promising techniques to detect particle dark matter. The idea
is very simple: if the galaxy is filled with WIMPs, 
then many of them should pass through the Earth, making it possible to look for the interaction of such 
particles with matter, e.g. by recording the recoil
energy of nuclei, as WIMPs scatter off them \cite{direarly1,direarly2, Wasserman:hh}. 

The key ingredients for the calculation of the signal
in direct detection experiments are the density and 
the velocity distribution of WIMPs in the solar neighborhood
and the WIMP-nucleon scattering cross section. 
With this information, it is then possible to evaluate the rate of events expected in an experiment ({\it i.e.} WIMP--nucleon scattering events) per
unit time, per unit detector material mass.

The rate is approximately given by 
\begin{equation}
R\approx \sum_i N_i n_\chi <\sigma_{i \chi}>,
\end{equation}
where the index, $i$, runs over nuclei species present in
the detector 
\[ N_i=\frac{\mbox{Detector mass}}{\mbox{Atomic mass of species }i} \]
is the number of target nuclei in the detector, 
\[ n_\chi \equiv \frac{\mbox{WIMP energy density}}{\mbox{WIMP mass}} \] 
is the local WIMP density and $<\sigma_{i \chi}>$ 
is the cross section for the scattering of WIMPs off 
nuclei of species $i$, averaged over the relative 
WIMP velocity with respect to the detector. For a more 
through discussion see, \eg, Ref.~\cite{Jungman:1995df}. 

\subsubsection{Scattering classifications}

The type of scattering processes considered can be classified by two important characteristics: elastic or inelastic scattering and spin-dependent or spin-independent scattering. 

\begin{itemize}
\item{Elastic and Inelastic Scattering}

The elastic scattering of a WIMP off of a nucleus in a detector is simply the interaction of the WIMP with a nucleus as a whole, causing it to recoil, ideally often enough to measure the recoil energy spectrum in the target. With a Boltzman velocity distribution of WIMPs, centered at 270 km/s, the spectrum of recoils is exponential with typical energies of $<E>\sim 50 \, \rm{keV}$. Current experiments can detect recoils of considerably lower energy, as low as 1-10 keV.   

Inelastic scattering, on the other hand, is not observed by the recoil of a target nuclei. Instead, the WIMP interacts with orbital electrons in the target either exciting them, or ionizing the target. Alternatively, the WIMP could interact with the target nuclei leaving it in an excited nuclear state. This process leaves the signature of a recoil followed by the emission of a photon a nanosecond, or so, later \cite{doublesig}. Such signatures have to compete with backgrounds of natural radioactivity, however.

\item{Spin-Dependent and Spin-Independent Scattering}

WIMP scattering off of nuclei is commonly discussed in the context of two classes of couplings. First, axial-vector (spin dependent) interactions result from couplings to the spin content of a nucleon. The cross sections for spin-dependent scattering are proportional to J(J+1) rather than the number of nucleons, so little is gained by using heavier target nuclei. For scalar (spin-independent) scattering, however, the cross section increases dramatically with the mass of the target nuclei, and typically dominates over spin-dependent scattering in current experiments which use heavy atoms as targets.

It should be pointed out that a WIMP which is not a Majorana particle could also scatter by vector interactions. Heavy Dirac neutrinos or MSSM sneutrinos are examples of particles which would scatter in this way. Neutralinos and Kaluza-Klein dark matter do not have such couplings, however.

For more on scalar, axial-vector and vector WIMP-nucleon scattering, see Appendix~\ref{dircalc}.

\end{itemize}

\subsubsection{Experimental Efforts}

More than 20 direct dark matter detection experiments are either now operating or are currently in development.  In these many experiments, numerous techniques have been developed to measure the nuclear recoil produced by dark matter scattering. Some of these methods include the observation of scintillation (used by DAMA, ZEPLIN-I, NAIAD, LIBRA), photons (used by CREST and CUORICINO) and ionization (used by HDMS, GENIUS, IGEX, MAJORANA and DRIFT). Some experiments use multiple techniques, such as CDMS and Edelweiss which use both ionization and photon techniques, CRESST-II and ROSEBUD which use both scintillation and photon techniques and XENON, ZEPLIN-II, ZEPLIN-III and ZEPLIN-MAX, which use both scintillation and ionization techniques. 

The use of such a large array of techniques and technologies is important not only to accelerate the progress of the field, but also to vary the systematic errors from experiment to experiment, allowing for a critical assessment of a positive signal.

Some experiments are also attempting to separate WIMP signatures from background by looking for an annual modulation in their rate. Such an effect would arise due to the Earth's annual motion around the Sun, resulting in a relative velocity relative to the galaxy's frame of reference \cite{Drukier:tm}. Under this effect, the Earth's velocity is given by
\begin{equation}
v_{\rm{E}} = 220 \,\rm{km/s}\, \bigg( 1.05 +0.07 \cos(2 \pi (t-t_m))  \bigg),
\end{equation}
where $t_m$ is roughly the beginning of June and the times are in units of years. The result of this effect is a $\cong 7 \%$ variation in the WIMP flux and direct detection rate over the course of the year. Since this variation is quite small, many events are needed to identify such a signature. For more on this technique and the status of direct detection techniques, see section~\ref{direxp}.


\subsection{Gamma-Ray Experiments} 
\label{gamexp}

In addition to detecting WIMPs directly, efforts are underway to attempt to observe the products of WIMP annihilations in the galactic halo, the center of the Sun or other regions. These annihilation products include neutrinos, positrons, anti-protons and gamma-rays.

To observe cosmic gamma-rays directly, observations must be made from space. This is because in the energy range we are most interested (GeV to TeV), photons interact with matter via $e^+ e^-$ pair production, which leads to an interaction length of approximately $38\,$g cm$^{-2}$, which is much shorter than the thickness of the Earth's atmosphere (1030 g cm$^{-2}$). Thus, at the energies we are considering, gamma-rays cannot reach ground based telescopes. Efforts have been developed, nevertheless, to observe gamma-rays indirectly with ground based experiments. In this section, we discuss the status of both ground and space-based gamma-ray telescopes.

\subsubsection{Ground-based telescopes}
\label{act}
When photons interact in the atmosphere, they
produce an electromagnetic cascade and thus a shower of secondary 
particles, allowing ground-based telescopes to {\it indirectly} observe 
gamma-rays through the detection of secondary particles and 
the Cerenkov light originating from their passage through the Earth's atmosphere.

It was P.~Blackett (winner of the Nobel prize in 1948) who first realized the
possibility of detecting Cerenkov light from cosmic air showers.
This realization was experimentally confirmed by W.~Galbraith and
J.~Jelley (1953). Cosmic gamma-rays can be difficult to observe in this way, however, as most of the observed Cerenkov 
light is due to cosmic-ray induced showers with isotropic
arrival directions. For detecting gamma-ray 
showers, an excess above the isotropic background of 
cosmic rays must be seen in the direction of a source. To accomplish this, the rejection of cosmic ray showers is of crucial importance.

\begin{figure}[h!]
\begin{center}
$\begin{array}{c@{\hspace{0.5in}}c}
\includegraphics[width=0.45\textwidth]{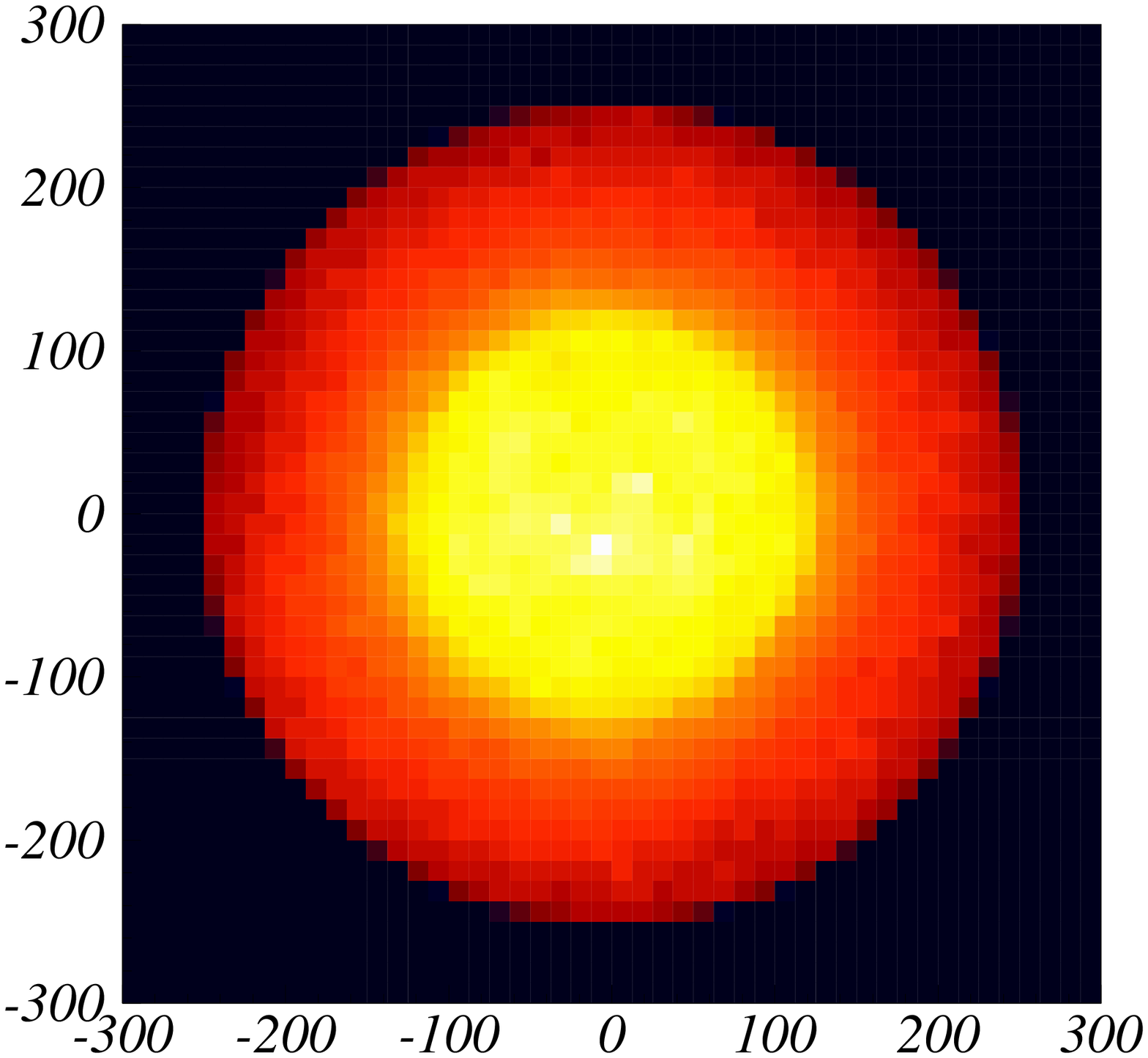} &
\includegraphics[width=0.45\textwidth]{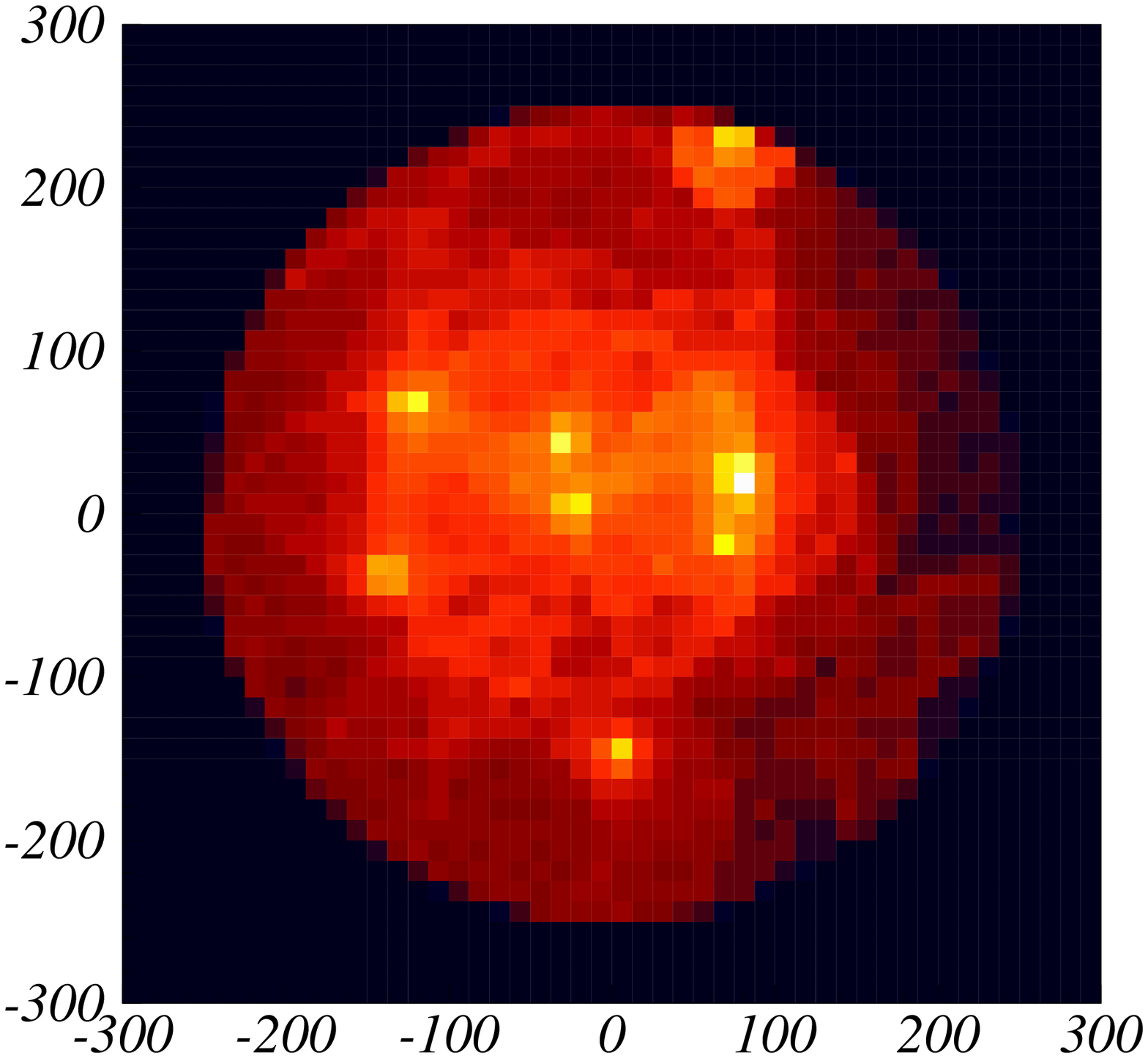} \\ [0.4cm]
\end{array}$
\end{center}
\caption[Simulated Cerenkov light from electromagnetic cascades]
{Simulations of Cerenkov light from electromagnetic cascades
initiated by a 1 TeV photon (left) and a 1 TeV proton (right). The figures
show the distribution of light to the ground in a $600 \times 600$~m$^2$ area. Figures kindly provided by I.~Perez.}
\label{chere}
\end{figure}

To distinguish between cosmic ray and gamma-ray induced air 
showers, the observed Cerenkov light is compared with numerical simulations of atmospheric showers (see Fig.~\ref{chere}). 
Apart from the difficulties in the treatment of interactions
at very high energies, numerical simulations are complicated by 
uncertainties associated with the density profile of the atmosphere 
and the Earth's magnetic field. Nevertheless, reliable codes 
for simulating atmospheric showers exist on the
market, for example \href{http://ik1au1.fzk.de/~heck/corsika/}{CORSIKA} 
and AIRES (see {\it e.g.} Knapp \etal~\cite{Knapp:2002vs}).

The methods of collecting Cerenkov light are quite varied, ranging from 
telescopes (and array of telescopes) to converted 
solar arrays. We show in table 6 a list of 
existing ground based experiments for the detection of gamma-rays.
These include imaging and non--imaging Air Cerenkov telescopes,
reconverted solar arrays as well as experiments
which detect secondary particles produced in showers.

\begin{figure}[t]
\includegraphics[width=\textwidth]{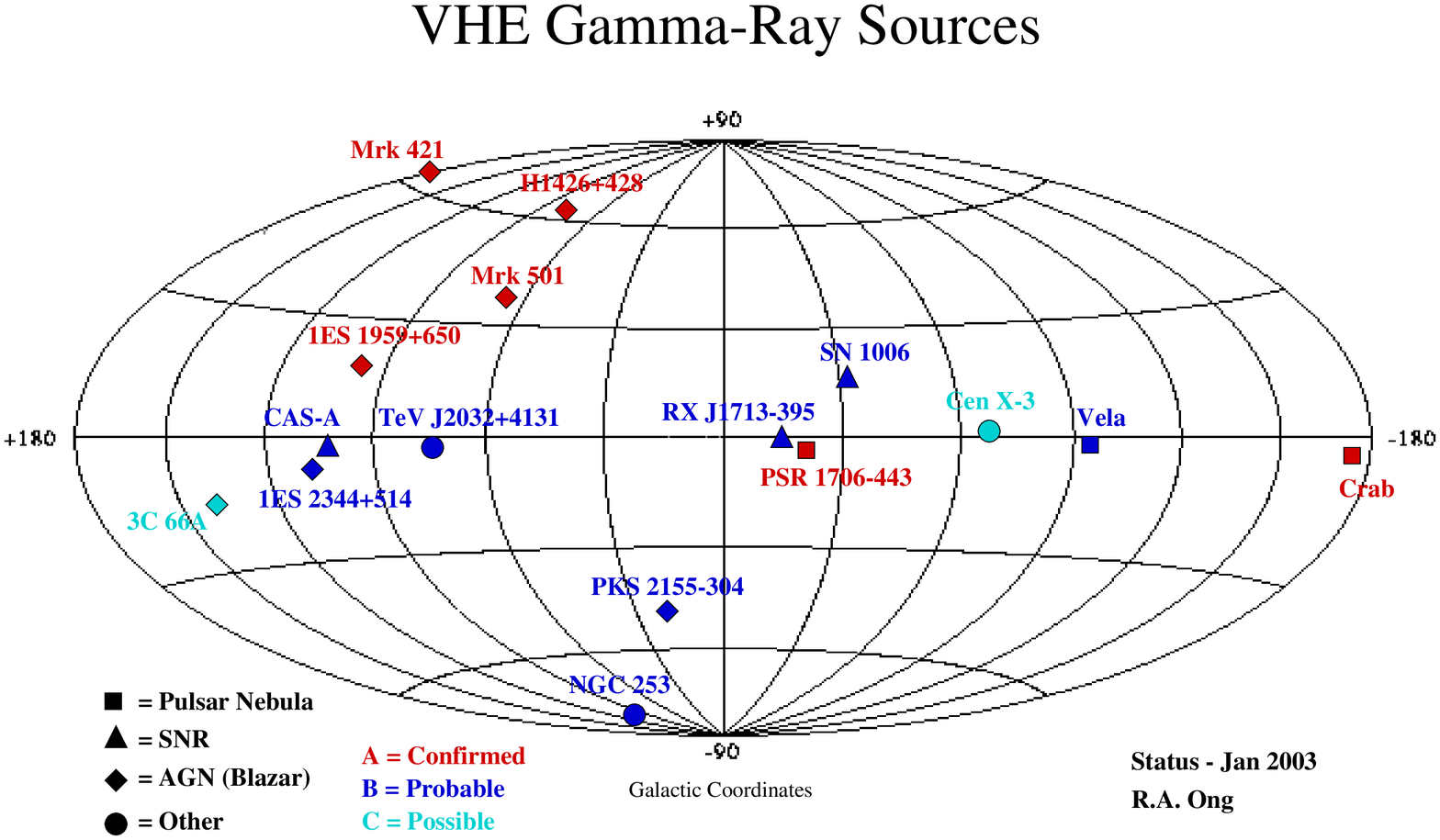}
\caption[Sky map of sources of TeV gamma-rays.]{Sky Map for sources of very high-energy (TeV) gamma-rays, 
as of January 2003 (Ref.~\cite{Ong:2003dz}).}
\label{status}
\end{figure}

The first observation of Cerenkov light due to gamma-ray emission 
from an astrophysical source was the detection of the Crab Nebula
(which today is regarded as the ``standard candle'' at these energies) with 
the Whipple Observatory 10m reflector ~\cite{weekes89}.
Currently, only six TeV gamma-ray sources have been {\it confirmed}, above 
10 GeV, having been detected
by multiple experiments at a high significance level (red symbols in 
Fig.~\ref{status}). Eight sources are {\it probable}, {\it i.e.}
detected at high significance by at least one group (blue symbols),
and two are ``possible'' (light blue symbols, see Ref.~\cite{Ong:2003dz} for more details). 

\begin{center}
\begin{table}
\label{IACT}
\begin{tabular}{llcccl}
\hline
\hline
{\it Imaging} \\
Group & Location & Telescope(s) & Threshold 
& Ref. \\
 &      & Num.$\times$Apert. & (TeV)  \\ 
\hline
\href{http://veritas.sao.arizona.edu/VERITAS_whipple.html}{Whipple} 
 &Arizona, USA & 10\,m &  0.4 &
\cite{WHIPPLE} \\
Crimea  & Ukraine & 6$\times$2.4\,m  & 1 
& \cite{CRIMEA} \\
SHALON  & Tien Shen, Ru & 4\,m  & 1.0 &
\cite{SHALON} \\
\href{http://icrhp9.icrr.u-tokyo.ac.jp/index.html}{CANG-II} 
& Woomera, Au & 10\,m & 0.5 
& \cite{CANGAROO}\\
\href{http://www.mpi-hd.mpg.de/hfm/CT/CT.html}{HEGRA} 
& La Palma, Es & 5$\times$5\,m  & 0.5 & \cite{HEGRA} \\
\href{http://lpnp90.in2p3.fr/~cat/index.html}{CAT}  &
Pyren\'ees, Fr & 4.5\,m & 0.25 & \cite{CAT}  \\
TACTIC & Mt.~Abu & 10\,m & 0.3 & \cite{TACTIC}
\\
\href{http://www.dur.ac.uk/~dph0www4/}{Durham} & Narrabri & 
3$\times$7\,m  & 0.25
& \cite{DURHAM99} \\
\href{http://www-ta.icrr.u-tokyo.ac.jp/}{7TA} & Utah, USA & 
7$\times$2\,m & 0.5 
& \cite{7TEL99} \\ \\
\hline
\hline
{\it Non-Imaging}\\
Group & Location & Type & Telescopes & Ref. \\
\hline 
Potchefstroom & South Africa & Lateral Array  & 4 & \cite{POTCH} \\
\href{http://tifrc1.tifr.res.in/~pnbhat/hegro.html}{Pachmarhi}
 & India & Lateral Array & 25 & \cite{PACHMARI99} \\ 
Beijing & China & Double & 2 & \cite{BEIJING} \\ \\
\hline
\hline
\it{Solar Arrays} \\
Group & Location & Heliostats & Threshold & Ref.
\\
     &    &    Now (future)  & GeV & \\
\hline                   
\href{http://www.astro.ucla.edu/~stacee/}{STACEE} & 
Albuq., USA &  32 (48)  & 180
& \cite{STACEE99} \\ 
\href{http://wwwcenbg.in2p3.fr/extra/Astroparticule/celeste/index.html}{CELESTE}
 & Themis, Fr & 40 (54) &     50$\pm10$ &
\cite{CELESTE99} \\ 
Solar-2 & Barstow, USA & 32 (64) &   20 & \cite{SOLAR-299} \\
\\
\hline
\hline
{\it Non - air Cerenkov} \\
Group  & Location & Telescope &  Threshold 
& Ref. \\
     &       &       & TeV &  \\
\hline                   
Milagro & Fenton Hill, US & Water Cher. & 0.5-1.0 &
\cite{MILAGRO} \\ 
Tibet HD  & Tibet & Scintillators  & 3 &
\cite{TIBET99}  \\ \\
\hline
\hline
\end{tabular}
\label{expe}
\caption[Atmospheric Cerenkov Imaging Observatories]
{Atmospheric Cerenkov Imaging Observatories {\it circa} 
September 2003.}
\end{table}
\end{center}
\normalsize

Although only a few TeV gamma-ray sources have been confirmed, many
more could be detected in next generation experiments. Among these experiments:
\begin{itemize}
\item
\href{http://hegra1.mppmu.mpg.de/MAGICWeb/}{MAGIC} is a 
17m imaging air Cerenkov telescope recently completed on the 
island of La Palma~\cite{magic}. It has already started taking data.
\item
\href{http://icrhp9.icrr.u-tokyo.ac.jp/index.html}{CANGAROO-III} 
is an array of four $10\,$m Cerenkov telescopes
being constructed in Woomera, Australia~\cite{canga}. It should start taking data in 2004.
\item
\href{http://www.mpi-hd.mpg.de/hfm/HESS/HESS.html}{HESS} 
consists of four 12$\,$m diameter Cerenkov telescopes,
at a site in the Gamsberg area of Namibia~\cite{hess}. The telescopes
are operational and started taking data.
\item
\href{http://veritas.sao.arizona.edu/index.html}{VERITAS} 
is an array of 7 telescopes in construction on Kitt Peak in Arizona, USA~\cite{veritas}. 
A preliminary version, VERITAS-4, with 4 telescopes should be operative 
around 2006.
\end{itemize}

\begin{figure}[t]
\includegraphics[width=\textwidth]{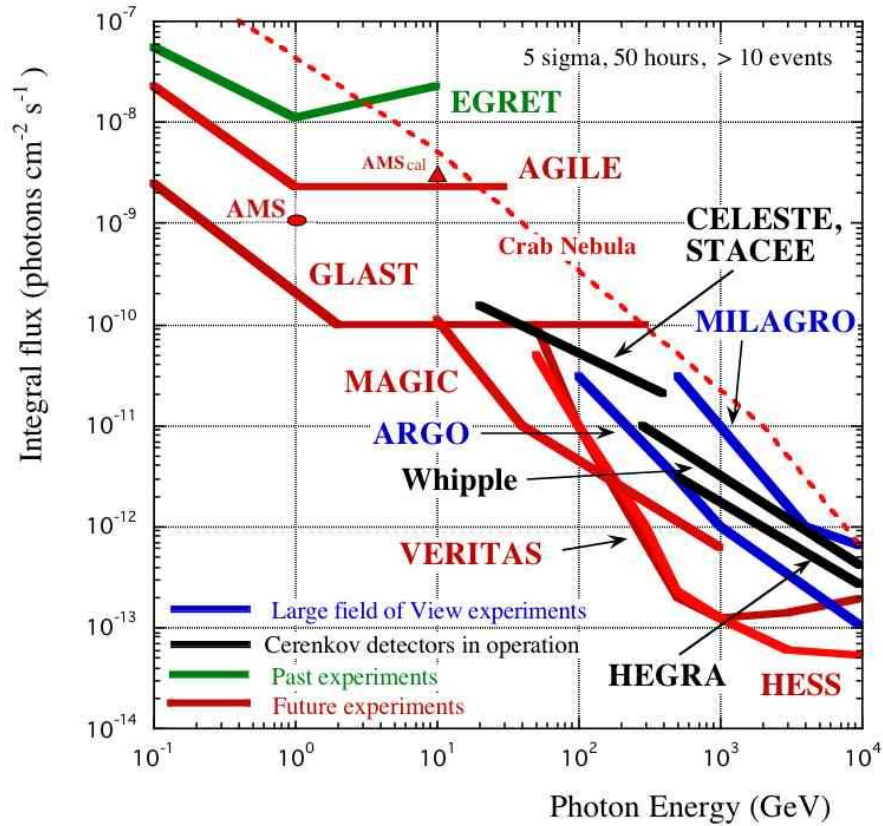}
\caption[Sensitivity of present and future gamma-ray detectors]
{Sensitivity of present and future detectors
in gamma-ray astrophysics
(from Ref.~\cite{Morselli:2002nw}).}
\label{sensi}
\end{figure}

\subsubsection{Space-based telescopes} 
\label{spacegamma}

The first high-energy (above GeV) gamma-ray space telescopes was EGRET (the Energetic Gamma-Ray Experiment Telescope), onboard the Compton gamma-ray observatory. Launched in 1991, 
EGRET has observed the universe in a range of energies 
extending up to approximately 30 GeV, amassing a large catalog of observed gamma-ray sources, although around
60\% of these sources remain unidentified.

The next space-based gamma-ray observatory will
be GLAST (Gamma-ray Large Area Space Telescope), which is scheduled for launch in 2007. As for its predecessor, GLAST will detect gamma-rays by
recording the characteristics of $e^+e^-$ pairs 
produced in the interaction of the incident gamma-ray
with a dense layer of tungsten. GLAST's effective area to gamma-rays will be a full square meter, considerably larger than with EGRET. GLAST will have an angular resolution on the order of arcminutes, compared to the degree level with EGRET (energy resolution varies with energy for both experiments). Unlike EGRET, GLAST will be sensitive to gamma-rays up to several hundred GeV in energy. 

GLAST is expected to be complementary to ground-based 
telescopes due to the lower range of energies 
observed, larger field of view, and higher duty
cycle. We show in Fig.\ref{sensi} the sensitivity of some
of the present and next generation ground-based and space-based 
gamma-ray experiments. We will use such information when
discussing the prospects for indirect detection of particle dark matter.

\subsection{Neutrino Telescopes}
\label{neuexp} \index{sensitivity of $\nu$~telescopes}

In addition to gamma-rays, neutrinos can be produced in the annihilations of dark matter particles. In this section, we review the status of high-energy neutrino telescopes, in particular, large volume Cerenkov detectors such as AMANDA, ANTARES and IceCube.

Neutrinos are considerably more difficult to observe than gamma-rays due to their weak interactions with ordinary matter. Neutrinos are not easily absorbed, however, allowing for their observation in underground, low background, experiments. In the GeV-TeV energy range, neutrinos are most easily observed by their ``muon tracks'' produced in charged current interactions inside of or nearby the detector volume. These muons travel through the detector emitting Cerenkov light which allows their trajectory to be reconstructed.

For a cosmic neutrino flux, $d\Phi_{\nu}/dE_{\nu}$, the rate of muon tracks in a detector is given by

\begin{equation}
\rm{rate} = \int_{E_{\mu}^{\rm thr}} dE_{\nu}
  \int_{0}^{1-\frac{E_{\mu}^{\rm thr}}{E_{\nu}}} dy\,
   A(E_{\mu})P_{\mu}(E_{\nu},y;\, E_{\mu}^{\rm thr})
    \, \frac{d\Phi_{\nu}}{dE_{\nu}},
\label{rate}
\end{equation}
where $E_{\mu}^{\rm{thr}}$ is the muon threshold energy of the experiment (generally between 10 and 100 GeV), $A(E_{\mu})$ is the effective area of the detector, typically in the range of 0.01 to 1.0 square kilometers (for further details and numerical values, see \eg Ref.~\cite{bailey:2003}) and $P_{\mu}(E_{\nu},y;\, E_{\mu}^{\rm thr})$ is the probability that a neutrino of energy 
$E_{\nu}$ interacts with a nucleon
producing a muon of energy $E_{\mu} \equiv (1-y)E_{\nu}$ 
above the detector threshold energy. 
As one would expect, this probability depends on the muon range, 
$R(E_{\mu},\, E_{\mu}^{\rm thr})$, {\it i.e.} the distance 
travelled by muons before their energy drops below 
$E_{\mu}^{\rm thr}$. The function $P_{\mu}(E_{\nu},y;\, E_{\mu}^{\rm thr})$
 is thus given by
\begin{equation}
P_{\mu}(E_{\nu},y;\, E_{\mu}^{\rm min}) =
 N_A\,R(E_{\mu},\, E_{\mu}^{\rm min}) \,
  \frac{d\sigma_{CC}^{\nu N}(E_{\nu},y)}{dy}
\label{prob}
\end{equation}
where $N_A=6.022\times 10^{23}\,g^{-1}$ is Avogadro's number and 
$d\sigma_{CC}^{\nu N}(E_{\nu},y)/dy$ is the differential
cross section for neutrino--nucleon charged--current scattering. 

The cross--section used in Eq.~\ref{prob} is described in high-energy physics textbooks, but carries uncertainties 
due to our limited knowledge of parton densities. It can be expressed as
\begin{equation}
\sigma^{\stackrel{(-)}{\nu}N}(s)=\int_0^1 dx\ \int_0^1 dy\
\frac{d^2\sigma^{\stackrel{(-)}{\nu}N}}{dx dy}
\end{equation}
with
\begin{equation}
\frac{d^2\sigma^{\stackrel{(-)}{\nu}N}}{dx dy} = \frac{G_F^2 s}{2 \pi}\
(1+xys/M_W^2)^{-2} \big[ (1-y) F_2^{\stackrel{(-)}{\nu}} + y^2 x
F_1^ {\stackrel{(-)}{\nu}}\pm y(1-\frac{y}{2}) x F_3^{\stackrel{(-)}{\nu}} \big]
\end{equation}
where $F_i=F_i(x,Q^2=xys)$ are the structure functions, 
$s=2M_N E_{\stackrel{(-)}{\nu}}$ and $G_F=1.1663\times
10^{-5}\ {\rm{GeV}}^{-2}$. For details regarding the calculation of high-energy neutrino-nucleon interactions including structure functions, see Ref.~\cite{Gandhi:1995tf}).


\begin{center}
\begin{figure}
\centering
\includegraphics{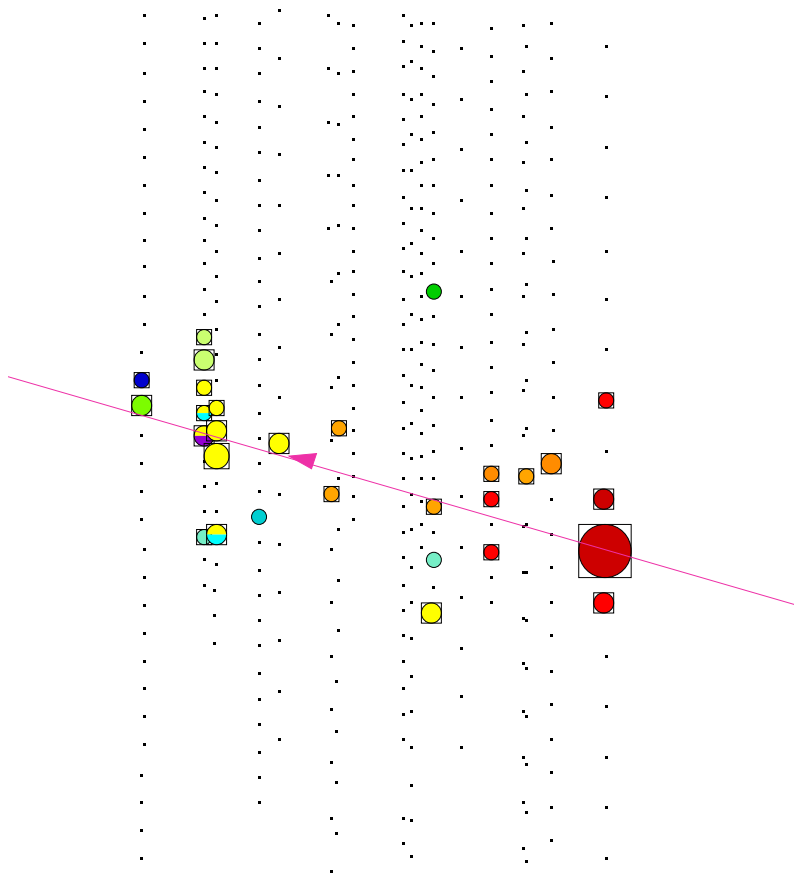}
\caption[Neutrino event in AMANDA]
{A muon neutrino event in AMANDA. Shown is the central 
part of the detector. The colorscale and symbol size correspond 
to hit time and amplitude~\cite{Wischnewski:2002wy}.}
\label{neutrevent}
\end{figure}
\end{center}

The muon range, $R(E_{\mu},\, E_{\mu}^{\rm min})$, appearing
in Eq.~\ref{prob}  follows from the energy--loss equation
\cite{Groom:in}
\begin{equation}
-dE_{\mu}/dX = \alpha_{\mu}(E_{\mu})
  +\beta_{\mu}(E_{\mu})\, E_{\mu},
\label{murange}
\end{equation}
with $X$ being the thickness of matter traversed
by the muon, and the quantities $\alpha_{\mu}(E_{\mu})$ 
and $\beta_{\mu}(E_{\mu})$ are the ionization loss 
and the fractional energy loss coefficients, respectively.
Integrating this result, we get the muon range
\begin{equation}
R(E_{\mu},\, E_{\mu}^{\rm min})\equiv
  X(E_{\mu}^{\rm min})-X(E_{\mu}) = 
   \frac{1}{\beta_{\mu}}\, \ln\, 
    \frac{\alpha_{\mu}+\beta_{\mu}E_{\mu}}
     {\alpha_{\mu}+\beta_{\mu}E_{\mu}^{\rm min}}
      \, .
\end{equation}
We adopt here the following values for the coefficients:
$\alpha_{\mu}=2.0\times 10^{-3}$ GeV (cm we)$^{-1}$
(${\rm cm we}\equiv g/{\rm cm}^2$) and 
$\beta_{\mu}=6.0\times 10^{-6}$ (cm we)$^{-1}$
~\cite{Giesel:2003hj}. 

We pass now to a brief description of existing and future
neutrino telescopes, focusing on kilometer--scale experiments.
The key idea is to detect muons, originating from neutrino 
fluxes as discussed above,  building large arrays of 
photo--multipliers deep in the ice, in a lake or in the sea,
to search for the Cerenkov light they are expected to emit 
as they move through these media.

The early pioneering effort made by the DUMAND collaboration 
~\cite{Grieder:iz} was followed by the deployment of the 
Lake Baikal experiment~\cite{Balkanov:1999rd} and of AMANDA 
~\cite{Andres:1999hm,amanda} at the South Pole. Although these experiments
have observed neutrinos produced in the Earth's atmosphere,
they have not, thus far, identified any extra--terrestrial neutrinos.

AMANDA, with approximately 50,000 square meters of effective area (at trigger
level) and a 30 GeV muon energy threshold, has been taking data for several years in its current configuration. ANTARES \cite{Aslanides:1999vq,antares}, currently under construction in the Mediterranean, will have a similar effective area and a lower energy threshold of about 10 GeV. Unlike with experiments at the South Pole, ANTARES will be sensitive in the direction of the galactic center.

IceCube\cite{Ahrens:2003ix,icecube}, beginning construction in 2005, and scheduled for completion in 2010, will be the first kilometer scale high-energy neutrino telescope. Using technology similar to AMANDA, IceCube will be considerably more sensitive to dark matter annihilations than current experiments. Even larger, and perhaps lower threshold, experiments may be needed beyond IceCube to further search for evidence of dark matter annihilations. For a review of high-energy neutrino astronomy, see Refs.~\cite{nureview1,nureview2}.

\subsection{Positron and Anti-Proton Experiments}
\label{posantipro}

Evidence for dark matter annihilations may also be observed in the spectra of cosmic positrons or anti-protons. Unlike gamma-rays and neutrinos, however, these charged particles do not point to their source due to the presence of galactic magnetic fields. Here we describe some of the experiments most important to these measurements.

The HEAT (High-Energy Antimatter Telescope) experiment made its first balloon flight in 1994-1995, measuring the spectrum of positrons between 1 and 30 GeV \cite{heat1995}. The results of this flight were very interesting, as they indicated an excess in the positron flux peaking at about 9 GeV and extending to higher energies. This excess could be a signature of dark matter annihilation in the local galactic halo (see section~\ref{positronsec}). A second HEAT flight in 2000 confirmed this observation \cite{heat2000,heat}.

The BESS (Balloon borne Experiment Superconducting Solenoidal spectrometer) experiment has had several successful balloon flights since 1993, providing the most detailed measurements of the cosmic anti-proton spectrum to date in the range of about 200 MeV to 3 GeV \cite{bess,bess2}. Above this energy, up to about 40 GeV, the CAPRICE experiment provides the best anti-proton measurements \cite{caprice}. There appears to be a mild excess in the anti-proton spectrum in the hundreds of MeV range, although it is very difficult to assess this result with any certainty.

\begin{center}
\begin{figure}
\centering
\includegraphics[width=1.0\textwidth]{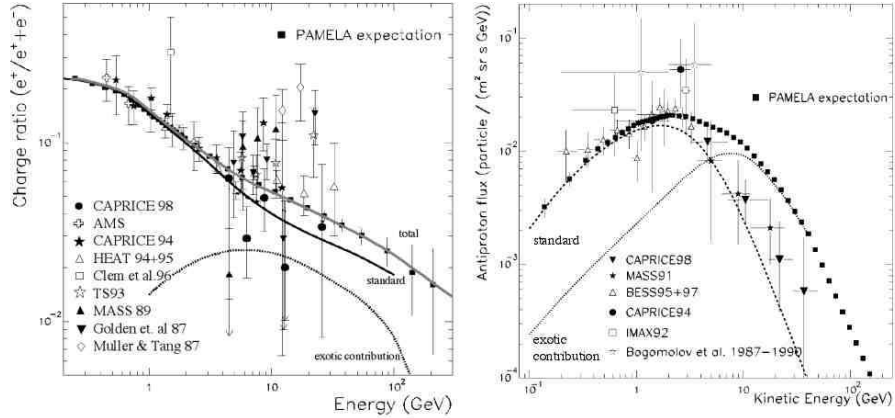}
\caption{The projected ability of the PAMELA experiment to measure the spectra of cosmic positrons (left) and anti-protons (right). A contribution from annihilating neutralino dark matter is included in the spectra shown. Notice, in comparison to the measurements made by HEAT, CAPRICE, BESS and other experiments, the dramatic improvement in precision. Also note the reach to higher energies made possible with PAMELA. From Ref.~\cite{Picozza:2002em}.}
\label{pamela}
\end{figure}
\end{center}

In the future, the experimental sensitivity to the cosmic positron and anti-proton spectra is likely to improve a great deal. Perhaps as early as 2005, the satellite borne PAMELA experiment, will begin its mission, measuring the spectra of both cosmic positrons and anti-protons with considerably improved precision. The primary objective of PAMELA is to the measure the cosmic anti-proton spectrum in the range of 80 MeV to 190 GeV and the cosmic positron spectrum in the range of 50 MeV to 270 GeV, far beyond the energies measured by HEAT, BESS or CAPRICE.  In Fig.~\ref{pamela}, we show the projected sensitivity of PAMELA to cosmic positrons (left) and anti-protons (right). The results are shown assuming a contribution from annihilating neutralino dark matter. It is clear that PAMELA will measure these spectra to far greater precision than previous experiments, especially at high energies (above $\sim$10 GeV).

AMS (the Alpha Magnetic Spectrometer) will considerably refine the measurement of the positron spectrum in its next manifestation, called AMS-02, onboard the International Space Station \cite{ams02}. AMS-02, with a 5000 $\rm{cm}^2 \, \rm{sr}$ aperture and a 1000 day duration, will provide exceptional precision in measuring the spectrum of cosmic positrons.

\subsection{Observations at Radio Wavelengths}
\label{radexp}

Observations at radio wavelengths belong to the realm
of ``classical''  astronomy. Radio emission from the galactic halo, particularly from the galactic center, can provide a valuable probe of particle dark matter. 

Electrons and protons produced in dark matter annihilations in the Galactic
Center region will emit synchotron radiation (at radio wavelengths) as they propagate through the galactic magnetic fields (see section~\ref{sec:synchro}). 

The observed Sgr A* (Galactic Center) radio emission could be explained 
in terms of synchrotron radiation emitted by 
shock-accelerated electrons (for more details
see Ref.~\cite{Liu:2004zi} and references therein).

Rather than reviewing the subject of radio observations of the galactic halo, we refer to Cane (Ref.~\cite{cane}) and to Brown (Ref.~\cite{radiobook}), which also
includes an interesting discussion of the absorption of
radio emission at different wavelengths. A complete catalog of observations of the Galactic center
at all frequencies, and in particular at radio wavelengths,
can be found in Ref.~\cite{narayan}. 
Additional information on specific measurements can be
found in Refs.~\cite{davies,ananta,lazio}. 


\section{Direct Detection}              
\label{direxp}

\begin{figure}[t]
\begin{center}
$\begin{array}{c@{\hspace{0.5in}}c}
\includegraphics[width=0.4\textwidth]{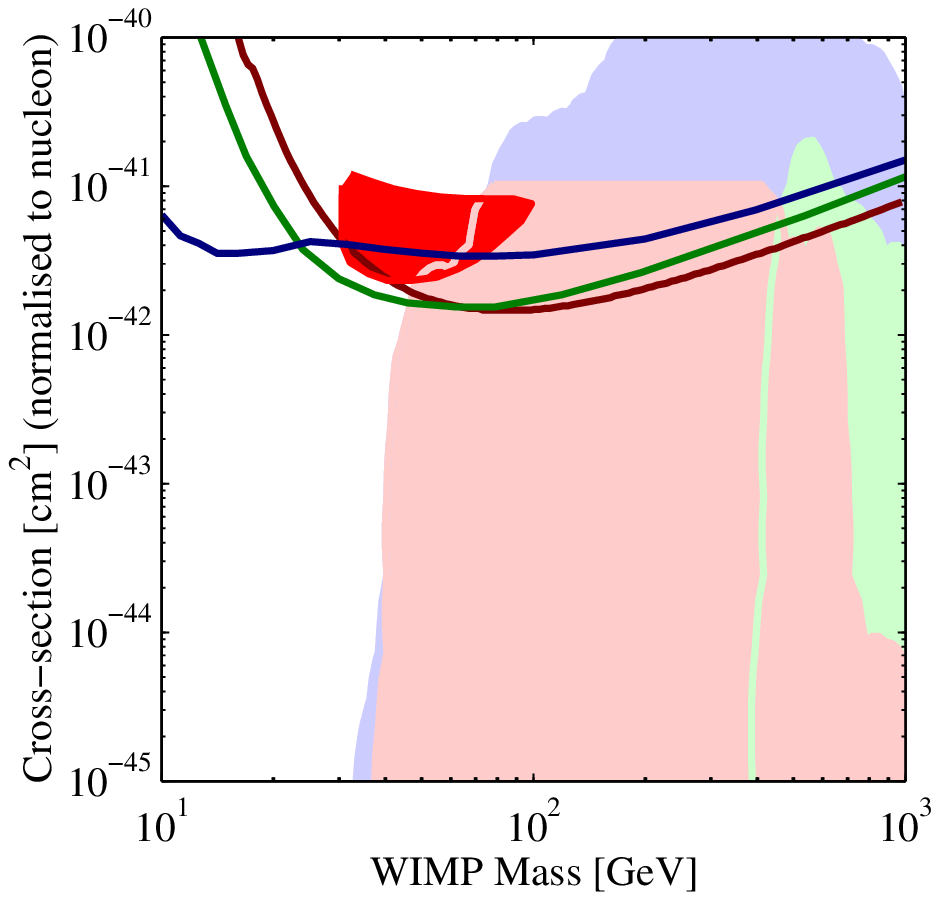} &
\includegraphics[width=0.4\textwidth]{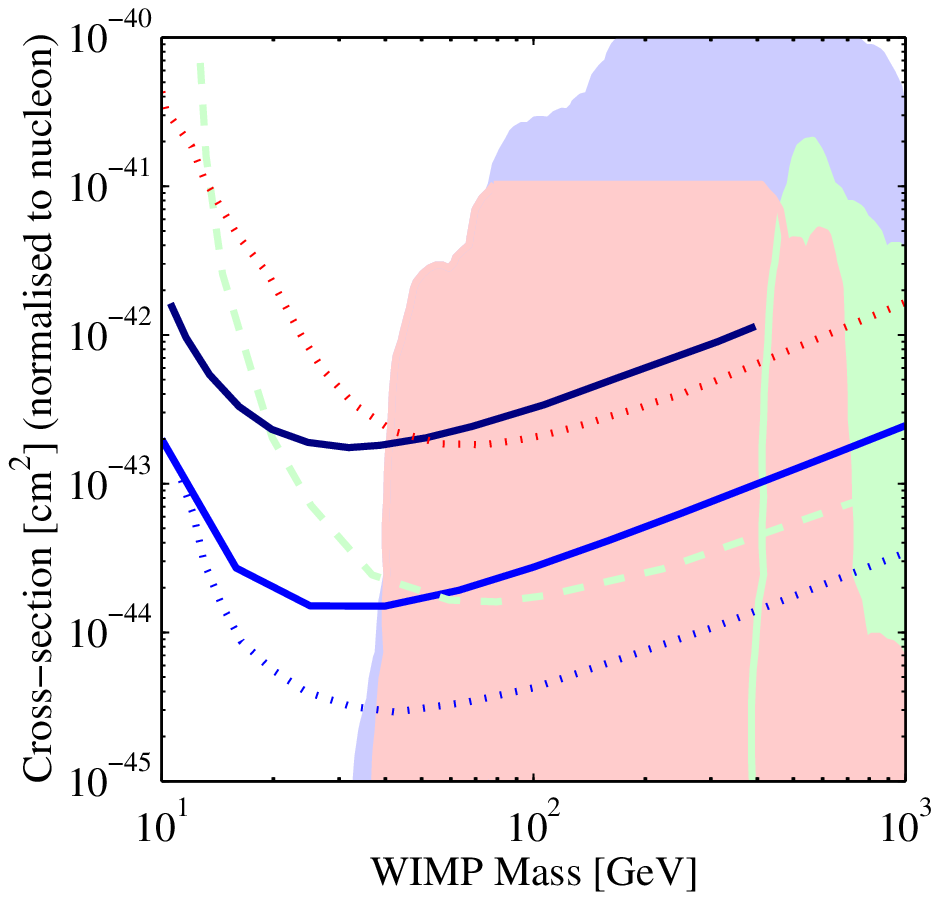} \\ [0.4cm]
\end{array}$
\caption{Current (left) and future (right) sensitivities of direct detection experiments. In the left frame, from top to bottom along the right side of the figure, the current limits from the CDMS, ZEPLIN-I and Edelweiss experiments are shown. The filled region near 30-100 GeV and $10^{-41}\,\rm{cm}^2$ is the parameter space favored by the DAMA experiment. In the right frame, from top to bottom along the right side of the figure, the projected reach of the GENIUS test facility (solid), CRESST-II (dots), CDMS-Soudan (solid), Edelweiss-II (dashed) and ZEPLIN-MAX (dots) are shown. In each frame, as filled regions, the space of models predicted by supersymmetry are shown \cite{gondolo}. The narrow region along the right side of the figure represents higgsino-like models, the region that reaches to the top of the figure represents mixed higgsino-gaugino models and the largest region represents gaugino-like models. These figures were made using the interface found at http://dendera.berkeley.edu/plotter/entryform.html.}
\label{direct}
\end{center}
\end{figure}

Many direct detection experiments have already produced quite strong limits on the elastic scattering cross section with protons or neutrons of potential dark matter candidates. Furthermore, experiments in the coming years will improve on current limits by several orders of magnitude making the prospects for discovery very great.

Presently, the best direct detection limits come from the CDMS, Edelweiss and ZEPLIN-I experiments, shown in the left frame of Fig.~\ref{direct}. These limits are for spin-independent (scalar) interactions.  With modern experiments, which use very heavy target nuclei, spin-dependent scattering experiments are not as sensitive to most dark matter candidates.

Also shown in the left frame of Fig.~\ref{direct} is the region in which the DAMA experiment claims a discovery (see \eg
Ref.~\cite{Bernabei:2003xg} for a recent review). DAMA, 
located at the INFN laboratories under the Gran Sasso
mountain in Italy and consisting of high purity NaI
crystals, has reported an annual modulation of their
event rate consistent with the detection of a
WIMP with a mass of approximately 60 GeV and a scattering
cross section on of the order of $10^{-41}$ cm$^2$.

Other experiments, such as EDELWEISS~\cite{Benoit:2002hf}
and CDMS \cite{Akerib:2003px} have explored the parameter space favored by DAMA
without finding any evidence of dark matter. A recent model independent
analysis has shown that it is difficult to reconcile the DAMA result with other
experiments \cite{Kurylov:2003ra} (see also Ref.~\cite{Ullio:2000bv})
although it may still be possible to find exotic particle candidates and halo 
models which are able to accommodate and explain the
data from all current experiments (for example, see Refs.~\cite{Prezeau:2003sv,Gelmini:2004gm,Smith:2001hy,Tucker-Smith:2004jv}).

Theoretical and experimental results on direct detection are usually 
obtained under some simplifying assumptions on the dark matter profile. In 
particular, an isothermal profile is often assumed, with $\rho \propto r^{-2}$
(thus, with a flat rotation curve), a local density of $\rho_0=0.3$ GeV cm$^{-3}$,
and a Maxwell-Boltzmann velocity distribution with a characteristic velocity of
$v_0=270$ km s$^{-1}$. Uncertainties in the
density and velocity distribution of dark matter lead to the enlargement of
the allowed region in the cross section-mass plane shown in Fig.~\ref{direct}, however, extending the mass range up to $\sim250$ GeV and the cross section range down
to $\sigma_{\chi-n}\sim 10^{-7}$pb~\cite{Belli:1999nz,Brhlik:1999tt}. If this results were confirmed, it could
explain the discrepancy between the observational findings of different
experiments. Unfortunately, subsequent analysises lead to different results
(see in particular Refs.~\cite{Copi:2002hm,Green:2003yh}), leaving the experimental
situation unclear.

Nevertheless, the DAMA collaboration
(whose raw data are not publicly available) insists
on the compatibility of their result with null searches 
of other experiments~\cite{Bernabei:2003xg}, questioning 
specific experimental issues like rejection procedures and
energy scale determination.  

Next generation experiments should clarify the 
experimental situation, thanks to the large 
improvement expected in
sensitivity, around two orders of magnitude in
scattering cross section for EDELWEISS II and 
even more for ZEPLIN-MAX (See Fig.~\ref{direct}, 
generated using the {\it dark matter limit plot 
generator} at \\ \href{http://dendera.berkeley.edu/plotter/entryform.html}
{http://dendera.berkeley.edu/plotter/entryform.html}).

\begin{figure}[h]
\begin{center}
\includegraphics[width=0.7\textwidth,clip=true]{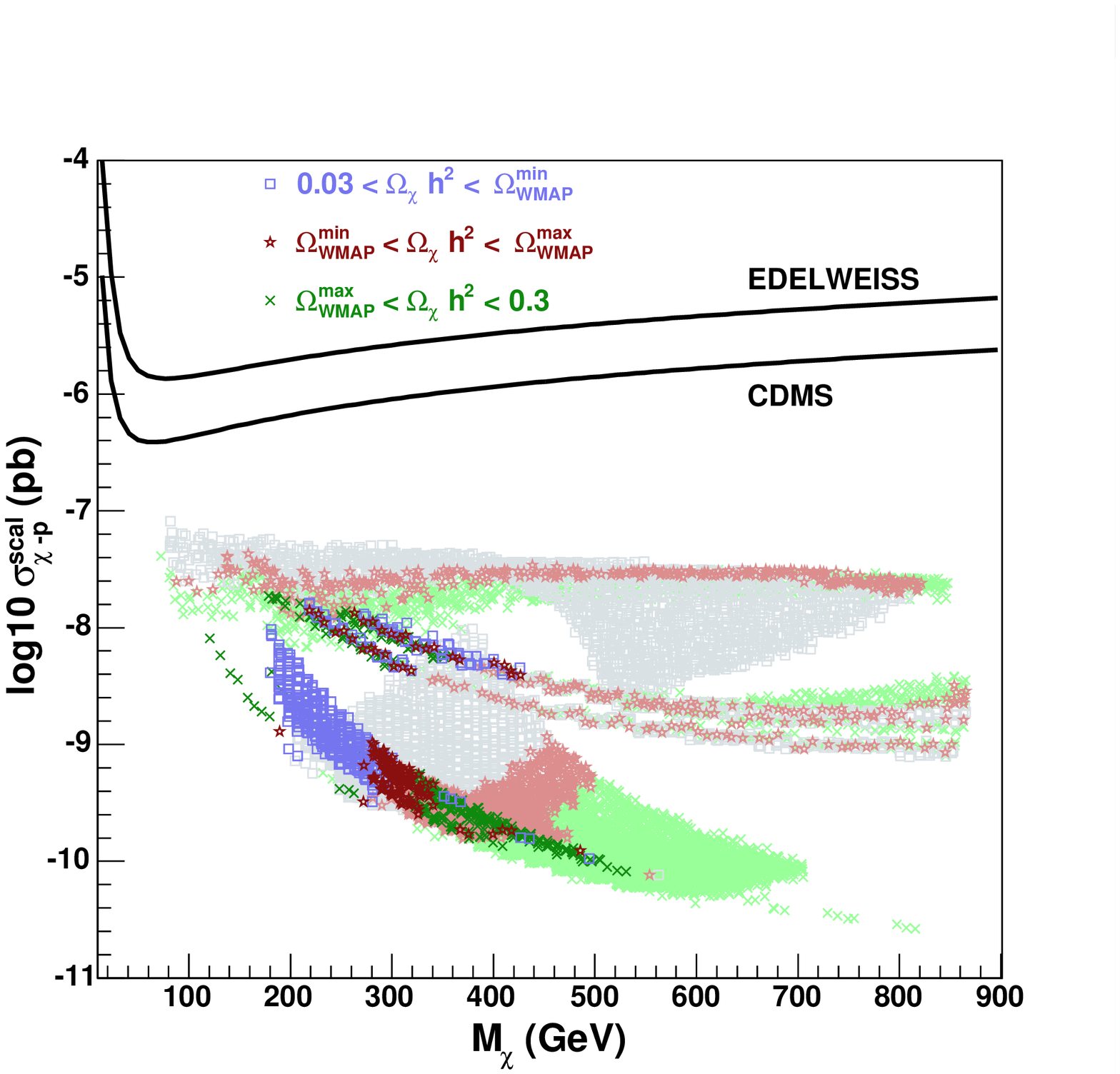}
\caption{Current experimental sensitivity of WIMP direct detection
experiments.Shown are the upper limits set by EDELWEISS ~\cite{Benoit:2002hf} 
and CDMS at Soudan ~\cite{Akerib:2004fq}. For comparison we also 
show predictions for different supersymmetric scenarios (see text).
Figure kindly provided by E.~Nezri.}
\label{DDmsugra}
\end{center}
\end{figure}

In Fig.~\ref{DDmsugra} we show the current constraints on neutralino 
dark matter in different supersymmetric scenarios. 
Shades paler than in the legend denote a value of the muon anomalous
magnetic moment outside of the $2\sigma$~ range \cite{gminus2current3} $
  8.1 \times 10^{-10}<\delta_{\mu}^{\mathrm{susy}} =
  \delta_{\mu}^{\mathrm{exp}}-\delta_{\mu}^{\mathrm{SM}}<44.1 \times 10^{-10}$
(see Ref.~\cite{Bertone:2004ag} for more details).
Future generation detectors will probe a wide portion of the 
supersymmetric parameter space and will give important insights
into the nature of dark matter particles. The $\bone$ particle (KK dark matter) should have
a scattering cross section with nucleons in the range of
$10^{-10}$--$10^{-12}$~pb, depending on its mass and on 
the mass difference with KK quark states~\cite{Servant:2002hb}.

\section{Indirect Detection}
\label{sec:annihi}

Indirect detection of dark matter is the technique of observing the 
radiation produced in
dark matter annihilations. The flux of such radiation
is proportional to the annihilation rate,
which in turn depends on the square of the dark
matter density, $\Gamma_A \propto \rho_{DM}^2$. Therefore, the ``natural'' places to look at, when searching for 
significant fluxes, are the regions where large dark matter densities 
accumulate. We will also refer to these regions or objects as {\it amplifiers}.

Dense regions of the galactic halo, such as the galactic center, may be excellent amplifiers for the purposes of detecting gamma-rays or neutrinos. Other astrophysical objects, such as the Sun or the Earth, could also as act as amplifiers for dark matter annihilations by capturing dark matter particles as they
lose energy through scattering with nucleons in the 
interiors of these objects. Only neutrinos can escape these dense objects, however. Annihilation products which are charged move under the influence of magnetic fields making it impossible to consider point sources of such radiation. Despite this, observations of cosmic positrons and anti-protons can be valuable tools in searching for particle dark matter.

In this section, we compare the predictions for gamma-ray, synchrotron, neutrino, positron and anti-proton fluxes from dark matter annihilation with current experimental
data and with the expected sensitivities of future 
experiments. We will show that by using these techniques, it is possible to constrain dark matter models and, in the future, potentially detect the presence of particle dark matter. 

\subsection{Gamma-rays and neutrinos from the Galactic center}
\label{sec:gammagc}

One of the most interesting regions for the indirect detection of dark matter
is the galactic center, where, according to the results of numerical simulations, the dark matter density 
profile is expected to grow as a power--law, $\rho \propto
r^{-\alpha}$. The possible values of $\alpha$, as well as alternative density profiles, have been 
discussed in chapter~\ref{obs}. Also recall from chapter~\ref{obs} that an additional 
enhancement of the density in this region could result from the process 
of adiabatic accretion onto the supermassive black hole at the galactic center.

Gamma-ray emission from the Galactic center has been discussed in the past 
by numerous authors (see \eg Bouquet, Salati and Silk~\cite{bousasi}, 
Stecker~\cite{Stecker:dz}, Berezinsky \etal~\cite{ber}, 
Bergstrom, Ullio and Buckley~\cite{Bergstrom:1997fj} for
neutralinos, Bertone, Servant and Sigl~\cite{Bertone:2002ms} for $\bone$ particles, Bertone and Sigl and Silk~\cite{Bertone:2002je} for the case of a density spike at the galactic center). Here, we will review these calculations and arguments for evaluating the prospects
for the indirect detection of dark matter near the Galactic center with present and 
next-generation experiments.

The flux of dark matter annihilation products is proportional to the number of annihilations per 
unit time, per unit volume, $\propto \sigma v \; n^2(r) \equiv
 \sigma v \rho^2(r)/m_{\rm{DM}}^2$, where $n(r)$ and $\rho(r)$ are the number and the mass density of a
dark matter particle, respectively. $m_{\rm{DM}}$ is the dark matter particle's mass and $\sigma v$ is its annihilation cross section multiplied by velocity. $r$ is the distance from the galactic center. The flux is also proportional to the spectrum of secondary particles of species, $i$, per 
annihilation, $dN_i/dE$. 
The flux observed is found by integrating the the density squared along the line-of-sight connecting the observer (the Earth) to the \GC. Including all factors, the observed flux can be written as
\begin{equation}
\label{fluxgamma2}
\Phi_i(\psi,E)=\sigma v \frac{dN_i}{dE} \frac{1}{4 \pi m_{\rm{DM}}^2}
\int_{\mbox{line of sight}}d\,s
\rho^2\left(r(s,\psi)\right)\label{flux},
\end{equation}
where the index $i$ denotes the secondary particle observed 
(in this section, $\gamma-$rays and neutrinos) and the
coordinate $s$ runs along the line of sight, in a
direction making an angle, $\psi$, from the direction
of the galactic center. If the dark matter particle is not its own anti-particle (particle-antiparticle annihilation), Eq.~\ref{fluxgamma2} is reduced by a factor of 2.

In order to separate the factors which depend on the halo profile 
from those which depend only on particle physics, we introduce,
following Ref.~\cite{Bergstrom:1997fj}, the quantity $J(\psi)$:
\begin{equation}
J\left(\psi\right) = \frac{1} {8.5\, \rm{kpc}} 
\left(\frac{1}{0.3\, \mbox{\small{GeV/cm}}^3}\right)^2
\int_{\mbox{\small{line of sight}}}d\,s\rho^2\left(r(s,\psi)\right)\,.
\label{gei}
\end{equation}
We define $\overline{J}(\Delta\Omega)$ as  the average of $J(\psi)$ over
a spherical region of solid angle, $\Delta\Omega$, centered on $\psi=0$.
The values of $\overline{J}(\Delta\Omega=10^{-3}\, \rm{str})$ are shown in the
last column of table 7 for the respective density profiles. 

We can then express the flux from a solid angle, $\Delta\Omega$, as
\be
\Phi_{i}(\Delta\Omega, E)\simeq5.6\times10^{-12}\,\frac{dN_i}{dE} 
\left( \frac{\sigma v}
{\rm{pb}}\right)\left(\frac{1\rm{TeV}} 
{m_{\rm{DM}}} \right)^2 \overline{J}\left(\Delta\Omega\right) 
 \; \Delta\Omega\,\,\rm{cm}^{-2} \rm{s}^{-1}\,.
\label{final}
\ee

\begin{center}
\begin{table}
\centering
\begin{tabular}{|c|ccccc|}
\hline 
&$\alpha$&$\beta$&$\gamma$&R (kpc)&$\overline{J}\left( 10^{-3}\right)$ \\
\hline 
Kra& 2.0& 3.0&0.4 & 10.0 &$ 2.166 \times 10^1$ \\
NFW& 1.0& 3.0& 1.0& 20& $1.352 \times 10^3$\\
Moore& 1.5& 3.0& 1.5& 28.0 &$ 1.544 \times 10^5$ \\
Iso& 2.0& 2.0& 0& 3.5&$2.868 \times 10^1$\\ 
\hline 
\end{tabular}
\label{tabjval}
\caption{Parameters of some widely used density profiles and corresponding value of $\overline{J}(10^{-3}\,\rm{str})$. For more on halo profiles, see chapter~\ref{obs}.}
\end{table}
\end{center}

\begin{figure}
\centering
\includegraphics[width=0.8\textwidth,clip=true]{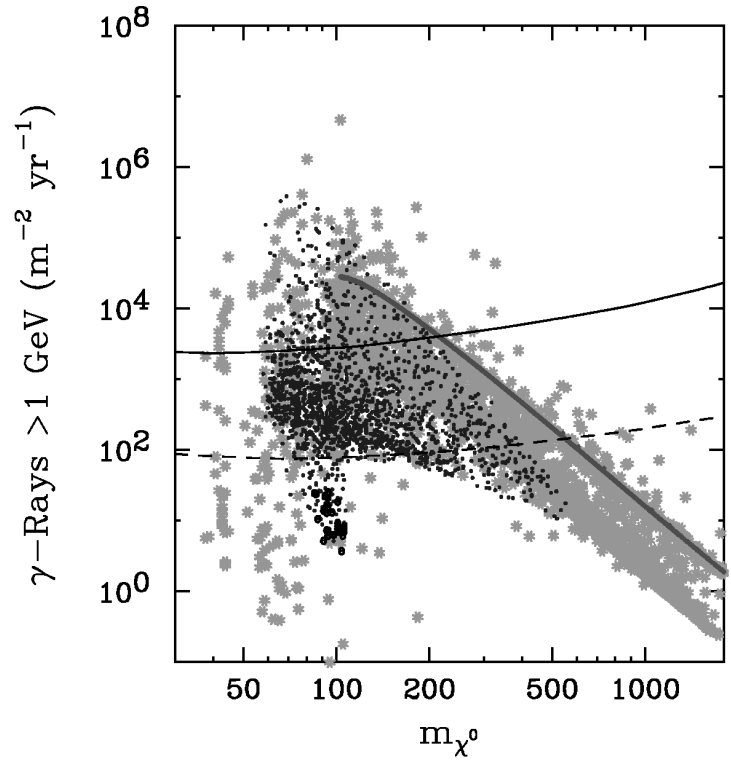}
\caption{The flux of gamma-rays above 1 GeV per square meter per year from the
Galactic center from annihilations of neutralino dark matter. A NFW halo profile has been used. For each point, the thermal relic density is below the maximum value allowed by WMAP. The solid and dashed lines are the limit from the EGRET experiment and predicted sensitivity for GLAST, respectively \cite{hooper}. The various shadings refer to different scenarios of supersymmetry breaking. For more information, see Ref.~\cite{wang}.}
\label{gammacont}
\end{figure}

\begin{figure}
\centering
\includegraphics[width=0.8\textwidth,clip=true]{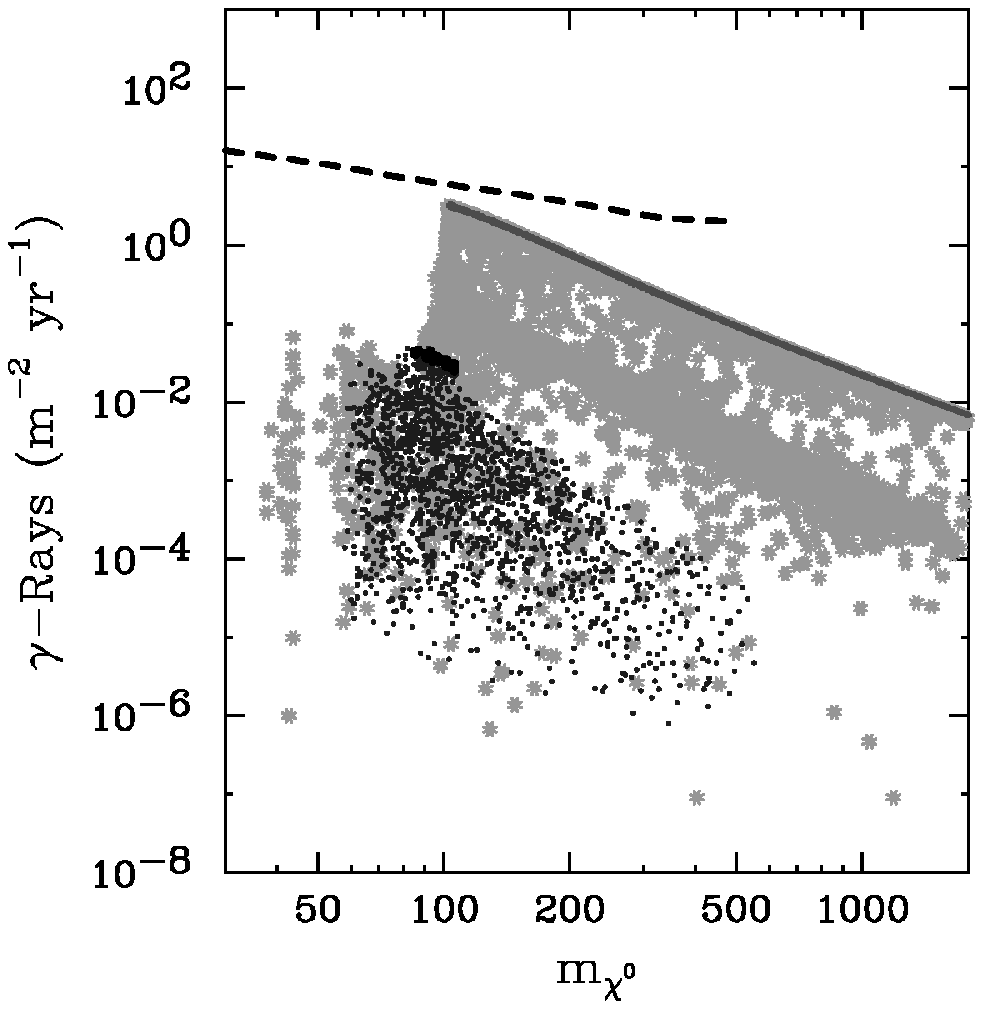}
\caption{As in the previous figure, but from the annihilation of neutralino dark matter to a $\gamma \gamma$ line. The dashed line represents the predicted sensitivity for GLAST. The flux for neutralino annihilation to $\gamma Z$ is similar. For more information, see Ref.~\cite{wang}. }
\label{gammaline}
\end{figure}

\subsubsection{Prospects for Neutralinos}

To study the detectability of gamma-ray fluxes from neutralino
annihilations, extensive scans of the MSSM are conducted, retaining only the small minority of models which are consistent 
with accelerator and cosmological constraints. We show in Fig.~\ref{gammacont}
the expected gamma-ray fluxes from the Galactic center for neutralino dark matter, considering an NFW halo profile. To adapt this spectrum to another profile, simply scale the flux by the value of $\overline{J}(\Delta\Omega=10^{-3}\, \rm{str})$ found in table 7. In Fig.~\ref{gammacont}, all continuum processes are included (typically dominated by annihilations to heavy quarks and gauge bosons for neutralino annihilation). Shown for comparison are the limit from EGRET and the projected reach of GLAST. For fairly heavy neutralinos, ACTs can also be effective. 

In addition to continuum gamma-ray emission, neutralinos can annihilate to mono-energetic gamma-ray lines via the processes $\chi \chi \rightarrow \gamma \gamma$ and $\chi \chi \rightarrow \gamma Z$~\cite{Rudaz:1990rt}. Such a line, if observed, would be a clear signature for dark matter annihilation (a ``smoking gun''). The flux of gamma-rays from such process is quite small, however, as no tree level feynman diagrams contribute to the process.  For the loop-level feynman diagrams which lead to gamma-ray line emission, see appendix~\ref{lines}.

\begin{figure}[h]
\centering
\includegraphics[width=\textwidth,clip=true]{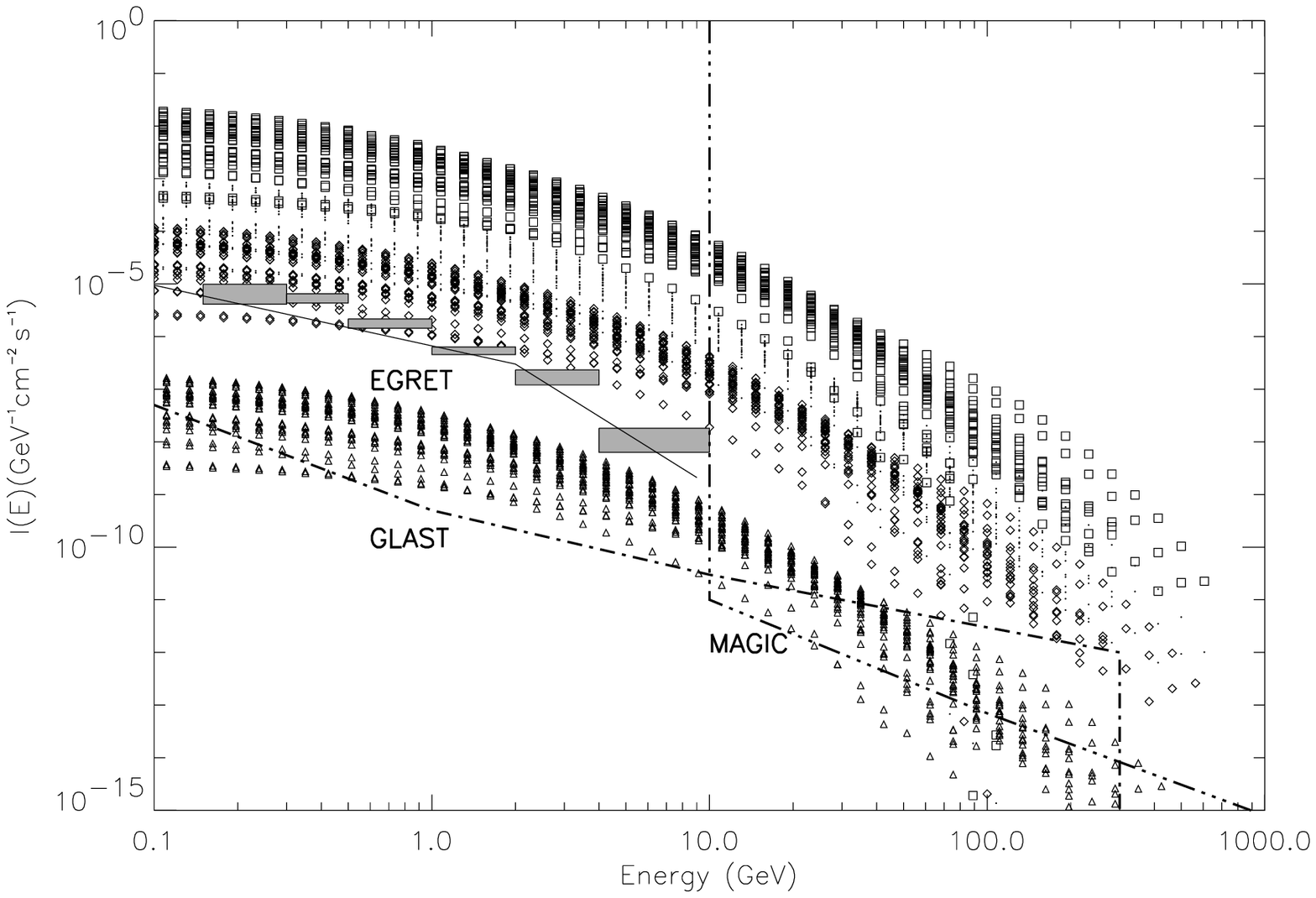}
\caption[Gamma-ray flux from the spike at the \GC (neutralinos)]
{Predicted gamma-ray fluxes for a large set of supersymmetric
models and halo profiles: $\gamma$=0.05(triangles), $\gamma$=0.12(diamonds), 
$\gamma$=0.2 (dots), $\gamma$=1.0 (squares). The flux observed by EGRET~\cite{narayan} is shown as grey boxes. Also shown are the projected sensitivities of for GLAST (1 month 
observation time) and MAGIC (50 hours).}
\label{gamGSallf}
\end{figure}

The gamma-ray fluxes predicted from the Galactic center can be considerably
enhanced if a density spike is considered. In Fig.~\ref{gamGSallf}, the
(continuum) gamma-ray flux from the Galactic center is shown for four different values of $\gamma$, the slope of the inner halo profile, in the presence of a density spike. For most of the models, a value of $\gamma$ between 0.05 and 0.1 can reproduce the scale of the gamma-ray flux observed by EGRET.


\subsubsection{Prospects for Kaluza-Klein dark matter}

\begin{figure}
\centering
\includegraphics[width=\textwidth,clip=true]{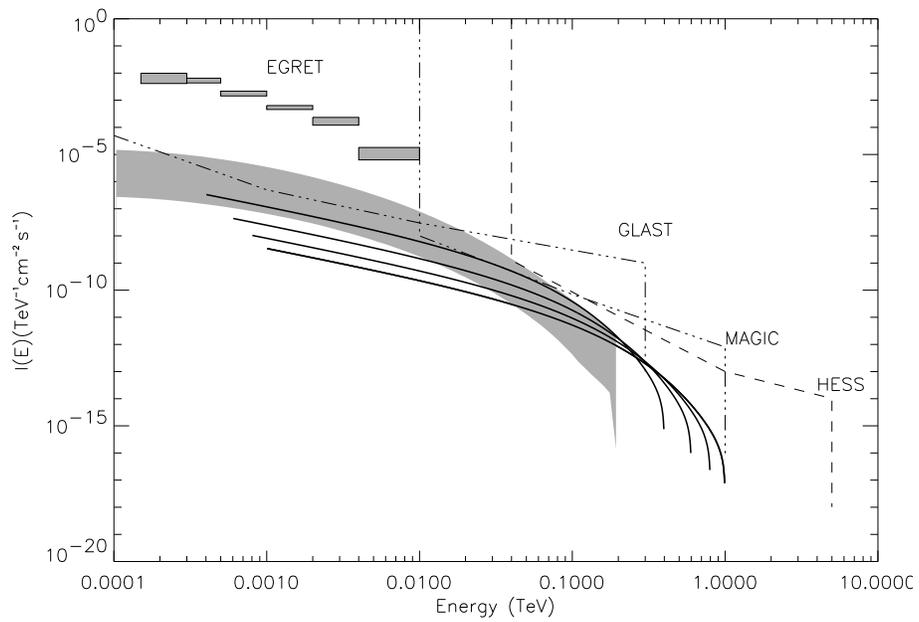}
\caption[Gamma ray flux from the spike at the \GC ($\bone$ particle)]
{Expected $\gamma-$ray fluxes for (top to bottom) $m_{\bone}=0.4$, 0.6, 0.8,
and 1 TeV and $\overline{J}\left(10^{-3}\right) = 500$. For
comparison shown are typical $\gamma-$ray fluxes predicted for
neutralinos of mass $\simeq200\,$GeV, (shadowed region) as well as EGRET \cite{mayer} data and expected 
sensitivities of the future GLAST\cite{Sadrozinski:wu}, MAGIC\cite{Petry:1999fm} and HESS\cite{Volk:2002iz} experiments.}
\label{gammas}
\end{figure}

Using the expression for the $\bone$ annihilation cross section found in 
Sec.~\ref{lkp}, the flux of annihilation products from the Galactic center can be simplified to
\be
\Phi_{i}(\Delta\Omega)  \simeq  3.4 \times 10^{-12}\frac{dN_i}{dE} 
\left( \frac{1\,\rm{TeV}} 
{m_{\bone}}\right)^4  \overline{J}\left(\Delta\Omega\right)
 \; \Delta\Omega \ \rm{cm}^{-2} \ \rm{s}^{-1}\,.
\ee
In Fig.~\ref{gammas}, the predicted $\gamma-$ray flux from KK dark matter
annihilations in the Galactic center is shown. Results for LKP masses of 0.4, 0.6, 0.8, and 1 TeV are shown. A halo profile with $\overline{J} \left( 10^{-3} \right) = 500$ has been used, although the effect of this choice is easily scaled with the values shown in table 7.

\begin{figure}
\centering
\includegraphics[width=\textwidth,clip=true]{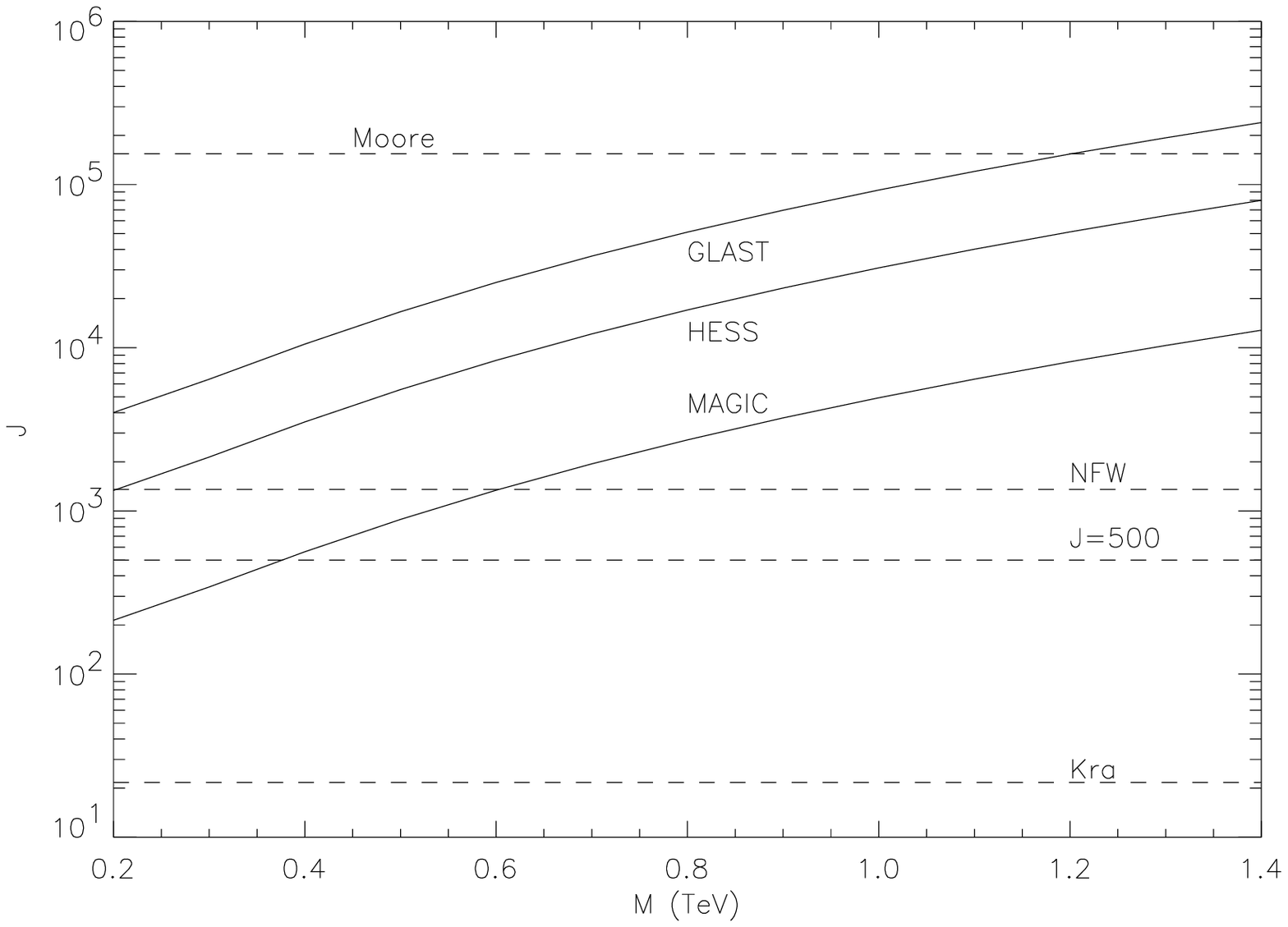}
\caption[Value of J required for the detectability of gamma rays]
{Value of $J=\overline{J}(10^{-3})$ required to produce $\gamma$ fluxes
observable by the future GLAST, MAGIC and HESS experiments, as a function of
the $\bone$ mass. For comparison we show the values of $J$ for some 
profiles discussed in the text.}
\label{constr}
\end{figure}

Unlike in the case of supersymmetry, with Kaluza-Klein dark matter, there are
few free parameters in calculating the gamma-ray spectrum from the Galactic
center ($m_{\bone}$ and $\overline{J}(\Delta\Omega)$). We can, therefore, easily place limits on the halo profile as a function of the LKP mass. We show in
Fig.~\ref{constr} the constraints on these parameters
based on the expected sensitivity of GLAST, MAGIC and HESS. 
For example, with an NFW profile, LKP masses below about 600 GeV will be excluded if MAGIC does not observe any radiation from the galactic center.

\begin{figure}
\centering
\includegraphics[width=\textwidth,clip=true]{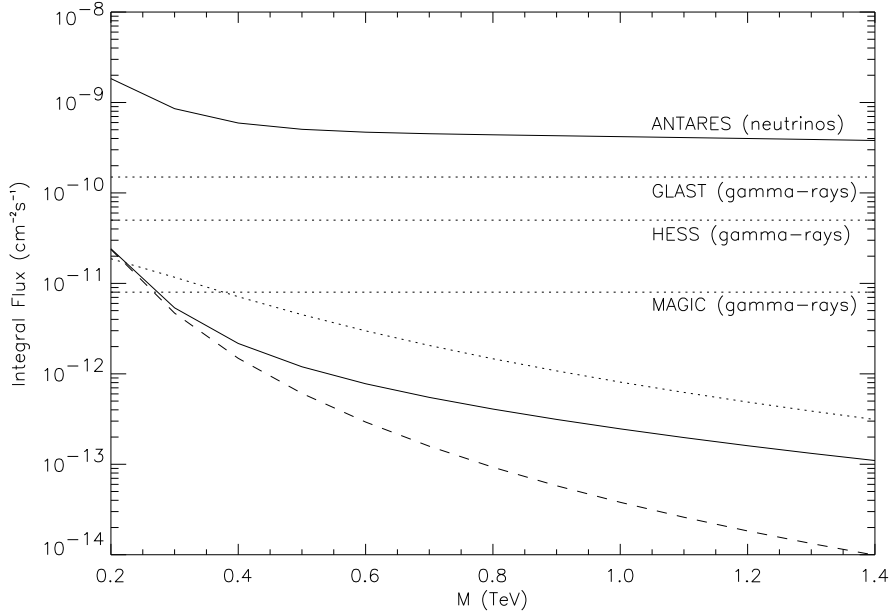}
\caption[Integral flux of $\gamma-$rays and neutrinos from the \GC]
{Integral flux of $\gamma-$rays (sloped dotted line) and 
muon neutrinos (solid) above 50 GeV, for 
$\overline{J}\left(10^{-3}\right) = 500$ . The dashed line shows 
the contribution of direct $\bone$ annihilation into neutrinos.
Horizontal lines are sensitivities 
of present and future experiments for $\gamma-$rays (dotted horizontal
lines) and neutrinos (upper solid line).}
\label{intflux}
\end{figure}

Neutrino telescopes will also be capable of searching for signals of dark
matter annihilation in the Galactic center (see Sec.~\ref{neuexp}), although these prospects are considerably poorer. In Fig.~\ref{intflux}, we plot the integral flux of muon neutrinos 
above $50$ GeV (solid line) as a function of the $\bone$ mass.
This result is obtained by adding the neutrino fluxes
from three different channels:
\begin{itemize}
\item Neutrinos produced directly in $\bone$ annihilations 
(dashed line), their spectrum being a line at energy $E_{\nu}=m_{\bone}$. 
\item Secondary neutrinos from decay of charged pions. This
spectrum can be evaluated using the expressions
for the charged pion decay found in Ref.~\cite{Lee:1996fp}.
\item Secondary neutrinos from ``prompt'' semi-leptonic
decay of heavy quarks (solid line). This spectrum is given, for example, in
Ref.~\cite{Jungman:1994jr}. 
\end{itemize}

We show in the same figure an estimate of
the sensitivity of the neutrino telescope ANTARES (upper 
solid line). To estimate this sensitivity, we first 
evaluated the rate of muons in ANTARES from the direction of 
the galactic center, which depends (see Eq.~\ref{rate}) 
on specific experimental quantities, such as the detector 
effective area and the threshold energy for the detection of 
muons. The rate is higher for more energetic neutrinos, being 
proportional to the muon range and to the neutrino-nucleon cross 
section, which are both increasing functions of energy. See Sec.~\ref{neuexp} for a discussion of neutrino telescopes.

Of these neutrino producing channels, ANTARES is most sensitive to those neutrinos produced directly in $\bone \bone \rightarrow \nu \nu$ annihilations. Although 
the branching ratio for this channel is an order of magnitude 
smaller than that into quark pairs, the neutrinos produced are 
emitted at the highest available energy, $E_{\nu}=m_{\bone}$, increasing their probability of being detected. Neutralinos, which do not annihilate directly to neutrinos, are more difficult to observe with neutrino telescopes.

In Fig.~\ref{intflux}, the integral flux of gamma-rays is shown for comparison, along with the projected sensitivity of future experiments GLAST, MAGIC and 
HESS. 


\subsubsection{The gamma-ray source at the Galactic center}
\index{gamma ray source at the GC}

The EGRET experiment (see Sec.~\ref{spacegamma}) has reported an excess 
of gamma-rays in the region of the galactic center, in an error circle 
of 0.2 degree radius including the position l=0(deg) and b=0(deg) 
\cite{mayer}. The name for this source, in the language of
high-energy astrophysicists, is 3EG J1746-2851. 
The radiation is well above the gamma-ray emission which would be expected 
from interactions of primary cosmic rays with the interstellar
medium (see, \eg, Refs.~\cite{Strong:1998fr,Cesarini:2003nr}).
%

It is intriguing to imagine that such excess emission 
could be the product of dark matter annihilations near the \GC.
However, it should be noted that some difficulties exist,
related to this interpretation. In fact, as shown in Refs.~\cite{hooper,Hooper:2002fx}, the EGRET source is not exactly coincident with the galactic center. This makes the interpretation of the EGRET signal as dark matter annihilation in a density spike (see Sec.\ref{smbh}) problematic.

Furthermore there is some evidence, though weak, that
the source could be variable. Such a result could rule out
completely the interpretation of the excess emission as
due to annihilation radiation from the galactic center. The 
variability of 3EG J1746-2851 has been recently 
discussed in Ref.~\cite{nolan}.

We briefly note here that multiple Atmospheric Cerenkov Telescopes (ACTs) have
recently reported an excess of gamma-rays from the Galactic center region. The
VERITAS collaboration, using the Whipple telescope, have reported a flux of
$1.6 \pm 0.5 \pm 0.3 \times 10^{-8} \, \rm{m}^{-2} \, \rm{s}^{-1}$ above 2.8
TeV \cite{Kosack:2004ri}. The CANGAROO collaboration has reported a flux of
approximately $2 \times 10^{-6} \, \rm{m}^{-2} \, \rm{s}^{-1}$  from this
region in the range of 250 GeV to 1 TeV \cite{Tsuchiya:2004wv}. We eagerly
await the resuls of HESS, which should be the most sensitive to the Galactic
center region.

It is certainly too soon to determine whether the fluxes observed by these experiments are the product of dark matter annihilations or are the result of another process, most likely astrophysical \cite{actoxford}. Improvements in the measurement of the gamma-ray spectrum, and improved angular resolution will be needed to resolve this issue.

\subsubsection{Upper Limit for the Neutrino Flux from the GC}

Despite the large uncertainties associated with the distribution of dark matter in the innermost 
regions of our galaxy, it is possible to set an upper limit 
on the neutrino flux by requiring that the associated gamma-ray 
emission does not exceed the flux observed by EGRET (see previous 
section)~\cite{Bertone:2004ag}.  

The maximum neutrino flux is obtained by normalizing the 
associated flux of gamma--rays to the EGRET data. 
This corresponds to fixing, for each model, the product, 
$\sigma v \, N_{\gamma}$, with  
$N_{\gamma}= \sum_i N_i \, R_i$.
Here $R_i$ is the branching ratio of all the channels, $i$,
contributing $N_i$ gamma-rays above a given threshold energy.

Having fixed the particle physics contents of our dark matter candidate,
the ratio between the number of photons and
the number of neutrinos emitted per annihilation is known.

The rescaled flux of muons, $\phi^{\rm{norm}}_{\mu}(>E_{th})$, 
will thus be given by
\begin{equation}
\phi^{\rm{norm}}_{\mu}(>E_{th})= \frac{\phi^{\rm{NFW}}_{\mu}(>E_{th})
\, \phi^{\rm{EGRET}}_{\gamma}(E_*)}{\phi^{\rm{NFW}}_{\gamma}(E_*)},
\end{equation}
where the label NFW indicates that NFW profiles have been 
used to compute profile-independent flux ratios
and $E_*$ is the energy at which we decide to normalize 
the flux to the gamma-ray data (in our case, $E_*=2$ GeV).

\begin{figure}[t]
\begin{center}
$\begin{array}{c@{\hspace{0.5in}}c}
\includegraphics[width=0.4\textwidth]{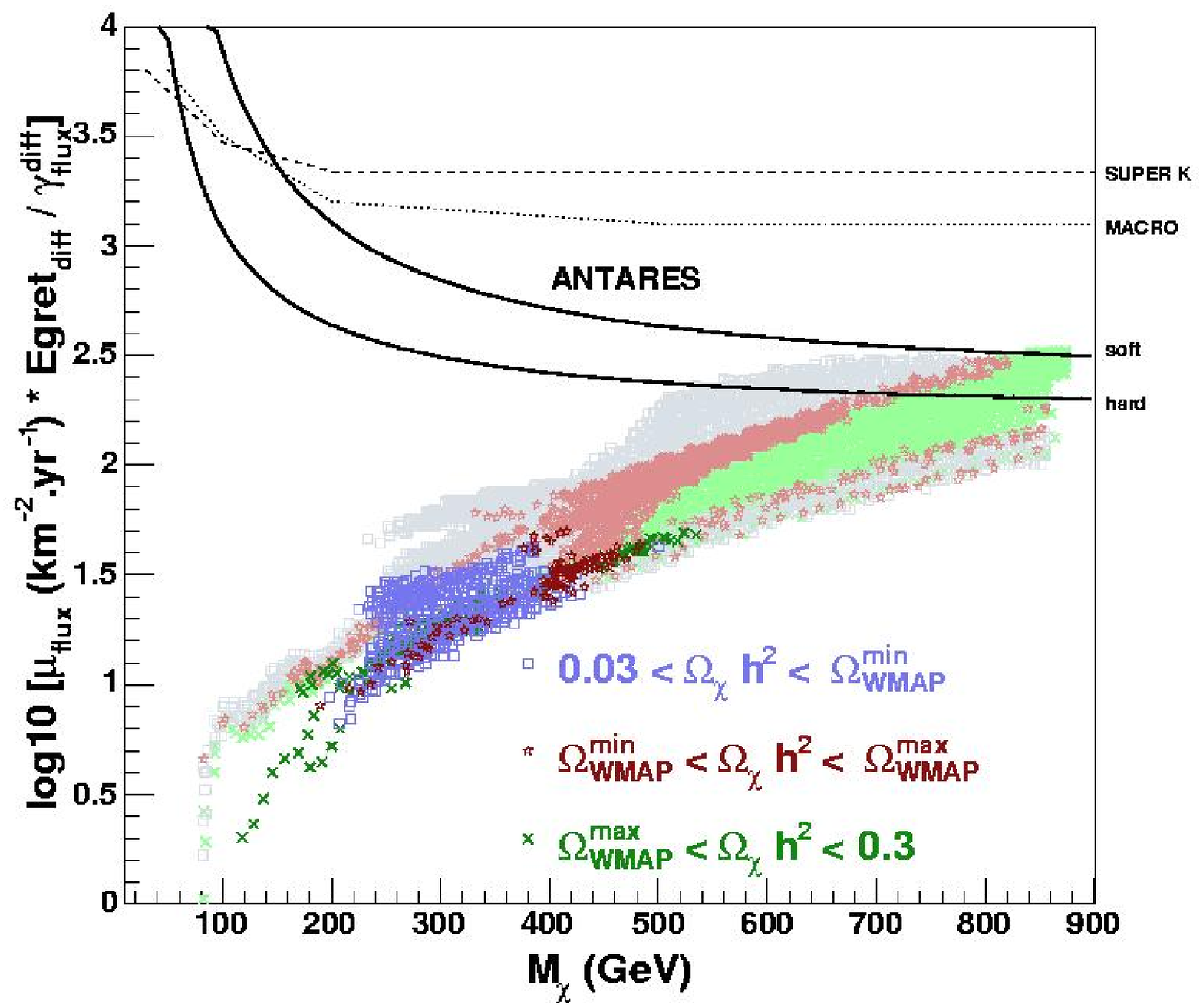} &
\includegraphics[width=0.4\textwidth]{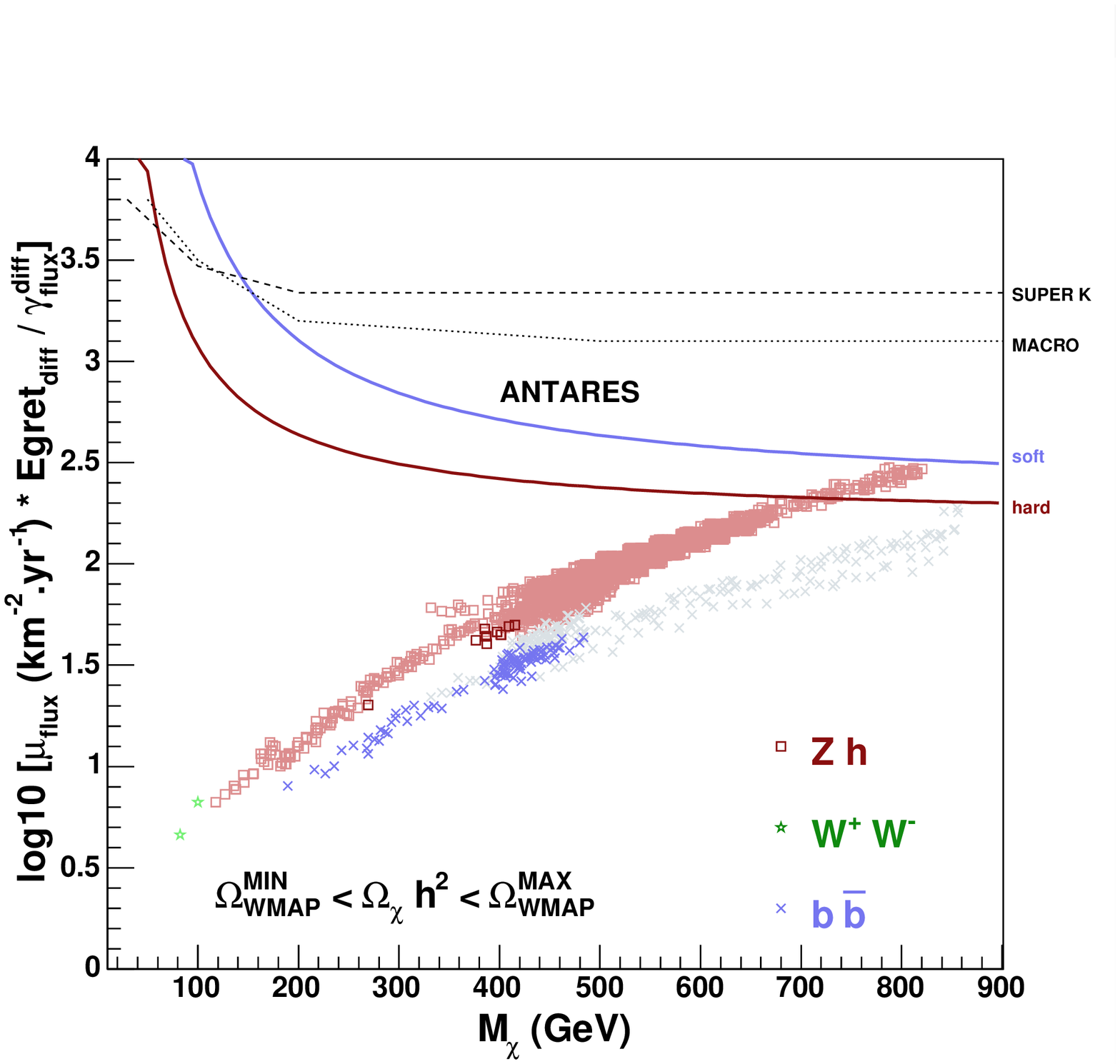} \\ [0.4cm]
\end{array}$
\caption{Neutrino-induced muon flux from the \GC normalized to EGRET. Models
are sorted by relic density (left) and leading annihilation channel (right).
Shades paler than in the legend denote a low $\delta_\mu^{\mathrm{susy}}$ value.}
\label{NeutrinosEGRETOh2}
\end{center}
\end{figure}

The results are shown in Fig.~\ref{NeutrinosEGRETOh2}, 
where shades paler than in the 
legend denote a low value for the muon's magnetic moment (see Ref.~\cite{Bertone:2004ag}
for more details). The neutrino induced 
muon flux normalized to the EGRET data represents an 
upper limit, as the observed gamma--ray emission certainly could 
be due to processes other than dark matter annihilation. 
The comparison with the sensitivity of ANTARES shows that
only the highest mass neutralinos can possibly be detected with neutrino from the
galactic center. In this case, conservatively assuming that 
the gamma-ray emission observed by EGRET is entirely due to 
neutralino annihilation, the upper limit on the neutrino flux
is barely above the minimum signal observable by ANTARES in 
3 years. 

If neutrinos are nevertheless observed above the given fluxes, then their
interpretation  as due to neutralino annihilation is problematic and
would actually require either the adoption of other dark matter candidates
annihilating dominantly into neutrino pairs or a different explanation,
e.g. in terms of astrophysical sources.

\subsection{Synchrotron Radiation from the Galactic Center}
\label{sec:synchro}

Another interesting means of indirect detection of dark matter
is observing the synchrotron radiation originating from
the propagation of secondary $e^{\pm}$'s in the galactic magnetic
fields. We will focus on what happens at the center of 
our galaxy, where most of the annihilation signal comes from.

\index{equipartition magnetic field}
The magnetic field around the Galactic center is thought to be at 
{\it equipartition}, \ie there is equipartition of 
magnetic, gravitational and kinetic energy of the plasma 
surrounding the central supermassive black hole (see Sec.~\ref{smbh}).

It is easy to derive the strength of the magnetic field
under a few simplifying assumptions. Let us consider the
existence of a galactic wind of particles with velocity, 
$v_{gw}$. These particles will be captured by the gravitational 
potential well of the black hole at the center of the
galaxy within the accretion radius, $r_a \equiv 2GM/v_{gw}^2$,
where $M$ is the mass of the central object.
Under the assumption of purely radial infall, the radial
dependence of the particle velocity is simply 
\be
v(r) = -c\left( \frac{r_S}{r} \right)^{1/2},
\ee
where $r_S\equiv GM/c^2$ is the Schwarzschild radius of the 
black hole. 
For a steady flow, the number of plasma particles through
a sphere of radius, $r$, around the black hole is 
\be
\dot N \equiv \frac{dN}{dt} = -4 \pi r^2 n(r) v(r),
\ee 
where $n(r)$ is the plasma number density. Solving for $n(r)$
we find
\be
n(r) = \frac{\dot N}{4 \pi c r_S^2} \left( \frac{r_S}{r} \right)^{3/2}.
\label{smden}
\ee
The accretion rate can be parametrised as follows
\be
\dot N m_p \approx 10^{22} \mbox{g s}^{-1} {\cal W} M_6^2,
\ee
where $m_p$ is the proton mass, $M_6$ is the mass of the central
black hole in units of $10^6 M_{\odot}$, and ${\cal W} \sim 1$
parameterizes the uncertainties of the physical parameters around the
black hole, namely the velocity and mass-loss rate of the
circum-nuclear wind (see Ref.~\cite{melia} for more 
details).

We now note that the infalling plasma is expected to be
highly ionized, and its energy density will reach equipartition with kinetic and gravitational energies.
After simple calculations, we find that the magnetic field 
under equipartition evaluates to~\cite{melia,shvartsman}
\be
B =  324 \mu \mbox{G} \left(\frac{r}{\mbox{pc}}\right)^
{-5/4}.
\ee

Far from the galactic center, the equipartition does not apply anymore
and we assume a flat profile for the magnetic field.
More specifically, the profile we adopt is 
\begin{equation}
B(r) = \mbox{max}\left[ 324 \mu \mbox{G} \left(\frac{r}{\mbox{pc}}\right)^
{-5/4}, \, 6\mu\mbox{G}\right]
\end{equation}
which means that the magnetic field is assumed to be in equipartition
with the plasma out to a galactocentric distance $r_c=0.23\,$pc
and equal to a typical value observed throughout the galaxy
at larger distances.

If the actual value of the magnetic field away from the central region
was smaller than the value we considered, this would imply a shift of
the radio spectrum
toward lower energies and thus, in the range of frequencies 
we are interested in,  a higher flux for a given frequency. 
This would also translate into stronger constraints for the mass and 
annihilation cross section. Nevertheless, we prefer to be conservative
and consider a quite high value of $B$. Note that magnetic fields
stronger than equipartition values are physically unlikely.


The mechanism of synchrotron emission is well known. We will now estimate 
the synchrotron luminosity produced by the propagation of secondary $e^{\pm}$'s
originating from dark matter annihilation in the galactic magnetic field

We recall that the critical synchrotron
frequency, $\nu_c(E)$, \ie the frequency around which the synchrotron
emission of an electron of energy, E, in a magnetic field of strength B,
peaks, can be expressed as
\begin{equation}
\nu_c(E) = \frac{3}{4 \pi} \frac{eB}{m_e c} \left(\frac{E}{m_e c^2}
\right)^2 \; ,
\end{equation}
where $m_e$ is the electron mass.
Inverting this relation, we determine the energy of the electrons 
which give the maximum contribution at that frequency,
\begin{equation}
E_m(\nu) = \left( \frac{4 \pi}{3} \frac{m_e^3 c^5}{e} \frac{\nu}{B}
\right)^{1/2} = 0.25 \left( \frac{\nu}{\mbox{MHz}} \right)^{1/2}  \left(
\frac{r}{\mbox{pc}} \right)^{5/8} \mbox{GeV}.
\label{em}
\end{equation} 

To compute the synchrotron luminosity we also need to know their
energy distribution, which in our case can be expressed as
(see, \eg, Ref.~\cite{Gondolo:2000pn})
\begin{equation}
 \frac{dn}{dE} = \frac{\Gamma Y_e(>E)}{P(E)} f_e(r),
\label{dn}
\end{equation}
where $\Gamma$ is the annihilation rate,
\begin{equation}
\Gamma = \frac{\sigma v}{m_{\rm{DM}}^2} \int_0^{\infty} \rho_{sp}^2 4 \pi r^2 \;\;
dr\,.
\end{equation}
The function $f_e(r)$ is given by
 \begin{equation}
f_e(r)= \frac{\rho_{sp}^2}{  \int_0^{\infty} \rho_{sp}^2 4 \pi r^2 \;\;
dr}
\end{equation}
and 
 \begin{equation}
P(E)= \frac{2 e^4B^2 E^2}{3 m_e^4 c^7}
\end{equation}
is the total synchrotron power spectrum.
Note that the general expression for $f_e(r)$ would have to take into
account spatial redistribution by diffusion (see, \eg, Ref.~\cite{Gondolo:2000pn}), but this is typically negligible~\cite{Bertone:2001jv}.

The quantity $Y_e(>E)$ is the number of $e^{\pm}$'s with energy above E
produced per annihilation, which depends on the annihilation modes,
and can be equivalently expressed as a function of the particle
mass, $m_{\rm{DM}}$, and the frequency considered, $\nu$. Actually,
Eq.~(\ref{em}) shows that for the frequencies we are interested in,
$E_m(\nu)<<m_{\rm{DM}}$, and thus the energy dependence of $Y_e(>E)$ can be neglected. 
We estimate $Y_e(>E)$ by the number of charged particles produced in
quark fragmentation (see below for further details).

For each electron the total power radiated in the frequency interval 
between $\nu$ and $\nu + d\nu$ is given
by
\be
P(\nu,E) = \frac{\sqrt{3} e^3}{m_e c^2} B(r) \frac{\nu}{\nu_c(E)}
\int_0^{\infty} K_{5/3}(y) dy  = \frac{\sqrt{3} e^3}{m_e
c^2} B(r) F\left( \frac{\nu}{\nu_c(E)} \right) \;,
\ee
where $ K_{n}(y)$ are the modified Bessel functions of order
$n$ (for definitions see \eg~\cite{rybi}) and
\begin{equation}
F\left( \frac{\nu}{\nu_c(E)} \right)=  \frac{\nu}{\nu_c(E)}
\int_0^{\infty} K_{5/3}(y) dy \;.
\end{equation}
Integrating this formula over the dark matter distribution, we obtain 
the total synchrotron luminosity
\begin{equation}
L_{\nu} = \int_0^{\infty} dr \; \;  4 \pi r^2  \int_{m_e}^{m_{\rm{DM}}} dE \;\;
\frac{dn_e}{dE} P(\nu,E) \;,
\label{lum}
\end{equation}
which by substitution becomes
\begin{equation}
L_{\nu} =\frac{\sqrt{3} e^3 \Gamma}{m_e c^2} \int_0^{\infty} dr 4 \pi r^2
f_e(r) B(r)  \int_{m_e}^{m_{\rm{DM}}} dE \frac{Y_e(>E)}{P(E)} F\left(
\frac{\nu}{\nu_c(E)} \right)\;.
\end{equation}

It is possible to simplify this formula by introducing the following
approximation (see
Rybicki \& Lightman~\cite{rybi}):
\begin{equation}
F\left( \frac{\nu}{\nu_c(E)} \right) \approx \delta(\nu / \nu_c(E) -
0.29)\;.
\label{approx}
\end{equation}
The evaluation of the integral then gives
\begin{equation}
L_{\nu}(\psi) \simeq \frac{1}{4 \pi}\frac{9}{8}
\left(\frac{1}{0.29 \pi} \frac{ m_e^3 c^5}{e} 
\right)^{1/2} \frac{\sigma v}{m_{\rm{DM}}^2} Y_e(m_{\rm{DM}},\nu)  \;\nu^{-1/2} \; I(\psi) 
\;,
\label{synchrolum}
\end{equation} 
where 
\begin{equation} 
I(\psi) = \int_0^{\infty} ds \;\; \rho^2\left(r(s,\psi)\right) B^{-1/2}
\left(r(s,\psi)\right)
\label{syn}
\end{equation} 
and $s$ is the coordinate running along the line-of-sight. 

For frequencies around $400\,$MHz, used below, and for the lowest value
of the magnetic field, we find that $E_m(400\,\mbox{MHz})
\lesssim 2\,\mbox{GeV}$.
In reality, for dark matter profiles with central cusps, e.g. the
NFW, Kravtsov, and Moore profiles, most of the annihilation 
signal comes from the inner region of the galaxy, where the magnetic field 
is probably higher. For $\nu=400\,\mbox{MHz}$ and $r < r_c$,
\begin{equation}
E_m(\nu)\simeq 0.3 \left(\frac{\nu}{400\,{\rm MHz}}\right)^{1/2}
\left(\frac{r}{\mbox{pc}}\right)^{5/8} \mbox{GeV}\,,\label{ecritic}
\end{equation}
which at the inner edge of the profile, corresponding to the Schwarzschild 
radius of the supermassive black hole at the galactic center, $R_S=1.3\times10^{-6}$ pc, takes the value of
$E_m(400\,\mbox{MHz})= 2.2 \times 10^{-5}$ GeV. We thus always have 
$E_m(400\,\mbox{MHz}) \ll m_{\rm{DM}}$, which means that most of the secondary
electrons are produced above this energy and contribute to the radio flux.

For a particle of mass $m_{\rm{DM}}$, the average electron multiplicity
per annihilation, $Y_e(m_{\rm{DM}})$, is
evaluated by adding the contribution of each annihilation channel
with cross section $(\sigma v)_i$, producing $Y_e^i(m_{\rm{DM}})$ electrons:
\be
\sigma v Y_e(m_{\rm{DM}})=\sum_i (\sigma v)_i Y_e^i(m_{\rm{DM}})\,,
\label{sigY}
\ee
where $\sigma v$ is the total annihilation cross section.

The main channels contributing to this flux are direct production of 
leptons and annihilation into quarks. 
The calculations are easily performed for Kaluza-Klein dark matter.  For direct production of leptons, 
\be
Y_e^{e^{\pm}}(M)=Y_e^{\mu^{\pm}}(M) \simeq 2
\ee 
in the relevant range of masses. In the quark channel, 
to count the number of electrons, $Y_e^{q\overline{q}}(M)$, we 
integrate the fragmentation functions for $e^{\pm}$'s from $\pi^{\pm}$'s. This results in 
\be
\sigma v Y_e(1\,\mbox{TeV})\simeq 6 \times 10^{-3}\,\mbox{TeV}^{-2}
\ee
and 
\be 
Y_e(1\,\mbox{TeV})\simeq 4.5,  
\ee
for $m_{\rm{DM}}=1$~TeV. The electron multiplicity in the hadronic 
channel alone would be much larger, roughly 20.

The case of neutralinos is much more complicated, as the dominating annihilation modes can vary from model to model (a discussion
of branching ratios in the framework of the mSUGRA models can
be found in Ref.~\cite{Bertin:2002ky}). Such calculations must, therefore, be conducted on a model-by-model basis.

One more step is necessary to calculate the observed radiation. We must multiply the synchrotron luminosity, $L_\nu$, with 
the synchrotron self-absorption coefficient, which we calculate next. 

\index{synchrotron self-absorption}
Synchrotron emission is accompanied by absorption, in which
a photon loses its energy due to the interaction with 
a charge in a magnetic field. 
The synchrotron self-absorption coefficient is by definition 
(see Rybicki and Lightman~\cite{rybi})
\begin{equation}
A_{\nu}= \frac{1}{a_{\nu}} \int_{0}^\infty (1-e^{-\tau(b)}) 2 \pi b \;
db,
\label{anu}
\end{equation}
where $\tau(b)$ is the optical depth as a function of the cylindrical
coordinate $b$,
\begin{equation}
\tau(b)= a_{\nu} \int_{d(b)}^{\infty} f_e(b,z) \;dz,
\label{la}
\end{equation}
and the coefficient, $a_{\nu}$, is given by
\begin{equation}
 a_{\nu} =\frac{e^3 \Gamma B(r)}{9 m_e \nu^2} \int_{m_e}^{m} E^2 \frac
{d}{dE} \left[ \frac{Y_e(>E)}{E^2 P(E)} \right] F\left( \frac{\nu}{\nu_c}
\right) \; dE.
\label{a2nu}
\end{equation}

The final luminosity is obtained by multiplying Eq.~(\ref{lum}) with
$A_{\nu}$, given by Eq.~(\ref{anu}). It is evident that in the limit of
small optical depth, the coefficient $A_{\nu} \rightarrow 1$, as can be
seen by expanding the exponential.

The lower limit of integration of Eq.~(\ref{la}) is 
\begin{equation}
d(b)= 0 \;\;\;\; for \;\;\;\;  b^2+z^2>(4Rs)^2, 
\end{equation}
\[ d(b)= \sqrt{(4Rs)^2-b^2} \;\;\;\;\; \mbox{elsewhere}. \]

Using the approximation introduced in Eq.~(\ref{approx}), we find
\begin{equation}
a_{\nu} =\frac{\Gamma Y}{4 \pi} \frac{c^2}{\nu^3},
\label{apnu}
\end{equation}
which can in turn be used to evaluate $\tau(b)$ in Eq.~(\ref{la}) and
$A_{\nu}$ in Eq.~(\ref{anu}).


If a density spike exists at the galactic center, the self-absorption coefficient 
cannot be neglected and can lead to a significant reduction
of the observed synchrotron flux by up to several orders of
magnitude.
The results of synchrotron emission in 
presence of a spike has been discussed in Ref.~\cite{Bertone:2001jv}. If a spike exists at
the galactic center, and if neutralinos are the dark matter particle,
only small values of $\gamma$ are compatible
with radio observations. Kaluza-Klein dark matter has annihilation
cross section typically larger than neutralinos. Thus Kaluza-Klein dark matter is very problematic in a scenario with a density spike.

Note that Eqs.~\ref{anu}--\ref{a2nu} are valid, strictly speaking,
only for position independent quantities. A rigorous treatment
of synchrotron emission and self-absorption would require the
solution of the radiative transport equation. 
Recently, Aloisio \etal 2004 Ref.~\cite{Aloisio:2004hy} have derived
the equilibrium distribution of electrons and positrons from neutralino 
annihilation at the Galactic center, and the resulting radiation considering 
adiabatic compression in the accretion flow, inverse Compton scattering of 
synchrotron photons (synchrotron self-Compton scattering), and synchrotron 
self-absorption. Such a detailed analysis allows a more precise estimate of
the radio emission and confirms that neutralino annihilation in a NFW profile 
with a spike would exceed the observed radio emission from the Galactic center.


If there is no spike at the galactic center, the optical depth is negligible 
and the self-absorption
coefficient is of the order of unity. In fact, using Eq.~\ref{apnu}, 
the optical depth in Eq.~\ref{la} can be expressed as
\begin{equation}
\tau\simeq\frac{\sigma v}{m_{\rm{DM}}^2} \frac{Y_e(M)}{4 \pi} \frac{1}{\nu^3} 
\int_0^{d_\odot} ds\rho^2(s)\,,
\end{equation}
where $d_\odot\simeq8\,$kpc is the distance of the Sun from the galactic center.
Using $m_{\rm{DM}}=1\,$TeV, $\sigma v = 1.6\times10^{-4}\,$TeV$^{-2}$ (for the cross
section for annihilation into right-handed up quarks) and a
NFW halo profile, we find 
\be
\tau = 1.78 \times 10^{-4} \left( \frac{\nu}{100\mbox{MHz}} 
\right)^{-3} \;.
\ee
We can thus neglect self-absorption unless
the frequency considered is very small. The absorption
on relativistic electrons from other sources is also 
negligible. Using 
\be
n(E) \lesssim 10^{-2}\,{\rm GeV}^{-1}{\rm cm}^{-2}
{\rm s}^{-1}{\rm sr}^{-1} 
\ee
for the locally observed differential electron
flux (see Ref.~\cite{Moskalenko:1997gh}) 
in the relevant energy range given by Eq.~(\ref{ecritic}),
one obtains an absorption coefficient per length
\be
\alpha_\nu \lesssim 6 \times 10^{-16} \mbox{pc}^{-1} 
\left( \frac{B}{\mu\mbox{G}} \right) 
\left( \frac{\nu}{\mbox{GHz}} \right) ^{-2}\;. 
\ee
Even if the
relativistic electron flux due to non-acceleration processes close
to the Galactic center is orders of magnitude larger, this effect would still be
negligible. However, for frequencies below
a few MHz, free-free absorption is important (see e.g. Ref.~\cite{cane}).

\begin{figure}[t]
\centering
\includegraphics[width=0.6\textwidth,clip=true]{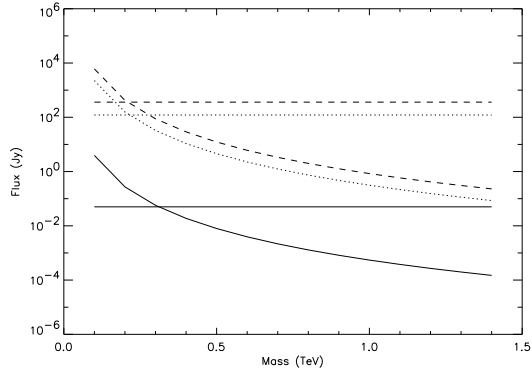}
\caption[Radio flux from regions close to the \GC]
{Predicted (curves) and observed (horizontal lines) radio flux 
from annihilating Kaluza-Klein dark matter from regions close to the galactic center. An NFW profile has been assumed. Three cases for the radio frequency and observed region are shown.}
\label{synchro}
\end{figure}

To compare with observations, we integrate over the relevant solid angle.
The comparison between predicted and observed fluxes can constrain the 
cross sections and masses of annihilating dark matter particles for a given halo profile. In particular, this method be used to provide a lower bound on the mass of a Kaluza-Klein dark matter particle. In Fig.~\ref{synchro} we show predicted and observed 
fluxes for Kaluza-Klein particles, for a NFW profile, as a function of the particle 
mass. Three cases are shown (see Ref.~\cite{Bertone:2002je} for more details). For each case the predicted and observed fluxes are plotted, the latter being represented by a horizontal line. The three cases are represented by solid, dashed and dotted lines. 
Case 1 is the most constraining, implying a lower bound on the
mass of about $0.3\,$TeV (assuming an NFW halo profile). 

The fluxes predicted at high latitudes can also be compared with 
observations (see Ref.~\cite{cane}). The strongest constraints result from
the lowest frequencies at which free-free and synchrotron 
self-absorption are not yet important, i.e. $\sim 10\,$MHz. Here, the observed background
emission between $0^\circ$ and $90^\circ$ from the galactic
anti-center is $\simeq6\times10^6\,$Jy. Comparing with the
predicted emission results in the limit
\be
\sigma v \lesssim10^{-22}\left(\frac{m_{\rm{DM}}}{\mbox{\small{TeV}}} \right)^2 
\frac{Y_e(1\mbox{\small{TeV})}}{Y_e(m_{\rm{DM}})}\,\mbox{cm}^{3} \mbox{s}^{-1}. 
\ee
While this is considerably weaker than the constraints above,
it is largely independent of the unknown halo profile near the galactic center.


The argument can also be turned the other way round and 
interpreted as a measure of the 
galactic magnetic field. One can in fact decide to select the 
values of $\gamma$ reproducing the normalization of the 
observed gamma-ray emission and look for the values of the
magnetic field that reproduce the correct normalization 
of the observed radio emission. Due to the uncertainties 
in the particle physics models, the precision obtained
is unfortunately poor (see Ref.~\cite{Bertone:2002je}).  

A comparison between the prospects for indirect detection 
at different 
wavelengths shows that constraints from synchrotron emission are 
more stringent than those obtained from gamma-ray
and neutrino fluxes (from current experiments),
although they are less robust, being derived under 
the additional assumptions made about the strength 
of the magnetic field at equipartition. 
In the future, however, the strongest constraints will likely result from gamma-ray experiments.

\subsection{Annihilation Radiation from External or Dwarf Galaxies} 

It also might be possible to observe annihilation radiation from galaxies outside of the Milky Way. In this case, although such galaxies are
far more distant than the galactic center, the observed emitting region is much larger. Dwarf galaxies within the Milky Way may also be observable regions of dark matter annihilation. 

Baltz \etal~\cite{Baltz:1999ra} studied the expected
flux of gamma-rays from M87 and several local dwarf spheroidal
galaxies. The predicted fluxes are calculated using the same 
formulae for indirect detection of secondary particles 
described above. Using a profile with a central core for the sources, they conclude that predicted 
fluxes are below the sensitivities of next-generation experiments,
unless the annihilation signal is boosted by a significant 
amount of dense clumps. A similar analysis was carried on for the
prospect of observing M31 with CELESTE (see Falvard \etal
~\cite{Falvard:2002ny}).

Tasitsiomi, Gaskins and Olinto~\cite{Tasitsiomi:2003vw}
focused on gamma-ray and synchrotron emission from the Large 
Magellanic Cloud (LMC). Fitting the LMC rotation curve with different 
profiles, they determined that although present data do not
constrain SUSY parameters, future experiments like GLAST (gamma-rays), and LOFAR (low frequencies emission), could
probe a significant portion of the SUSY parameter space.
Similar, though more optimistic, conclusions have been 
obtained by  Pieri and Branchini~\cite{Pieri:2003cq}.

Finally, particularly interesting are the prospects for dark matter observations in the 
Draco and Sagittarius dwarf galaxies~\cite{ferrer,Tyler:2002ux}.

\subsection{High-Energy Neutrinos from the Sun or Earth}
\label{nufromsun}


In addition to gamma-rays, neutrinos can be produced in dark matter annihilations. Unlike gamma-rays, however, neutrinos can escape from dense media in which such annihilations may take place. For example, WIMPs which are captured in deep gravitational wells such as the Sun or Earth, can annihilate at great rates. Although gamma-rays cannot escape these objects, neutrinos often can, providing an interesting signature to search for with high-energy neutrino telescopes \cite{indirectneutrino1,indirectneutrino2,indirectneutrino3,indirectneutrino4,indirectneutrino5,Bergstrom:1998xh,indirectneutrino6,indirectneutrino7,indirectneutrino8}

\subsubsection{Capture and annihilation in the Sun}

In order to provide an observable flux of neutrinos, dark matter particles must be gathered in high concentrations. In the following calculation, we will focus on WIMP capture in the Sun, as these prospects are more promising than for capture in the Earth.

The rate at which WIMPs are captured in the Sun depends on the nature of the interaction the WIMP undergoes with nucleons in the Sun. For spin-dependent interactions, the capture rate is given by \cite{capture,Jungman:1995df} 
\begin{equation} 
C_{\mathrm{SD}}^{\odot} \simeq 3.35 \times 10^{20} \, \mathrm{s}^{-1} 
\left( \frac{\rho_{\mathrm{local}}}{0.3\, \mathrm{GeV}/\mathrm{cm}^3} \right) 
\left( \frac{270\, \mathrm{km/s}}{\bar{v}_{\mathrm{local}}} \right)^3  
\left( \frac{\sigma_{\mathrm{H, SD}}} {10^{-6}\, \mathrm{pb}} \right)
\left( \frac{100 \, \mathrm{GeV}}{m_{\rm{DM}}} \right)^2 ,
\label{c-eq}
\end{equation} 
where $\rho_{\mathrm{local}}$ is the local dark matter density, 
$\sigma_{\mathrm{H,SD}}$ is the spin-dependent, WIMP-on-proton (hydrogen)
elastic scattering cross section, $\bar{v}_{\mathrm{local}}$ 
is the local rms velocity of halo dark matter particles and 
$m_{\rm{DM}}$ is the dark matter particle's mass. 

\newpage

The analogous formula for the capture rate from spin-independent (scalar) scattering is \cite{capture,Jungman:1995df}
\ba
C_{\mathrm{SI}}^{\odot} \simeq 1.24 \times 10^{20} \, \mathrm{s}^{-1} 
\left( \frac{\rho_{\mathrm{local}}}{0.3\, \mathrm{GeV}/\mathrm{cm}^3} \right) 
\left( \frac{270\, \mathrm{km/s}}{\bar{v}_{\mathrm{local}}} \right)^3
\left( \frac{100 \, \mathrm{GeV}}{m_{\rm{DM}}} \right)^2 \times \nonumber\\ 
\times \left( \frac{2.6 \, \sigma_{\mathrm{H, SI}}
+ 0.175 \, \sigma_{\mathrm{He, SI}}}{10^{-6} \, \mathrm{pb}} \right) 
 \; .
\ea
Here, $\sigma_{\mathrm{H, SI}}$ is the spin-independent, WIMP-on-proton
elastic scattering cross section and $\sigma_{\mathrm{He, SI}}$ is the 
spin-independent, WIMP-on-helium, elastic scattering cross section. 
Typically, $\sigma_{\mathrm{He, SI}} \simeq 16.0 \, \sigma_{\mathrm{H, SI}}$.
The factors of $2.6$ and $0.175$ include information on the solar 
abundances of elements, dynamical factors and form factor suppression.

Although these two rates appear to be comparable in magnitude, the spin-dependent and spin-independent cross sections can differ radically. For example, for Kaluza-Klein dark matter, the spin-dependent cross section is typically three to four orders of magnitude larger than the spin-independent cross section \cite{Servant:2002aq,Cheng:2002ej} and solar accretion by spin-dependent 
scattering dominates. Spin-dependent capture also dominates for most neutralino models. On the other hand, for scalar dark matter candidates (such as sneutrinos, or candidates from theory space little Higgs models), the spin-independent cross section can dominate.

If the capture rates and annihilation cross sections are sufficiently
high, equilibrium may be reached between these processes.  
For $N$ WIMPs in the Sun, the rate of change of this
quantity is given by
\begin{equation}
\dot{N} = C^{\odot} - A^{\odot} N^2  ,
\end{equation}
where $C^{\odot}$ is the capture rate and $A^{\odot}$ is the 
annihilation cross section times the relative WIMP velocity per volume.  
$C^{\odot}$ was given in Eq.~\ref{c-eq}, while $A^{\odot}$ is 
\begin{equation}
A^{\odot} = \frac{\langle \sigma v \rangle}{V_{\mathrm{eff}}}, 
\end{equation}
where $V_{\mathrm{eff}}$ is the effective volume of the core
of the Sun determined roughly by matching the core temperature with 
the gravitational potential energy of a single WIMP at the core
radius.  This was found in Refs.~\cite{equ1,equ2} to be
\begin{equation}
V_{\rm eff} = 5.7 \times 10^{27} \, \mathrm{cm}^3 
\left( \frac{100 \, \mathrm{GeV}}{m_{\rm{DM}}} \right)^{3/2} \;.
\end{equation}
The present WIMP annihilation rate is given by
\begin{equation} 
\Gamma = \frac{1}{2} A^{\odot} N^2 = \frac{1}{2} \, C^{\odot} \, 
\tanh^2 \left( \sqrt{C^{\odot} A^{\odot}} \, t_{\odot} \right) \;, 
\end{equation}
where $t_{\odot} \simeq 4.5$ billion years is the age of the solar system.
The annihilation rate is maximized when it reaches equilibrium with
the capture rate.  This occurs when 
\begin{equation}
\sqrt{C^{\odot} A^{\odot}} t_{\odot} \gg 1 \; .
\end{equation}
For many of the particle physics models which are most often considered (most supersymmetry or Kaluza-Klein models, for example), the WIMP capture and annihilation rates reach, or nearly reach, equilibrium in the Sun. This is often not the case for the Earth. This is true for two reasons. First, the Earth is less massive than the Sun and, therefore, provides fewer targets for WIMP scattering and a less deep gravitational well for capture. Second, the Earth accretes WIMPs only by scalar (spin-independent) interactions. For these reasons, it is unlikely that the Earth will provide any observable neutrino signals from WIMP annihilations in any planned experiments (for a recent analysis of WIMP capture in the Earth, see Ref.~\cite{edsjoearth}).


The flux of neutrinos produced in WIMP annihilations is highly model dependent as the annihilation fractions to various products can vary a great deal from model to model. We will attempt to be as general in our discussion as possible while still considering some specific cases as well.

In supersymmetry, there are no tree level diagrams for direct neutralino annihilation to neutrinos. Many indirect channels exist, however. These include neutrinos from heavy quarks, gauge bosons, tau leptons and Higgs bosons. These processes result in a broad spectrum of neutrinos, but with typical energies of 1/2 to 1/3 of the neutralino mass. For experimental (muon) energy thresholds of 10-100 GeV, lighter WIMPs can be very difficult or impossible to detect for this reason.

For neutralinos lighter than the $W^{\pm}$ mass ($80.4\,\,$GeV), annihilation to $b \bar{b}$ typically dominates, with a small admixture of $\tau^+ \tau^-$ as well. In these cases, neutrinos with less than about 30 GeV energy are produced and detection is difficult. For heavier neutralinos, annihilation into gauge bosons, top quarks and Higgs bosons are important in addition to $b \bar{b}$ and $\tau^+ \tau^-$. In particular, gauge bosons can undergo two body decays ($Z \rightarrow \nu \nu\,\,$ or $W^{\pm} \rightarrow l^{\pm} \nu$) producing neutrinos with an energy of about half of the WIMP mass. Neutralinos with a substantial higgsino component often annihilate mostly into gauge bosons. 

For Kaluza-Klein dark matter, the picture is somewhat different. Kaluza-Klein dark matter particles annihilate directly to a pair of neutrinos about 3-4\% of the time \cite{Servant:2002aq,Cheng:2002ej,Hooper:2002gs}. Although this fraction is small, the neutrinos are of higher energy and are, therefore, easier to detect. The more frequent annihilation channels for Kaluza-Klein dark matter are charged leptons (60-70\%) and up-type quarks (20-30\%). Of these, the $\tau^+ \tau^-$ mode contributes the most to the neutrino flux. Unlike in supersymmetry, a large fraction of lightest Kaluza-Klein particles annihilate into long lived particles, such as up quarks, electrons and muons, which lose their energy in the Sun long before decaying. Bottom and charm quarks lose some energy before decaying, but not as dramatically.

Neutrinos which are produced lose energy as they travel through the Sun \cite{edsjo,Jungman:1994jr,crotty}. The probability of a neutrino escaping the sun without interacting is given by
\begin{equation} 
P = e^{-E_{\nu}/E_k}
\end{equation}
where $E_k$ is $\simeq130$ GeV for $\nu_{\mu}$, $\simeq160$ GeV for $\nu_{\tau}$, $\simeq200$ GeV for $\bar{\nu_{\mu}}$ and $\simeq230$ GeV for $\bar{\nu_{\tau}}$. Thus we see that neutrinos above a couple hundred GeV are especially depleted. For a useful parameterization of solar effects, see Ref.~\cite{edsjo}. Note that neutrino oscillations can also play an important role in calculating the flux of muon neutrinos in a detector \cite{crotty}.


\subsubsection{Detection of high-energy neutrinos from the Sun}


Several experiments are potentially able to detect the flux of high
energy neutrinos from dark matter annihilations in the solar core.
The AMANDA experiment is currently the largest operating neutrino
telescope. The AMANDA B-10 array, due to its ``soda can'' geometry, was not very sensitive in the direction of the Sun (the horizon), although the current version of the experiment, AMANDA-II, does not have this problem and can place limits on dark matter annihilations from 
the center of the Sun and Earth. ANTARES, with a lower energy threshold (10 GeV) and IceCube, with a much greater effective area, will each function as effective dark matter experiments (see section~\ref{neuexp} for a description of neutrino telescopes).

The background for this class of experiments consists of atmospheric
neutrinos \cite{atmback} and neutrinos generated in cosmic ray
interactions in the Sun's corona \cite{Bergstrom:1998xh,sunback1}.  In the direction of the
Sun (up to the angular resolution of a neutrino telescope), tens of events
above 100 GeV and on the order of 1 event per year above 1 TeV, per square
kilometer are expected from the atmospheric neutrino flux. The rate of events from neutrinos generated by cosmic ray 
interactions in the Sun's corona is predicted to be less than a few events per
year per square kilometer above 100 GeV.

\begin{figure}
\centering
\includegraphics[width=0.6\textwidth,clip=true]{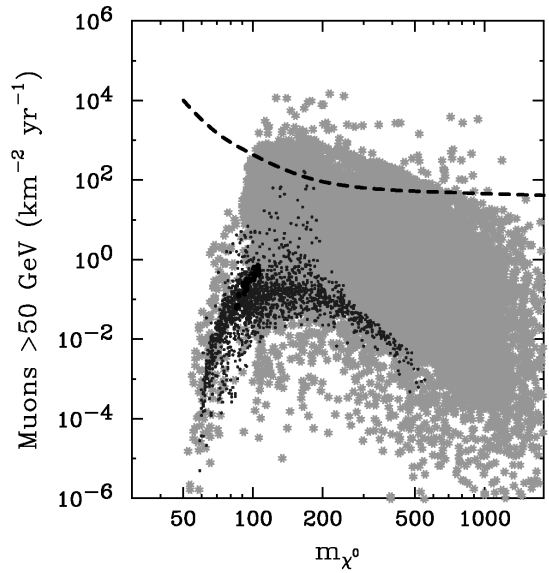}
\caption{The number of events from neutralino annihilation in the Sun per year in a neutrino telescope with an effective area equal to one square kilometer and a 50 GeV muon threshold \cite{wang}. The lightly shaded region represents the general Minimal Supersymmetric Standard Model (MSSM), the darker region corresponds to mSUGRA models, a subset of the MSSM. For each point shown, the relic density is below the maximum value allowed by the WMAP data ($\Omega_{\chi} h^2 \le 0.129$). The sensitivity projected for IceCube is shown as a dashed line \cite{edsjolimits}.}
\label{figsusy1}
\end{figure}

The sensitivity of a square kilometer neutrino detector with a moderate 
muon energy threshold (50 GeV) to supersymmetric dark matter is shown in 
Fig.~\ref{figsusy1}. From this figure, it is clear that high-energy neutrinos 
will be an observable signature in only a small fraction of possible 
supersymmetry models, although such experiments are still certainly an 
important probe.

\begin{figure}[t]
\centering\leavevmode
\includegraphics[width=0.4\textwidth]{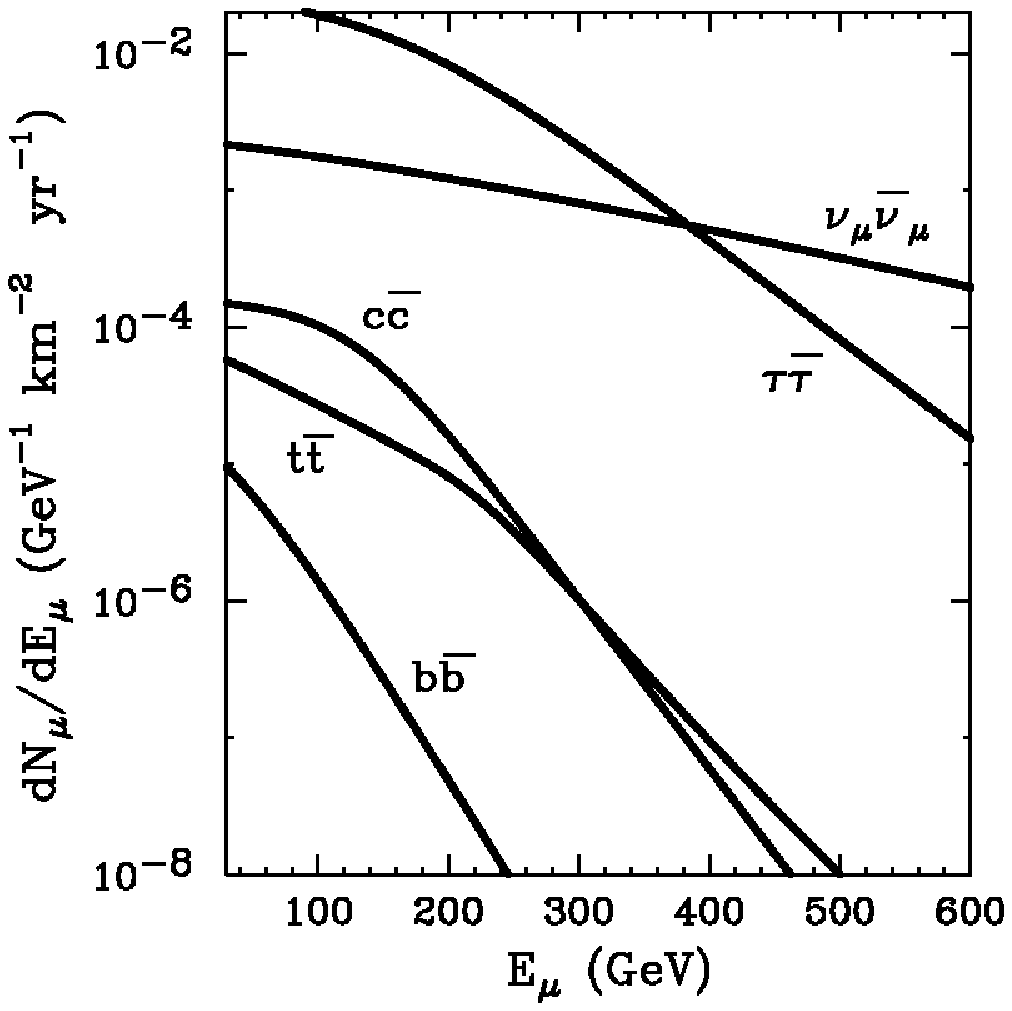} 
\includegraphics[width=0.4\textwidth]{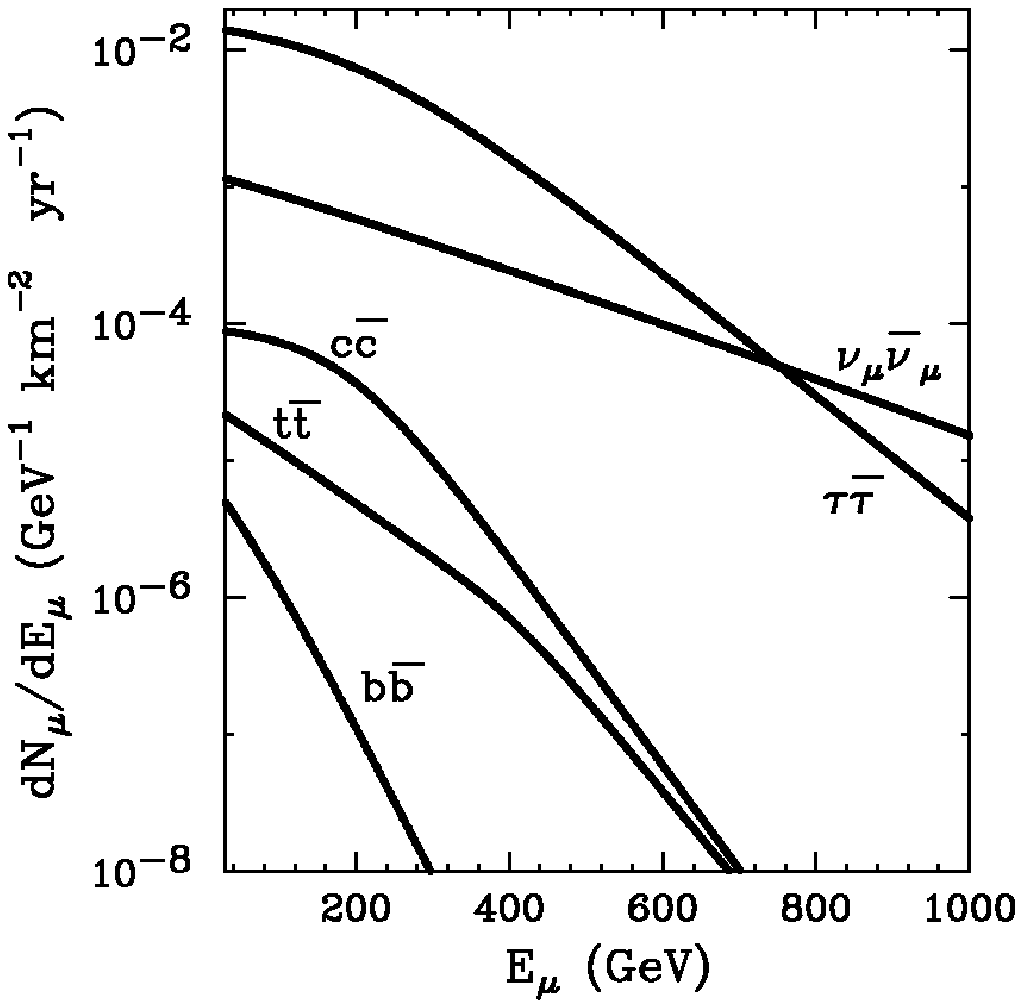} \\ [0.4cm]
\caption{The spectrum of muons at the Earth generated in charged-current 
interactions of muon neutrinos generated in the annihilation of 
$600$ GeV (left side) and $1000$ GeV (right side)
Kaluza-Klein dark matter particles in the Sun \cite{Hooper:2002gs}.  The elastic scattering cross section used for capture in the Sun was fixed at $10^{-6}$ pb for both graphs.  
The rates are proportional to that cross section.}
\label{figkk1}
\end{figure}

For Kaluza-Klein dark matter, the prospects for detection via high-energy neutrinos are substantially better. This is largely due to the dominating annihilation modes. The spectrum of muons in a detector due to LKP annihilations in the Sun is shown in Fig.~\ref{figkk1} for various annihilation channels and for two choices of LKP mass. Unlike in the case of supersymmetry, annihilation to neutrinos and taus dominates the neutrino spectrum. In supersymmetry, b quarks and gauge bosons dominate, producing fewer observable neutrinos.

\begin{figure}[t]
\centering\leavevmode
\includegraphics[width=0.6\textwidth]{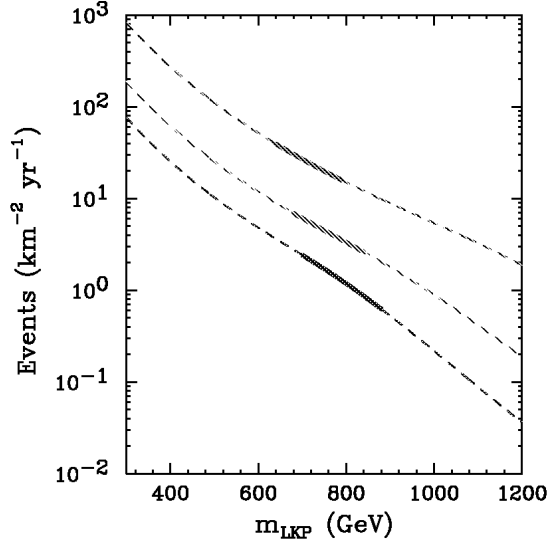}
\caption{The number of events per year in a detector with effective 
area equal to one square kilometer and a muon energy threshold of 50 GeV \cite{Hooper:2002gs}.  Contours are shown, from bottom to top, for 
$r_{\qR}=0.1$, $0.2$, and $0.3$, where $r_{\qR}$ is the mass splitting of the LKP and KK quarks over the LKP mass. The 
expected size of the one-loop radiative corrections predict $0.1 \lsim r_{\qR} \lsim 0.2$, therefore, the $r_{\qR}=0.3$ contour
is shown merely for comparison.
The relic density of the LKP's lies within the range
$\Omega_{\B} h^2 = 0.16 \pm 0.04$ for the solid sections 
of each line.  The relic density is smaller (larger) for 
smaller (larger) LKP masses.}
\label{figkk2}
\end{figure}

In Fig.~\ref{figkk2}, the event rates from Kaluza-Klein dark matter annihilation in the Sun are shown for a square kilometer detector with a threshold of 50 GeV. Each of the three lines correspond to variations in the Kaluza-Klein spectrum. For the spectrum predicted in Ref.~\cite{Cheng:2002iz},
a kilometer scale neutrino telescope could be sensitive to a 
LKP with mass up to about 800 GeV.  The relic density
of the LKP varies from low to high values from 
left to right in the graph.  The range of masses of the LKP that
gives the appropriate relic density was estimated from Refs.~\cite{Cheng:2002ej,Servant:2002aq} and shown in the
figure by the solid sections of the lines. Combining the expected
size of the one-loop radiative corrections with a relic density
appropriate for dark matter, we see that IceCube should see 
between a few events and tens of events per year.  

For detectors with smaller effective areas one simply has to scale the 
curves down by a factor $A/(1 \; \mathrm{km}^2)$ to obtain the event rate.  
In particular, for the first generation neutrino telescopes including AMANDA and 
ANTARES, with effective areas up to $0.1$ km$^2$,
the event rate could be as high as ten events per year for a
500 GeV LKP. The current limits from AMANDA-II (with data up to 2001) is roughly 3000 muons per square kilometer per year from the Sun \cite{Ackermann:2004um}. This sensitivity is expected to improve significantly with the analysis of more recent and future data.


\subsection{$e^+$ and $\bar{p}$ from Annihilations in
the Galactic Halo} 
\label{positronsec}

Charged particles, such as positrons and anti-protons, which are generated in
dark matter annihilations do not travel in straight lines. Therefore, rather
than observing a single region, such as the Galactic center or the Sun, the entire galactic halo can contribute to the flux of such particles.  In this section, we will discuss the impact on dark matter annihilations in the galactic halo on the cosmic positron and anti-proton spectrum.

\subsubsection{The positron excess} 

In 1994 and 1995, the High Energy Antimatter Telescope (HEAT) observed
a flux of cosmic positrons well in excess of the predicted rate, peaking around $\sim10\,$ GeV and extending to higher energies \cite{heat1995}. This result was confirmed by another HEAT flight in 2000 \cite{heat2000,heat}. Although the source of these positrons is not known, it has been suggested in numerous articles that this signal could be the product of dark matter annihilations, particularly within the context of supersymmetry \cite{positrons1,positrons2,Baltz:2001ir,positrons3,positrons4,positrons5,positrons6} and Kaluza-Klein dark matter \cite{Cheng:2002ej,kribskkpos}.

If the dark matter is evenly distributed in our local region (within a few kpc), the rate of annihilations may be insufficient to produce the observed excess. It has been suggested, however, that if sufficient clumping were present in the galactic halo, that the rate at which such particles annihilate could be enhanced enough to accommodate the data.

Positrons can be produced in a variety of dark matter annihilation
modes. Direct annihilation to $e^+ e^-$ is suppressed for neutralinos, but
occurs frequently for Kaluza-Klein dark matter \cite{Cheng:2002ej}. Also,
annihilations to $ZZ$ or $W^{+}W^{-}$ can produce positrons with energy of half
of the WIMP mass ~\cite{positrons2}. A continuum of positrons, extending to much lower energies, will in most cases also be produced in the cascades of annihilation products such as heavy leptons, heavy quarks, Higgs bosons and gauge bosons. The spectrum of positrons produced in dark matter annihilations can vary significantly depending on the mass and annihilation modes of the WIMP.

\begin{figure}
\centering\leavevmode
\includegraphics[width=0.4\textwidth,clip=true]{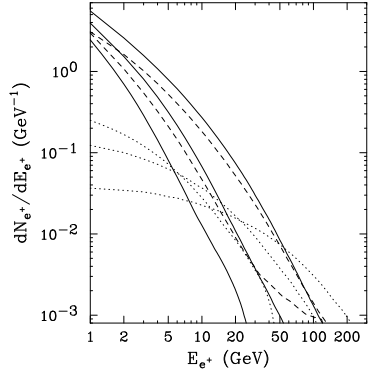}
\caption[]
{The positron spectrum from neutralino annihilations for the most important annihilation modes. Solid lines represent the positron spectrum, per annihilation, for $\chi^0 \chi^0 \rightarrow b \bar{b}$, for LSPs with masses of 50, 150 and 600 GeV. The dotted lines are the same, but from the process $\chi^0 \chi^0 \rightarrow \tau^{+} \tau^{-}$. Dashed lines represent positrons from the process $\chi \chi \rightarrow W^{+}W^{-}$ for LSPs with masses of 150 and 600 GeV. The spectrum from $\chi \chi \rightarrow ZZ$ is very similar.}
\label{positronspec1}
\end{figure}

As positrons propagate, they move under the influence of the tangled galactic magnetic fields, travelling in a random walk, and losing energy via inverse Compton and synchrotron processes. The diffusion-loss equation describing this process is given by

\begin{eqnarray}
\frac{\partial}{\partial t}\frac{dn_{e^{+}}}{dE_{e^{+}}} &=& \vec{\bigtriangledown} \cdot \bigg[K(E_{e^{+}},\vec{x})  \vec{\bigtriangledown} \frac{dn_{e^{+}}}{dE_{e^{+}}} \bigg] \nonumber\\
&+& \frac{\partial}{\partial E_{e^{+}}} \bigg[b(E_{e^{+}},\vec{x})\frac{dn_{e^{+}}}{dE_{e^{+}}}  \bigg] + Q(E_{e^{+}},\vec{x}),
\label{dif}
\end{eqnarray}
where $dn_{e^{+}}/dE_{e^{+}}$ is the number density of positrons per unit energy, $K(E_{e^{+}},\vec{x})$ is the diffusion constant, $b(E_{e^{+}},\vec{x})$ is the rate of energy loss and $Q(E_{e^{+}},\vec{x})$ is the source term.

The diffusion constant \cite{diffusion} and rate of energy loss can be parameterized by

\begin{equation}
K(E_{e^{+}}) = 3 \times 10^{27} \bigg[3^{0.6} + E_{e^{+}}^{0.6} \bigg] \,\rm{cm}^2 \, \rm{s}^{-1}
\label{k}
\end{equation}
and
\begin{equation}
b(E_{e^{+}}) = 10^{-16} E_{e^{+}}^2 \,\, \rm{s}^{-1},
\label{b}
\end{equation}
respectively. $b(E_{e^{+}})$ is the result of inverse Compton scattering on both starlight and the cosmic microwave background \cite{lossrate}. The diffusion parameters are constrained from analyzing stable nuclei in cosmic rays (primarily by fitting the boron to carbon ratio) \cite{L1,L2}.

In equations~\ref{k} and~\ref{b}, there is no dependence on location.  This is due to the assumption of a constant diffusion zone. For our galaxy, the diffusion zone is best approximated as a slab of thickness $2L$, where $L$ is chosen to be 4 kpc, the best fit to observations \cite{diffusion,L1,L2}. The radius of the slab is unimportant, as it is much larger than the distances which positrons can propagate at these energies. Outside of the diffusion zone, the positron density is assumed to be (nearly) zero (free escape boundary conditions). For detailed descriptions of two zone diffusion models, see Refs.~\cite{positrons6,L1,L2,2zonediffusion1,2zonediffusion2}.

The effect of propagation on the positron spectrum depends strongly on the distance from the source. To compare to the data recorded by HEAT, a quantity called the ``positron fraction'' is typically considered. The positron fraction is the ratio of the positron flux to the combined positron and electron fluxes. The spectra for secondary positrons, secondary electrons and primary electrons can be found in Ref.~\cite{Moskalenko:1997gh}. 

\begin{figure}[t]
\begin{center}
$\begin{array}{c@{\hspace{0.5in}}c}
\includegraphics[width=0.4\textwidth]{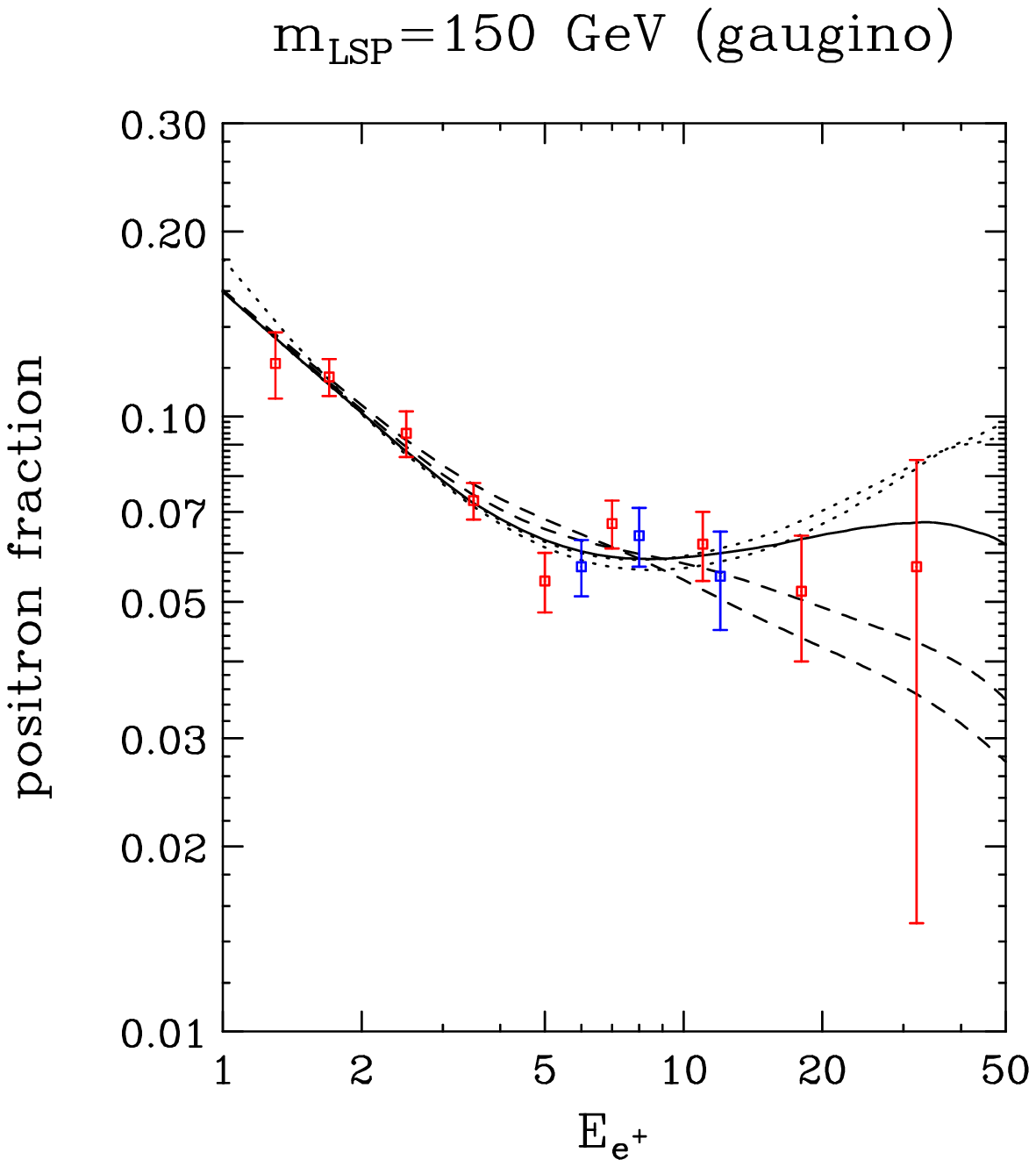} &
\includegraphics[width=0.4\textwidth]{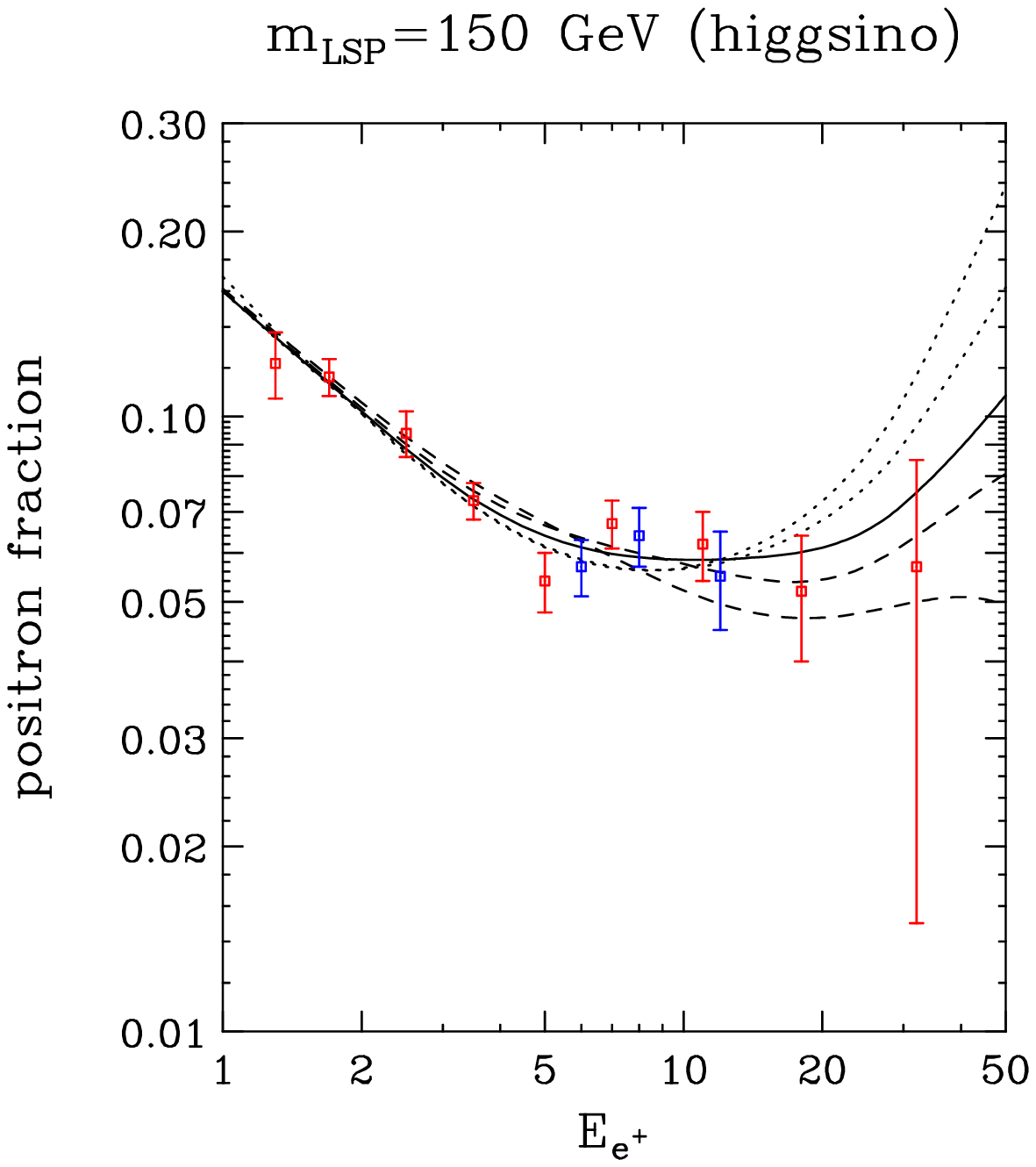} \\ [0.4cm]
\end{array}$
\end{center}
\caption{The predicted positron fraction, as a function of positron energy (in GeV), for a 150 GeV neutralino which annihilates 96\% to $b \bar{b}$ and 4\% to $\tau^+ \tau^-$ (left) or 58\% to $W^+ W^-$ and 42\% to $ZZ$ (right). The solid line represents the distance to the source (the dark matter clump) at which the predicted spectrum best fit the data (0.42 and 0.62 kpc for the left and right panels, respectively). Dotted lines represent the spectra for a source at a distance less than found for the best fit (0.23 and 0.19 kpc or 0.20 and 0.19 kpc for the left and right panels, respectively). For these two lines, the $\chi^2$ is larger by 1 and 4, respectively (1 and 2-$\sigma$). The dashed lines are the same, but for distances greater than found for the best fit (0.85 and 1.3 kpc or 1.1 and 1.6 kpc for the left and right panels, respectively) . The normalization was considered a free parameter. The error bars shown are for the HEAT experiment. Red (lighter) error bars are from the 94-95 flight. The three blue (darker) errors bars between 6 and 12 GeV are from the 2000 flight.}
\label{fit2}
\end{figure}

Figure~\ref{fit2} shows the positron fraction, as a function of positron energy, for two scenarios with supersymmetric dark matter candidates. The various lines represent clumps of dark matter at different distances from Earth. Note the substantial variation in the positron spectrum which results. In all cases, the normalization was considered a free parameter. The predicted spectrum is compared to the error bars of the 1995 and 2000 HEAT data.

These results show that the spectral shape of the observed positron excess can be fit well by dark matter annihilation models. This neglects the issue of the annihilation rate (normalization), however. To produce the observed excess, a very high annihilation rate is required in the local region (within a few kpc). For supersymmtric dark matter, this requires very dramatic dark matter substructure \cite{Hooper:2003ad}. For Kaluza-Klein dark matter, with larger cross sections and more favorable annihilation models, it may be more natural to accommodate the observed positron excess \cite{kribskkpos}.

In the future, new experiments, such as AMS-02 \cite{ams02}, PAMELA \cite{pamela} and Bess Polar \cite{besspolar}, will refine the positron spectrum considerably. See section~\ref{posantipro} for more details.

\subsubsection{Anti-protons}

Anti-protons travel much greater distances than positrons before losing their energy as they propagate through the galactic magnetic fields. Therefore, the dark matter distribution throughout much of the galaxy can contribute to the observed anti-proton spectrum~\cite{Stecker:jc,Rudaz:1987ry}. The measurement of the BESS experiment finds a cosmic anti-proton flux of $1.27^{+0.37}_{-0.32} \times 10^{-6} \rm{cm}^{-2} \rm{s}^{-1} \rm{sr}^{-1} \rm{GeV}^{-1}$ in the range of 400 to 560 MeV. This measurement is difficult to interpret in the context of dark matter annihilations due to large uncertainties in the size of the diffusion zone and other propagation characteristics \cite{Donato:2003xg}. Future experiments, especially those with sensitivity at greater energies, will be needed to identify signatures of dark matter in the cosmic anti-proton spectrum. For more information on anti-protons from dark matter annihilations, see Refs.~\cite{antiproton1,Donato:2003xg,Bergstrom:1999jc}.

\subsection{The Role of Substructures} 

Annihilation radiation could be enhanced by the presence of substructures
in the galactic halo. The actual effect depends crucially on the 
prescription of the profile and the spatial distribution of substructures.

Several groups focused on the signal enhancement due to the
presence of ``clumps'' in dark matter distribution, a common feature of 
N-body simulations. The effect of the enhancement of the annihilation 
radiation on the gamma-ray flux has been studied by, \eg, Bergstrom 
\etal~\cite{Bergstrom:1998zs}, Calcaneo-Roldan and Moore~\cite{Calcaneo-Roldan:2000yt},
Tasitsiomi and Olinto~\cite{Tasitsiomi:2002vh},
Berezinsky, Dokuchaev and Eroshenko 2003 \cite{Berezinsky:2003vn}
and Stoehr \etal~\cite{Stoehr:2003hf}). 

Recently the problem has been carefully investigated by 
Koushiappas, Zentner and Walker~\cite{Koushiappas:2003bn}, 
by means of a semi-analytic model of structure formation calibrated 
on high-resolution N-body simulations.
The authors concluded that previous estimates were optimistic, and that
it may be possible for the upcoming experiments GLAST and VERITAS
to detect gamma-rays from dark matter clumps only if the neutralino 
is relatively light, \ie $m_{\chi} \lsim 100\,$ GeV.

Blasi \etal\cite{Blasi:2002ct} studied the synchrotron emission
produced by secondary electron-positron pairs, produced by neutralino
annihilations, in the galactic magnetic field. If confirmed, their 
results would imply a microwave emission observable over CMB 
anisotropies, which is potentially identifiable by its spatial
structure or its radio spectrum. As we mentioned before, the presence of substructure is also a possible explanation
for the positron excess observed by HEAT \cite{Baltz:2001ir,Hooper:2003ad}.

Clumps are not the only substructures that can potentially
increase the annihilation flux. Among other structures considered 
in the literature, are the so-called caustics. In fact, continuous 
infall of dark matter on our galaxy should give rise to ring shaped 
overdensities, called caustics (see \eg Ref.~\cite{Sikivie:1999jv}). 
Unfortunately the prospects for the 
detection of annihilation radiation from these substructures do not 
appear promising (see Bergstrom, Edsjo and Gunnarsson~\cite{Bergstrom:2000bk}).

Apart from galactic substructures, one could ask what
the annihilation flux from all structures and substructures in 
the Universe is, \ie what extra-galactic background would be
produced by dark matter annihilation. This problem has been investigated
by Bergstrom, Edsjo and Ullio~\cite{Bergstrom:2001jj},
Taylor and Silk~\cite{Taylor:2002zd}, and Ullio \etal
~\cite{Ullio:2002pj}. In particular, the authors of the last 
reference stressed the possibility of observing, for some
specific regions of the SUSY parameter space, and sufficiently dense substructures, a spectacular feature in the gamma-ray spectrum produced by cosmological redshift and 
absorption along the line-of-sight of the gamma-ray line 
from dark matter annihilation.

\subsection{Constraints from Helioseismology}
\label{sec:helio}

The seismic diagnostics of the Sun's interior puts important constraints on the internal thermodynamic
structure of the Sun. Indeed, such research has led to
significant improvements in our understanding of microphysics such as the
equation of state and the opacity calculations, and to a better
determination of specific cross-sections in the $pp$ chain
(see \eg Ref.~\cite{Lopes:2002gp} and references therein). 

It is intriguing to investigate whether the fact that 
the Sun evolves in a halo of WIMPs affects its internal 
structure and the details of its evolution.
Modifying an existing numerical code for the Solar
structure, Lopes \etal~\cite{Lopes:2002gp} estimated the 
influence of the WIMP halo on the evolution and structure of 
the Sun, and calculated the deviations of the ``modified Sun'' with respect 
to the Solar Standard Model and to helioseismic
data. They then rejected the portions of the WIMP parameter space
leading to Solar models in conflict with Helioseismic
observations. 

\begin{figure}
\centering
\includegraphics[width=0.7\textwidth,clip=true]{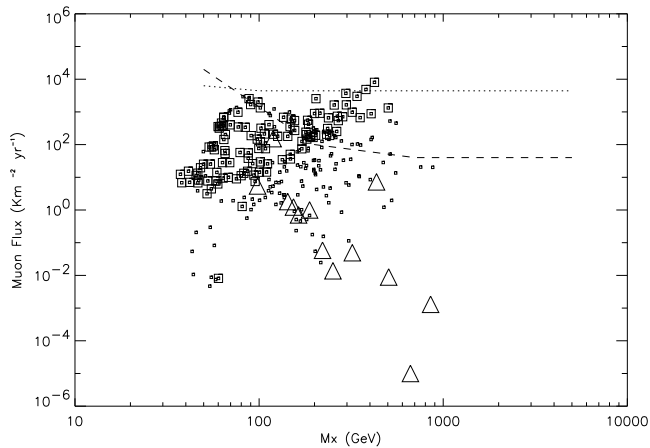}
\caption[Effects of helioseismic constraints on the muon flux from the Sun]
{Predicted neutrino-induced muon flux produced by neutralino annihilation
in the Sun. Small squares correspond to models within the phenomenological MSSM, triangles correspond
to selected benchmark points within the mSUGRA framework.
Big squares are used to highlight models leading to a local variation
of luminosity of the solar core larger than $10^{-5}$ (which could
thus be potentially probed by upcoming solar seismic observations).
The dotted and dashed curves represent the current limit sensitivity
of MACRO and the expected sensitivity of IceCube.}
\label{helio3}
\end{figure}

Although current measurement do not appear to impose strong constraints on dark matter particles (see also
Lopes, Silk and Hansen~\cite{Lopes:2001ra}
and Bottino \etal~\cite{Bottino:2002pd}), 
it is expected that future helioseismic experiments will be sensitive to
luminosities from WIMP annihilations in the solar core larger than $10^{-5}$
times the solar core luminosity (see figure~\ref{helio3}). Furthermore, if the
dark matter density increases toward the central region of our galaxy, as is
suggested by N-body simulations, stars nearer to the Galactic center would  evolve in a WIMP halo
which is much more dense, where the effects of dark matter on
the stellar structure could be of enormous importance.


Preliminary calculations~\cite{lopespc} suggest that evolution times of stars
evolving in dense dark matter halos are significantly shorter.
If confirmed, these results would change our understanding
of stellar evolution and shed new light on
the stellar population near the center of our galaxy.

\subsection{Constraints on Superheavy Dark Matter}

In our discussion thus far, we have focused on the detection of weakly interacting dark matter particles with masses near the electroweak scale. Here, we will extend that discussion to include particles which are much more massive.

Recently, Albuquerque and Baudis~\cite{Albuquerque:2003ei}
have studied the prospects for the direct detection of supermassive dark matter particles. They find that if such particles are strongly interacting (simpzillas), masses below $\sim 10^{15} \,$GeV can be probed by current experiments. A superheavy, weakly interacting particle is not constrained by this method, however.

The prospects for the observation of supermassive dark matter annihilation from the galactic center
are not very promising (see \eg Ref.~\cite{Bertone:2001jv}). Nevertheless, portions of the relevant 
parameter space can be efficiently probed by gamma-ray experiments. 

The prospects for the observation of high-energy neutrinos from
the Sun are potentially interesting~\cite{Albuquerque:2000rk,crotty}.
For simpzillas, kilometer scale neutrino telescopes, such as IceCube can test a broad range of masses. A signature unique to this scenario would 
be a predominance of tau neutrinos with energies above $\sim$TeV. 

The compilation of
results in the work of Starkman \etal~\cite{Starkman:1990nj} considers the constraints on superheavy dark matter found from double-$\beta$ decays, cosmic-ray detectors, galactic-halo
stability, cooling of molecular clouds, proton-decay detectors and
longevity of neutron stars. 

The constraints derived from old neutron stars is particularly
interesting. The argument goes as follows: if WIMPs 
exist they would accrete on neutron stars, the same way as 
they do on the Sun (see Sec.~\ref{nufromsun} and Sec.~\ref{sec:helio})~\cite{Goldman:nd}.
For certain regions of the WIMP parameter space, the accretion
can be so efficient that WIMPs become self-gravitating,
then collapse to a mini black hole, which finally destroys
the star. However, a large portion of the parameter space of modern 
superheavy candidates would escape these constraint, since 
the collapse could be prevented by self-annihilations.

\section{Conclusions}

There is compelling evidence for the existence of dark matter. Although our
understanding of its nature and distribution is still
incomplete, many independent observations suggest that about 30\% of 
the total energy density of the Universe is made of some sort of
non-baryonic dark matter. We have reviewed such observations and discussed
how they compare with theoretical predictions, and in particular with 
the results of N-body simulations. 


The dark matter problem is not only relevant to astrophysicists but also to 
the particle and high-energy physics community. In fact, some of the best 
dark matter candidates come from possible extensions of the Standard Model of
particle physics. There is certainly no shortage of particle dark matter candidates found in such models. Among those
proposed in literature, we have focused on the dark matter particles found in models of supersymmetry (the lightest neutralino) and models with Universal Extra Dimensions (Kaluza-Klein dark matter). Although many simple models of supersymmetry, extra dimensions or other scenarios are widely discussed by the particle and astroparticle
communities, the phenomenology of the actual physical theory could be
more rich and complex. Collider experiments are probing significant regions 
of the parameter space of these hypothetical particles. Conversely, a positive
astrophysical detection of dark matter would provide invaluable information regarding the physics ``beyond the 
Standard Model''.

The astroparticle community has started a vigorous and broad  program of experiments
that may be able to shed new light on the physics and astrophysics 
of dark matter. Before discussing the results of direct and indirect searches,
we have reviewed the present and future experiments on which they are based. 

Among the most promising dark matter searches appears to be direct detection. The current situation is complicated by the claim of a positive detection by the DAMA experiment, which have been contradicted by several other experiments. It is unclear, but more
and more improbable, whether it is possible to find a theoretical 
scenario that accommodates all the experimental findings. The much 
higher (several orders of magnitude) sensitivity of future experiments
should be able to solve this controversy.

Indirect dark matter detection via annihilations in the Galactic center region is also an exciting possibility, although 
 the prospects for the observation of gamma-rays, neutrinos and synchrotron
radiation from that direction strongly depend on astrophysical  
parameters, such as the profile of dark matter in the innermost regions, which
unfortunately are poorly known. Nevertheless, the development of next-generation
gamma-ray and neutrino telescopes will allow us to test many scenarios,
especially if effects such as the adiabatic accretion onto the central black
hole significantly enhance the dark matter density and corresponding
annihilation signal. If the Galactic center turns out to contain less dark matter, observations of dwarf galaxies, external galaxies and local dark substructure may play an important role for indirect searches. 

Indirect searches for dark matter through the observation of 
high energy neutrinos produced in dark matter annihilations in the Sun are also promising. These rates do not depend strongly on the dark matter halo distribution and are thus fairly unambiguous probes of particle dark matter models. Measurement of the positron and anti-proton spectra, which are soon to improve dramatically, can also provide an opportunity to observe products of dark matter annihilations in the galactic halo.

Collectively, the direct, indirect and collider searches for particle dark matter have incredible prospects for discovery in the coming years. We hope that this review can be a useful tool in guiding members of the scientific community closer to the goal of dark matter identification which has eluded us for so long.

\section{Acknowledgements}

We wish to thank G.Sigl for earlier collaboration and countless 
stimulating discussions. We thank K.~Abazajian, J~.Beacom, A.~Birkedal-Hansen, B.~Dobrescu, S.~Hansen, I.~Liubarsky, I.~Lopes, E.~Nezri, J.~Orloff, I.~Perez, G.~Servant, C.~Skordis, P.~Salati, C.~Spiering, F.~Stoehr, T.~Tait,
J.~Taylor and A.~Zentner for illuminating comments
and discussions. Special thanks to P.~Salati for careful reading of an 
earlier version of the manuscript and T.~Plehn and L.~Wang for numerous useful comments.
The work of GB 
was supported at an earlier stage by an ``Allocation de Recherche'', PhD 
program Universite Paris 7 at the Institut d'Astrophysique de Paris, and is 
now supported by the DOE and the NASA grant NAG 5-10842 at Fermilab. DH is supported by the Leverhulme trust.

\newpage

\begin{appendix}

\section{Neutralino Mass Eigenstates}
\label{diagonal}

In the Minimal Supersymmetric Standard Model (MSSM), the neutral electroweak gauginos ($\tilde{B}, \tilde{W^3}$) and higgsinos ($\tilde{H^0_1}, \tilde{H^0_2}$) have the same quantum numbers and, therefore, mix into four mass eigenstates called neutralinos. The neutralino mass matrix in the $\widetilde{B}$-$\widetilde{W}^3$-$\widetilde{H}_1^0$-$\widetilde{H}_2^0$ basis is given by
\begin{equation}
\arraycolsep=0.01in
{\cal M}_N=\left( \begin{array}{cccc}
M_1 & 0 & -M_Z\cos \beta \sin \theta_W^{} & M_Z\sin \beta \sin \theta_W^{}
\\
0 & M_2 & M_Z\cos \beta \cos \theta_W^{} & -M_Z\sin \beta \cos \theta_W^{}
\\
-M_Z\cos \beta \sin \theta_W^{} & M_Z\cos \beta \cos \theta_W^{} & 0 & -\mu
\\
M_Z\sin \beta \sin \theta_W^{} & -M_Z\sin \beta \cos \theta_W^{} & -\mu & 0
\end{array} \right)\;,
\end{equation}
where $M_1$, $M_2$ and $\mu$ are the bino, wino and higgsino mass parameters, respectively, $\theta_W$ is the Weinberg angle and $\tan \beta$ is the ratio of the vacuum expectation values of the Higgs bosons. This matrix can be diagonalized by the matrix, $N$.
\begin{equation}
M_{\chi^0}^{\rm{diag}} = N^{\dagger}  M_{\chi^0} N.
\end{equation}
The masses of the four mass eigenstates are then given by
\cite{ElKheishen,BBO}
%
\begin{eqnarray}
\epsilon _1M_{\chi _1^0}&=&-({1\over 2}a-{1\over 6}C_2)^{1/2}
+\left [-{1\over 2}a-{1\over 3}C_2+{{C_3}\over {(8a-{8\over 3}C_2)^{1/2}}}
\right ]^{1/2}+{1\over 4}(M_1+M_2)\;, \nonumber \\ \\
\epsilon _2M_{\chi _2^0}&=&+({1\over 2}a-{1\over 6}C_2)^{1/2}
-\left [-{1\over 2}a-{1\over 3}C_2-{{C_3}\over {(8a-{8\over 3}C_2)^{1/2}}}
\right ]^{1/2}+{1\over 4}(M_1+M_2)\;, \nonumber \\ \\
\epsilon _3M_{\chi _3^0}&=&-({1\over 2}a-{1\over 6}C_2)^{1/2}
-\left [-{1\over 2}a-{1\over 3}C_2+{{C_3}\over {(8a-{8\over 3}C_2)^{1/2}}}
\right ]^{1/2}+{1\over 4}(M_1+M_2)\;, \nonumber \\ \\
\epsilon _4M_{\chi _4^0}&=&+({1\over 2}a-{1\over 6}C_2)^{1/2}
+\left [-{1\over 2}a-{1\over 3}C_2-{{C_3}\over {(8a-{8\over 3}C_2)^{1/2}}}
\right ]^{1/2}+{1\over 4}(M_1+M_2)\;, \nonumber \\
\end{eqnarray}
%
where $\epsilon _i$ is the sign of the $i$th eigenvalue
of the neutralino mass matrix, and
%
\begin{eqnarray}
C_2&=&(M_1M_2-M_Z^2-\mu ^2)-{3\over 8}(M_1+M_2)^2\;, \\
C_3&=&-{1\over 8}(M_1+M_2)^3
+{1\over 2}(M_1+M_2)(M_1M_2-M_Z^2-\mu^2)+(M_1+M_2)\mu ^2
\nonumber \\
&&+(M_1\cos ^2\theta _W^{}
+M_2\sin ^2\theta _W^{})M_Z^2+\mu M_Z^2\sin 2\beta\;, \\
C_4&=&-(M_1\cos ^2\theta _W^{}+M_2\sin ^2\theta _W^{})
M_Z^2\mu\sin 2\beta -M_1M_2\mu^2\nonumber \\
&&+{1\over 4}(M_1+M_2)
[(M_1+M_2)\mu ^2+(M_1\cos ^2\theta _W^{}+M_2\sin ^2\theta _W^{})
M_Z^2+\mu M_Z^2\sin 2\beta] \nonumber \\
&&+{1\over 16}(M_1M_2-M_Z^2-\mu^2)(M_1+M_2)^2
-{3\over 256}(M_1+M_2)^4\;, \\
a&=&{1\over {2^{1/3}}}{\rm Re}\left [-S+i(D/27)^{1/2}\right ]^{1/3}\;, \\
D&=&-4U^3-27S^2\;, \quad U=-{1\over 3}C_2^2-4C_4, \quad S=-C_3^2
-{2\over 27}C_2^3  +{8\over 3}C_2C_4\;.
\end{eqnarray}
%
The four masses above are not generally in the order $M_{\chi_1^0}<M_{\chi_2^0}<M_{\chi_3^0}<M_{\chi_4^0}$, although it is conventional to relabel the states, from lightest to heaviest.

The mixing matrix, $N$, is then given by~\cite{ElKheishen,BBO}
\begin{eqnarray}
{{N_{i2}}\over {N_{i1}}}&=&-{1\over {\tan \theta _W^{}}}
{{M_1-\epsilon _iM_{\chi _i^0}}\over {M_2-\epsilon _iM_{\chi _i^0}}}\;,
\\
{{N_{i3}}\over {N_{i1}}}&=&{{-\mu [M_2-\epsilon _iM_{\chi _i^0}]
[M_1-\epsilon _iM_{\chi _i^0}]-M_Z^2\sin \beta \cos \beta
[(M_1-M_2)\cos ^2\theta _W^{}+M_2-\epsilon _iM_{\chi _i^0}]}\over
{M_Z[M_2-\epsilon _iM_{\chi _i^0}]\sin \theta _W^{}[-\mu \cos \beta+
\epsilon _iM_{\chi _i^0} \sin \beta ]}}\;, \nonumber \\ \\
{{N_{i4}}\over {N_{i1}}}&=&{{-\epsilon _iM_{\chi _i^0}
[M_2-\epsilon _iM_{\chi _i^0}]
[M_1-\epsilon _iM_{\chi _i^0}]-M_Z^2\cos ^2\beta
[(M_1-M_2)\cos ^2\theta _W^{}+M_2-\epsilon _iM_{\chi _i^0}]}\over
{M_Z[M_2-\epsilon _iM_{\chi _i^0}]\sin \theta _W^{}[-\mu \cos \beta+
\epsilon _iM_{\chi _i^0} \sin \beta ]}}\;, \nonumber \\
\end{eqnarray}
and
\begin{eqnarray}
N_{i1}=\left [1+\left ({{N_{i2}}\over {N_{i1}}}\right )^2
+\left ({{N_{i3}}\over {N_{i1}}}\right )^2+\left ({{N_{i4}}
\over {N_{i1}}}\right )^2
\right ]^{-1/2}\;.
\end{eqnarray}
The lighest neutralino ($\chi^0_1$) is a mixture of gauginos and higgsinos:
\begin{equation}
\chi^0_1 = N_{11}\tilde{B}     +N_{12} \tilde{W}^3
          +N_{13}\tilde{H}^0_1 +N_{14} \tilde{H}^0_2.
\end{equation}
The gaugino fraction of $\chi^0_1$ is defined as
\begin{equation}
f_G = N_{11}^2 +N_{12}^2
\label{eq:fG}
\end{equation}
and its higgisino fraction as
\begin{equation}
f_H = N_{13}^2 +N_{14}^2.
\label{eq:fH}
\end{equation}

\pagebreak

\section{Neutralino Annihilation Cross Sections in the Low Velocity Limit}
\label{annappendix}

In this appendix, we give the amplitudes and cross sections for the most important neutralino annihilation channels in the low velocity limit (the first term in the expansion $\sigma v = a + b v^2 + ...$). This is sufficient for indirect detection but generally insufficient for relic density calculations in which velocity dependent contributions are important. For a more complete list, with all S and P-wave tree level annihilation amplitudes, see Refs.~\cite{ann1,Jungman:1995df,Nihei:2002ij,Nihei:2001qs,Birkedal-Hansen:2002sx}.

\subsection{Annihilation Into Fermions}

\begin{figure}[t!]
\centering
\includegraphics[width=0.7\textwidth,clip=true]{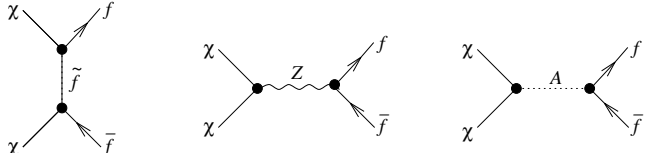}
\caption[]
{Tree level diagrams for neutralino annihilation into fermion pairs. From Ref.~\cite{Jungman:1995df}.}
\label{fermionfeyn}
\end{figure}

Neutralinos can annihilate to fermion pairs by three tree level diagrams~\cite{ann1,ellislsp,ann2,ann3}. These processes consist of s-channel exchange of pseudoscalar Higgs and $Z^0$-bosons and t-channel exchange of sfermions (see Fig.~\ref{fermionfeyn}).  

The amplitude for pseudoscalar Higgs exchange is given by
\begin{equation}
{\cal A}_A = 4\sqrt{2}\, g\, T_{A\, 11}\, h_{Aff} \;
		{1\over 4- (m_A/m_\chi)^2 + i\, \Gamma_A m_A / m_\chi^2}.
\label{Aannamp}
\end{equation}
Here, $m_A$ is the pseudoscalar Higgs mass and $\Gamma_A$ is the pseudoscalar Higgs width. $T_{A\, 11}$ is the $A^0$-neutralino-neutralino coupling and is given by  
\begin{equation}
T_{A\, 11}=-\sin \beta Q^{\prime\prime}_{1,1} + \cos \beta S^{\prime\prime}_{1,1},
\end{equation}
where $Q^{\prime\prime}_{1,1} = N_{3,1} (N_{2,1} - \tan \theta_W N_{1,1})$ and $S^{\prime\prime}_{1,1} = N_{4,1} (N_{2,1} - \tan \theta_W N_{1,1})$. $N$ is the matrix which diagonalizes the neutralino mass matrix in the $\widetilde{B}$-$\widetilde{W}^3$-$\widetilde{H}_1^0$-$\widetilde{H}_2^0$ basis, $M_{\chi^0}^{\rm{diag}} = N^{\dagger}  M_{\chi^0} N$ (see Appendix~\ref{diagonal}). $\theta_W$ is the Weinberg angle and $\tan \beta$ is the ratio of the Higgs vacuum expectation values. $h_{Aff}$ is the $A^0$-fermion-fermion Yukawa coupling. For up-type fermions, this is given by
\begin{equation}
h_{Aff} = -\frac{g m_f \cot \beta}{2 m_{W^{\pm}}}.
\end{equation}
For down-type fermions, it is
\begin{equation}
h_{Aff} = -\frac{g m_f \tan \beta}{2 m_{W^{\pm}}}.
\end{equation}


The amplitude for neutralino annihilation via sfermion exchange to a pair of fermions, $f_i \bar{f}_i$, is given by
\begin{equation}
{\cal A}_{\tilde{f}} = \sqrt{2}\,  \sum_{j=1}^6 \frac{1}{P_j} \bigg(\left[(X^{'}_{ij1})^2+(W^{'}_{ij1})^2  \right] \frac{m_{f_i}}{m_{\chi}} + 2 X^{'}_{ij1} W^{'}_{ij1}   \bigg),
\end{equation}
where $P_j=1+(m_{\tilde{f_j}}/m_{\chi})^2-(m_{f_i}/m_{\chi})^2$ and the sum is over the six sfermion states which couple to the final state fermion. The fermion-sfermion-neutralino couplings, $X^{'}_{ij1}$ and $W^{'}_{ij1}$, are given by
\begin{equation}
X^{'}_{ij1} = X_{1} (\Pi_L \Theta_f)_{i,j} + Z_{i,k,1} (\Pi_R \Theta_f)_{k,j}
\end{equation}
and
\begin{equation}
W^{'}_{ij1} = Y_{1} (\Pi_R \Theta_f)_{i,j} + Z_{i,k,1} (\Pi_L \Theta_f)_{k,j},
\end{equation}
where
\begin{equation}
X_1 = -g \sqrt{2} \,[\,T_3(f_i) N_{2,1}^* - \tan \theta_W \,(\,T_3(f_i)- e(f_i)\,) \, N_{1,1}^* \,],
\end{equation}
and
\begin{equation}
Y_1 = g \sqrt{2} \, \tan \theta_W \, e(f_i)  N_{1,1}^*.
\end{equation}
For final state up-type quarks,
\begin{equation}
Z_{i,j,1} =  - \frac{g}{\sqrt{2} \, m_{W^{\pm}} \, \sin \beta}  \Theta_{i,j} \, N_{4,1}^*.
\end{equation}
For final state down-type quarks,
\begin{equation}
Z_{i,j,1} =  - \frac{g}{\sqrt{2} \, m_{W^{\pm}} \, \cos \beta}  \Theta_{i,j} \, N_{3,1}^*.
\end{equation}
And for final state leptons,
\begin{equation}
Z_{i,j,1} =  - \frac{g}{\sqrt{2} \, m_{W^{\pm}} \, \cos \beta}  \Theta_{i,j} \, N_{3,1}^*.  
\end{equation}
Here, $T_3(f_i)$ and $e(f_i)$ are the weak hypercharge and electric charge of the final state fermion. $N$, again, is the matrix which diagonalizes the neutralino mass matrix. $\Theta_f$'s are the appropriate 6 x 6 sfermion mass matrices and $\Pi_{L,R}$ are left and right projection operators:
\begin{equation}
\Pi_L =  
\left( \begin{array}{cccccc}
1 & 0 & 0 & 0 & 0 & 0 \\
0 & 1 & 0 & 0 & 0 & 0 \\
0 & 0 & 1 & 0 & 0 & 0 \\
\end{array} \right),
\end{equation}
\begin{equation}
\Pi_R =  
\left( \begin{array}{cccccc}
0 & 0 & 0 & 1 & 0 & 0 \\
0 & 0 & 0 & 0 & 1 & 0 \\
0 & 0 & 0 & 0 & 0 & 1 \\
\end{array} \right).
\end{equation}
%
%
%
%
%
Lastly, the amplitude for neutralino annihilation to fermions via $Z$ exchange is given by
\begin{equation}
{\cal A}_{Z} = 2 \sqrt{2} \frac{g^2}{\cos^2\theta_{\rm{W}}} O^{\prime\prime\, L}_{1,1} T_3(f_i) \frac{m_{f_i} m_{\chi}}{m_Z^2},
\end{equation}
where $T_3(f_i)$ is the weak hypercharge of the fermion. The coupling $O^{\prime\prime\, L}_{1,1}$ is given by $\frac{1}{2}(-N_{3,1} N_{3,1}^*  + N_{4,1} N_{4,1}^* )$. 

Summing these three contributions to the amplitude, we can calculate the cross section for this process:
\begin{equation}
\sigma v(\chi\chi\rightarrow \bar{f}_i f_i)_{v\rightarrow0} = 
		{c_f \beta_f\over 128 \pi  m_\chi^2} 
		|{\cal A}_A(\chi\chi\rightarrow \bar{f}_i f_i)+{\cal A}_{\tilde{f}}(\chi\chi\rightarrow \bar{f}_i f_i)+{\cal A}_Z(\chi\chi\rightarrow \bar{f}_i f_i)|^2,
\label{fermanncross}
\end{equation}
where $\beta_f = \sqrt{1 - m_f^2/m_\chi^2}$. $c_f$ is a color factor which is equal to three for quark final states and one otherwise.

We emphasize that all tree level (low velocity) neutralino annihilation diagrams to fermion pairs have amplitudes which are proportional to the final state fermion mass. For sfermion and $Z^0$ exhange, this is because the $Z^0$-fermion-fermion and neutralino-fermion-fermion couplings preserve chirality. For psuedoscalar Higgs exchange, the amplitude introduces an explicit factor of the fermion mass in the Yukawa coupling. We also note that the Yukawa coupling which appears in the psuedoscalar Higgs exchange amplitude is proportional to $\tan \beta$ for down-type quarks and $\cot \beta$ for up-type quarks. The net result of these observations is that neutralino annihilation into fermions will be dominated by heavy final states, $b \bar{b}$, $\tau^- \tau^+$ and, if kinematically allowed, $t \bar{t}$. Furthermore, if $\tan \beta$ is large, bottom-type fermions may dominate over up-type fermions, even if less massive. For example, annihilations to $b \bar{b}$ may dominate over $t \bar{t}$, even for heavy neutralinos.

\subsection{Annihilation Into Gauge Bosons}

\begin{figure}[t]
\begin{center}
$\begin{array}{c@{\hspace{0.5in}}c}
\includegraphics[width=0.4\textwidth]{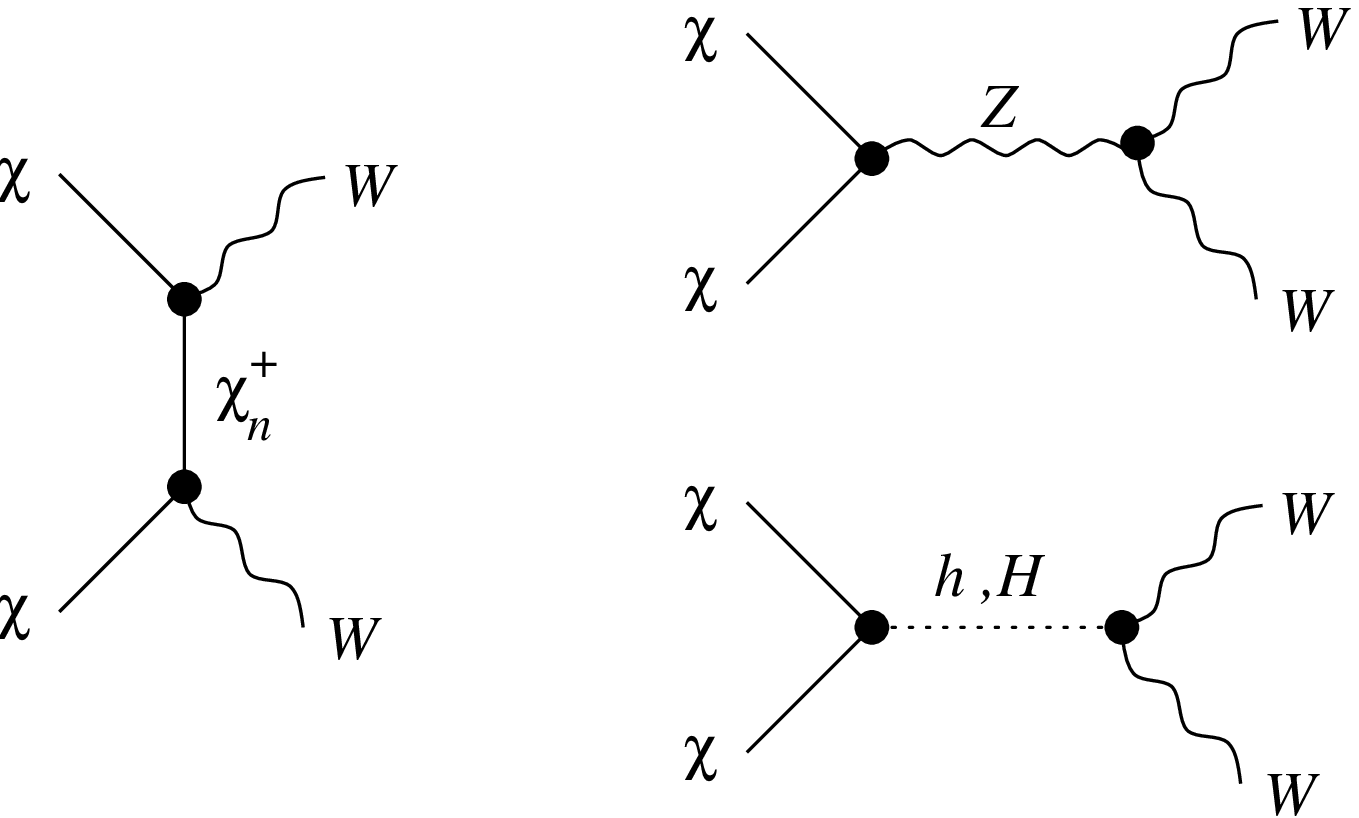} &
\includegraphics[width=0.4\textwidth]{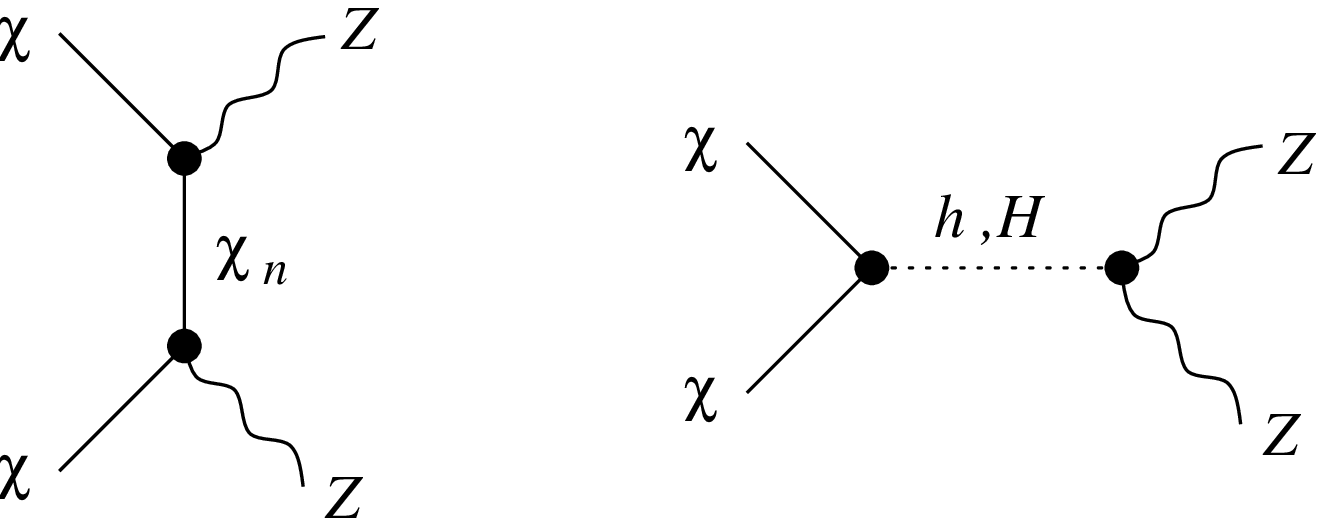} \\ [0.4cm]
\end{array}$
\end{center}
\caption[]
{Tree level diagrams for neutralino annihilation into gauge boson pairs. From Ref.~\cite{Jungman:1995df}.}
\label{gaugefeyn}
\end{figure}

Generally, neutralinos can annihilate into gauge boson pairs via several processes (see Fig.~\ref{gaugefeyn}) \cite{ann1,ann4,ann5,ann6}. In the low velocity limit, however, only t-channel processes via chargino or neutralino exchange are non-vanishing. 

In the low velocity limit, the amplitude for neutralino annihilation to $W^{\pm}$-pairs is given by
\begin{equation}
{\cal A}(\chi\chi\rightarrow W^+ W^-)_{v\rightarrow0} = 
2 \sqrt{2}\, \beta_W g^2 \sum_{n=1}^2 
\left[ (O^L_{1,n})^2 + (O^R_{1,n})^2 \right] {1\over P_n},
\end{equation}
where $\beta_W = \sqrt{1 - m_W^2 / m_\chi^2}$ and $P_n = 1 + (m_{\chi_n^{\pm}} / m_\chi)^2 - (m_W / m_\chi)^2$. The sum is over chargino states. $O^L_{1,n}$ and $O^R_{1,n}$ are the neutralino couplings to charginos given by $\frac{-1}{\sqrt{2}} N_{4,1} V_{2,n}^* +N_{2,1} V_{1,n}^*$ and $\frac{1}{\sqrt{2}} N_{3,1}^* U_{2,n} +N_{2,1}^* U_{1,n}$, respectively, where $N$, again, is the matrix which diagonalizes the neutralino mass matrix. The $V$'s and $U$'s are components of the chargino mass matrix, in the basis

\begin{equation}
U = \left(\matrix{ \cos\phi_-  -\sin\phi_- \cr
		\sin\phi_-  \cos\phi_+ \cr}\right) 
\end{equation}
and
\begin{equation}
V = \left(\matrix{ \cos\phi_+  -\sin\phi_+ \cr
		\sin\phi_+  \cos\phi_- \cr} \right),
\end{equation}
where
\begin{equation}
\tan 2\phi_- = 2\sqrt{2} m_W 
	{{(\mu \sin\beta + M_2 \cos\beta)}
	\over {(M_2^2 - \mu^2 + 2m_W^2 \cos 2\beta)}} 
\end{equation}
and
\begin{equation}
\tan 2\phi_+ = 2\sqrt{2} m_W
	{{(\mu\cos\beta + M_2 \sin\beta)}
	\over {(M_2^2 - \mu^2 - 2m_W^2 \cos\beta)}}.
\end{equation}

The amplitude for annihilations to $Z^0$-pairs is similar:
\begin{equation}
{\cal A}(\chi\chi\rightarrow Z^0Z^0)_{v\rightarrow0} = 
4 \sqrt{2}\, \beta_Z {g^2\over \cos^2\theta_W}  \sum_{n=1}^4
\left( O^{\prime\prime\, L}_{1,n}\right)^2 {1\over P_n}.
\end{equation}
Here, $\beta_Z = \sqrt{1 - m_Z^2 / m_\chi^2}$, and $P_n = 1 + (m_{\chi_n} / m_\chi)^2 - (m_Z / m_\chi)^2$. The sum is over neutralino states. The coupling $O^{\prime\prime\, L}_{1,n}$ is given by $\frac{1}{2}(-N_{3,1} N_{3,n}^*  + N_{4,1} N_{4,n}^* )$.

The low velocity annihilation cross section for this mode is then given by
\begin{equation}
\sigma v(\chi\chi\rightarrow GG)_{v\rightarrow0} =
{1\over S_G} {\beta_G\over 128 \pi  m_\chi^2} 
|{\cal A}(\chi\chi\rightarrow GG)|^2,
\end{equation}
where $G$ indicates which gauge boson is being considered. $S_G$ is a statistical factor equal to one for $W^+ W^-$ and two for $Z^0 Z^0$.

It is useful to note that pure-gaugino neutralinos have a no S-wave annihilation amplitude to gauge bosons. Pure-higgsinos or mixed higgsino-gauginos, however, can annihilate efficiently via these channels, even at low velocities.

\subsection{Annihilation Into Higgs Bosons}

\begin{figure}[t]
\begin{center}
$\begin{array}{c@{\hspace{0.5in}}c}
\includegraphics[width=0.4\textwidth]{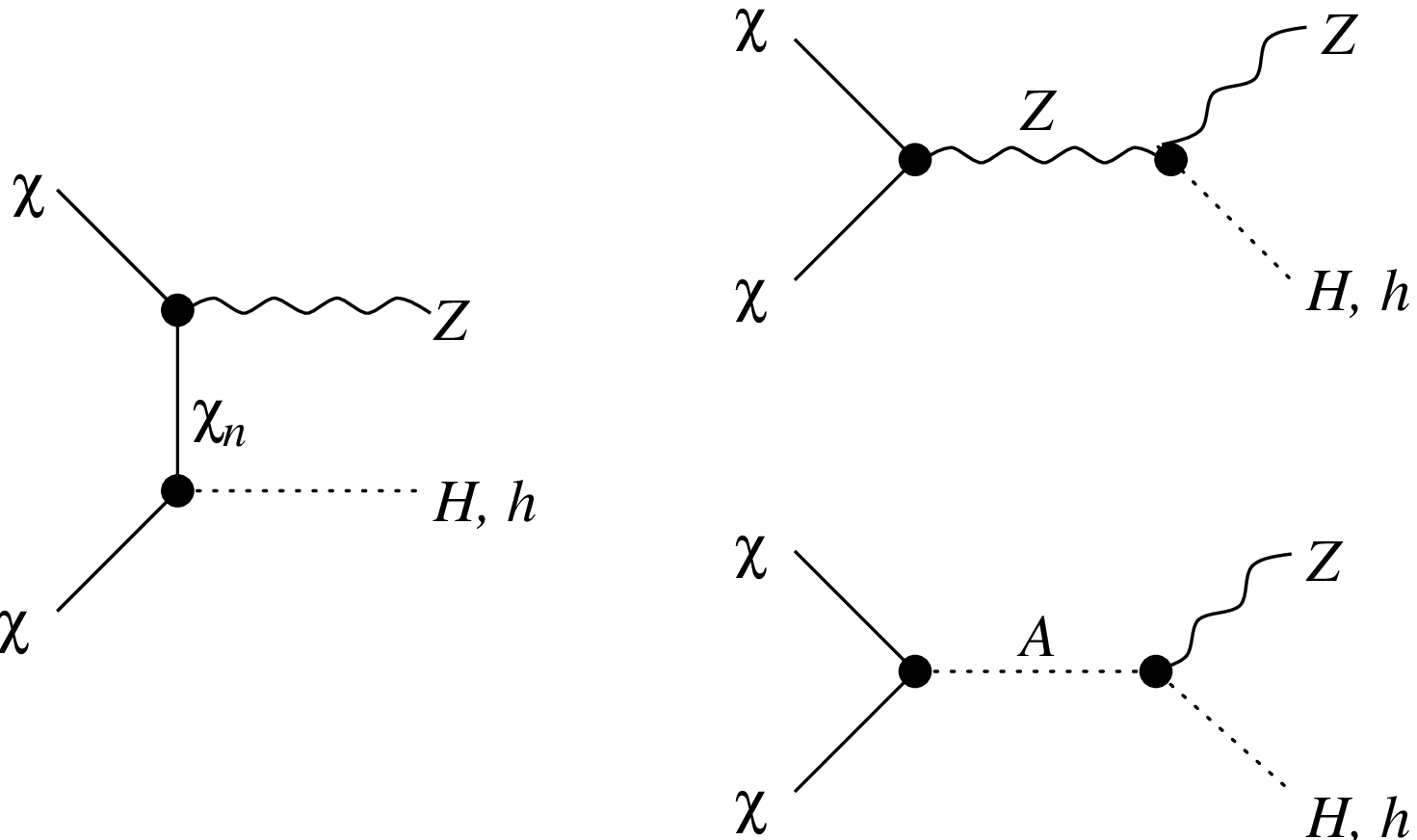} &
\includegraphics[width=0.4\textwidth]{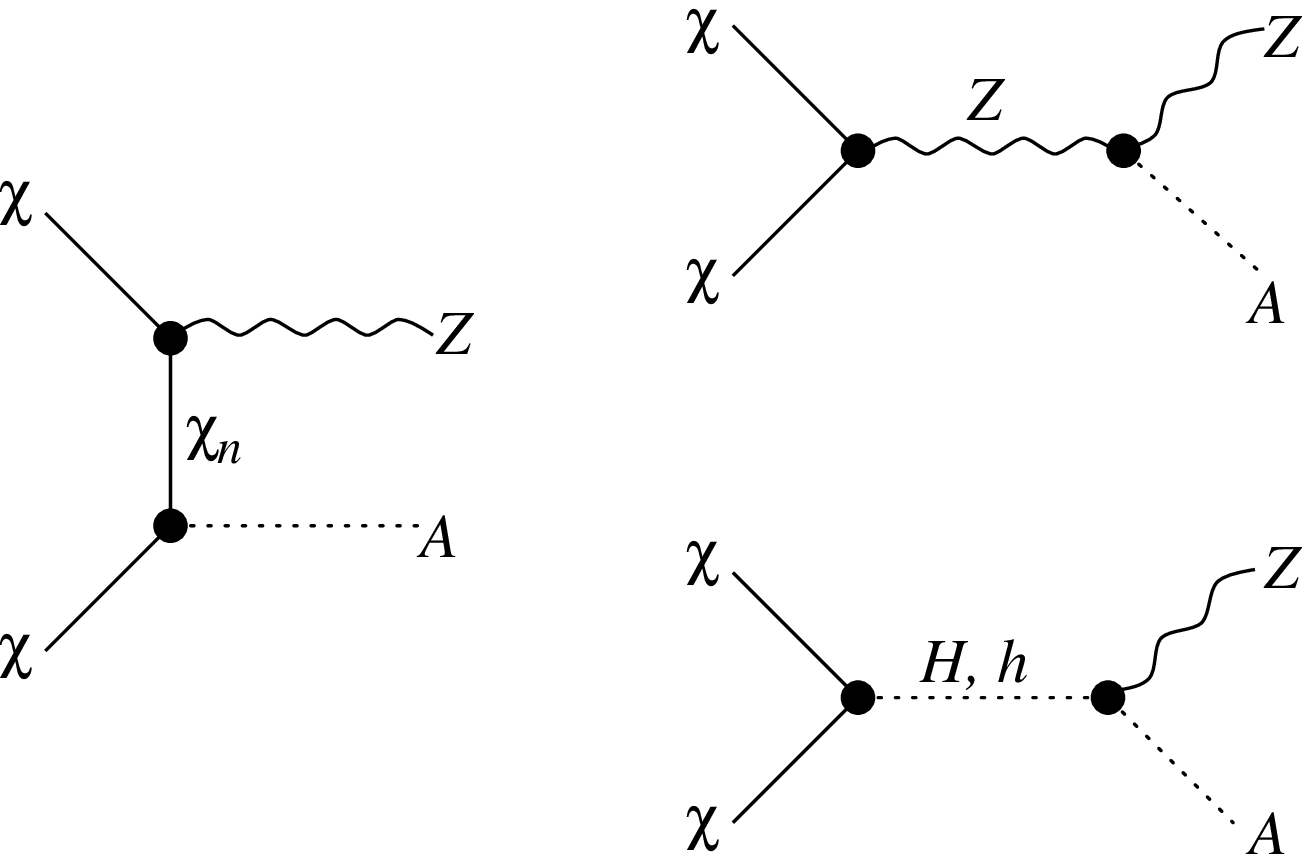} \\ [0.4cm]
\end{array}$
\end{center}
\caption[]
{Tree level diagrams for neutralino annihilation into a $Z$ and a Higgs boson. From Ref.~\cite{Jungman:1995df}.}
\label{zhiggsfig}
\end{figure}

\begin{figure}[t]
\begin{center}
$\begin{array}{c@{\hspace{0.5in}}c}
\includegraphics[width=0.4\textwidth]{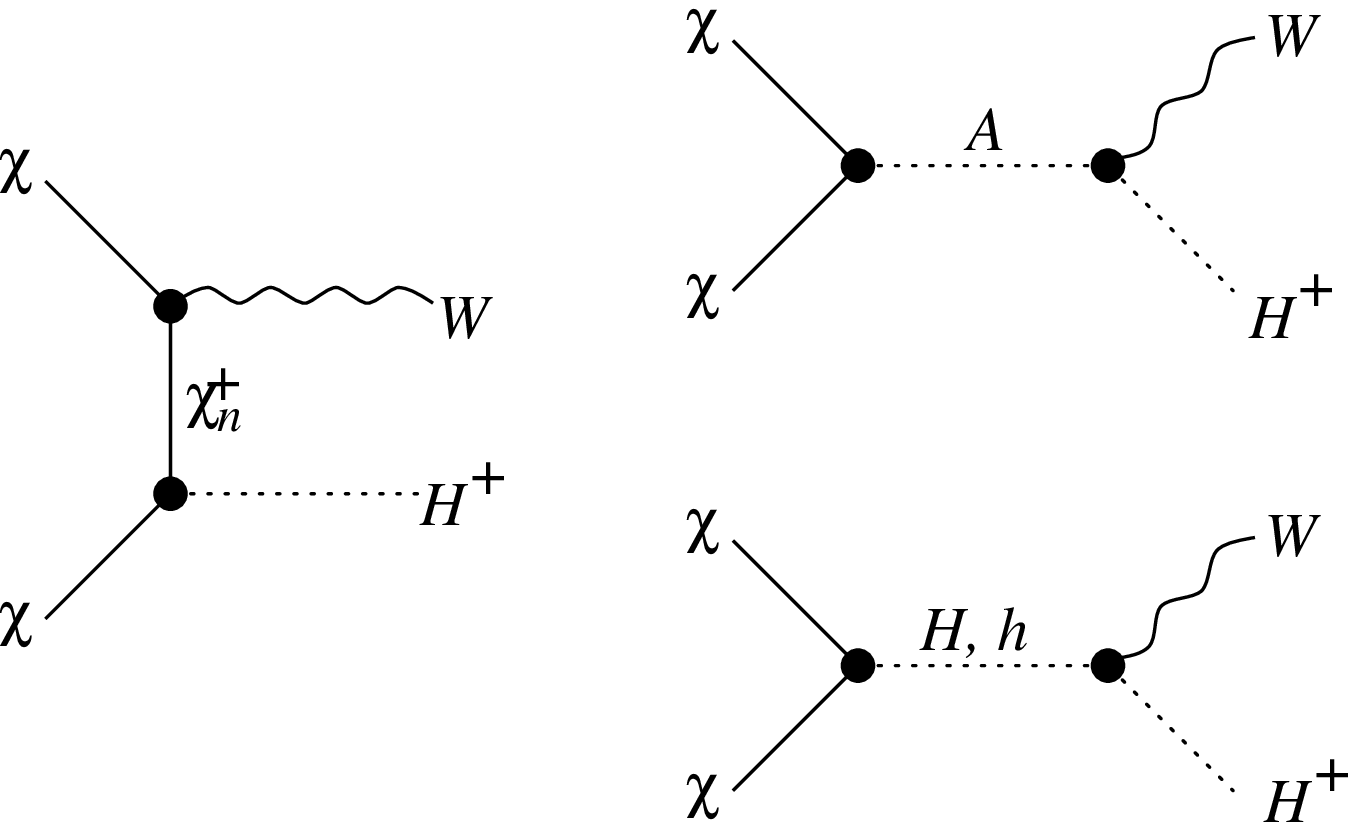} &
\includegraphics[width=0.4\textwidth]{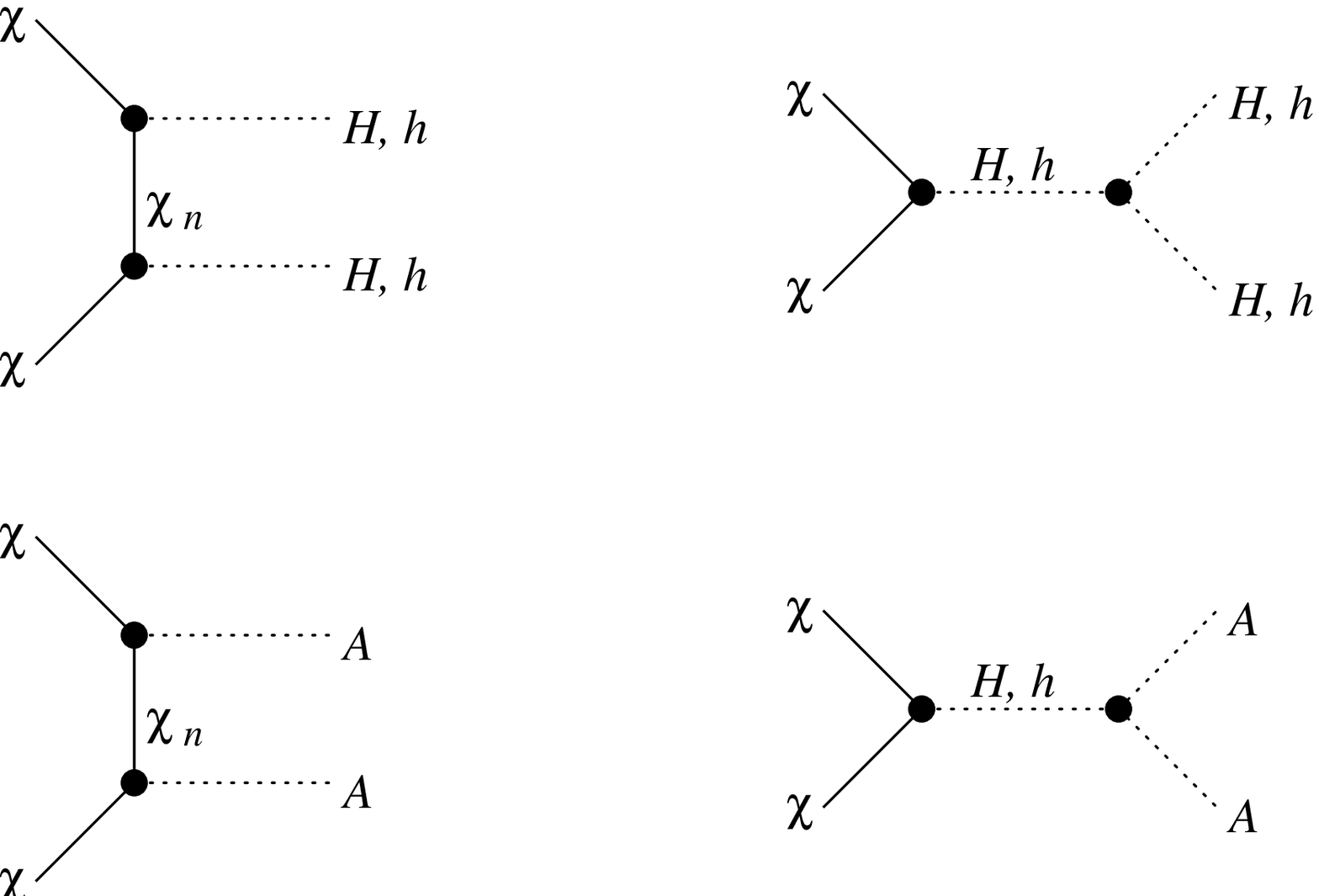} \\ [0.4cm]
\end{array}$
\end{center}
\caption[]
{Tree level diagrams for neutralino annihilation into a $W^{\pm}$ and a Higgs boson or a pair of Higgs bosons. From Ref.~\cite{Jungman:1995df}.}
\label{whiggsfig}
\end{figure}

\begin{figure}[t]
\begin{center}
\includegraphics[width=0.4\textwidth]{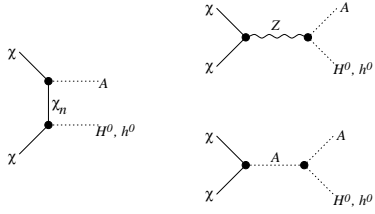} 
\end{center}
\caption[]
{Tree level diagrams for neutralino annihilation into a neutral Higgs boson and a pseudoscalar Higgs boson. From Ref.~\cite{Jungman:1995df}.}
\label{higgshiggsfig}
\end{figure}

There are many tree level diagrams which contribute to neutralino annihilation into Higgs boson pairs or a Higgs boson and a gauge boson (see Figs.~\ref{zhiggsfig}, \ref{whiggsfig} and \ref{higgshiggsfig})\cite{ann1,ann4,ann5,ann6,ann7,ann8,ann9}.

In the low velocity limit, the amplitude for neutralino annihilation to a $Z^0$ and a light neutral Higgs, $h^0$, is given by
\begin{eqnarray} 
\label{zhiggs}
{\cal A}(\chi\chi\rightarrow Z^0 h^0)_{v\rightarrow0} = 
- 2 \sqrt{2}\, \beta_{Zh}\, {m_\chi \over m_Z} {g^2\over \cos\theta_W} 
\biggl[ - 2\sum_{n=1}^4  O^{\prime\prime L}_{1,n} T_{h\, 1,n} \\ \nonumber
\times {m_{\chi_n} - m_\chi \over m_\chi P_n} + O^{\prime\prime L}_{1,1} 
		{m_\chi \sin(\beta-\alpha)\over m_Z \cos\theta_W}
	- {2\cos(\alpha-\beta) T_{A\, 1,1} \over 4 - m_A^2/m_\chi^2 
		+ i\, \Gamma_A m_A / m_\chi^2} \biggr],
\end{eqnarray}
where $\Gamma_A$ is the pseudoscalar Higgs width and $T_{h\, 1,n}$ is the $h^0-\chi_0-\chi_n$ Yukawa coupling (see below).  The couplings, $O^{\prime\prime L}_{1,n}$, is given by $N_{3,1} (N_{2,n}- \tan \theta_W N_{1,n})/2 + N_{3,n} (N_{2,1}-\tan \theta_W N_{1,1})/2$ and $P_n = 1 + (m_{\chi_n} / m_\chi)^2 - {1\over 2}(m_Z / m_\chi)^2 - {1\over 2}(m_h/m_\chi)^2$. $\tan \beta$ is the ratio of the Higgs vacuum expectation values and the mixing angle, $\alpha$, is related to $beta$ by
\begin{equation}
\sin 2\alpha =  -\sin 2\beta
	\left({{m_H^2 + m_h^2}\over{m_H^2 - m_h^2}} \right)
\end{equation}
and
\begin{equation}
\cos 2\alpha = -\cos 2\beta
	\left({{m_A^2 - m_Z^2}\over{m_H^2 - m_h^2}}\right).	
\end{equation}
The three terms of Eq.~\ref{zhiggs} correspond to neutralino, $Z^0$ and $A^0$ exchange, from first to last.

The expression for neutralino annihilations to a $Z^0$ and a heavy Higgs boson, $H^0$, is the same, but with $\sin(\beta - \alpha)$ and $\cos(\alpha - \beta)$ exchanged, and the couplings and masses of $h^0$ replaced by the couplings and masses of $H^0$. These Yukawa couplings are given by
\begin{equation}
T_{h\, 1,n} = \sin \alpha \, Q^{\prime \prime}_{1,n} + \cos \alpha \, S^{\prime \prime}_{1,n}
\end{equation}
and
\begin{equation}
T_{H\, 1,n} = -\cos \alpha \, Q^{\prime \prime}_{1,n} + \sin \alpha \, S^{\prime \prime}_{1,n}.
\end{equation}
Here, $S^{\prime \prime}_{1,n} =  N_{4,1} (N_{2,n}-\tan \theta_W N_{1,n})/2+ N_{4,n} (N_{2,1}-\tan \theta_W N_{1,1})/2$. $Q^{\prime \prime}_{1,n}$ is defined above.

The amplitude for annihilations to a $W^{\pm}$ and a charged Higgs boson is given by
\begin{eqnarray}
\label{wchhiggs}
{\cal A}(\chi\chi\rightarrow W^\pm H^\mp)_{v\rightarrow0} = 
4\sqrt{2}\,\beta_{WH} g^2 \biggl[
	-{1\over 2} \sum_{n=1}^2
	{m_\chi\over m_W}
	{O^R_{1,n}Q^{\prime R}_{1,n} - O^L_{1,n} Q^{\prime L}_{1,n} 
			\over P_n} \\ \nonumber
+ {1\over 2} \sum_{n=1}^2{m_{\chi^+_n} \over m_W}
	{O^R_{1,n}Q^{\prime L}_{1,n} - O^L_{1,n} Q^{\prime R}_{1,n} 
			\over P_n}
	- {m_\chi \, T_{A\, 11}\over m_W (4 - m_A^2 / m_\chi^2)}
\biggr],
\end{eqnarray}
where $P_n = 1 + (m_{\chi_n^\pm} / m_\chi)^2
- {1\over 2}(m_{H^\pm} / m_\chi)^2 - {1\over 2}(m_W/m_\chi)^2$. $O^R_{1,n}$ and $O^L_{1,n}$ are couplings given earlier in this appendix. $Q^{\prime L}_{n,m}$ and $Q^{\prime R}_{n,m}$ are the chargino-neutralino-charged Higgs couplings, given by
\begin{equation}
	Q^{\prime\, L}_{nm} = \cos\beta \left[
		N_{4n}^* V_{1m}^* 
		+ \sqrt{{1\over 2}} (N_{2n}^* 
				+ \tan\theta_W N_{1n}^*) V_{2m}^*
		\right] 
\end{equation}
and
\begin{equation}
	Q^{\prime\, R}_{nm} = \sin\beta \left[
		N_{3n} U_{1m} 
		- \sqrt{{1\over 2}} (N_{2n} 
				+ \tan\theta_W N_{1n}) U_{2m}
		\right],
\end{equation}
where each of the quantities used have been defined earlier in this appendix. The first and second terms of Eq.~\ref{wchhiggs} correspond to chargino exchange. The third term corresponds to pseudoscalar Higgs exchange (there is no low velocity contribution from scalar Higgs exchange).

Finally, the low velocity amplitude for neutralino annihilation into one neutral Higgs boson and one pseudoscalar Higgs boson is given by

\begin{eqnarray}
	{\cal A}(\chi\chi& \rightarrow h^0 A^0)_{v\rightarrow0} =
	\sqrt{2}\, g^2
	\Biggl\{
		-4\sum_{n=1}^4 {T_{h 0n} T_{A 0n}}
		\biggl[ 
		{m_{\chi_n} \over m_\chi P_n}
	-  {m_A^2 - m_h^2 \over m_\chi^2}\biggr]\cr
	&\hskip-40pt - 2 {m_Z\over m_\chi}
		{\sin(\alpha+\beta) \cos2\beta \over 2\cos\theta_W}	
		{ T_{A\, 00} \over 4- m_A^2/m_\chi^2}
	- {\cos(\alpha-\beta) O^{\prime\prime}_{L00} \over 2 \cos^2\theta_W}
		{ m_A^2 - m_a^2 \over m_Z^2}
	\Biggr\}.
\end{eqnarray}
Here, $P_n = 1 + (m_{\chi_n} / m_\chi)^2 - {1\over 2}(m_A / m_\chi)^2 - {1\over 2}(m_h/m_\chi)^2$. The other quantities have been defined earlier in this appendix. Again, the amplitude for the analogous process with a heavy rather than light Higgs boson in the final state is the same, but with $\sin(\alpha + \beta)$ and $\cos(\alpha - \beta)$ exchanged and the light Higgs couplings and masses replaced with those for the heavy Higgs boson.

In the low velocity limit, there is no amplitude for neutralino annihilations to $H^+ H^-$, $h^0 h^0$, $H^0 H^0$, $A^0 A^0$ or $Z^0 A^0$. 

The low velocity cross section for neutralino annihilation via any of these modes is 
\begin{equation}
\sigma v(\chi\chi\rightarrow XY)_{v\rightarrow0}\; = 
	{\beta_{XY} \over 128 \pi m_\chi^2}
	|{\cal A}(\chi\chi\rightarrow
	XY)_{v\rightarrow0}|^2,
\end{equation}
where $X$ and $Y$ are labels referring to the final state particles.

\subsection{Annihilation Into Photons}
\label{lines}

Although there are no tree level processes for neutralino annihilation into photons, loop level processes to $\gamma \gamma$ and $\gamma Z^0$ are very interesting, as they may provide a spectral line feature observable in indirect detection experiments.

\begin{figure}[h]
 \centering
 \mbox{\subfigure{\epsfig{file=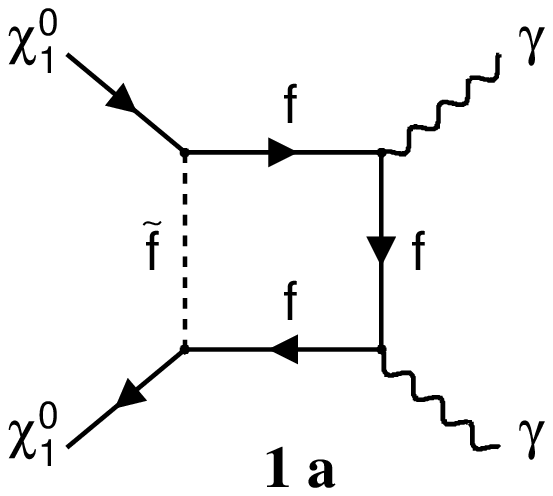,width=3cm}}\quad
       \subfigure{\epsfig{file=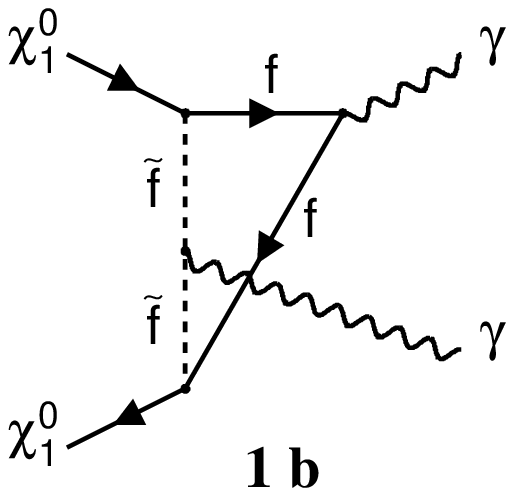,width=3cm}}\quad
       \subfigure{\epsfig{file=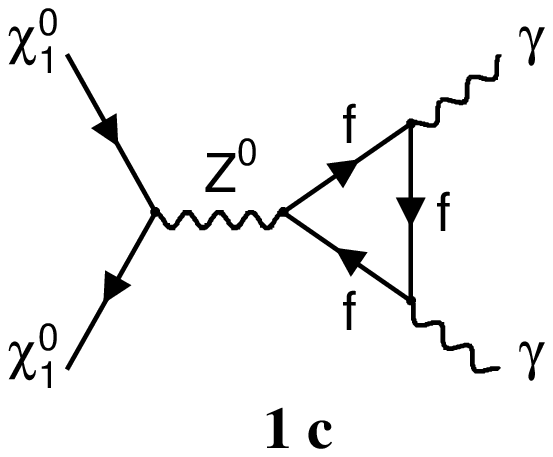,width=3cm}}\quad
       \subfigure{\epsfig{file=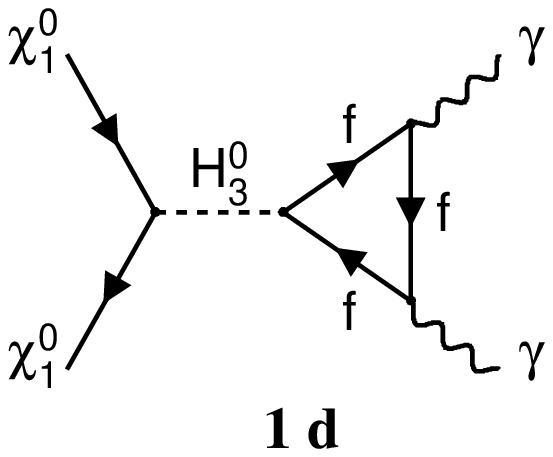,width=3cm}}}
 \mbox{\subfigure{\epsfig{file=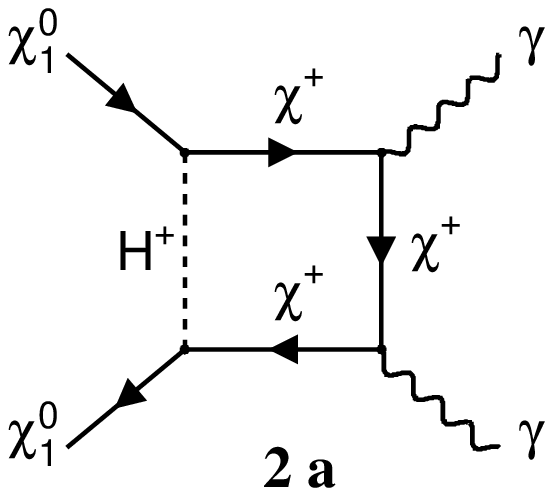,width=3cm}}\quad
       \subfigure{\epsfig{file=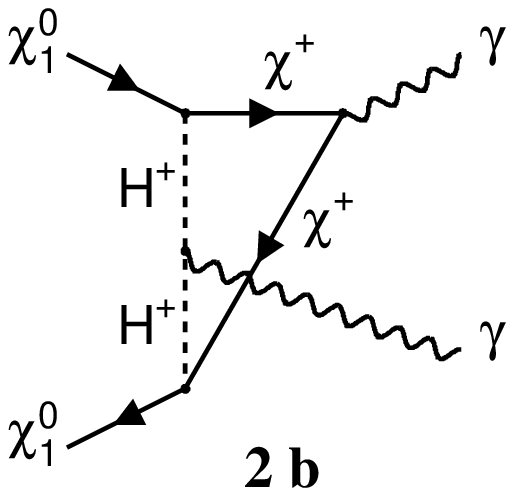,width=3cm}}\quad
       \subfigure{\epsfig{file=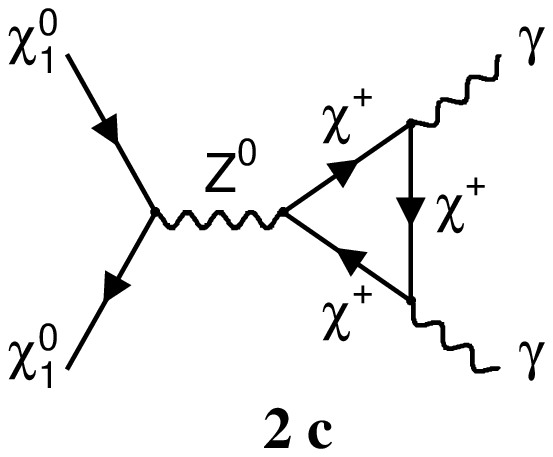,width=3cm}}\quad
       \subfigure{\epsfig{file=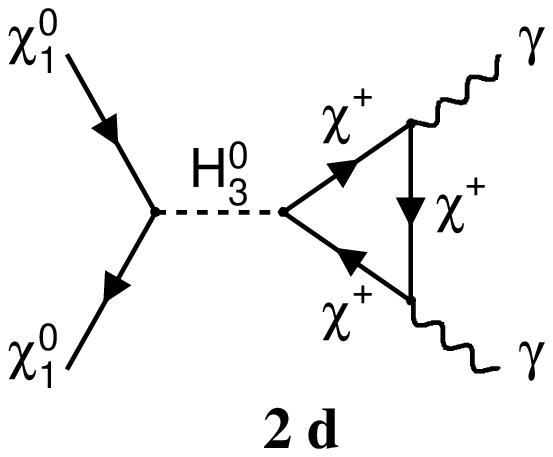,width=3cm}}}
 \mbox{\subfigure{\epsfig{file=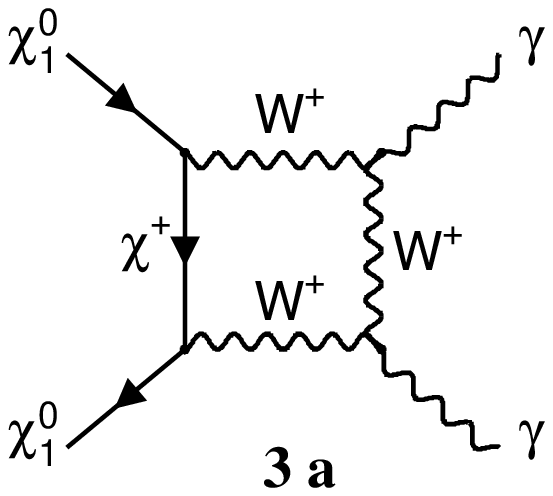,width=3cm}}\quad
       \subfigure{\epsfig{file=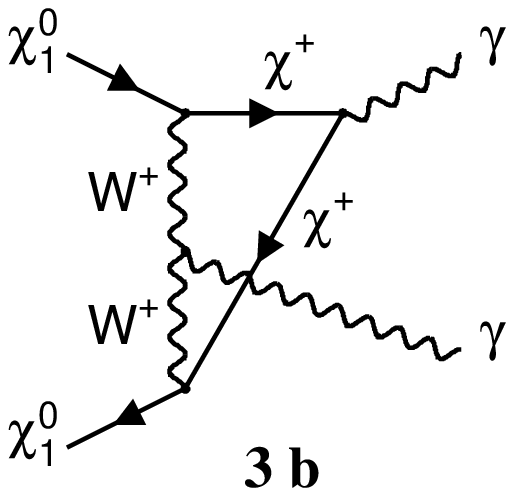,width=3cm}}\quad
      \subfigure{\epsfig{file=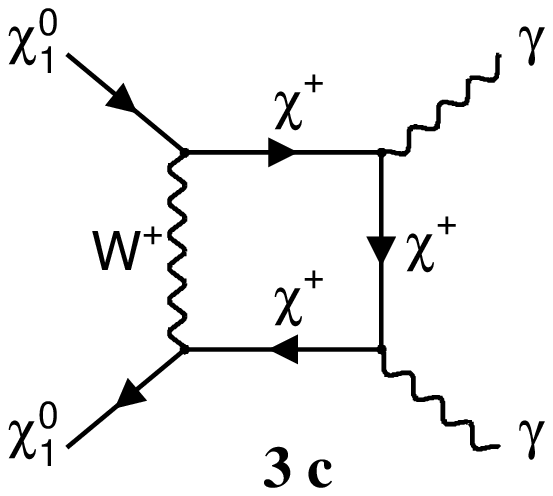,width=3cm}}}
 \mbox{\subfigure{\epsfig{file=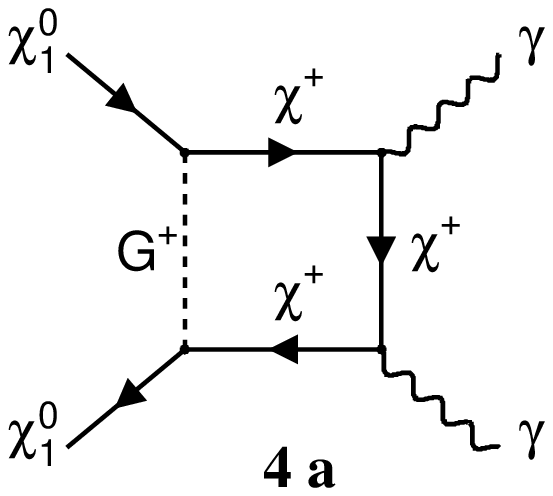,width=3cm}}\quad
       \subfigure{\epsfig{file=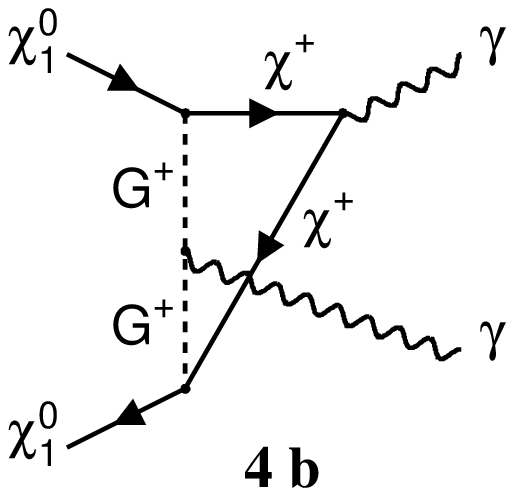,width=3cm}}}
\caption{Diagrams contributing, at one loop level, to neutralino 
annihilation into two photons. From Ref.~\cite{bergstromgammagamma}.}\label{fig:gammagamma}
\end{figure}

In Fig.~\ref{fig:gammagamma}, all of the one-loop diagrams are shown for neutralino annihilation to a pair of photons. In Fig.~\ref{fig:gammaz}, the corresponding diagrams to a photon and a $Z^0$ are shown. We do not include the corresponding amplitudes or cross sections here. For those results, see Ref.~\cite{bergstromgammagamma} and Ref.~\cite{bergstromgammaz} for $\gamma \gamma$ and $\gamma Z^0$ final states, respectively. Also see Ref.~\cite{Gounaris:2003jw}.

\pagebreak

\pagebreak

\begin{figure}[h]
 \centering
 \mbox{\subfigure{\epsfig{file=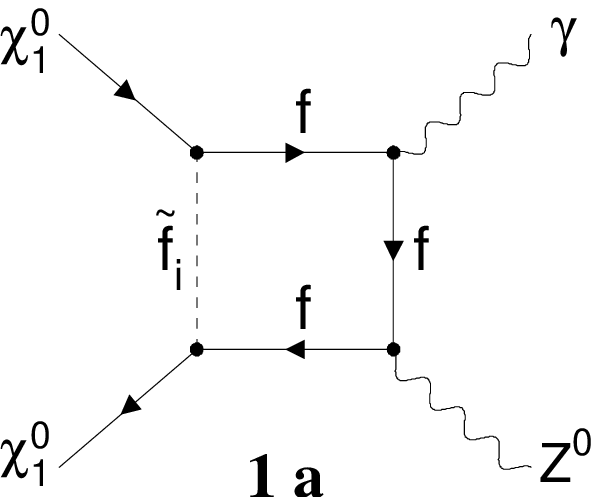,width=3cm}}\quad
       \subfigure{\epsfig{file=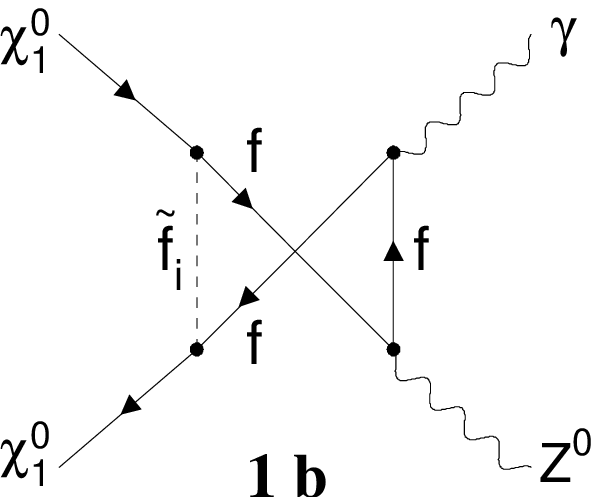,width=3cm}}\quad
       \subfigure{\epsfig{file=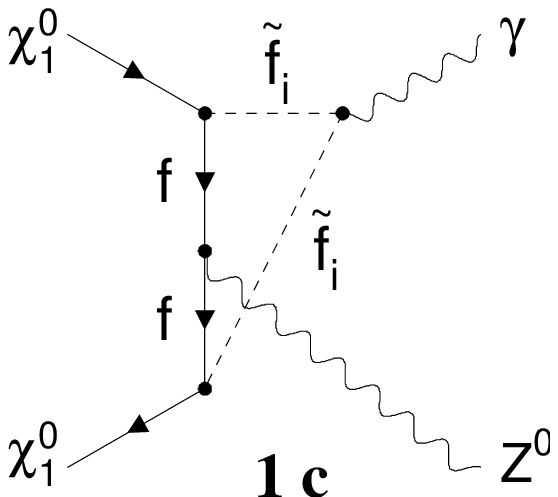,width=3cm}}\quad
       \subfigure{\epsfig{file=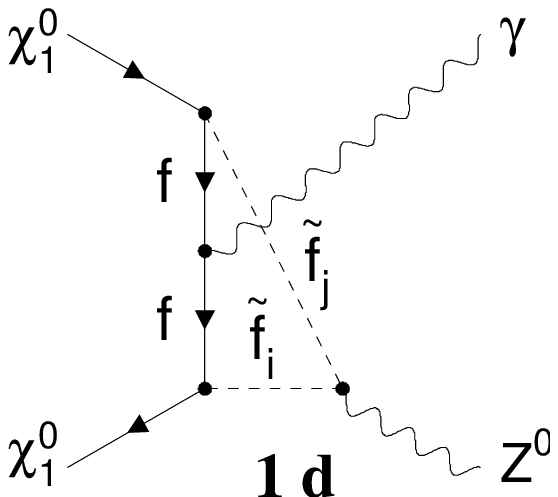,width=3cm}}}
 \mbox{\subfigure{\epsfig{file=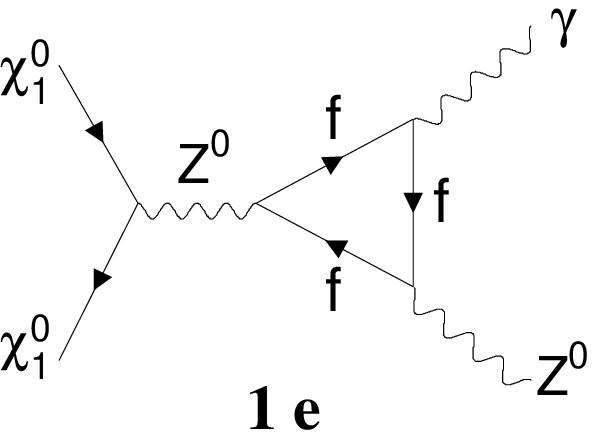,width=3cm}}\quad
       \subfigure{\epsfig{file=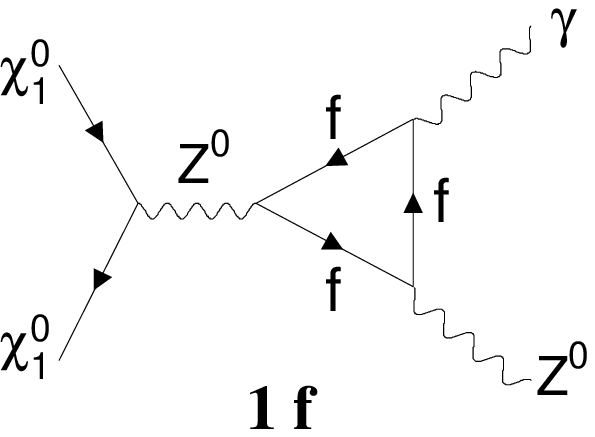,width=3cm}}\quad
       \subfigure{\epsfig{file=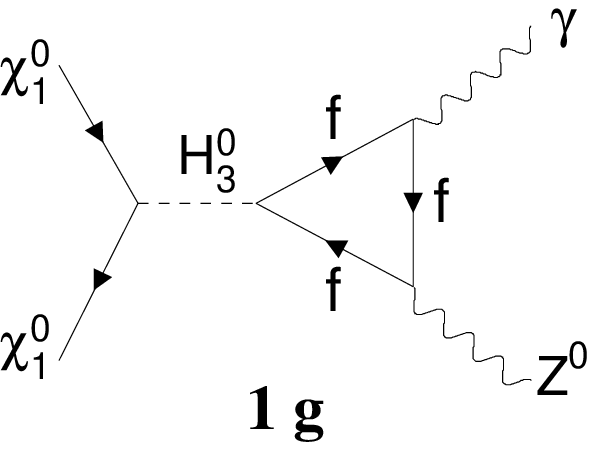,width=3cm}}\quad
       \subfigure{\epsfig{file=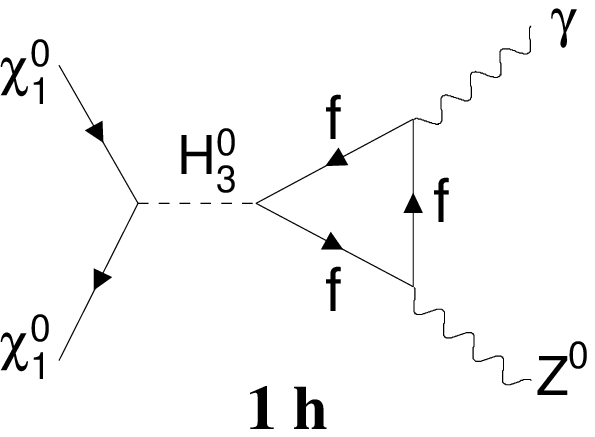,width=3cm}}}
 \mbox{\subfigure{\epsfig{file=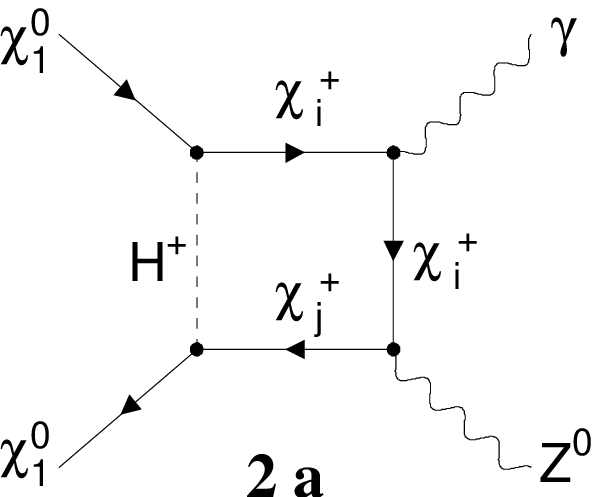,width=3cm}}\quad
       \subfigure{\epsfig{file=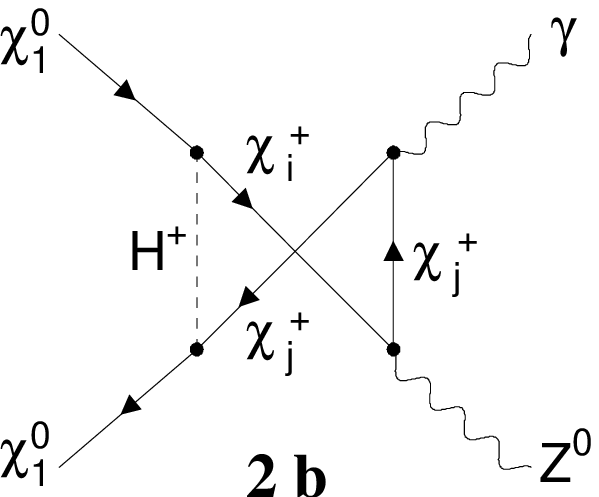,width=3cm}}\quad
       \subfigure{\epsfig{file=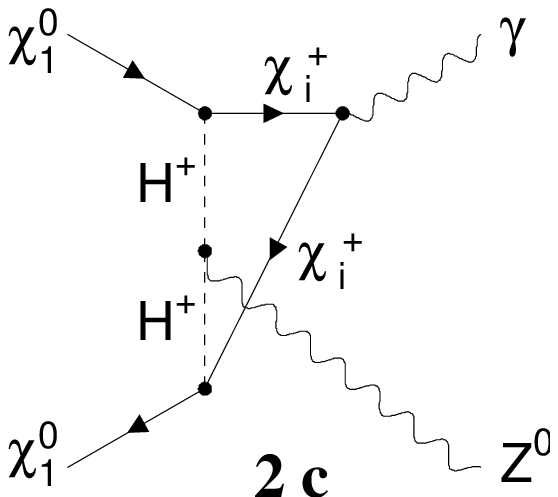,width=3cm}}\quad
       \subfigure{\epsfig{file=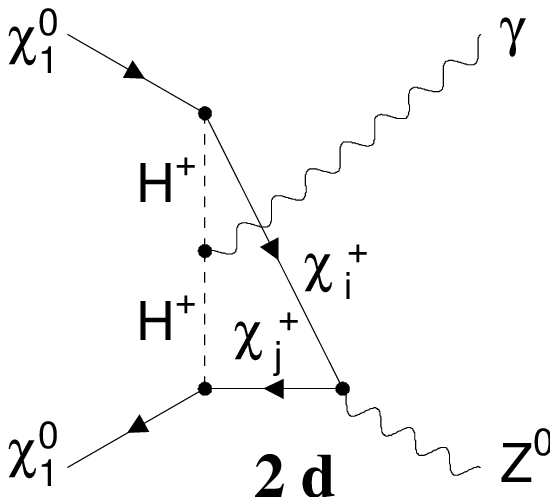,width=3cm}}}
 \mbox{\subfigure{\epsfig{file=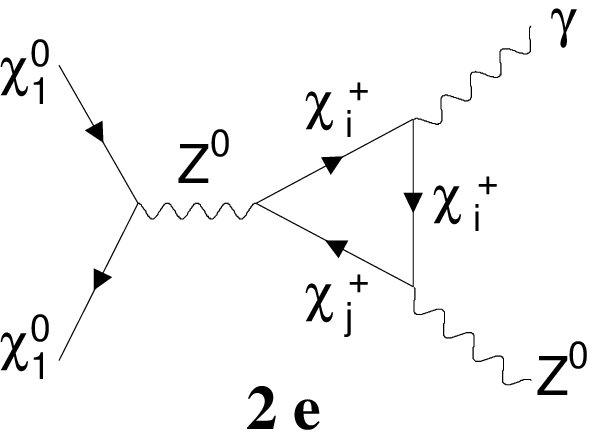,width=3cm}}\quad
       \subfigure{\epsfig{file=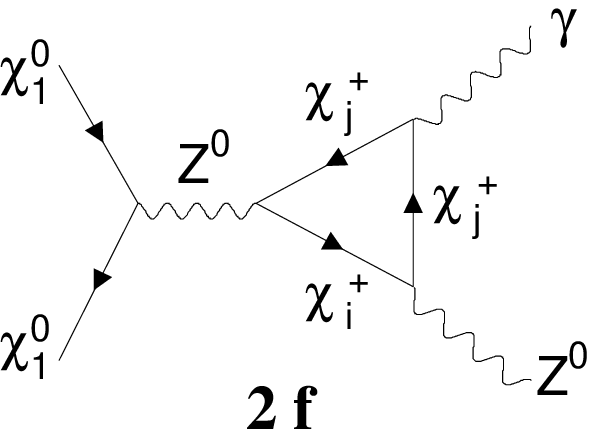,width=3cm}}\quad
       \subfigure{\epsfig{file=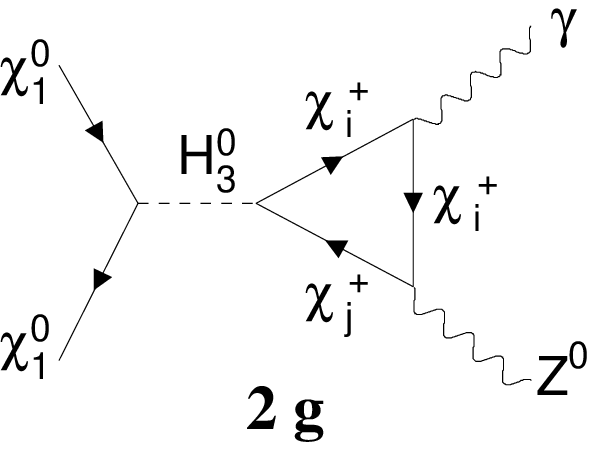,width=3cm}}\quad
       \subfigure{\epsfig{file=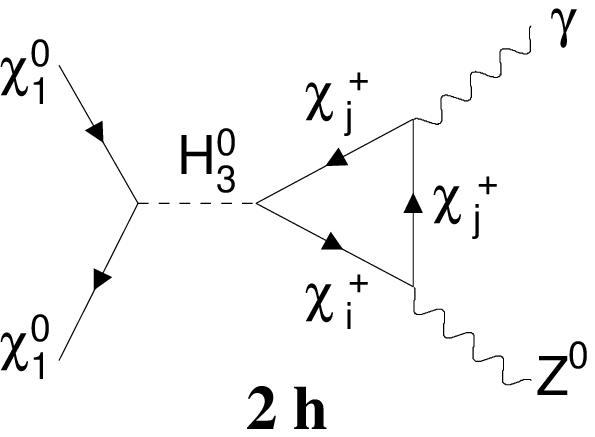,width=3cm}}}
\end{figure}

\pagebreak

\begin{figure}[h]
\centering
 \mbox{\subfigure{\epsfig{file=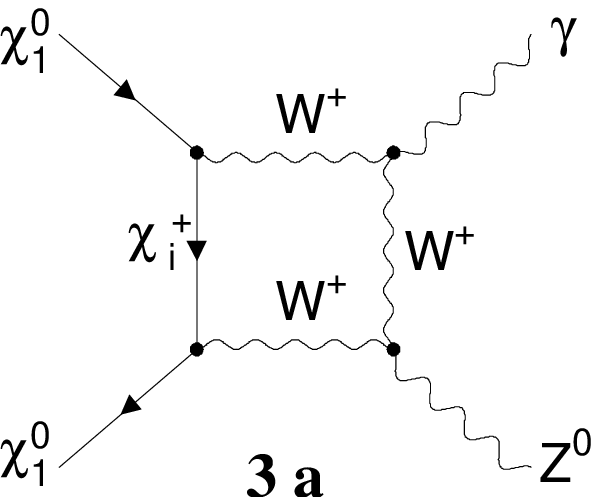,width=3cm}}\quad
       \subfigure{\epsfig{file=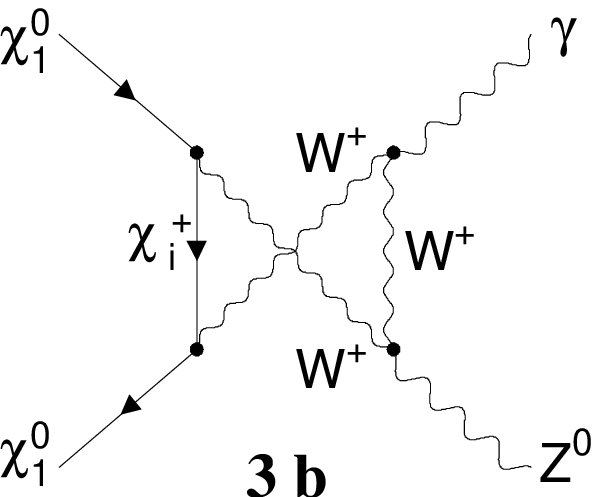,width=3cm}}\quad
       \subfigure{\epsfig{file=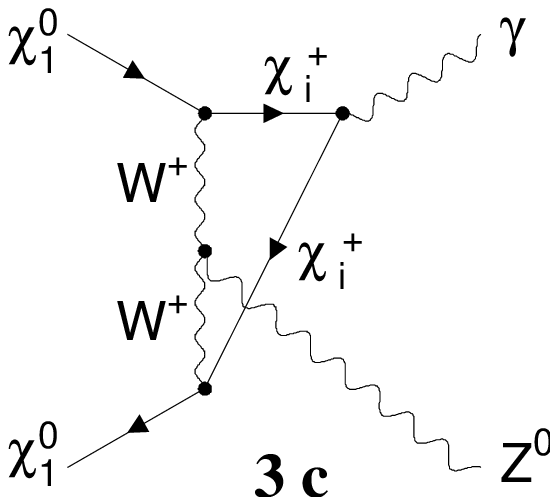,width=3cm}}}
 \mbox{\subfigure{\epsfig{file=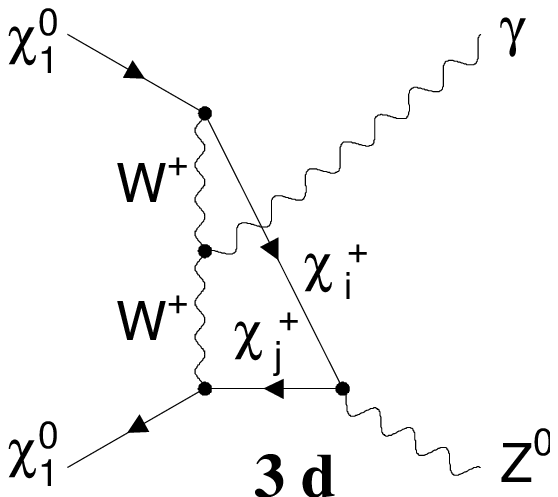,width=3cm}}\quad
       \subfigure{\epsfig{file=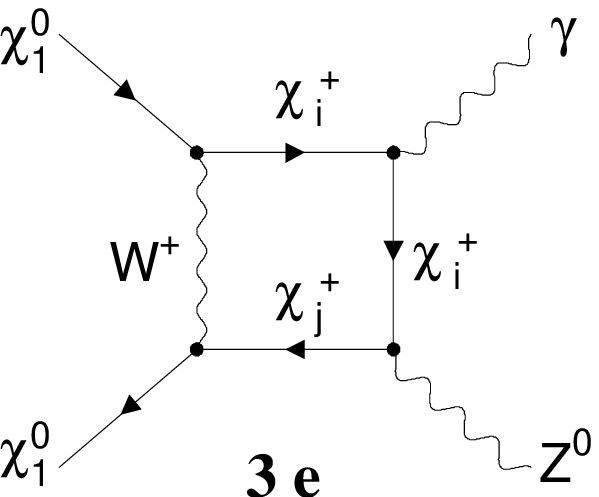,width=3cm}}\quad
       \subfigure{\epsfig{file=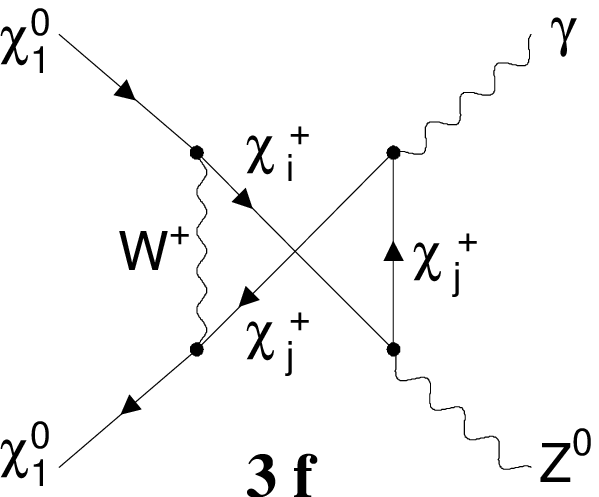,width=3cm}}}
 \mbox{\subfigure{\epsfig{file=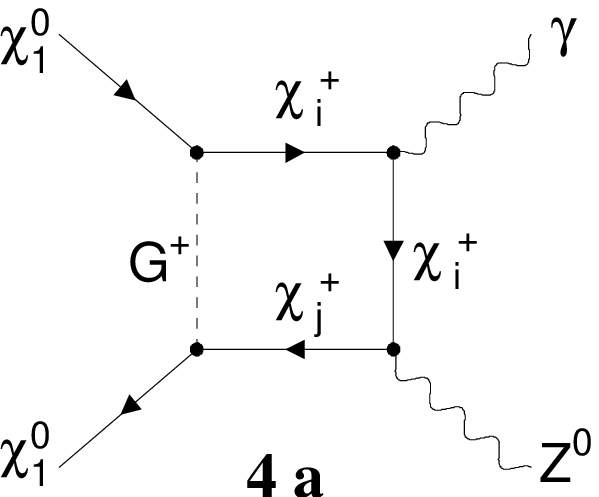,width=3cm}}\quad
       \subfigure{\epsfig{file=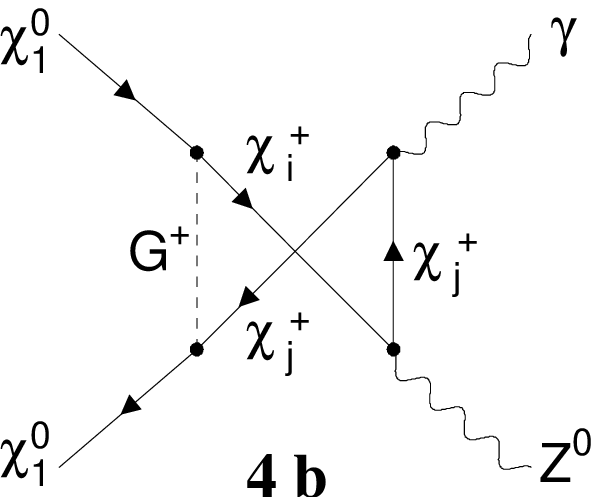,width=3cm}}\quad
       \subfigure{\epsfig{file=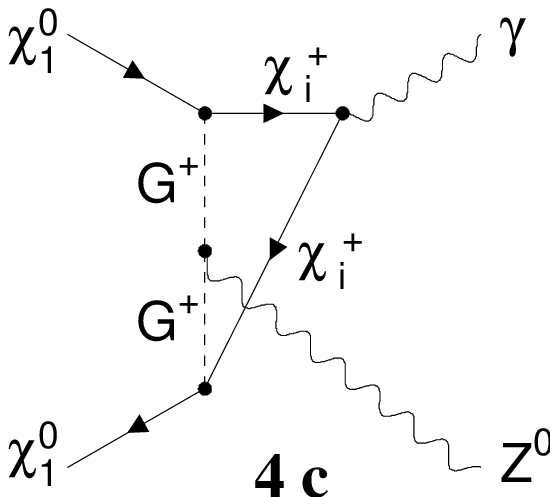,width=3cm}}}
 \mbox{\subfigure{\epsfig{file=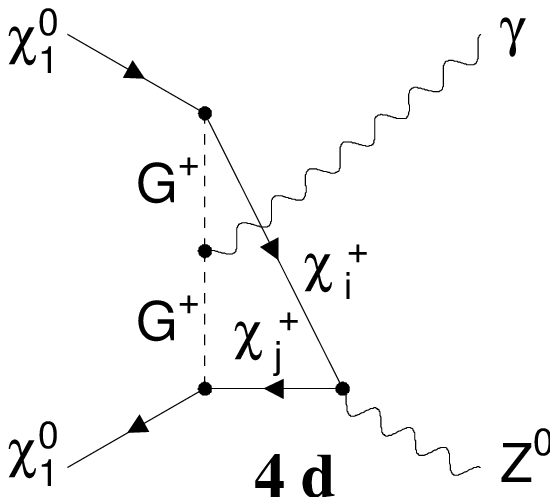,width=3cm}}\quad
       \subfigure{\epsfig{file=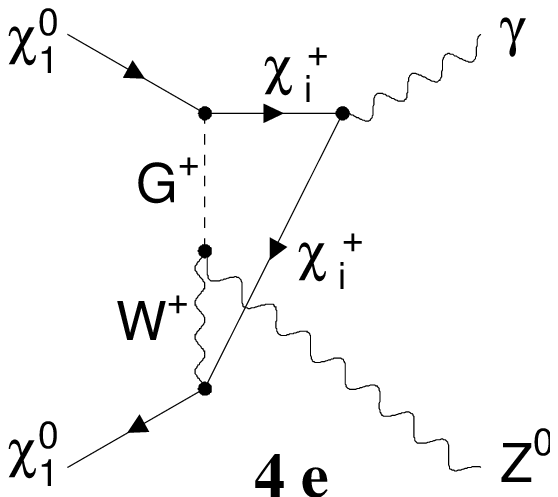,width=3cm}}\quad
       \subfigure{\epsfig{file=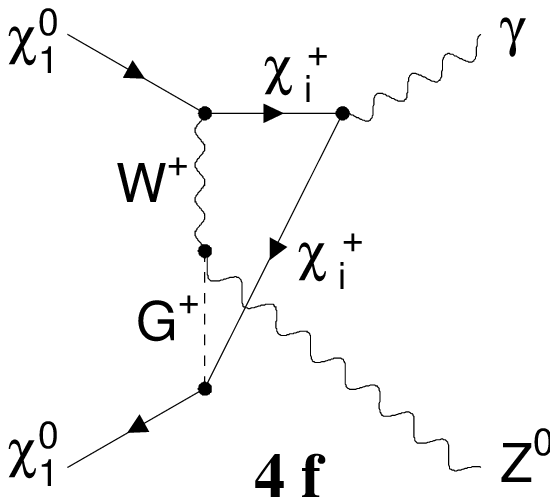,width=3cm}}}
\caption{Diagrams contributing, at one loop level, to neutralino 
annihilation into a photon and a $Z^0$.  From Ref.~\cite{bergstromgammaz}.}\label{fig:gammaz}
\end{figure}

\pagebreak

\begin{figure}[h]
\centering
\includegraphics[width=0.5\textwidth,clip=true]{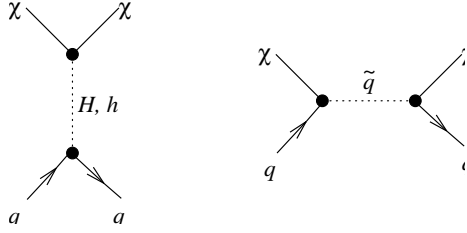}
\caption[]
{Tree level Feynman diagrams for neutralino-quark scalar (spin-independent) elastic scattering. From Ref.~\cite{Jungman:1995df}.}
\end{figure}

\section{Elastic Scattering Processes}
\label{dircalc}

\subsection{Scalar Interactions}
\label{scalarint}

\begin{figure}
\centering
\includegraphics[width=0.7\textwidth,clip=true]{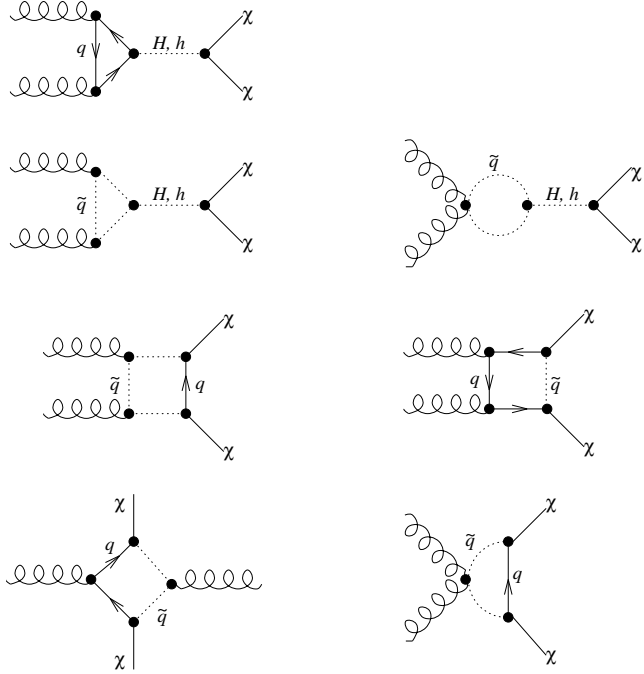}
\caption[]
{Feynman diagrams for neutralino-gluon scalar (spin-independent) elastic scattering. Notice that no tree level processes exist. From Ref.~\cite{Jungman:1995df}.}
\label{gluonloops}
\end{figure}

Consider a WIMP with scalar interactions with quarks given by
\begin{equation}
{\cal L}_{scalar} =  a_q \bar{\chi} \chi \bar{q} q,
\end{equation}
where $a_q$ is the WIMP-quark coupling. Then the scattering cross section for the WIMP off of a proton or neutron is given by
\begin{equation}
\label{scacross1}
\sigma_{scalar} = 
\int_0^{4m_r^2v^2} {d\sigma(v=0) \over d|\vec{v}|^2} =
{4m_r^2\over \pi} f_{p,n}^2,
\end{equation}
where $v$ is the relative velocity of the WIMP, $m_r$ is the reduced mass of the nucleon ($m_r \simeq m_{p,n}$ for WIMPs heavier than $\sim 10\,$ GeV) and $f_{p,n}$ is the WIMP coupling to protons or neutrons, given by 
\begin{equation}
\label{scalarterms}
f_{p,n} = \sum_{q=u,d,s}  f_{Tq}^{(p,n)} a_q  \frac{m_{p,n}}{m_q}  + \frac{2}{27}f_{TG}^{(p,n)} \sum_{q=c,b,t} a_q \frac{m_{p,n}}{m_q},
\end{equation}
where $f_{Tu}^{(p)}=0.020 \pm 0.004, f_{Td}^{(p)}=0.026 \pm 0.005, f_{Ts}^{(p)}=0.118 \pm 0.062, f_{Tu}^{(n)}=0.014 \pm 0.003, f_{Td}^{(n)}=0.036 \pm 0.008$ and $f_{Ts}^{(n)}=0.118 \pm 0.062$ \cite{fvalues}. $f_{TG}^{(p,n)}$ is related to these values by
\begin{equation}
f_{TG}^{(p,n)} = 1 - \sum_{q=u,d,s} f_{Tq}^{(p,n)}.
\end{equation}
The term in Eq.~\ref{scalarterms} which includes $f_{TG}^{(p,n)}$ results from the coupling of the WIMP to gluons in the target nuclei through a heavy quark loop. The couplings of squarks and Higgs bosons to heavy quarks leads to a loop level coupling of the WIMP to gluons \cite{ann3,Barbieri:zs,ann7}. Such diagrams are shown in Fig.~\ref{gluonloops}.
%

To attain the scalar cross section for a WIMP scattering off of a target nucleus, one should sum over the protons and neutrons in the target:
\begin{equation}
\label{scal}
\sigma =  \frac{4 m_r^2}{\pi}  \bigg(Z f_p + (A-Z) f_n \bigg)^2, 
\end{equation}
where $Z$ and $A-Z$ are the numbers of protons and neutrons in the nucleus, respectively.

The above expression is valid only at zero momentum transfer between the WIMP and the nucleon. For finite momentum transfer, the differential cross section must be multiplied by a nuclear form factor. The appropriate factor, called the {\it Woods-Saxon} form factor, is given by~\cite{woodssaxon}
\begin{equation}
F(Q) = \bigg(\frac{3 j_1(q R_1)}{q R_1}\bigg)^2 \, \rm{exp}[-(qs)^2],
\end{equation}
where $j_1$ is the first spherical bessel function and the momentum transferred is $q=\sqrt{s m_N Q}$. $R_1$ is given by $\sqrt{R^2 - 5 s^2}$, where $R$ and $s$ are approximately equal to $1.2 \, \rm{fm}\, A^{1/3}$ and $1\, \rm{fm}$, respectively. 

Although less accurate than the Woods-Saxon form factor, the following simple form factor is sometimes used in its place~\cite{Ahlen:mn,Freese:1987wu}:
\begin{equation}
F(Q) = \rm{exp}[-Q/2Q_0].
\end{equation}
Here, $Q$ is the energy tranferred from the WIMP to the target and $Q_0=1.5/(m_N R_0^2)$ where $R_0 = 10^{-13} \, \rm{cm} \, [0.3+0.91 (m_N/\rm{GeV})^{1/3}]$.

In the context of neutralino scattering, the value of $a_q$ can be calculated from the parameters of the MSSM~\cite{Gelmini:1990je,Srednicki:1989kj,Drees:1992rr,Drees:1993bu}. Following Ref.~\cite{fvalues}, $a_q$ is in this case given by:
\begin{eqnarray}
\label{scalcou}
a_q & = & - \frac{1}{2(m^{2}_{1i} - m^{2}_{\chi})} Re \left[
\left( X_{i} \right) \left( Y_{i} \right)^{\ast} \right] 
- \frac{1}{2(m^{2}_{2i} - m^{2}_{\chi})} Re \left[ 
\left( W_{i} \right) \left( V_{i} \right)^{\ast} \right] \nonumber \\
& & \mbox{} - \frac{g m_{q}}{4 m_{W} B} \left[ Re \left( 
\delta_{1} [g N_{12} - g' N_{11}] \right) D C \left( - \frac{1}{m^{2}_{H}} + 
\frac{1}{m^{2}_{h}} \right) \right. \nonumber \\
& & \mbox{} +  Re \left. \left( \delta_{2} [g N_{12} - g' N_{11}] \right) \left( 
\frac{D^{2}}{m^{2}_{H}}+ \frac{C^{2}}{m^{2}_{h}} 
\right) \right],
\end{eqnarray}
where
\begin{eqnarray}
X_{i}& \equiv& \eta^{\ast}_{11} 
        \frac{g m_{q}N_{1, 5-i}^{\ast}}{2 m_{W} B} - 
        \eta_{12}^{\ast} e_{i} g' N_{11}^{\ast}, \nonumber \\
Y_{i}& \equiv& \eta^{\ast}_{11} \left( \frac{y_{i}}{2} g' N_{11} + 
        g T_{3i} N_{12} \right) + \eta^{\ast}_{12} 
        \frac{g m_{q} N_{1, 5-i}}{2 m_{W} B}, \nonumber \\
W_{i}& \equiv& \eta_{21}^{\ast}
        \frac{g m_{q}N_{1, 5-i}^{\ast}}{2 m_{W} B} -
        \eta_{22}^{\ast} e_{i} g' N_{11}^{\ast}, \nonumber \\
V_{i}& \equiv& \eta_{22}^{\ast} \frac{g m_{q} N_{1, 5-i}}{2 m_{W} B}
        + \eta_{21}^{\ast}\left( \frac{y_{i}}{2} g' N_{11},
        + g T_{3i} N_{12} \right)
\label{xywz}
\end{eqnarray}
where $y_i$ and $T_{3i}$ denote hypercharge and isospin, and
\begin{eqnarray}
\delta_{1} = N_{13} (N_{14}),\,\,\,\, \delta_{2} = N_{14}
(-N_{13}), \nonumber \\
B = \sin{\beta} (\cos{\beta}),\,\,\,\, C = \sin{\alpha} (\cos{\alpha}),
\nonumber \\
D = \cos{\alpha} (-\sin{\alpha}). \,\,\,\,\,\,\,\,\,\,\,\,\,\,\,\,\,\,\,\,\,
\label{moredefs}
\end{eqnarray}
Here, $i=1$ for up-type and 2 for down-type quarks. $\alpha$ is the Higgs mixing angle. $m_{1i}, m_{2i}$ denote elements of the appropriate 2 x 2 squark mass matrix. $N_{1n}$ are elements of the matrix which diagonalizes the neutralino mass matrix in the $\widetilde{B}$-$\widetilde{W}^3$-$\widetilde{H}_1^0$-$\widetilde{H}_2^0$ basis (see Appendix~\ref{diagonal}). $\eta$ is the matrix which diagonalizes the appropriate squark mass matrices.

To crudely estimate what scale we expect for the scalar cross section between a neutralino and nucleon, we can carry out a back-of-the-envelope estimate. Considering a gaugino-like neutralino, we see that as $\delta_1$ and $\delta_2$ both vanish, so do most of the terms in Eq.~\ref{scalcou}. We are left with a neutralino-quark coupling of $a_q \sim A / m_{\tilde{q}}^2$, where $A$ is the product of the various order 1 couplings, mixing matrix parameters, etc. which contribute. For a typical case, $A$ might be $\sim 10^{-3}$ or so, although it can vary a great deal. Inserting this coupling into Eqs.~\ref{scacross1} and~\ref{scalarterms}, we estimate a neutralino-nucleon scalar cross section of $\sim A^2 m_p / m_{\tilde{q}}^4$, which is roughly $10^{-9} \, \rm{picobarns}$, for TeV mass squarks. These results can vary dramatically, however, depending on the characteristics of the model being considered (see Figs.~\ref{direct} and~\ref{DDmsugra}). 

We can contrast this with the much larger neutralino annihilation cross sections. Considering again a gaugino-like neutralino, its amplitude for annihilations into $b \bar{b}$ via psuedoscaler Higgs exchange (see Eq.~\ref{Aannamp}) is roughly ${\cal A}_A \sim m_b \tan \beta \sqrt{f_h} /m_{W^{\pm}}$ where $f_h$ is the higgsino fraction of the WIMP. The annihilation cross section (Eq.~\ref{fermanncross}) is then roughly $\sigma \sim 3\, m_b^2 \tan^2 \beta f_h /128 \pi  m_{\chi}^2 m_{W^{\pm}}^2$. For even a very small higgsino fraction, say $1\%$, and a 200 GeV neutralino, we find a cross section of $\sim 10^{-3} \,$picobarns for small values of $\tan \beta$ and a few picobarns for $\tan \beta = 30$.


\begin{figure}
\centering
\includegraphics[width=0.5\textwidth,clip=true]{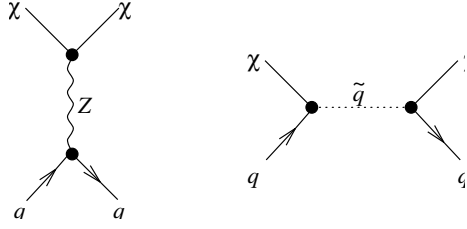}
\caption[]
{Tree level Feynman diagrams for neutralino-quark axial-vector (spin-dependent) elastic scattering. From Ref.~\cite{Jungman:1995df}.}
\end{figure}

\subsection{Axial-Vector Interactions}

Next, we consider a WIMP with axial-vector interactions with quarks given by
\begin{equation}
{\cal L}_A = d_q\; \bar{\chi} \gamma^\mu \gamma_5 \chi \;\,
	\bar{q} \gamma_\mu \gamma_5 q,
\end{equation}
where $d_q$ is the generic coupling. 

For such a WIMP, the spin-dependent scattering cross section can be written as \cite{gondolodirect}
\begin{equation}
{d\sigma \over d |\vec{v}|^2} = {1 \over 2 \pi  v^2} \overline{|T(v^2)|^2},
\end{equation}
where $v$, again, is the relative velocity of the WIMP, and $T(v^2)$ is the scattering matrix element. This expression can be integrated over the Boltzman velocity distribution of halo WIMPs to arrive at an average elastic scattering cross section. At zero momentum, the matrix element, $T(v^2)$, is given by
\begin{equation}
\label{matrixelement}
\overline{|T(0)|^2} = \frac{4 (J+1)}{J} |(d_u \Delta_u^p+ d_d \Delta_d^p+d_s \Delta_s^p)<S_p> + (d_u \Delta_u^n+ d_d \Delta_d^n+d_s \Delta_s^n)<S_n>|^2,
\end{equation}
where $J$ is the nuclear spin and the $\Delta$'s are the fraction of the nucleon spin carried by a given quark. Their values are measured to be $\Delta_u^p =\Delta_d^n=0.78 \pm 0.02, \Delta_d^p=\Delta_u^n=-0.48 \pm 0.02$ and $\Delta_s^p=\Delta_s^n=-0.15 \pm 0.02$. $<S_p>$ and $<S_n>$ are the expectation values of the total spin of protons and neutrons, respectively. Notice that for target nuclei with even numbers of protons and neutrons, there is zero total spin, and the cross section vanishes.

The values of $<S_p>$ and $<S_n>$ depend on the nucleus being considered. For $^{73}\rm{Ge}$, the interacting shell model finds $<S_p>$ and $<S_n>$ to be 0.011 and 0.468, respectively. For $^{28}\rm{Si}$, they are given by -0.0019 and 0.133. For $^{27}\rm{A}$, they are 0.3430 and 0.269. And for $^{39}\rm{K}$, they are -0.184 and 0.054 \cite{mallot}.

For non-zero momenta, a more complex form of equation~\ref{matrixelement} is needed. This equation is given by
\begin{eqnarray} \nonumber
\overline{|T(v^2)|^2} = \frac{(J+1)}{J} |(d_u \Delta_u^p+ d_d \Delta_d^p+d_s \Delta_s^p+d_u \Delta_u^n+ d_d \Delta_d^n+d_s \Delta_s^n) \\ \nonumber
<S_p+S_n>  F^0(v^2) + (d_u \Delta_u^p+ d_d \Delta_d^p+d_s \Delta_s^p-d_u \Delta_u^n+ d_d \Delta_d^n+d_s \Delta_s^n) \\
<S_p-S_n>  F^1(v^2) |^2,
\end{eqnarray}
where the $F$'s are nuclear form factors given by
\begin{equation}
F^0(v^2) \simeq \exp(- r_0^2 v^2/\hbar^2)
\end{equation}
and
\begin{equation}
F^1(v^2) \simeq \exp( - r_1^2 v^2 / \hbar^2 +i c v/\hbar ),
\end{equation}
where $r_0$ and $r_1$ are parameters which depend on the nucleus being considered, with typical values of $1.3-2.1\,\,\rm{fm}^{-1}$.

Again, within the context of neutralino scattering, the value of $d_2$ can be calculated from the parameters of the MSSM~\cite{direarly2,Engel:bf,ann2,ann3,Ellis:1991ef,lambdas1}. Following Ref.~\cite{fvalues}, $d_2$ is in this case given by:
\begin{eqnarray}
d_2 & = & \frac{1}{4(m^{2}_{1i} - m^{2}_{\chi})} \left[
\left| Y_{i} \right|^{2} + \left| X_{i} \right|^{2} \right] 
+ \frac{1}{4(m^{2}_{2i} - m^{2}_{\chi})} \left[ 
\left| V_{i} \right|^{2} + \left| W_{i} \right|^{2} \right] \nonumber \\
& & \mbox{} - \frac{g^{2}}{4 m_{Z}^{2} \cos^{2}{\theta_{W}}} \left[
\left| N_{13} \right|^{2} - \left| N_{14} \right|^{2}
\right] \frac{T_{3i}}{2},
\label{alpha2}
\end{eqnarray}
where the quantities used are defined in Appendix~\ref{scalarint}.

\subsection{Vector Interactions}

As a third case, consider a WIMP with vector interactions with quarks, given by
\begin{equation}
{\cal L}_{vec}^q =  b_q\; \bar{\chi} \gamma_\mu \chi\;\, \bar{q}
\gamma^\mu q.
\end{equation}
Here, $b_q$ is the WIMP-quark vector coupling. In this case, the contributions of each quark in the nucleus add coherently and large cross sections result for large nuclei. The WIMP-nucleus cross section in this case is straight forward~\cite{direarly2}
\begin{equation}
\sigma_{0\,vec} = { m_\chi^2 m_N^2 b_N^2 \over 64 \pi (
     m_\chi + m_N)^2  },
\end{equation}
where $b_N$ is simply $b_N= 2Z b_p + (A-Z) b_n$.

As an example of a WIMP with vector interactions, consider a Dirac neutrino. In this case, $b_q= G_F (T^3_q - 2e_q \sin^2 \theta_{\rm{W}})/\sqrt{2}$, where $G_F$ is the Fermi constant, $T^3_q$ and $e_q$ are the weak isospin and electric change of the quark $q$, respectively, and $\theta_{\rm{W}}$ is the Weinberg angle. Summing over the quarks in a proton or neutron, we get $b_p=G_F (1-4 \sin^2 \theta_{\rm{W}})/(2\sqrt{2})$ and $b_n= -G_F/(2\sqrt{2})$. Since $4 \sin^2 \theta_{\rm{W}} \cong 1$, the neutron-neutrino cross section is much larger than the analogous proton-neutrino interaction. The Dirac neutrino-neutron cross section is then given by $\sigma_{\nu, n} = G^2_F  m_\nu^2 m_n^2 / (512 \pi (
     m_\nu + m_n)^2)$. A cross section of this size has been ruled out by direct scattering experiments, except perhaps in the case of a very light WIMP. Similar calculations show that other similar particles, such as sneutrinos, are also excluded by this method \cite{falksneutrino}.

Neutralinos, being Majorana fermions, do not have vector interactions with quarks.



\end{appendix}



\end{document}